\documentclass{aa}

\usepackage{graphics}
\usepackage{graphicx}
\usepackage{multicol}
\usepackage{txfonts}
\usepackage{amsmath}
\usepackage{verbatim}
\usepackage{pdflscape}
\usepackage{afterpage}

\usepackage{natbib,twoopt}
\usepackage{amsmath} 
\usepackage{hyperref} 
\bibpunct{(}{)}{;}{a}{}{,} 
\makeatletter
\newcommandtwoopt{\citeads}[3][][]{\href{http://adsabs.harvard.edu/abs/#3}%
{\def\hyper@linkstart##1##2{}%
\let\hyper@linkend\@empty\citealp[#1][#2]{#3}}}
\newcommandtwoopt{\citepads}[3][][]{\href{http://adsabs.harvard.edu/abs/#3}%
{\def\hyper@linkstart##1##2{}%
\let\hyper@linkend\@empty\citep[#1][#2]{#3}}}
\newcommandtwoopt{\citetads}[3][][]{\href{http://adsabs.harvard.edu/abs/#3}%
{\def\hyper@linkstart##1##2{}%
\let\hyper@linkend\@empty\citet[#1][#2]{#3}}}
\newcommandtwoopt{\citeyearads}[3][][]%
{\href{http://adsabs.harvard.edu/abs/#3}
{\def\hyper@linkstart##1##2{}%
\let\hyper@linkend\@empty\citeyear[#1][#2]{#3}}}
\makeatother

\hypersetup{colorlinks=true,linkcolor=blue,citecolor=blue,urlcolor=blue}

\def\NtotalCEP{456}

\def\NtotalRRL{789}
\def\NnearRRL{22}
\def\NboundRRL{7}

\begin{document}

\title{Multiplicity of Galactic Cepheids and RR Lyrae stars from Gaia DR2}
\subtitle{II. Resolved common proper motion pairs
\thanks{Tables \ref{cepheids-bound-table1} to \ref{various-bound-table} are available in electronic form
at the CDS via anonymous ftp to \url{cdsarc.u-strasbg.fr} (130.79.128.5)
or via \url{http://cdsweb.u-strasbg.fr/cgi-bin/qcat?J/A+A/}}}
\titlerunning{Multiplicity of Galactic Cepheids and RR Lyrae stars from Gaia DR2 - II}
\authorrunning{P. Kervella et al.}
\author{
Pierre~Kervella\inst{1}
\and
Alexandre Gallenne\inst{2}
\and
Nancy Remage Evans\inst{3}
\and
Laszlo Szabados\inst{4}
\and
Fr\'ed\'eric Arenou\inst{5}
\and
Antoine M\'erand\inst{6}
\and
Nicolas Nardetto\inst{7}
\and
Wolfgang Gieren\inst{8}
\and
Grzegorz Pietrzynski\inst{9}
}
\institute{
LESIA, Observatoire de Paris, Universit\'e PSL, CNRS, Sorbonne Universit\'e, Univ. Paris Diderot, Sorbonne Paris Cit\'e, 5 place Jules Janssen, 92195 Meudon, France, \email{pierre.kervella@obspm.fr}.
\and
European Southern Observatory, Alonso de C\'ordova 3107, Casilla 19001, Santiago, Chile.
\and
Smithsonian Astrophysical Observatory, MS 4, 60 Garden Street, Cambridge, MA 02138, USA.
\and
Konkoly Observatory, MTA CSFK, Konkoly Thege M. \'ut 15-17, H-1121, Hungary.
\and
GEPI, Observatoire de Paris, Universit\'e PSL, CNRS, 5 Place Jules Janssen, 92190 Meudon, France.
\and
European Southern Observatory, Karl-Schwarzschild-Str. 2, 85748 Garching, Germany.
\and
Universit\'e C\^ote d'Azur, OCA, CNRS, Lagrange, France
\and
Universidad de Concepci{\'o}n, Departamento de Astronom\'{\i}a, Casilla 160-C, Concepci{\'o}n, Chile.
\and
Nicolaus Copernicus Astronomical Centre, Polish Academy of Sciences, Bartycka 18, PL-00-716 Warszawa, Poland
}
\date{Received ; Accepted}
\abstract {The multiplicity of classical Cepheids (CCs) and RR Lyrae stars (RRLs) is still imperfectly known, particularly for RRLs.}
{In order to complement the close-in short orbital period systems presented in Paper~I, our aim is to detect the wide, spatially resolved companions of the targets of our reference samples of Galactic CCs and RRLs.}
{Angularly resolved common proper motion pairs were detected using a simple progressive selection algorithm to separate the most probable candidate companions from the unrelated field stars.}
{We found 27 resolved, high probability gravitationally bound systems with CCs out of 456 examined stars, and one unbound star embedded in the circumstellar dusty nebula of the long-period Cepheid RS Pup.
We found seven spatially resolved, probably bound systems with RRL primaries out of 789 investigated stars, and 22 additional candidate pairs.
We report in particular new companions of three bright RRLs: OV And (companion of F4V spectral type), RR Leo (M0V), and SS Oct (K2V).
In addition, we discovered resolved companions of 14 stars that were likely misclassified as RRLs.} 
{The detection of resolved non-variable companions around CCs and RRLs facilitates the validation of their GDR2 parallaxes. The possibility to conduct a detailed analysis of the resolved coeval companions of CCs and old population RRLs will also be valuable to progress on our understanding of their evolutionary path.}
\maketitle

\section{Introduction}

Classical Cepheids (CCs) and RR Lyrae stars (RRLs) are essential standard candles for Galactic (e.g., \citeads{2013ApJ...763...32D}) and extragalactic (e.g., \citeads{2016ApJ...826...56R}) distance determinations.
These pulsating stars are the subject of several studies following the first \citepads{2017A&A...605A..79G, 2018MNRAS.474.2142I} and second \citepads{2018arXiv180502079C, 2018MNRAS.481.1195M, 2018pas6.conf..101R} Gaia data releases.
Their multiplicity fraction has a particular importance, as the presence of companions may bias their apparent brightness and affect their evolutionary path. The recent discovery of the binary evolution pulsators \citepads{2012Natur.484...75P, 2013MNRAS.428.3034S} is an example of the potential impact of binarity on the properties of oscillating stars. Coeval, gravitationally bound companions are also valuable for conducting comparative evolutionary modeling.

In Paper~I \citepads{Kervella2018a}, we searched for companions of CCs and RRLs from the signature of the presence of a companion on their proper motion (PM). For this purpose, we used the positions measured by the Hipparcos and Gaia spacecrafts at two epochs separated by 24.25 years to determine the PM of their center of mass.
The presence of a companion results in a ``virtual orbit'' of the photocenter around the center of mass of the system  (see, e.g., \citeads{1999AJ....118.1086B,2000AJ....120.1106B,2013A&A...556A.133S}).
As the astrometric missions measured the position of the photocenter, we detected the presence of companions from a comparison of the two PM vectors (Hipparcos and Gaia DR2) to the mean PM vector.
The difference between the photocenter's PM vector and that of the center of mass is referred to as the  proper motion anomaly (PMa) in the following.
The time baseline of more than two decades between Hipparcos and Gaia provides a sensitivity to orbital periods of up to several hundred years, depending on the distance of the target and the mass ratio of the CC to its companion. However, its sensitivity decreases for longer orbital periods (i.e., millennia), as the principle of the determination of the mean PM of the center of mass (difference of positions) cancels the signature of the orbital PM of the photocenter if it is very slow.

In the present paper, we search for spatially resolved, common proper motion or gravitationally bound companions of our samples of CCs and RRLs. 
Despite the high contrast that makes their detection difficult, several spatially resolved CC companions have already been found using optical interferometry \citepads{2013A&A...552A..21G, 2014A&A...561L...3G, Gallenne2018}, adaptive optics \citepads{2014A&A...567A..60G} or  Hubble Space Telescope (HST) imaging \citepads{2008AJ....136.1137E, 2018ApJ...863..187E, 2016AJ....151..129E}.
The database of the binary and multiple Galactic CCs maintained at Konkoly Observatory\footnote{\url{http://www.konkoly.hu/CEP/intro.html}} \citepads{2003IBVS.5394....1S} provides a list of the known CCs in multiple systems.
A database of candidate binaries with an RRL component is provided by \citetads{2016MNRAS.459.4360L}\footnote{\url{ http://rrlyrbincan.physics.muni.cz}}. No RRL resolved companion is currently known, and only \object{TU UMa} has been convincingly demonstrated to be a binary and has accurate orbital parameters (\citeads{2016A&A...589A..94L}; Paper~I). Its companion is likely a white dwarf orbiting at a small angular separation ($\approx 10$~mas; Paper~I).

In Sect.~\ref{identificationcriteria}, we present the selection criteria that we adopted to discriminate the unrelated field stars from the physical companions. The resulting detections are presented in Sect.~\ref{detectedcompanions} for CCs (Sect.~\ref{CEPresolvedcompanions}) and RRLs (Sect.~\ref{RRLresolvedcompanions}). We also present the candidate companions of variable stars that were incorrectly classified as RRLs in Sect.~\ref{VARresolvedcompanions}.
We discuss in Sect.~\ref{notesonstars} the results we obtained on selected individual stars, including the detections of PM anomalies presented in Paper~I.

\section{Selected samples\label{samples}}

We  adopted the list of 455 CCs from \citetads{2000A&AS..143..211B} plus \object{Y Car}, together with their listed photometric distances, that are based on multicolor period-luminosity relations. This catalog is tied to an LMC distance modulus of $\mu_\mathrm{LMC} = 18.25$, that is too short compared to recent measurements. We therefore renormalized the listed distances using the distance modulus established by \citetads{2013Natur.495...76P}. This correction increases the distances of all Cepheids in the \citetads{2000A&AS..143..211B} catalog by $\approx 11\%$.
The distance of \object{Y Car} is taken from \citetads{1992ApJ...385..680E}.
The RRL sample was extracted from the General Catalogue of Variable Stars (GCVS) \citepads{2017ARep...61...80S} that comprises 8509 stars listed as RR type.
We adopt for the RRLs the GDR2 parallaxes $\varpi_\mathrm{G2}$.
The catalog uncertainty of $\varpi_\mathrm{G2}$ for the RRLs in the magnitude range of our sample is typically  $30-100\,\mu$as.
For the targets with $\varpi_\mathrm{G2} \lesssim 0.5$~mas the uncertainty on the GDR2 parallaxes of the faint candidate companions becomes too large to enable an efficient selection process.
We therefore limited our sample to the RRLs with a GDR2 parallax $\varpi _\mathrm{GDR2}> 0.5$~mas.
This results in a sample of {\NtotalRRL} stars classified as RRLs within 2\,kpc.

The GDR2 parallaxes and PMs were corrected following the procedure described in Paper~I, and we set a uniform uncertainty of 15\% on all the adopted {\NtotalCEP} CC (photometric) and {\NtotalRRL} RRL (GDR2) parallaxes. It corresponds to the range of accepted parallaxes in the companion selection process (Sect.~\ref{parallax}).

\section{Companion identification criteria\label{identificationcriteria}}

We defined criteria for the selection of wide companion candidates based on 1) the similarity of their parallax, 2) their tangential differential velocity, 3) their projected linear separation. We also tested them to determine whether they are gravitationally bound.

\subsection{Parallax \label{parallax}}

The GDR2 parallax is our primary criterion for the selection of potential companions.
The first step in our selection is based on the compatibility of the GDR2 parallax of the field stars with the adopted value of the target stars (CCs or RRLs).
The candidates whose GDR2 parallax is outside  the range of expected values ($\pm 15\%$ around the adopted parallax of the target) are rejected. 
We do not use a variable weight depending on the proximity in parallax; the candidate companions are uniformly rejected if they are outside  the expected $\varpi$ uncertainty range. This choice is intended to reject the statistical outliers, reduce the number of false positive detections and ensure that we do not bias our detections toward the prior adopted value of the target parallax.

We also reject the field stars that have a parallax S/N < 3 to reduce the number of inconclusive detections.
The improvement of the uncertainties of the Gaia parallaxes in the future data releases will result in the inclusion of additional companion candidates.

\subsection{Proper motion}

We determined the difference in tangential velocity $dv_\mathrm{tan}$ (perpendicular to the line of sight) between the field stars and the tested CC and RRL. We excluded the candidates that exhibit a differential tangential velocity larger than $dv_\mathrm{tan, max} = 20$\,km\,s$^{-1}$.
This limit corresponds to an increase in projected separation of $\approx 20$\,pc in ten million years.
This is much larger than the usually considered limit for common proper motion pairs (see, e.g., \citeads{2016A&A...587A..51S}), even if \citetads{2017arXiv170903532P} report the existence of comoving pairs up to a separation of 10\,pc.
A field star is removed from the list of candidate companions if its $dv_\mathrm{tan}$ is more than 1$\sigma$ above $dv_\mathrm{tan, max}$.
The candidates that exhibit a low differential velocity $dv_\mathrm{tan} < 5$\,km\,s$^{-1}$ are flagged as \texttt{LowV}.

We also set as a condition for field star selection that the position angle of its PM vector is within $\pm 15^\circ$ of that of the PM vector of the CC or RRL target star, if the candidate is located within a projected radius of 10\,kau.
This range of permitted angles  accounts for a possible orbital motion. For wider separation candidates, the acceptance range for the PM vector position angle is reduced to $\pm 5^\circ$ to limit the number of false positives.
When available, we adopted the mean proper motion $\mu_\mathrm{HG}$ computed from the Hipparcos and GDR2 positions (see Paper~I), to mitigate the effect of an orbiting close-in companion on the adopted PM vector. This is particularly important for the stars showing a strong PMa,  for example the short-period Cepheid \object{V1334 Cyg} (Paper~I; \citeads{2013A&A...552A..21G, Gallenne2018}).

\subsection{Projected proximity}

The search radius around the CCs and RRLs is set to 1\,pc at the distance of each target, with a minimum angular radius of 1\,arcminute.
The probability that candidate companions are bound to the target star increases as their projected separation with the target becomes smaller.
We therefore allocated a continuous linear weight in the ranking of the candidate companions to their linear projected distance to the target.
The field stars that are within $50\,000$\,au for CCs and $30\,000$\,au from RRLs are flagged as \texttt{Near}.
These different maximum radii take into account the difference in mass between CCs and RRLs, and also the fact that the young CCs can be found in open clusters (whose populations we wish to probe) contrary to old RRLs (except in the case of chance associations).
These radii are comparable to the widest binary systems \citepads{2017AJ....153..257O,2013ARA&A..51..269D,2017A&A...598L...7K}, and they also incorporate the binaries  possibly formed from adjacent prestellar cores with slow relative proper motion \citepads{2017MNRAS.468.3461T}.

\subsection{Gravitationally bound candidates\label{boundcandidates}}

We test the possibility that the candidate companions are gravitationally bound to the target CC or RRL by comparing the differential tangential velocity $dv_\mathrm{tan}$ with the escape velocity $v_\mathrm{esc}$ at their projected separation (see, e.g., \citeads{2017A&A...598L...7K}). This is a first-order approach as the third, radial component of the differential velocity is usually unknown for the field stars. It should be accounted for in a precise comparison of the relative velocity with the escape velocity. The expression of the escape velocity $v_\mathrm{esc}$ for two bodies of total mass $m_\mathrm{tot}$, located at a distance $r$ from each other is
\begin{equation}
v_\mathrm{esc} = \sqrt{\frac{2 G m_\mathrm{tot}}{r}}.
\end{equation}

We adopt the predicted masses of the CC and RRL targets as described in Paper~I.
A limitation of this approach is that the true total mass $m_\mathrm{tot}$ of the systems is often higher than that of the CC or RRL alone. For instance, the system of $\delta$\,Cep contains at least one close-in component of a mass of $\approx 1\,M_\odot$ \citepads{2015ApJ...804..144A,2016MNRAS.461.1451G}, and a distant component \object{HD 213307} \citepads{2002AJ....124.1695B, 2010ApJ...725.2392M} that is likely an early-type, relatively massive binary (Sect.~\ref{delcep}). The determination of the escape velocity for the different components of this quadruple system should be based on the total mass of the considered components that exceeds the mass of the Cepheid alone.
To account for the mass of the secondary star (and possible additional components), we assumed an identical mass for the companion in the computation of the escape velocity by considering $m_\mathrm{tot} = 2\, m_\mathrm{target}$.

The candidate companions are flagged as \texttt{Bound} when they are already flagged as \texttt{Near} and their tangential differential velocity  $dv_\mathrm{tan}$ is within $3\sigma$ of the escape velocity $v_\mathrm{esc}$ (they are also normally flagged as \texttt{LowV}). This relatively permissive criterion accounts for the significant uncertainties in the physical parameters used in the computation.

\section{Candidate companion parameters}

For CCs, individual color excess estimates $E(B-V)$ are available in the literature, and we adopted these values for their candidate companions. For the candidate companions of RRLs, we determined $E(B-V)$ from the 3D extinction maps by \citetads{2018MNRAS.478..651G}, \citetads{2018A&A...616A.132L}, or \citetads{2017AstL...43..472G}, depending on the sky coverage of each map. For the bright RR Lyrae stars \object{RR Leo} and \object{SS Oct}, we adopted the individual estimates by \citetads{2008MNRAS.386.2115F}.

For the candidates that have $JHK_s$ magnitudes from the 2MASS catalog \citepads{2003tmc..book.....C, 2006AJ....131.1163S}, we derived their linear radius $R$ and effective temperature $T_\mathrm{eff}$ from their broadband magnitudes using the visible-infrared surface brightness--color relations from \citetads{2004A&A...426..297K} and the GDR2 parallax.
For completeness, we also cross-identified the candidate companions in the WISE catalog \citepads{2010AJ....140.1868W,2012yCat.2311....0C}.

When $JHK_s$ infrared magnitudes were not available, we used Eq. (1) in \citetads{2018A&A...616A..10G} to compute the extinction in the $G$, $G_\mathrm{BP}$, and $G_\mathrm{RP}$ bands from the adopted value of the color excess $E(B-V)$. This provided the dereddened color excess $C_\mathrm{XP} = E(G_\mathrm{BP}-G_\mathrm{RP})_0$ and the absolute dereddened $M_G$ magnitude of the candidates in the $G$ band. We then converted the dereddened color to an effective temperature using the polynomial expression by \citetads{2010A&A...523A..48J} (their Eq.~2):
\begin{equation}
\log T_\mathrm{eff} = 3.999 - 0.654\,C_\mathrm{XP} + 0.709\,C_\mathrm{XP}^2 - 0.316\,C_\mathrm{XP}^3.
\end{equation}
For the stars with $C_\mathrm{XP}>1.4$, this polynomial expression is unreliable, and we adopt the following linear correction:
\begin{equation}
        \log T_\mathrm{eff,corr} = \left( \frac{\log T_\mathrm{eff} - 3.3}{1.5 - 6}\right) \, (C_\mathrm{XP} - 6) + 3.3.
\end{equation}

For main sequence  candidate companions, we converted the effective temperature and absolute magnitude into spectral type using the grid by \citetads{2013ApJS..208....9P}\footnote{\url{http://www.pas.rochester.edu/~emamajek/EEM_dwarf_UBVIJHK_colors_Teff.txt}} (see also \citeads{2012ApJ...746..154P}).
In some cases, when the candidates are located close to the primary star, the derived $(T_\mathrm{eff},M_G)$ combination is inconsistent due to the contamination of the $G_\mathrm{BP}$ and $G_\mathrm{RP}$ magnitudes by the bright CCs or RRLs.
The $G$ magnitude and astrometry can be correct while color photometry is contaminated, given the different sizes of the pixel windows: 18 pixels along scan for bright stars in the astrometric field ($G$ band) against 60 pixels ($3.5\arcsec$) for $G_\mathrm{BP}$ and $G_\mathrm{RP}$.
When there was an inconsistency, we assumed that the candidate is a main sequence star and based our provisional spectral type determination on the absolute magnitude $M_G$ alone.

\section{Detected companions\label{detectedcompanions}}

\subsection{Cepheid companions\label{CEPresolvedcompanions}}

We individually examined   the fields of the \texttt{Bound} or \texttt{Near} candidates by eye to check the dubious cases. The resulting list of candidates is presented in Table~\ref{cepheids-candidate-list}.
We present in Figures~\ref{cepheid-field-1} to \ref{cepheid-field-5} the Second Generation Digitized Sky Survey Red (DSS2-Red) fields around the CCs that show \texttt{Bound} candidate companions. Their properties and those of their companions are listed in Tables~\ref{cepheids-bound-table1} and \ref{cepheids-bound-table2}.
Selected CCs with \texttt{Near} candidates are shown in Figs.~\ref{cepheids-near1} and \ref{cepheids-near2}.
The full list of detected \texttt{Near} candidate companions of CCs is given in Tables~\ref{cepheids-near-table1} and \ref{cepheids-near-table2}.

\begin{figure}[]
\centering
\includegraphics[width=\hsize]{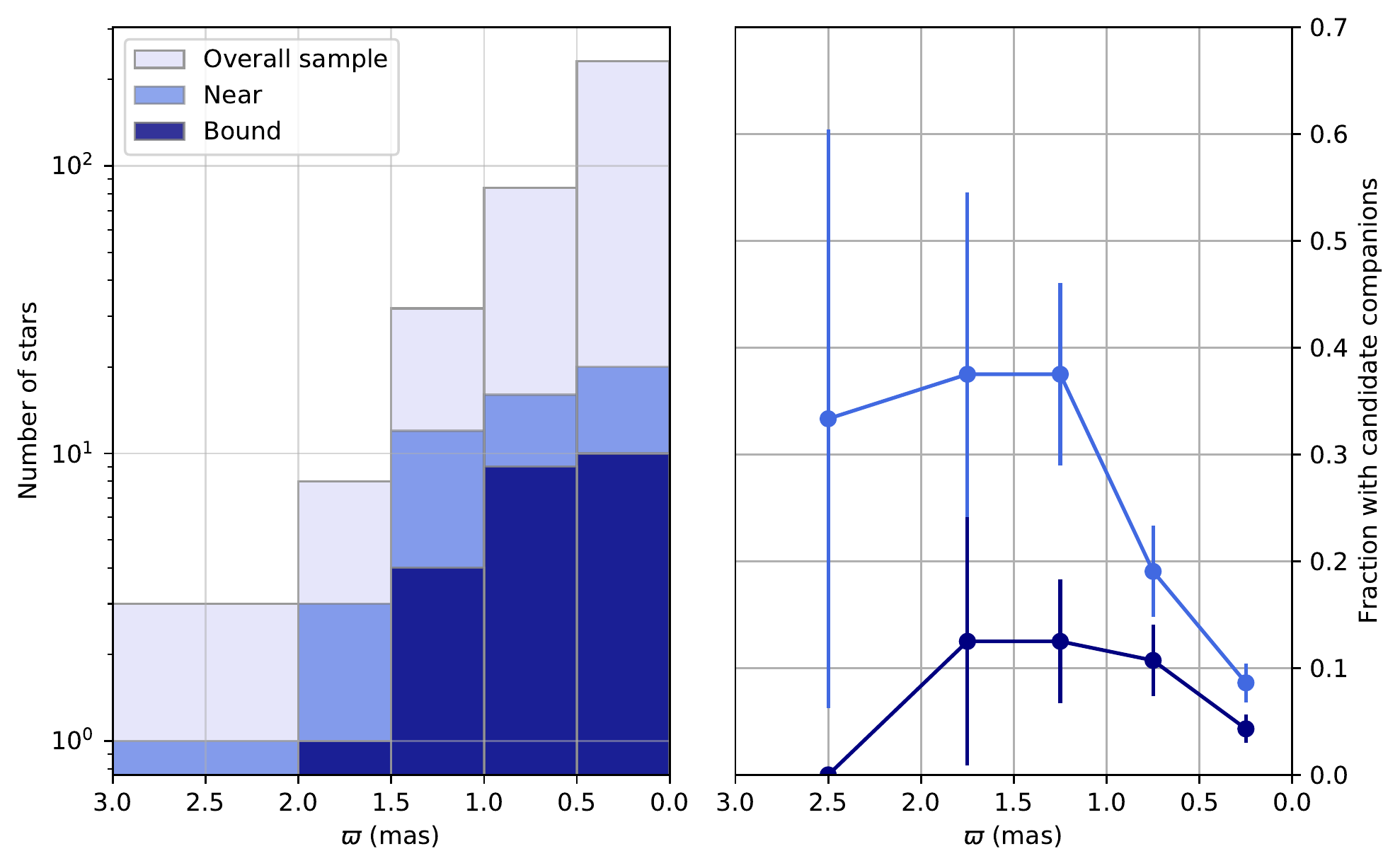}
\caption{\textit{Left:} Histogram of the CCs with candidate \texttt{Near} and \texttt{Bound} companions as a function of parallax.
\textit{Right:} Fraction of the observed CCs with candidate companions.
The error bars represent the binomial proportion 68\% confidence interval.
\label{CC-histo}}
\end{figure}

For the more distant CCs of our sample, the probability of chance associations for the detected candidate companions is increased compared to the closer CCs and the RRLs (which are on average located at shorter distances than CCs).
Figure~\ref{CC-histo} shows the histogram of the CCs with detected \texttt{Near} and \texttt{Bound} candidates. The decrease in the sensitivity of the algorithm with distance is clearly visible in the right panel. We observe a peak fraction of $\approx 10\%$ of CCs with \texttt{Bound} candidate companions.
A validation of the candidate companions for the distant CCs will be possible using the future Gaia data releases, which will provide more accurate estimates of their relative parallaxes and tangential velocities. The measurement of radial velocities for the CCs and their candidate companions will provide their full 3D relative velocity and therefore enable a stringent test of their gravitational bind.

\begin{table*}
 \caption{Wide companions of Cepheids for the targets with \texttt{Near} (\texttt{N}) or \texttt{Bound} (\texttt{B}) resolved candidate companions. The ``$\varpi_\mathrm{exp.}$'' column gives their expected parallax from the renormalized period--luminosity distances by \citetads{2000A&AS..143..211B}. The ``Total'' column gives the number of GDR2 field stars with compatible parallaxes that were examined as candidates. The ``Vis'' column indicates the result of a visual inspection of the field of the considered targets, with $\checkmark$ indicating a likely comoving system and $-$ a dubious association.
The observational properties and field charts of the Cepheids and their associated candidate companions are presented in Appendix~\ref{cep-tables-appendix}.}
 \label{cepheids-candidate-list}
 \centering
 \small
  \begin{tabular}{lrrrcccll}
  \hline
  \hline
         Target    & $\varpi_\mathrm{exp.}$ & Period & Total   & \texttt{N}   & \texttt{B} & Vis. & Comment \\ 
             & (mas) & (d) &  &    &  \\ 
  \hline  \noalign{\smallskip}
\multicolumn{8}{c}{\textit{Cepheids with \texttt{Bound} candidates}} \\
  \noalign{\smallskip}
\object{TV CMa}  &  0.477  &  4.67      &   4  &   1  &   1 & $\checkmark$ & Tight, high probability bound companion \\ 
 \object{ER Car}  &  0.959  &  7.72      &  41  &   2  &   1 & $\checkmark$ & Very wide comoving candidate \\ 
 \object{CE Cas B}  &  0.343  &  4.48      &  15  &   1  &   1 & $\checkmark$ & Known physical Cepheid companion (\object{CE Cas A}) \\ 
 \object{DF Cas}  &  0.462  &  3.83      &   7  &   1  &   1 & $\checkmark$ & Slow PM, possible association\\ 
 \object{V0659 Cen}  &  1.287  &  5.62      &  66  &   1  &   1 & $\checkmark$ & Very wide comoving candidate \\ 
 \object{delta Cep}  &  3.755  &  5.37      &  53  &   1  &   1 & $\checkmark$ & Known physical companion ($\delta$\,Cep B) \\ 
 \object{AX Cir}  &  1.917  &  5.27      &  79  &   2  &   1 & $\checkmark$ & Wide, probably bound companion\\ 
 \object{BP Cir}  &  1.700  &  2.40      & 101  &   2  &   1 & $\checkmark$ & Very wide comoving companion, possible group \\ 
 \object{R Cru}  &  1.170  &  5.83      &  49  &   1  &   1 & $\checkmark$ & Tight, probably bound companion\\ 
 \object{X Cru}  &  0.678  &  6.22      &  36  &   2  &   1 & $\checkmark$ & Wide, possibly bound companion \\ 
 \object{VW Cru}  &  0.710  &  5.27      &  22  &   1  &   1 & $\checkmark$ & Possible comoving group \\ 
 \object{V0532 Cyg}  &  0.727  &  4.68      &  19  &   1  &   1 & $\checkmark$ & Wide, probably bound companion \\ 
 \object{V1046 Cyg}  &  0.372  &  4.94      &   6  &   1  &   1 & $\checkmark$ & Tight, high probability bound companion \\ 
 \object{CV Mon}  &  0.601  &  5.38      &  27  &   3  &   2 & $\checkmark$ & High probability bound candidates \\ 
\object{RS Nor}  &  0.487  &  6.20      &  36  &   2  &   1 & $\checkmark$ & Tight, probably bound companion\\ 
 \object{SY Nor}  &  0.429  & 12.65      &  14  &   2  &   1 & $\checkmark$ & Tight, probably bound companion \\ 
 \object{QZ Nor}  &  0.556  &  5.41      &  69  &   3  &   1 & $\checkmark$ & Wide, possible chance association \\ 
 \object{AW Per}  &  1.218  &  6.46      &  10  &   1  &   1 & $\checkmark$ & Tight bound companion \\ 
 \object{U Sgr}  &  1.669  &  6.75      & 165  &   3  &   2 & $\checkmark$ & Very wide, possible comoving cluster \\ 
 \object{V0350 Sgr}  &  1.141  &  5.15      &  57  &   2  &   1 & $\checkmark$ & High probability bound companion \\ 
 \object{V0950 Sco}  &  1.073  &  4.82      &  92  &   2  &   1 & $\checkmark$ & Likely bound candidate \\  
 \object{CM Sct}  &  0.518  &  3.92      &  21  &   1  &   1 & $\checkmark$ & Wide comoving association \\ 
 \object{EV Sct}  &  0.556  &  4.40      &  43  &   1  &   1 & $\checkmark$ & Wide, probably bound candidate, possible group \\ 
 \object{Polaris} ($\alpha$\,UMi) &  7.540  &  5.67      &  15  &   1  &   1 & $\checkmark$ & Known physical companion (Polaris B) \\ 
 \object{SX Vel}  &  0.490  &  9.55      &  17  &   1  &   1 & $\checkmark$ & Tight, high probability bound companion \\  
 \object{CS Vel}  &  0.292  &  5.90      &  10  &   1  &   1 & $\checkmark$ & Possibly bound candidate and comoving group \\ 
 \object{DK Vel}  &  0.431  &  2.48      &  17  &   1  &   1 & $\checkmark$ & Wide, possibly bound candidate \\ 
  \hline  \noalign{\smallskip}
\multicolumn{8}{c}{\textit{Cepheids with \texttt{Near} candidates}} \\
  \noalign{\smallskip}
\object{FF Aql}  &  2.048  &  6.40      &  60  &   1  &   0 & $-$ & PM divergence, unlikely association \\  
\object{V0916 Aql}  &  0.325  & 13.44      &   3  &   1  &   0 & $\checkmark$ & Tight companion, small parallax \\ 
 \object{eta Aql}  &  3.755  &  7.18      &  33  &   1  &   0 & $-$ & PM divergence, unlikely association \\  
\object{CK Cam}  &  1.959  &  3.29      &  27  &   1  &   0 & $-$ & PM divergence, unlikely association \\  
\object{Y Car}  &  0.695  &  3.64      &  32  &   1  &   0 & $\checkmark$ & Comoving group, possible cluster \\ 
\object{UX Car}  &  0.710  &  3.68      &  29  &   1  &   0 & $-$ & Very wide, unlikely association \\   
\object{UZ Car}  &  0.437  &  5.20      &  30  &   1  &   0 & $-$ & Very wide, PM divergence, unlikely association \\   
\object{XY Car}  &  0.366  & 12.43      &   9  &   1  &   0 & $-$ & Very wide, possible comoving star \\ 
 \object{EY Car}  &  0.446  &  2.88      &  27  &   1  &   0 & $-$ & Tangential velocity difference, possible association \\ 
\object{DD Cas}  &  0.322  &  9.81      &   3  &   1  &   0 & $-$ & Tangential velocity difference, unlikely association \\  
\object{VW Cen}  &  0.256  & 15.04      &   5  &   1  &   0 & $\checkmark$ & Very wide, possibly comoving companion \\  
\object{XX Cen}  &  0.589  & 10.95      &  23  &   1  &   0 & $-$ & PM difference, unlikely association \\ 
\object{AY Cen}  &  0.639  &  5.31      &  45  &   1  &   0 & $\checkmark$ & Slight PM divergence, possible association \\
\object{AV Cir}  &  1.802  &  3.07      &  57  &   1  &   0 & $-$ & Very wide, unrelated field stars \\ 
\object{T Cru}  &  1.306  &  6.73      &  45  &   1  &   0 & $-$ &  Wide, PM divergence, unlikely association \\ 
\object{SU Cyg}  &  1.234  &  3.85      &  74  &   1  &   0 & $\checkmark$ & Probably comoving unbound companion \\ 
\object{SZ Cyg}  &  0.419  & 15.11      &   5  &   1  &   0 & $\checkmark$ & Possible comoving companion \\  
\object{V1334 Cyg}  &  1.502  &  4.75      &  24  &   1  &   0 & $\checkmark$ & Probable comoving companion \\ 
\object{RR Lac}  &  0.515  &  6.42      &  12  &   1  &   0 & $-$ &  Wide, unlikely association \\
\object{S Nor}  &  1.126  &  9.75      & 104  &   2  &   0 & $\checkmark$ & Very wide, one possibly bound and one unbound \\ 
\object{U Nor}  &  0.699  & 12.64      &  35  &   1  &   0 & $-$ &  Wide, unlikely association \\ 
 \object{Y Oph}  &  1.186  & 17.12      &  20  &   1  &   0 & $\checkmark$ & Very wide, possibly comoving candidate \\ 
\object{AQ Pup}  &  0.312  & 30.10      &   6  &   1  &   0 & $\checkmark$ & Possible comoving companion \\  
\object{Y Sgr}  &  2.146  &  5.77      &  90  &   2  &   0 & $\checkmark$ &  Wide, possible association \\
\object{XX Sgr}  &  0.733  &  6.42      &  39  &   1  &   0 & $-$ & Tangential velocity difference, unlikely association \\ 
 \object{AP Sgr}  &  1.287  &  5.06      &  91  &   2  &   0 & $\checkmark$ & Wide, possibly comoving candidate\\ 
\object{Y Sct}  &  0.581  & 10.34      &  18  &   1  &   0 & $-$ &  Wide, unlikely association \\  
 \object{R TrA}  &  1.733  &  3.39      &  66  &   1  &   0 & $\checkmark$ & Wide, uncertain companion, slight PM offset \\ 
\object{LR TrA}  &  1.048  &  3.44      &  40  &   1  &   0 & $-$ &  Wide, unlikely association, PM difference \\ 
  \hline
\end{tabular}
\end{table*}

\subsection{RR Lyrae star companions\label{RRLresolvedcompanions}}

\begin{figure}[]
\centering
\includegraphics[width=\hsize]{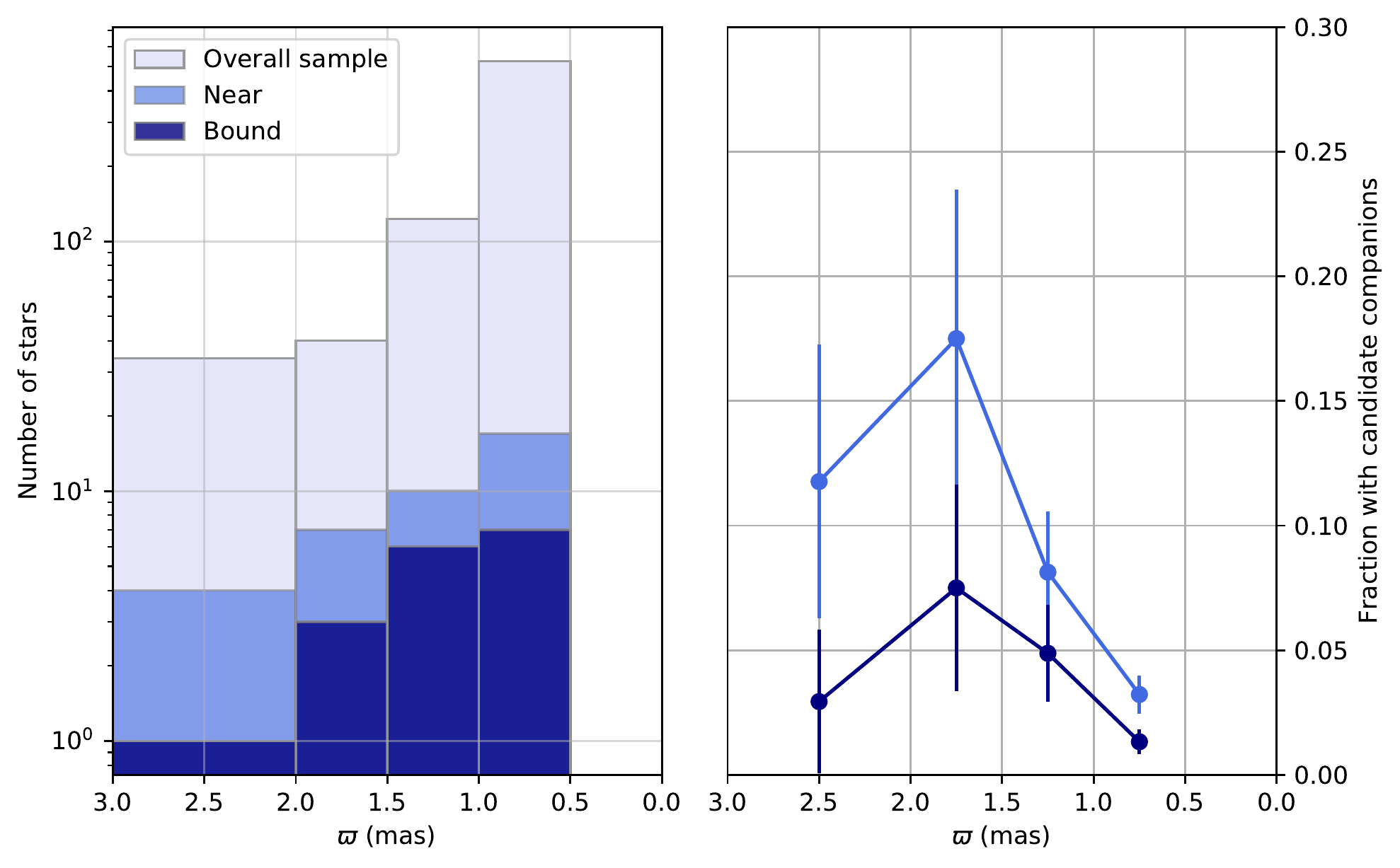}
\caption{\textit{Left:} Histogram of the RRLs and various variables with candidate \texttt{Near} and \texttt{Bound} companions as a function of parallax. The lower parallax limit is set at $\varpi = 0.5$~mas for our RRL sample.
\textit{Right:} Fraction of stars with candidate companions.
The error bars represent the binomial proportion 68\% confidence interval.
\label{RRL-histo}}
\end{figure}

RR Lyrae stars are old population stars, with a typical age of 10\,Ga, that usually belong to the thick disk of the Galaxy.
On average, nearby RRLs have significantly faster proper motions than the surrounding field stars, making their common proper motion companions easier to identify.
We applied our companion search criteria to {\NtotalRRL} RRLs, and this resulted in {\NboundRRL} targets with candidates flagged as \texttt{Bound} candidates and {\NnearRRL} flagged as  \texttt{Near}. The {\NboundRRL} RRLs with \texttt{Bound} candidates are listed in Table~\ref{rrlyr-candidate-list}, and their observational properties are given in Table~\ref{rrlyrae-bound-table}.
The full list of detected \texttt{Near} candidate companions of RRLs is given in Tables~\ref{rrl-near-table1} and \ref{rrl-near-table2}.
The histogram of the RRLs and various variable stars with detected candidate companions is presented in Fig.~\ref{RRL-histo}.
For the nearby stars, we detect \texttt{Bound} candidate companions around $\approx 5-10\%$ of the examined targets.
The fields surrounding the RRLs with \texttt{Bound} candidates are shown in Figs.~\ref{rrlyr-field-1} and \ref{rrlyr-field-2}, and a selection of the RRLs with \texttt{Near} candidates is shown in Fig.~\ref{rrlyr-near}.

\begin{table*}
 \caption{RR Lyrae stars and other types of variable stars with \texttt{Near} (\texttt{N}) or \texttt{Bound} (\texttt{B}) resolved candidate companions.
 A mention of ``/ 0'' after the period indicates a fundamental mode pulsator as identified in the GDR2 catalog (\texttt{rrlyrae} table).
The observational properties and field charts of the RRLs and their associated candidate companions are presented in Appendix~\ref{rrl-tables-appendix}.
}
 \label{rrlyr-candidate-list}
 \centering
 \small
 \begin{tabular}{lrlrcccl}
  \hline
  \hline
         Target    & $\varpi_\mathrm{G2}$ & Period & Total   & \texttt{N}   & \texttt{B} & Vis. & Comment \\ 
             & (mas) & (d) &  &    &  \\ 
  \hline  \noalign{\smallskip}
\multicolumn{8}{c}{\textit{RR Lyrae with \texttt{Bound} candidates}} \\
  \noalign{\smallskip}
 \object{OV And}  &  0.938  &  0.471      &   9  &   1  &   1 & $\checkmark$ & Tight, high probability bound companion \\ 
  \object{CS Del}  &  0.635  &  0.366      &   7  &   1  &   1 & $\checkmark$ & Very tight, high probability bound companion \\ 
 \object{V0893 Her}  &  2.679  &   0.492      &  13  &   1  &   1 & $\checkmark$ & High probability bound, slight PM difference \\ 
 \object{RR Leo}  &  1.032  &  0.452      &   1  &   1  &   1 & $\checkmark$ & Tight, high probability bound system \\ 
 \object{SS Oct}  &  0.862  &  0.622 / 0  &   3  &   1  &   1 & $\checkmark$ & Very tight bound companion \\ 
 \object{EY Oph}  &  1.850  &     NA      &  36  &   2  &   1 & $\checkmark$ & Slight PM divergence, possibly bound companion \\ 
 \object{V0487 Sco}  &  1.069  &  0.329      &  59  &   1  &   1 & $\checkmark$ & Fast PM, probably bound companion \\ 
 \hline  \noalign{\smallskip}
\multicolumn{8}{c}{\textit{RR Lyrae with \texttt{Near} candidates}} \\
  \noalign{\smallskip}
\object{V0830 Cyg}  &  0.618  &  0.401 / 0  &  27  &   1  & 0 & $-$ & Probable chance association \\ 
\object{CZ Lac}  &  0.852  &  0.432 / 0  &  25  &   1  &   0 & $\checkmark$ & PMa, probably bound, triple system \\ 
\object{V0424 Lyr}  &  1.579  &  0.580 / 0  &  33  &   1  &   0 & $-$ & Rich star field, probable chance association \\ 
\object{AG Nor}  &  0.978  &  0.505 &  68  &   3  &   0 & $\checkmark$ & Close parallax value, possibly comoving \\ 
\object{KP Nor}  &  0.520  &NA &  25  &1  &0 & $-$ & PM difference, chance association \\ 
\object{IT Oph}  &  6.206  &NA & 578  &1  &0 & $-$ & Dense star field, chance association \\ 
\object{MS Oph}  &  0.658  &NA &  26  &1  &0 & $-$ & Probable chance association \\  
 \object{V1693 Oph}  &  0.677  &  0.522 &  27  &1  &0 & $-$ & Uncertain association \\ 
 \object{UY Ori}  &  2.840  &NA &  18  &1  &0 & $-$ & Divergent PM vectors \\ 
\object{V1154 Ori}  &  1.087  &NA &  51  &1  &0 & $\checkmark$ & Probable comoving association \\  
\object{V0701 Sgr}  &  2.000  &  0.627 & 127  & 2  &0 & $-$ & PM difference, chance association \\ 
 \object{V2481 Sgr}  &  0.910  &NA & 105  &1  &0 & $-$ & Divergent PM, chance association \\ 
\object{V2626 Sgr}  &  1.363  &  0.462 &  88  &1  &0 & $\checkmark$ & Slight PM position angle difference, uncertain  \\ 
 \object{V3531 Sgr}  &  0.765  &  0.542 &  26  &1  &0 & $-$ & Very wide, unlikely association \\ 
 \object{V4107 Sgr}  &  1.265  &NA & 224  &1  &0 & $-$ & Dense stellar field, association unclear \\ 
 \object{V4313 Sgr}  &  1.700  &NA & 279  &2  &0 & $\checkmark$ & Possible comoving group \\ 
\object{V4355 Sgr}  &  2.433  &NA & 361  &5  &0 & $-$ & Probable chance association  \\ 
\object{V4591 Sgr}  &  1.837  &NA & 276  &1  &0 & $-$ & PM difference, chance association \\  
\object{IY Sco}  &  2.530  &NA & 309  &2  &0 & $-$ & Rich star field, chance association \\
\object{KN Sco}  &  0.959  &NA &  43  &1  &0 & $-$ & Unlikely association \\ 
\object{V0828 Sco}  &  0.877  &NA &  33  &   1  &   0 & $-$ & PM difference, chance association \\
\object{V0348 Sct}  &  1.249  &NA      &  45  &   1  &   0 & $-$ & Chance association \\ 
  \hline  \noalign{\smallskip}
\multicolumn{8}{c}{\textit{Other variable types with \texttt{Bound} candidates}} \\
\noalign{\smallskip}
\object{HM Aql}  &  1.829  &  0.345      &  28  &   1  &   1 & $\checkmark$ & Relatively wide companion, likely bound \\
 \object{EN CMi}  &  0.569  &  0.540      &   8  &   1  &   1 & $\checkmark$ & Wide, high probability bound candidate \\ 
\object{NQ Cyg}  &  0.938  &  0.312      &  25  &   1  &   1 & $\checkmark$ & Eclipsing binary, fast PM, probably bound companion \\ 
\object{V1391 Cyg}  &  1.232  &  0.596      &  17  &   1  &   1 & $\checkmark$ & Eclipsing binary, tight companion \\  
\object{V2121 Cyg}    &   25.597  &  0.800$^a$  &   3  &     1  &     1 & $\checkmark$ & $\gamma$\,Dor pulsator, PMa, fast proper motion \\ 
\object{UU Dor}  &  0.856  &     NA      &   5  &   1  &   1 & $\checkmark$ & Eclipsing binary, tight high probability bound companion \\  
\object{IW Lib}    &    3.571  &  1.783$^b$  & 22  &     1  &     1 & $\checkmark$ & W UMa eclipsing, high probability bound companion \\ 
 \object{AZ Men}  &  1.088  &  0.318      &   8  &   1  &   1 & $\checkmark$ & High probability bound companion \\ 
\object{V1171 Oph}  &  1.099  &     NA      &  16  &   2  &   1 & $\checkmark$ & Uncertain variable class, high probability bound companion  \\  
 \object{V1330 Sgr}  &  1.229  &  0.427      & 193  &   2  &   1 & $-$ & Wide, uncertain companion \\ 
\object{V1382 Sgr}  &  1.611  &  0.493      & 316  &   1  &   1 & $\checkmark$ & Comoving candidate, slight PM difference  \\ 
\object{V2248 Sgr}  &  1.948  &     0.315      &  22  &   1  &   1 & $\checkmark$ & W UMa eclipsing, high probability bound companion \\  
\object{V3166 Sgr}    &    6.646  & NA  & 192  &     3  &     2 & $\checkmark$ & Uncertain variable class, two high probability bound \\\object{HR Sco}  &  7.370  &     NA      & 425  &   2  &   1 & $-$ & Uncertain variable class, dense field, likely chance association \\
\hline
\end{tabular}
\tablebib{(a): \citetads{2009A&A...499..967C}; (b): \citetads{2002AcA....52..397P}.}
\end{table*}

\subsection{Variables of other classes \label{VARresolvedcompanions}}

Fourteen targets that were incorrectly classified as RRLs have \texttt{Bound} candidate companions. The results of the visual inspection of the corresponding fields are given in Table~\ref{rrlyr-candidate-list} and the detailed candidate properties are listed in Table~\ref{various-bound-table}.
Their respective field charts are presented in Figs.~\ref{various-field1} to \ref{various-field3}.

\section{Notes on individual stars\label{notesonstars}}

In this section we present a selection of individual stars from our sample that host highly probable resolved physical companions.
We also discuss the stars for which a significant PMa was identified in Paper~I. 

\subsection{Cepheids}

\subsubsection{U Aql}

The quest for masses and luminosities for CCs has benefitted greatly from the availability of the ultraviolet spectrum using  the International Ultraviolet Explorer (IUE)  and HST telescopes. Through these studies, the picture of multiplicity has become increasingly complex. The Cepheid \object{U Aql} ($P=7.02$\,d; \object{HD 183344}) is a good case in point. Its substantial orbital motion was only recognized in 1979 \citepads{1979PASP...91..840S}. The discussion of \citetads{1987PASP...99..610W} provides an orbit and a summary of previous velocity information. The spectrum of the hottest star in the system dominates in the ultraviolet below about $\lambda=200$\,nm, and IUE observations have been discussed by \citetads{1985ApJ...296..175B} and \citetads{1992ApJ...384..220E}. These studies provide a temperature of $9300 \pm 100$\,K and a spectral type of B9.8V, respectively.

We detect a very strong PMa on \object{U Aql}. From its combination with the spectroscopic orbital parameters by \citetads{Gallenne2018b}, we determined in Paper~I that its companion has a mass of $1.9 \pm 0.3\,M_\odot$ and is orbiting on a 5.6\,au orbit. This mass is slightly lower than expected from the effective temperature of $T_\mathrm{eff}=9300 \pm 100$\,K determined from IUE spectroscopy by \citetads{1985ApJ...296..175B} and the spectral type of B9.8V from \citetads{1992ApJ...384..220E}.
We do not detect any resolved companion of \object{U Aql}.

\subsubsection{FF Aql}

\object{FF Aql} is a known binary system whose spectroscopic orbital parameters were determined by \citetads{1990AJ.....99.1598E}.
The astrometric observations by \citetads{2007AJ....133.1810B} using the Hubble Space Telescope Fine Guidance Sensor (FGS) revealed the orbital shift of the photocenter of the system.
In Paper~I, we identified a strong PMa on \object{FF Aql} with a S/N from the GDR2 measurement of $\Delta_\mathrm{G2} = 5.6$.
Combining the PM anomalies with the spectroscopic orbital parameters by \citetads{Gallenne2018b}, we derived the orbital parameters of the system and the mass of the secondary component ($m_2 \approx 0.8 \pm 0.1\,M_\odot$; Paper~I), which is lower than the estimate of $1.5\,M_\odot$ by \citetads{1990AJ.....99.1598E} from IUE spectroscopy.
Future Gaia data releases will provide an accurate determination of the parameters of the photocenter astrometric orbit, as the orbital period of the system ($P=1433$\,d) is a good match to the observing lifetime of the satellite.

\citetads{2014A&A...567A..60G} searched for resolved companions of the  short-period Cepheid \object{FF Aql} using adaptive optics up to a separation of $1.7\arcsec$, but did not find any. \citetads{Gallenne2018b} detected a possible close-in companion using near-infrared interferometry.
The visual companion found by \citetads{1963icvd.book.....J} and recovered by \citetads{2011MNRAS.413.1200R} and \citetads{2016AJ....151..129E} is \object{Gaia DR2 4514145288240592512} at a separation of $6.98\arcsec$ from the CC. We find that its GDR2 parallax ($\varpi = 1.971 \pm 0.041$~mas) is compatible with that of \object{FF Aql} ($\varpi = 1.839 \pm 0.107$~mas). However, its PM vector ($\mu = [-8.52 \pm 0.07, -19.93 \pm 0.07]$~mas\,a$^{-1}$) is very different from the mean Hipparcos-Gaia PM vector of the CC ($\mu_\mathrm{HG} = [-0.14 \pm 0.01, -9.34 \pm 0.01]$~mas\,a$^{-1}$). This corresponds to a projected relative velocity of $\approx 30$\,km\,s$^{-1}$. This in principle excludes that the stars are gravitationally bound, confirming the conclusion by \citetads{1990AJ.....99.1598E}.  However, the coincidence in terms of position and parallax is quite remarkable from a statistical point of view. This candidate could still be comoving with the CC, and may be bound, if it is a close binary and its measured PM is affected by its orbital motion.

\subsubsection{RW Cam}

\object{RW Cam} exhibits a strong PMa signal-to-noise ratio ($\Delta_\mathrm{G2} = 9.3$) (Paper~I). It is a  long-period CC ($P=16.41$\,d) that is a known binary \citepads{1977MNRAS.178..505M,1985ApJ...296..175B,1994ApJ...436..273E}.
As reported in Paper~I, we estimate that its orbital period is probably on the order of a few hundred years, for a semimajor axis of $\approx 200$~mas. We do not detect any additional resolved companion of \object{RW Cam}.

\subsubsection{Y Car}

 \citetads{1997AJ....114.1176B} found that the spectral type of the companion of \object{Y Car} is B9V, and \citetads{2005AJ....130..789E} showed that \object{Y Car} is actually a triple system. This companion was detected by \citetads{Gallenne2018b} via near-infrared interferometry.
 We reported in Paper~I a very strong PMa S/N ($\Delta_\mathrm{G2}=14.8$) that confirms the multiplicity of this short-period double-mode Cepheid ($P=3.64$\,d for the fundamental mode).
We did not detect any \texttt{Bound} candidate companion of \object{Y Car}, but we identified several nearby stars with comparable PM vectors that could constitute a comoving group (Fig.~\ref{cepheids-near1}).

\subsubsection{YZ Car}

The combination of the spectroscopic orbital elements of \citetads{Gallenne2018b} (which are in agreement with those determined by \citeads{2016ApJS..226...18A}) and the PMa vectors results in a companion mass of $m_2 \approx 1.9 \pm 0.3\,M_\odot$ (Paper~I) for the companion of the long-period pulsator \object{YZ Car} ($P=18.2$\,d). The interferometric companion search by \citetads{Gallenne2018b} did not reveal the faint secondary, but established an upper limit of B3V on its spectral type, in agreement with the estimate of B8V-A0V from \citetads{1993PASP..105..915E} and the mass derived in Paper~I, which corresponds to an A-type dwarf.

We did not detect any \texttt{Bound} candidate companions, but two nearby stars present comparable proper motions and parallaxes (Fig.~\ref{cepheids-near1}).

\subsubsection{CE Cas AB}

This visual binary is composed of two CCs, labeled  components A ($P=5.14$\,d) and B ($P=4.48$\,d).
This is the only such configuration known in the Milky Way.
The projected separation of the two stars is 7.2\,kau, corresponding to an orbital period on the order of 5000\,years.
\object{CE Cas} A and B are members of the open cluster \object{NGC 7790} \citepads{2013A&A...560A..22M}, together with the Cepheid \object{CF Cas} ($P=4.875$\,d).
The two components of \object{CE Cas} are present in the GDR2, but with statistically different parallaxes ($\varpi_A = 0.317 \pm 0.031$~mas; $\varpi_B = 0.262 \pm 0.030$~mas).
From their PM vectors, we  confirm, however, that the two Cepheids are gravitationally bound (Fig.~\ref{cepheid-field-1}).
The GDR2 parallax of component B is likely biased, possibly due to light contamination from the nearby component A.
We did not find any additional resolved candidate companions of \object{CE Cas}.

\subsubsection{SU Cas}

The short-period pulsator \object{SU Cas} ($P=1.95$\,d) is a member of a cluster of young stellar objects \citepads{2012MNRAS.421.1040M}, and its distance ($d=414 \pm 11$\,pc; $\varpi = 2.42 \pm 0.06$~mas) was estimated by \citetads{2012ApJ...753..144M}.
This value is compatible with the GDR2 parallax ($\varpi = 2.15 \pm 0.08$~mas).
\object{SU Cas} is located close to a reflection nebula \citepads{1984ApJ...283..254T, 2003A&A...399..141M}.
It exhibits a strong PMa with a S/N of $\Delta_\mathrm{G2} = 5.6$ (Paper~I).
The combination of the spectroscopic orbital parameters determined by \citetads{2008A&A...488...25G} and the PM anomalies point to the presence of a very low mass companion on a tight orbit of only 1.7\,au.
This companion is different from the $2.4\,M_\odot$ candidate with spectral type B9.5V spectral type listed by \citetads{2013AJ....146...93E} that orbits the Cepheid at a much larger distance of $\approx 100$\,au.
No detection is reported by \citetads{2015A&A...579A..68G} from near-infrared interferometric observations, setting an upper limit of A0 to the spectral type of close-in companions.
We did not detect any \texttt{Bound} resolved stellar companion of \object{SU Cas} in the GDR2.

\subsubsection{V0659 Cen}

\citetads{2013AJ....146...93E} identified a close companion of the short-period Cepheid \object{V0659 Cen} ($P=5.62$\,d) at a separation of $0.63\arcsec$ using HST/WFC3 imaging.
We found a marginal PMa at the Hipparcos epoch (S/N $\Delta_\mathrm{Hip} = 2.7$), but it is not significant at the GDR2 epoch ($\Delta_\mathrm{G2} = 1.6$). We note, however, that the error bars of the GDR2 PM vector are larger than usual, possibly due to the presence of the close companion at a separation of $0.63\arcsec$ that is not listed separately in the GDR2 catalog.

We found a resolved common proper motion candidate star of spectral type M3V at a very wide projected separation of 48\,kau (Fig.~\ref{cepheid-field-1}) that does not correspond to the candidates observed by \citetads{2016AJ....151..129E} (see also \citeads{2016AJ....151..108E}).

\subsubsection{$\delta$ Cep\label{delcep}}

\begin{figure*}[]
\centering
\includegraphics[width=9cm]{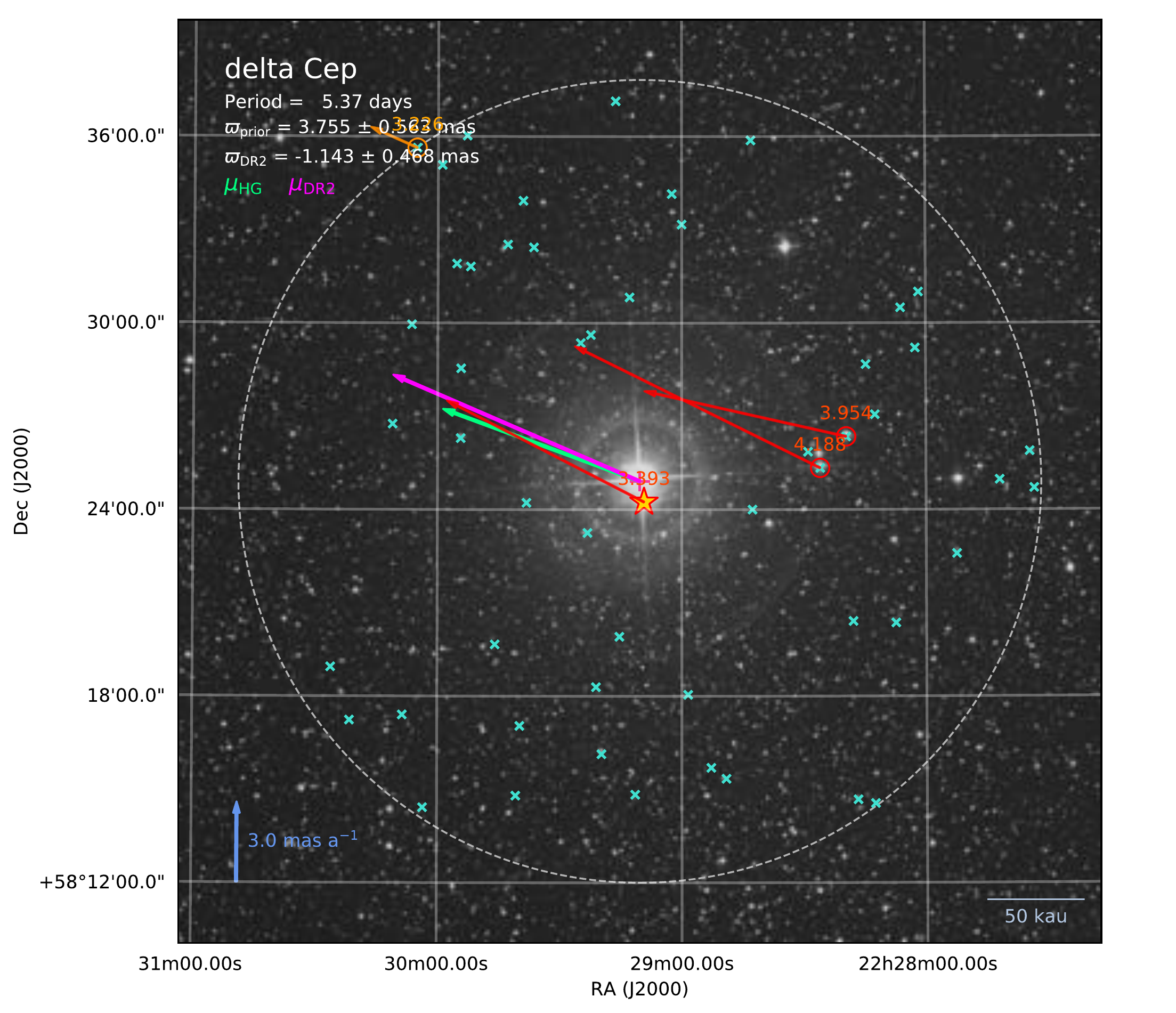}
\includegraphics[width=9cm]{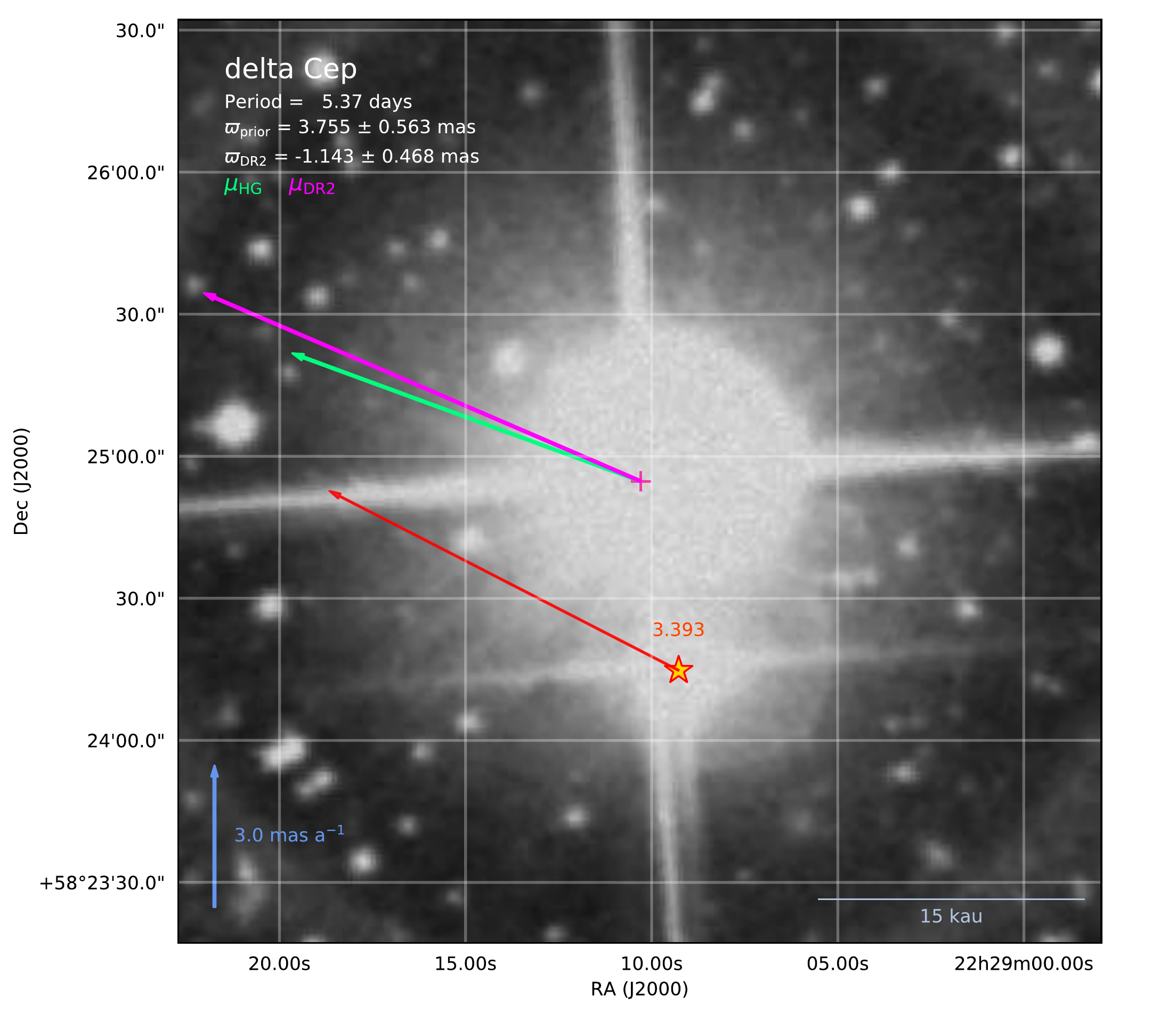}
\caption{Field around $\delta$\,Cep (Left panel: wide field; Right panel: narrow field), with its companion $\delta$\,Cep B (yellow star) and other comoving stars. The background image is from the DSS2-Red. \label{delta-cep-wide}}
\end{figure*}

As shown by \citetads{2012ApJ...747..145M}, the prototype Cepheid $\delta$\,Cep is a member of a cluster that also includes the K1.5b supergiant $\zeta$\,Cep.
This Cepheid has shown a periodic X-ray variability \citepads{2017ApJ...838...67E} and evidence of a significant mass loss \citepads{2012ApJ...744...53M}.
Figure~\ref{delta-cep-wide} shows the field around $\delta$\,Cep, with some of the stars in the cluster that show a comparable proper motion.

The bright visual companion $\delta$\,Cep~B (\object{HD 213307}) is associated with its surrounding nebula \citepads{2010ApJ...725.2392M, 2012ApJ...744...53M} as confirmed by the presence of a bow shock.
The GDR2 parallax of $\delta$\,Cep~B ($\varpi_\mathrm{G2,B} = 3.393 \pm 0.049$~mas) is noticeably smaller than the HST FGS determination by \citetads{2002AJ....124.1695B} ($\varpi_\mathrm{FGS} = 3.66 \pm 0.15$~mas), as well as the Hipparcos parallax ($\varpi_\mathrm{Hip} = 3.77 \pm 0.16$~mas) and the combined cluster distance from \citetads{2012ApJ...747..145M} ($\varpi_\mathrm{comb} = 3.68 \pm 0.08$~mas).
The GDR2 parallax of $\delta$\,Cep~B is consistent with the distance derived by \citetads{Borgniet2018} using the HR-SPIPS spectral analysis technique.
Rescaling the SPIPS fit of \citetads{2015A&A...584A..80M}, this larger distance corresponds to a spectroscopic projection factor of $p = 1.39 \pm 0.03$.
The spectral type of $\delta$\,Cep~B was estimated in the B7-B8 III-IV range by \citetads{2002AJ....124.1695B}, with a probable F0V companion. This corresponds to a mass of $\approx 5.6\,M_\odot$ for the pair Ba+Bb.
The PMa vector of $\delta$\,Cep~B exhibits S/N levels of $\Delta_\mathrm{Hip} = 3.8$ and $\Delta_\mathrm{G2} = 2.1$, respectively for Hipparcos and GDR2 (Table~\ref{delCepB}), confirming its binarity.
Its differential tangential velocity with respect to A is $dv_\mathrm{tan} = 1.18 \pm 0.03$\,km\,s$^{-1}$ (Table~\ref{delCepB}).
This is comparable to the escape velocity at a separation of 12\,kau, i.e., $v_\mathrm{esc} \approx 0.90$\,km\,s$^{-1}$ for a total system mass of $11\,M_\odot$ (for the four components of the system, see below).
Measuring the radial velocity of $\delta$\,Cep~B is unfortunately a difficult task as it is a fast rotating star ($v_\mathrm{rot}\,\sin i \approx 140$\,km\,s$^{-1}$; \citeads{1970CoAsi.239....1B}). Moreover, the presence of the F0V companion $\delta$\,Cep~Bb with an estimated orbital period of 390\,days and an induced orbital velocity amplitude of $K_1 \approx 15$\,km\,s$^{-1}$ \citepads{2002AJ....124.1695B} also complicates the measurement of the $\gamma$-velocity of Ba+Bb (i.e., its center-of-mass radial velocity).
So at the moment, we cannot firmly conclude that $\delta$\,Cep~B is gravitationally bound to A.
However, the low differential tangential velocity of the two objects, their proximity in space, and their high masses are statistically strong indications that they are gravitationally bound.

A variation in the $\gamma$-velocity of $\delta$\,Cep A reported by \citetads{2015ApJ...804..144A} revealed the presence of the close-in orbiting companion $\delta$\,Cep~Ab.
$\delta$ Cep exhibits a strong PMa in the Hipparcos and GDR2 catalogs, at S/N levels of $\Delta_\mathrm{Hip} = 7.0$ and $\Delta_\mathrm{G2} = 4.7$ (see Table~\ref{delCepB} and Paper~I), respectively.
Adopting the GDR2 parallax of $\delta$\,Cep~B as that of the system, they correspond to tangential velocity anomalies of $\Delta v_\mathrm{tan,Hip} = 2.1 \pm 0.4$\,km\,s$^{-1}$ and $\Delta v_\mathrm{tan,G2} = 5.3 \pm 1.5$\,km\,s$^{-1}$.
These velocities are too large to be induced by the orbital motion of the distant companion $\delta$\,Cep~B.
In Paper~I, the mass of the close-in companion $\delta$\,Cep~Ab was estimated to $m_{Ab}=0.72 \pm 0.11\,\,M_\odot$, assuming a mass of $m_A = 4.80 \pm 0.72\,\,M_\odot$ for the CC. This corresponds to a red dwarf between the spectral types K3V and M0V \citepads{2013ApJS..208....9P}.
This late spectral type explains why this companion could not be detected by \citetads{2016MNRAS.461.1451G} using optical interferometry.

Based on the observed PMa, we confirm that $\delta$\,Cep is probably a quadruple system, pending the confirmation that component B is gravitationally bound to A.

\begin{table}[]
\caption{Absolute and relative motion of $\delta$\,Cep A and $\delta$\,Cep B (HD 213307).
The absolute and linear proper motions ($\mu$ and $\Delta\mu$) are expressed in mas\,a$^{-1}$ and the differential tangential velocity of B relative to A ($dv_\mathrm{tan}$) in km\,s$^{-1}$.
The angular proper motion was converted to velocity using the GDR2 parallax of component B ($\varpi_\mathrm{G2,B} = 3.393 \pm 0.049$~mas).}
\centering
\small
\begin{tabular}{lcccc}
\hline \hline \noalign{\smallskip}
 & \multicolumn{2}{c}{$\delta$\,Cep A} & \multicolumn{2}{c}{$\delta$\,Cep B}  \\
 & $\alpha$ & $\delta$ & $\alpha$ & $\delta$ \\
\hline \noalign{\smallskip}
$\vec{\mu}_\mathrm{HG}$ & $+14.069_{0.009}$ & $+2.703_{0.016}$ & $+14.134_{0.009}$ & $+3.548_{0.015}$ \\
$dv_\mathrm{tan}$ & & & $+0.089_{0.013}$ & $+1.181_{0.021}$ \\
 \noalign{\smallskip} \hline \noalign{\smallskip}
$\vec{\mu}_\mathrm{H}$ & $+15.35_{0.22}$ & $+3.52_{0.18}$ & $+16.19_{0.59}$ & $+4.28_{0.50}$ \\
$\Delta \vec{\mu}_\mathrm{H}$ & $+1.28_{0.23}$ & $+0.82_{0.19}$ & $+2.06_{0.59}$ & $+0.73_{0.51}$ \\
$\Delta_\mathrm{H}$ & \multicolumn{2}{c}{7.0} & \multicolumn{2}{c}{3.8} \\
 \noalign{\smallskip} \hline \noalign{\smallskip}
$\vec{\mu}_\mathrm{G2}$ & $+17.64_{0.81}$ & $+3.98_{0.73}$ & $+14.09_{0.09}$ & $+3.79_{0.09}$ \\
$\Delta \vec{\mu}_\mathrm{G2}$ & $+3.57_{0.82}$ & $+1.27_{0.72}$ & $-0.04_{0.11}$ & $+0.24_{0.12}$ \\
$\Delta_\mathrm{G2}$ & \multicolumn{2}{c}{4.7} & \multicolumn{2}{c}{2.1} \\
 \noalign{\smallskip} \hline
\end{tabular}
\label{delCepB}
\end{table}

\subsubsection{CP Cep}

\citetads{1977MNRAS.178..505M} argued that the long-period pulsator \object{CP Cep} is a binary with a B3 spectral type companion.
From Paper~I, \object{CP Cep} ($P = 17.86$\,d) exhibits a significant PMa with a S/N of $\Delta_\mathrm{Hip}= 3.2$. We therefore confirm the binarity of this CC with a close-in orbiting companion.
We do not detect additional resolved companions in the GDR2 catalog.

\subsubsection{AX Cir}

We reported in Paper~I the presence of a strong PMa on \object{AX Cir} induced by its close-in orbiting companion \citepads{1960PASP...72..500J,2004MNRAS.350...95P} of spectral type B6V \citepads{1994ApJ...436..273E}. This companion was resolved using optical interferometry by \citetads{2014A&A...561L...3G}.
The combination of the GDR2 astrometric PM vectors and the spectroscopic orbit enabled us in Paper~I to determine that the close-in companion of \object{AX Cir} has a mass of $\approx 5.2\,M_\odot$, larger than that of the Cepheid.
This value, compared with the mass from a STIS spectrum (in prep.) of $3.5\,M_\odot$ (B9V), suggests that the companion is itself a binary.

We also identified a very faint \texttt{Bound} candidate (Fig.~\ref{cepheid-field-1}) at a large projected separation of 42\,kau, whose spectral type is M3.5V (Table~\ref{cepheids-bound-table1}).

\subsubsection{BP Cir}

The short-period \object{BP Cir} ($P=2.398$\,d) is a known binary star through its changing $\gamma$-velocity \citepads{2004MNRAS.350...95P}, although its orbital elements are still uncertain. In Paper~I, we did not detect any significant PMa, possibly due to the very long orbital period.
\citetads{Gallenne2018b} detected a close-in companion of \object{BP Cir} from near-infrared interferometry, at a separation corresponding to an orbital period of $\approx 14680$\,d.

We identified a very low mass common proper motion candidate companion of spectral type M2V at a large projected separation of 39\,kau, with a parallax of $\varpi_\mathrm{G2} = 1.53 \pm 0.19$~mas). We note that the GDR2 parallax of \object{BP Cir} ($\varpi_\mathrm{G2} = 1.02 \pm 0.04$~mas) is significantly different from the value derived from the Leavitt law ($\varpi_\mathrm{PL} = 1.70$~mas).

\subsubsection{R Cru}

\object{R Cru} is a suspected binary from its $\gamma$-velocity drift \citepads{1982MNRAS.199..925L}.
We detected a tight, \texttt{Bound} comoving companion of \object{R Cru} (Fig.~\ref{cepheid-field-2}) of spectral type G8V at a projected separation of $7.70\arcsec$ (6.6\,kau) from the CC (Table~\ref{cepheids-bound-table1}). This companion was also identified by \citetads{2016AJ....151..129E} from HST imaging at a separation of $7.64\arcsec$. The slight change in separation between their observing epoch (2014.0) and the GDR2 epoch (2015.5) is indicative of a possible orbital motion.
The spectral type that we determine for \object{R Cru B} is compatible with this star being the source of the X-ray emission detected by \citetads{2016AJ....151..108E} using XMM imaging.

In addition, we found a marginal PMa ($\Delta_\mathrm{G2}=2.2$; visible in Fig.~\ref{cepheid-field-2}) that indicates the possible presence of another close-in companion.

\subsubsection{X Cru}

The resolved companion of \object{X Cru} (Fig.~\ref{cepheid-field-2}) is a main sequence solar-type dwarf ($T_\mathrm{eff} = 6000 \pm 200$\,K, $R = 1.0 \pm 0.1\,R_\odot$) of likely spectral type G1V (Table~\ref{cepheids-bound-table1}). It is located at a large projected separation of 40\,kau from the CC.

\subsubsection{SU Cru}

\object{SU Cru} is a known spectroscopic binary \citepads{1996A&A...311..189S} that shows a high PMa ($\Delta_\mathrm{G2}=7.4$; Paper~I).
However, this signal is possibly biased by a significant error in the Hipparcos PM or position values, as the PM vector in the GDR2 catalog is much lower. We did not detect a resolved companion.

\subsubsection{VW Cru}

\object{VW Cru} ($P=5.265$\,d) is not a known binary CC \citepads{2003IBVS.5394....1S}. Its resolved candidate \texttt{Bound} companion is located at a projected separation of 29\,kau (Fig.~\ref{cepheid-field-2}). It is a hot dwarf of spectral type A2V (Table~\ref{cepheids-bound-table1}).

\subsubsection{SU Cyg}

\object{SU Cyg} is a 3.845\,d period CC that exhibits a very strong PMa of $\Delta_\mathrm{G2}=15.9$.
It is a member of the open cluster \object{Turner 9} \citepads{2013MNRAS.434.2238A}, and a known binary, whose orbital parameters were determined by \citetads{1988ApJS...66..343E}.
The close-in main sequence companion is a HgMn star \citepads{1998A&A...332L..33W} for which we derived a provisional mass of $m_2 = 4.7 \pm 0.7\,M_\odot$ in Paper~I.

\citetads{1997AJ....113.2104T} suggested that the close A2V nearby field star (\#1 in their Table 1) may be physically associated with the CC.
This source (\object{Gaia DR2 2031776202584173952}) is located at a separation of $24.8\arcsec$, west of the CC (Fig.~\ref{cepheids-near1}).
Its GDR2 parallax ($\varpi_\mathrm{G2} = 1.157 \pm 0.034$~mas) is within $1\sigma$ of the GDR2 parallax of \object{SU Cyg} ($\varpi_\mathrm{G2} = 1.198 \pm 0.052$~mas).
The difference in GDR2 tangential velocity between the two stars reaches $dv_\mathrm{tan} = 4.5 \pm 0.9$\,km\,s$^{-1}$, which indicates that this field star is not gravitationally bound to the CC (the escape velocity is $v_\mathrm{esc} \approx 1$\,km\,s$^{-1}$ at the projected separation).
The difference is even larger if we consider the difference between the proper motion $\mu_\mathrm{HG}$ of the CC (mean PM between the Hipparcos and GDR2 epochs) and the GDR2 PM vector of the A2V field star.
However, as argued by \citetads{1997AJ....113.2104T}, the GDR2 parallax confirms its small physical separation from the CC, which makes it a good fiducial to estimate the color excess $E(B-V)$ of the CC.

\subsubsection{V0532 Cyg}

The short-period Cepheid \object{V0532 Cyg} is a suspected binary star \citepads{1996AstL...22...33G} with a possible period of around 400\,days.
We did not detect any significant PMa, but a common proper motion companion of spectral type F0V (Table~\ref{cepheids-bound-table1}) is present at a projected separation of 29\,kau (Fig.~\ref{cepheid-field-2}).

\subsubsection{V1046 Cyg}

We identified a close resolved common proper motion companion of \object{V1046 Cyg} at a projected separation of 6.7\,kau (Fig.~\ref{cepheid-field-2}).
It is a hot dwarf of spectral type B8V, with a probable mass around $3.5\,M_\odot$ (Table~\ref{cepheids-bound-table1}).

\subsubsection{V1334 Cyg}

This short-period, first overtone CC is a known spectroscopic \citepads{1995ApJ...445..393E, 2000AJ....119.3050E} and interferometric \citepads{2013A&A...552A..21G, Gallenne2018} binary star, that exhibits the strongest PMa of our sample ($\Delta_\mathrm{G2} = 31.0$).
A detailed discussion of the PMa of \object{V1334 Cyg}, and its use to determine the physical parameters of the system, is presented in Paper~I.

A common proper motion companion (\object{Gaia DR2 1964855939153629312}) is present at a separation of $66\arcsec$ (44\,kau, Fig.~\ref{cepheids-near2}). Its parallax of $\varpi_\mathrm{G2} = 1.409 \pm 0.318$~mas is in good agreement with the value $\varpi_\mathrm{G18} = 1.388 \pm 0.015$~mas determined by \citetads{Gallenne2018}.
Adopting the same color excess $E(B-V)=0.025 \pm 0.009$ as the CC \citepads{2008MNRAS.389.1336K}, the $G$ band magnitude of the candidate companion ($m_G = 19.7$) is $M_G=10.4$. This star is thus likely a low mass red dwarf of early M spectral type. This hypothesis is compatible with its relatively large $G - G_\mathrm{RP} =1.1$ color index.

\subsubsection{$\zeta$ Gem}

The association of the nearby visual companion \object{HD 268518} (\object{WDS J07041+2034 B}) with the 10-day period CC $\zeta$\,Gem (\object{HD 52973}) is disproved by its very different GDR2 parallax ($\varpi_\mathrm{G2} = 28.65 \pm 0.06$~mas) and proper motion. It was found to be a spectroscopic binary with an F4V spectral type, and proposed to be a physical companion of the CC by \citetads{2012ApJ...748L...9M}. We did not detect any other resolved candidate companions to $\zeta$\,Gem.
However, $\zeta$\,Gem displays a marginal PMa at a level of $\Delta_\mathrm{G2}=2.3$ (Paper~I), which indicates the possible presence of a close-in orbiting companion, although it shows no sign of orbital motion \citepads{2015AJ....150...13E}.

\subsubsection{T Mon}

\object{T Mon} is a rare long-period CC ($P=27.02$\,d). It exhibits a moderate PMa ($\Delta_\mathrm{G2}=2.6$), from which we derived a high companion mass of $8.4\,M_\odot$ (Paper~I), comparable to the mass of the CC. This is in line with the conclusions by \citetads{1999ApJ...524..379E} who proposed that component B is a chemically peculiar star of the magnetic Ap $\alpha$ CVn type, and likely to be a binary itself, as chemically peculiar stars often are.
However, no detection was  reported from near-infrared interferometry by \citetads{Gallenne2018b}. 
We did not identify resolved common proper motion candidates.

\subsubsection{CV Mon}

\object{CV Mon} is a member of the open cluster \object{van den Bergh 1} \citepads{2015MNRAS.446.1268C}, with a color excess $E(B-V) = 0.75 \pm 0.02$\,mag \citepads{1998AJ....115.1958T}. This value is consistent with that of \citetads{2013A&A...550A..70G}, $E(B-V)=0.722$, which we adopt here (Table~\ref{cepheids-bound-table1}).
It shows a PMa at a level of $\Delta_\mathrm{Hip} = 2.7$, but of only $\Delta_\mathrm{G2}=2.0$ (Paper~I).

As shown in Fig.~\ref{cepheid-field-3}, we identified two nearby stars, \object{Gaia DR2 3127142224816361600} and \object{Gaia DR2 3127142327895572352,} that are potentially gravitationally bound to the CC.
The position angle of the PM vector of the latter differs by $15.7^\circ$ from that of the CC. This is slightly more than our acceptable limit for \texttt{Bound} companions ($15^\circ$ at this separation); nevertheless, we   have included it in our \texttt{Bound} list in Table~\ref{cepheids-bound-table1}.
These two stars were already identified by \citetads{1994AJ....108..653E} as probable companions; they were measured at slightly larger angular separations from the CC, which may be due to their orbital motion.
Their tangential velocities relative to the target are below 1\,km\,s$^{-1}$, therefore compatible with being gravitationally bound to the CC at their respective projected linear separations of 18 and 24\,kau.
The spectral type of both companions is around B8V, corresponding to a mass of $\approx 3.5\,M_\odot$ \citepads{2013ApJS..208....9P}.

\subsubsection{S Mus}

\object{S Mus} ($P=9.66$\,d) is a known spectroscopic binary \citepads{1990PASP..102..551E,1990AJ.....99..353B}, and a member of the open cluster \object{ASCC 69} \citepads{2013MNRAS.434.2238A,2014ApJ...785L..25E,2017MNRAS.464.1119C}. We detected a moderate PMa (Paper~I), which we used in combination with the spectroscopic orbit of \citetads{Gallenne2018b} to determine a mass of $m_2 = 2.2 \pm 0.3\,M_\odot$ for the secondary star. This is significantly lower than  expected from the B3.5V spectral type estimated by \citetads{1994AJ....108.2251E} from IUE observations, but it should be noted that the short orbital period of \object{S Mus} ($P_\mathrm{orb} = 505$\,d) results in a strong smearing of the PMa, particularly for the Hipparcos epoch. The low mass we determined could therefore be significantly biased and should be considered preliminary.
\citetads{Gallenne2018b} detected the companion of \object{S Mus} from near-infrared interferometry, and established the orbital elements of the system.
Epoch astrometry from future Gaia data releases will provide a very accurate astrometric orbit of the photocenter of the system.

The resolved candidate companion reported by \citetads{2016AJ....151..108E} and Evans et al.~(2019, in prep.) is not present in the GDR2, possibly due to its angular proximity to the bright CC. Evans et al.~concluded that it is not a young companion, but that X-rays are probably produced by the spectroscopic binary companion. We did not detect other \texttt{Bound} candidate companions.

\subsubsection{S Nor}

\object{S Nor} ($P=9.75$\,d) is known to host a resolved companion at a separation of $0.90\arcsec$ (equivalent to $\approx 800$\,au) discovered by \citetads{2013AJ....146...93E}. We did not detect this companion in the GDR2 catalog as it is located too close to the CC.
We did not find other, more distant \texttt{Bound} candidate companions, and we do not confirm the companion proposed by \citetads{1994AJ....108..653E}.
However, \object{S Nor} is a member of the open cluster \object{NGC 6087} \citepads{2013MNRAS.434.2238A}, and is therefore located in a rich stellar field where many \texttt{Near} candidates are present with comparable PM vectors (Fig.~\ref{cepheids-near2}).

We observed a significant PMa in the Hipparcos ($\Delta_\mathrm{Hip} = 3.8$) and moderate in the GDR2 ($\Delta_\mathrm{G2} = 2.2$) that confirm the presence of a close-in orbiting component.
Combining them with the spectroscopic orbital parameters by \citetads{2008A&A...488...25G} (who assumed a circular orbit), we derived a companion mass of $1.5 \pm 0.2\,M_\odot$ (Paper~I).
The PMa is an important confirmation of a close orbit, which has been difficult to identify. 
There are three well-covered seasons of CORAVEL data from \citetads{1987A&AS...70..389M} and \citetads{1994A&AS..108...25B} showing little evidence of orbital motion.
No signature of a close-in companion was found by \citetads{Gallenne2018b} in radial velocity at a level of $\pm 1$\,km\,s$^{-1}$.

\subsubsection{RS Nor}

\object{RS Nor} ($P=6.20$\,d) does not show a significant PMa (Paper~I) and is not a known binary.
We detected a relatively massive \texttt{Bound} candidate companion of spectral type A1V (Table~\ref{cepheids-bound-table1}) at a projected separation of 15\,kau (Fig.~\ref{cepheid-field-3}).

\subsubsection{SY Nor}

\object{SY Nor} ($P=12.64$\,d) is a known binary system \citepads{2003IBVS.5394....1S} with a close-in B4.5V companion identified from IUE observations by \citetads{1994ApJ...436..273E}.
We detected a strong PMa induced by this companion ($\Delta_\mathrm{G2} = 8.0$).

We also detected two resolved, gravitationally bound candidates (Fig.~\ref{cepheid-field-3}).
The resolved companions have respective (projected separation of 5.7\,kau) and  (42\,kau).
The closer resolved candidate (separation 5.7\,kau), \object{Gaia DR2 5884729035245399424} (\object{SY Nor B}), was  identified by \citetads{1981A&AS...44..179P} and is discussed in Appendix A of \citetads{1994ApJ...436..273E} as potentially bound to the CC.
Its B2V spectral type corresponds to a mass of $\approx 7\,M_\sun$, which is comparable to the expected CC mass ($\approx 6.2\,M_\odot$). This high mass may indicate that B is itself a binary.
The AB pair is present in the WDS catalog (\object{WDS J15548-5434}; \citeads{2001AJ....122.3466M}), and we list in Table~\ref{SYNor-positions} the relative position of component B.
We confirm that \object{SY Nor B} is gravitationally bound to the CC, and although its orbital period is likely on the order of 100\,ka, its astrometric orbital displacement may be detectable in the future Gaia data releases or, for example,  by using the astrometric dual field mode of the GRAVITY instrument \citepads{2017A&A...602A..94G}.

\begin{table}[htp]
\caption{Relative positions of SY Nor B with respect to SY Nor A.}
\centering
\begin{tabular}{lllc}
\hline \hline
Epoch & PA $(^\circ)$ & Separation $(\arcsec)$ \\
\hline
1934 & 315 & 2.1 \\
1967 & 315 & 2.2 \\
2015.5 & $314.7862 \pm 0.0015$ & $2.46341 \pm 0.00007$ \\
\end{tabular}
\label{SYNor-positions}
\end{table}%

The wider candidate companion (42\,kau), of approximate spectral type F6V,  marginally satisfies the gravitational bound condition on the tangential velocity. However, considering that the inner system is already composed of at least three massive stars with a total mass on the order of $20\,M_\odot$, we classify this star as a \texttt{Bound} candidate. \object{SY Nor} is therefore likely a hierarchical quadruple system, possibly quintuple if B is a binary.

\subsubsection{QZ Nor}

The short-period, first overtone \citepads{2009A&A...504..959K} Cepheid \object{QZ Nor} ($P=3.786$\,d) is not known to be a binary system \citepads{2003IBVS.5394....1S}. It probably belongs to the open cluster \object{NGC 6067} \citepads{1983AJ.....88..379E, 2013MNRAS.434.2238A}, together with the Cepheid \object{V0340 Nor}.
We identified a low mass (spectral type K1V, Table~\ref{cepheids-bound-table2}), common proper motion companion at a large projected separation of 30\,kau (Fig.~\ref{cepheid-field-3}).
Its GDR2 parallax is compatible with that of the CC within their uncertainties.

\subsubsection{Y Oph}

\object{Y Oph} is a long-period CC ($P=17.12$\,d) with peculiar sinusoidal photometric and radial velocity curves.
It shows phase jumps in its $O-C$ diagram \citepads{1989CoKon..94....1S}, and exhibits a strong reddening, that may be partly of circumstellar origin.
\citetads{1978PASP...90..188A} and \citetads{1978A&A....62...75P} proposed that \object{Y Oph} is a spectroscopic binary star, which was not confirmed by \citetads{1986AJ.....92..436E}. 
IUE observations by \citetads{1992ApJ...384..220E} excluded the possibility of a close-in companion hotter than spectral type A0, but \citetads{1989CoKon..94....1S} determined a possible binary orbital period of 1223\,d.
We did not detect any significant PMa in \object{Y Oph} ($\Delta_\mathrm{G2} = 0.7$; Paper~I), and we thus do not confirm its binarity.

The GDR2 parallax of \object{Y Oph} ($\varpi_\mathrm{G2} = 1.39 \pm 0.08$~mas) is different from the measurements obtained using the classical Baade-Wesselink technique (e.g., $\varpi = 1.79 \pm 0.09$~mas from \citeads{2013A&A...550A..70G}, $\varpi = 1.81 \pm 0.13$~mas from \citeads{2007A&A...476...73F}) or its interferometric version ($\varpi = 2.04 \pm 0.07$~mas; \citeads{2007ApJ...664.1093M}).
Considering this uncertainty on the GDR2 parallax, we considered a broad parallax range from $\varpi = 1.2$ to 2.5\,mas as prior for the wide companion search. No companion was detected assuming this prior condition on the parallax.
The only field star showing a proper motion comparable to that of \object{Y Oph} is \object{Gaia DR2 4175017934690530688}, a faint red dwarf. With a parallax of $\varpi_\mathrm{G2} = 1.726 \pm 0.395$~mas, it is compatible with our adopted range of acceptable parallaxes. Future Gaia data releases will provide more accurate parallax values of both this star and \object{Y Oph} and confirm if they are comoving companions.
We note that \citetads{2014A&A...567A..60G} did not find any companion of \object{Y Oph} from adaptive optics imaging up to a projected separation of $1.7\arcsec$.

\subsubsection{AW Per\label{awper}}

\object{AW Per} is a  fundamental mode CC with a pulsation period of $P_\mathrm{puls} = 6.46$\,d.
It is a known spectroscopic binary system \citepads{1989AJ.....97.1153W}.
\citetads{2000AJ....120..407E} determined accurate orbital parameters of the inner \object{AW Per} Aab system and in particular an orbital period of $P_\mathrm{orb} \approx 40$\,years.
They determined a minimum dynamical mass of $6.6\,M_\odot$ for the CC's spectroscopic companion Ab, which is significantly higher than the mass inferred from its spectral type ($4.0\,M_\odot$). They therefore proposed that the Ab component is itself a binary system.
\citetads{2008MNRAS.383..139M} measured the position angle and separation of the companion from the CC at epoch November 2001 (MJD 52235): $\theta = 13.74 \pm 0.26$~mas, $\mathrm{PA} = 184 \pm 2^\circ$.
\citetads{2015A&A...579A..68G} detected the companion of \object{AW Per} using near-infrared interferometry.
We detected a very strong PMa ($\Delta_\mathrm{G2}=21.4$; Paper~I), that we attribute to the photocenter displacement of the inner Aab system.
The combination of the spectroscopic orbital parameters from \citetads{2016Obs...136..209G} with the PM anomalies (Paper~I) confirm the high mass of the companion ($8.8 \pm 1.3\,M_\odot$), and consequently its binarity.

We detected an additional high probability \texttt{Bound} candidate companion at a projected linear separation of 8.4\,kau (Fig.~\ref{cepheid-field-3}). This is a low mass star whose spectral type is likely around K3.5V. \object{AW Per} is therefore likely a hierarchical quadruple system.

\subsubsection{RS Pup\label{rspup}}

The long-period Cepheid \object{RS Pup} ($P \approx 41.5$\,d) is one of the most luminous Galactic CCs. It is remarkable as it is embedded in a large dusty nebula that scatters the light from the pulsating star creating light echoes. This phenomenon was discovered by \citetads{1961PASP...73...72W} and first studied by \citetads{1972A&A....16..252H}. \citetads{Kervella:2014lr} determined the distance to the CC ($d_\mathrm{echoes} = 1910 \pm 80$\,pc) using polarimetric imaging of the light echoes with the Hubble Space Telescope ACS camera.
The presence of light echoes implies that the nebula is associated with the CC, but it is too massive to have been created by mass loss from the star itself \citepads{2012A&A...541A..18K}. The extension of the nebula ($\approx 1$\,pc) is negligible with respect to the distance to the CC ($\approx 2$\,kpc).
\citetads{2017A&A...600A.127K} determined the spectroscopic projection factor of \object{RS Pup} $p=1.25 \pm 0.06$ and its color excess $E(B-V)=0.496 \pm 0.006$.

\begin{figure}[]
\centering
\includegraphics[width=8cm]{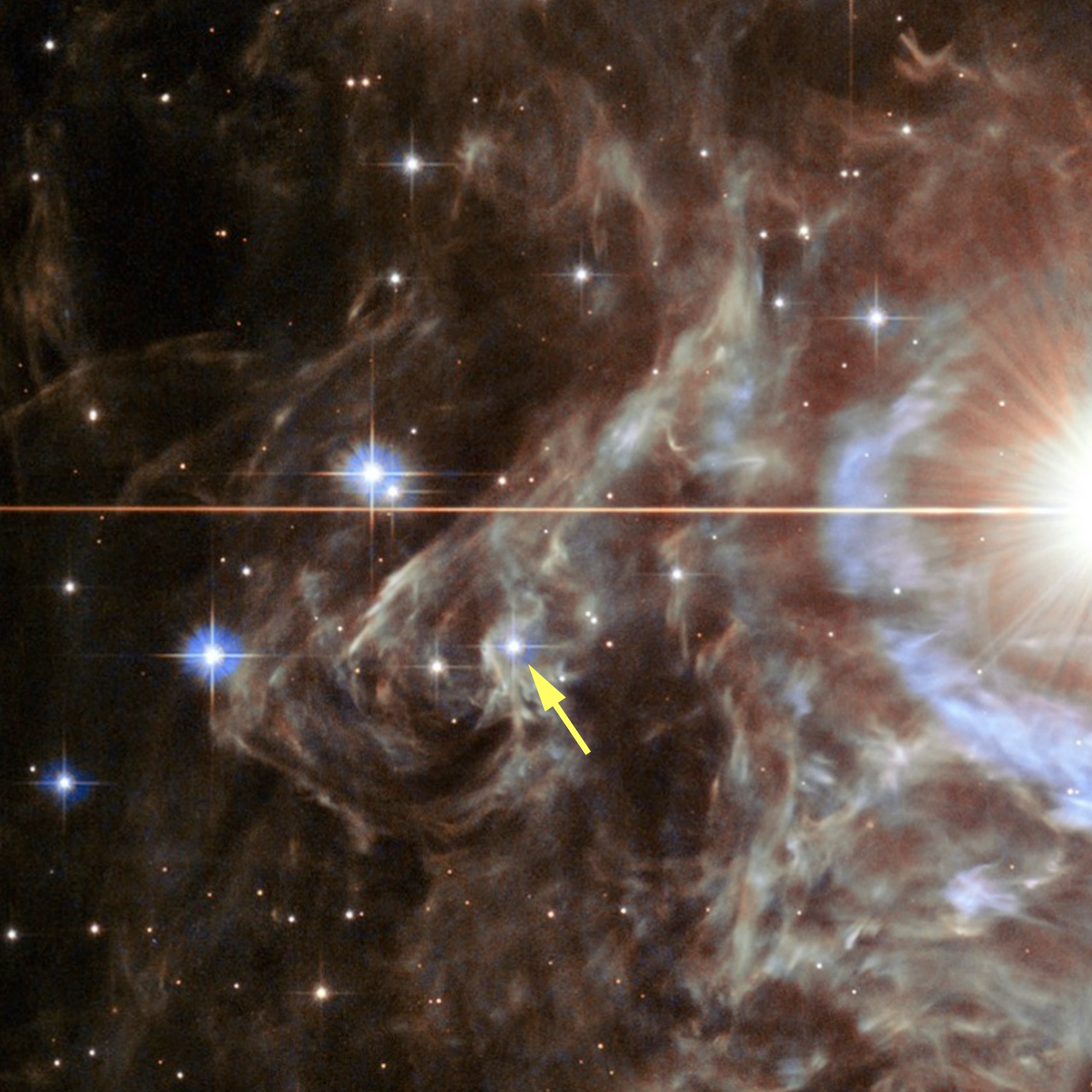}
\caption{Star S1 (yellow arrow) associated with the dusty nebula surrounding the long-period Cepheid RS Pup (the bright star at the right edge of the image).
Color rendition: NASA, ESA, Z. Levay, and the Hubble Heritage Team STScI/AURA-Hubble/Europe Collaboration.\label{RSPupS1}}
\end{figure}

The star \object{Gaia DR2 5546476755539995008} (hereafter S1) appears to be embedded within the nebula surrounding RS\,Pup because it is interacting with the dust cloud (Fig.~\ref{RSPupS1}).
The differential proper motion with respect to RS\,Pup is $\mu[\mathrm{S1\,-\,RS\,Pup}] = (-0.15, +3.26)$~mas\,a$^{-1}$.
This corresponds to a tangential velocity difference of $\approx 30$\,km\,s$^{-1}$, and S1 is therefore not gravitationally bound to the CC.
Its GDR2 parallax is $\varpi[\mathrm{S1}] = 0.532 \pm 0.048$~mas, and its spectral type is F6V (Table~\ref{cepheids-bound-table2}).
The presence of this star in the nebula makes it an interesting proxy for the GDR2 parallax of \object{RS Pup}, without the possible bias due to the large amplitude photometric and color variability of the CC.
We note that the parallax of S1 is compatible within the error bars with that of the CC determined from polarimetric imaging of the light echoes $\varpi_\mathrm{echoes} = 0.524 \pm 0.022$~mas \citepads{Kervella:2014lr}.

\object{RS Pup} shows a moderate PMa $\Delta_\mathrm{G2} = 1.8$ (Paper~I), which indicates the possible presence of a close-in orbiting companion. However, the presence of the bright light echoes close to the CC may perturb the Gaia astrometric measurements.

\subsubsection{S Sge}

We detected a very strong PMa of $\Delta_\mathrm{G2} = 20.0$ (Paper~I) on \object{S Sge} ($P=8.38$\,d).
Combined with the spectroscopic orbital parameters from \citetads{2008A&A...488...25G}, we derived an inclination of the orbital plane of $i=75 \pm 10\,\deg$, and a companion mass of $3.0 \pm 0.5\,M_\odot$. This mass is still higher than that from the IUE spectrum (1.7 to $1.5\,M_\odot$; \citeads{1993AJ....106.1599E}), making it likely that the
companion is itself a binary; however, the short orbital period of 676\,days  results in a temporal smearing of the Hipparcos PMa vector, and this mass should be considered provisional.
We did not detect any resolved \texttt{Bound} candidate companions.

\subsubsection{U Sgr}

\object{U Sgr} ($P=6.75$\,d) does not display a significant PMa.
It is a member of the open cluster \object{Messier 25} (\object{IC 4725}; \citeads{1955MNSSA..14...38I, 2013MNRAS.434.2238A}).
We identified a number of field comoving stars of the cluster that are visible in Fig.~\ref{cepheid-field-3}, two of which are \texttt{Bound} candidates.
These are early A-type dwarf or subgiant stars (Table~\ref{cepheids-bound-table2}).

\subsubsection{V0350 Sgr}

We detected a very strong PMa in \object{V0350 Sgr} ($\Delta_\mathrm{G2} = 8.7$; Paper~I) induced by its known close-in companion \citepads{2015AJ....150...13E,2018ApJ...866...30E}.
The combination of the astrometric PM anomalies and the spectroscopic orbit by \citetads{Gallenne2018b} results in a mass of $m_2 = 3.8 \pm 0.6\,M_\odot$ for the secondary.
This mass is compatible with the B9.5V spectral type determined by \citetads{2013AJ....146...93E} from IUE observations, within the error bars.
A possible interferometric detection of the companion was reported by \citepads{Gallenne2018b}.
We also found a \texttt{Bound} candidate at a projected separation of 26\,kau (Fig.~\ref{cepheid-field-4}) that has a spectral type of A2V (Table~\ref{cepheids-bound-table2}).
\object{V350 Sgr} is therefore probably a triple system.

\subsubsection{W Sgr}

\object{W Sgr} is a $P=7.6$\,d CC that is a member of a triple system, with one wide (A0V spectral type; \citeads{1978MNRAS.183..701M, 1985ApJ...296..175B, 1991ApJ...372..597E}) and one spectroscopic companion (discovered by \citeads{1989A&A...216..125B}).
\citetads{Gallenne2018b} determined the spectroscopic orbital parameters of the inner system.
Combined with the moderate PMa of the star, the spectroscopic companion mass was estimated to $m_2 = 1.1 \pm 0.2\,M_\odot$ (Paper~I).
The known wide component of \object{W Sgr} is too close to the CC for Gaia ($<0.2\arcsec$), and we did not detect other \texttt{Bound} candidate companions.

\subsubsection{RV Sco}

The very strong PMa that we detected ($\Delta_\mathrm{G2} = 9.7$; Paper~I) confirms that this star is a close binary.
\citetads{1989CoKon..94....1S} proposed an orbital period of $\approx 8000$\,d from the residuals of the $O-C$ diagram.
\citetads{1992ApJ...384..220E} established an upper limit of A3 for possible main sequence companions of \object{RV Sco}.
We note that the Hipparcos and GDR2 PM anomalies are significantly different (time separation of 8850\,d), and the future Gaia data releases will provide the necessary astrometry to improve the determination of the companion properties.
We did not identify any resolved \texttt{Bound} candidates from the GDR2 catalog.

\subsubsection{V0636 Sco}

From its strong PMa ($\Delta_\mathrm{G2} = 11.9$; Paper~I) and the spectroscopic orbital parameters from \citetads{Gallenne2018b}, we determined that the close-in companion of this $P=6.80$\,d pulsator has a mass of $m_2 = 2.3 \pm 0.3\,M_\odot$. This estimate is in good agreement with the determination by \citetads{1998ApJ...505..903B} who found a spectral type of B9.5V ($m_2 \approx 2.5\,M_\odot$) for \object{V0636 Sco B} using UV spectroscopy from the HST.
A possible interferometric detection of this companion is reported by \citepads{Gallenne2018b}.
We did not identify any resolved \texttt{Bound} candidate from the GDR2 catalog.

\subsubsection{V0950 Sco}

We did not find a PMa ($\Delta_\mathrm{G2}=1.8$; Paper~I) on \object{V0950 Sco}.
We isolated one resolved \texttt{Bound} candidate, \object{Gaia DR2 5960623340819000192} (Fig.~\ref{cepheid-field-4}) located at a projected separation of 15\,kau from the CC and of spectral type G1V (Table~\ref{cepheids-bound-table2}).

\subsubsection{CM Sct}

\object{CM Sct} is a short-period CC ($P=3.92$\,d) that is not classified as a known spectroscopic binary, in line with our non detection of any significant PMa (Paper~I).

We detected a resolved \texttt{Bound} candidate companion of spectral type A1V (Table~\ref{cepheids-bound-table2}), \object{Gaia DR2 4253603428053877504} (Fig.~\ref{cepheid-field-4}), located at a projected separation of 49\,kau ($26.8\arcsec$). The GDR2 parallax of this companion ($\varpi = 0.547 \pm 0.050$~mas) is slightly larger than that of the CC ($\varpi = 0.405 \pm 0.065$~mas). Considering that the expected parallax from the Leavitt law is $\varpi = 0.518$~mas, this difference may be due to an offset in the GDR2 parallax of the CC.

\subsubsection{EV Sct}

\object{EV Sct} is a member of the open cluster \object{NGC 6664} \citepads{1976AJ.....81.1125T,1987A&AS...70..389M,2003MNRAS.345..269H,2013MNRAS.434.2238A}, and the GDR2 parallax value ($\varpi = 0.526 \pm 0.054$~mas) confirms this association.
As a consequence, the stellar density is high in the field around \object{EV Sct}, and we detected a number of \texttt{LowV} sources with a similar proper motion as the CC (Fig.~\ref{cepheid-field-4}).
We identified one candidate \texttt{Bound} companion of spectral type B9V (Table~\ref{cepheids-bound-table2}), \object{Gaia DR2 4156513016572003840}, that is located at a projected separation of  44\,kau ($22.6\arcsec$) from the primary.

\citetads{1999A&A...350L..55K} proposed that \object{EV Sct} is a binary comprising two CCs, the companion being itself a very short-period CC ($P=1.2$\,d). We detected only a $\Delta_\mathrm{G2} = 1.1$ PMa on \object{EV Sct}, and we therefore cannot confirm the presence of a close-in orbiting companion. Future Gaia data releases will provide a stringent test of the presence of this companion.

\subsubsection{SZ Tau}

\object{SZ Tau} is a suspected spectroscopic binary star of low radial velocity amplitude \citepads{1996AstL...22...33G,1996AstL...22..175G}. It is also a member of the open cluster \object{NGC 1647} \citepads{1964PZ.....15..242E,1992AJ....104.1865T, 2013MNRAS.434.2238A}.
\citetads{1992ApJ...384..220E} placed an upper limit of A1 on the spectral type of a main sequence companion, and \citetads{2016AJ....151..108E} found little evidence of orbital motion (to about $\pm 1$\,km\,s$^{-1}$).
The moderate PMa that we observed ($\Delta_\mathrm{G2}=1.9$; Paper~I) is not conclusive, but further Gaia observations are needed to confirm the detection.

\subsubsection{Polaris ($\alpha$ UMi)\label{polaris}}

\object{Polaris} Aa ($\alpha$\,UMi, \object{HD 8890}, \object{HIP 11767}) is the nearest CC, but with a surprising uncertainty in its distance.
Its parallax has been measured by Hipparcos, and was found to be $\varpi_\mathrm{Hip} = 7.54 \pm 0.11$~mas \citepads{2007ASSL..350.....V, 2007A&A...474..653V}.
Unfortunately, \object{Polaris A} is too bright for Gaia, but the parallax of its physical companion \object{Polaris B} has been measured at $\varpi_\mathrm{G2} = 7.321 \pm 0.028$~mas (value corrected for the 29$\,\mu$as ZP offset).
This value is larger by $4.4\sigma$ than the parallax $\varpi_\mathrm{FGS} = 6.26 \pm 0.24$~mas for the same star measured by \citetads{2018ApJ...853...55B} using the HST FGS.
With the agreement between the Hipparcos and Gaia results, the FGS parallax is an outlier, but the reasons for this are unclear.
A discussion of the parallaxes of \object{Polaris A} and B is presented in Appendix B of \citetads{2018A&A...619A...8G} and in \citetads{2515-5172-2-3-126}.
\object{Polaris B} is a fast rotating main sequence F3V star \citepads{2008MNRAS.387L...1U}.
\object{Polaris} is part of a tight binary system with the secondary component \object{Polaris Ab} \citepads{2002ApJ...567.1121E, 2008AJ....136.1137E, 2018ApJ...863..187E}.
Using the Gaia distance, \citetads{2018ApJ...863..187E} found a preliminary mass of $3.45 \pm 0.75\,M_\odot$.
Using the FGS parallax, the likely mass is  6.5 to $7\,M_\odot$, and the Cepheid is pulsating in the first overtone and crossing the instability strip for the first time \citepads{2018A&A...611L...7A, 2515-5172-2-3-126}.

\begin{figure}[]
\centering
\includegraphics[width=9cm]{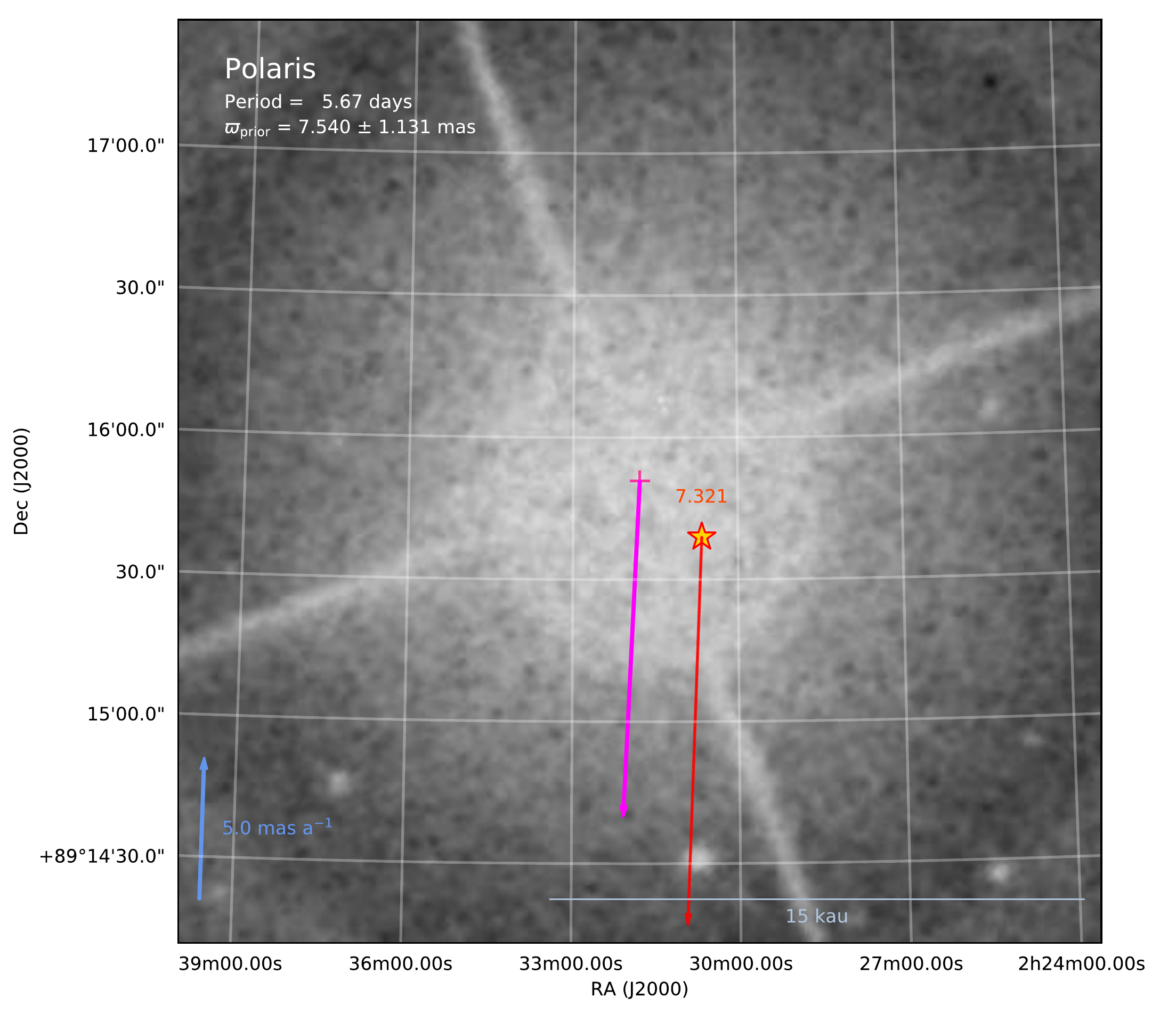}
\caption{Field around Polaris from the DSS2-Red showing its companion Polaris~B (yellow star).\label{polaris-wide}}
\end{figure}

For the present companion search, we adopt the Hipparcos parallax $\varpi_\mathrm{Hip} = 7.54$~mas as our prior estimate, with a 15\% relative uncertainty ($\pm 1.1$~mas).
This parallax range is compatible with the GDR2 parallax of \object{Polaris B}, whose properties are listed in Table~\ref{cepheids-bound-table2}.
This choice of prior parallax has a low influence on the efficiency of the search algorithm. 
This is particularly true as Polaris is located far from the plane of the Milky Way, and therefore in a low star density field.
We also adopt the Hipparcos proper motion vector $\mu_\mathrm{Hip} = [+44.48 \pm 0.11, -11.85 \pm 0.13]$~mas\,a$^{-1}$ \citepads{2007ASSL..350.....V}.
The field around Polaris is presented in Fig.~\ref{polaris-wide}.
The differential tangential velocity of \object{Polaris B} with respect to A is $dv_\mathrm{tan} = 1.96 \pm 0.32$~km\,s$^{-1}$.
The radial velocities determined by \citetads{1996JRASC..90..140K} of Polaris~A ($\gamma$-velocity of the inner binary system $v_\mathrm{rad} = -16.42 \pm 0.03$\,km\,s$^{-1}$) and Polaris~B ($v_\mathrm{rad} = -14.7 \pm 1.2$\,km\,s$^{-1}$) correspond to a differential radial velocity $dv_\mathrm{rad} = 1.7 \pm 1.2$\,km\,s$^{-1}$. This results in a space velocity of Polaris B with respect to A of $dv = 2.6 \pm 1.2$\,km\,s$^{-1}$.
The escape velocity at the projected separation of 2.3\,kau is $v_\mathrm{esc} \approx 3.2$\,km\,s$^{-1}$, assuming a total mass of $6.7\,M_\odot$ for the three components.
We thus conclude that the Polaris B is likely gravitationally bound to A, thus forming a triple stellar system.

\subsubsection{SX Vel}

\object{SX Vel} ($P=9.55$\,d) is not classifed as a known binary, but it shows a PMa at a level of $\Delta_\mathrm{Hip} = 2.9$. It is however not present in the GDR2 ($\Delta_\mathrm{G2} = 1.0$).
It was tested for being a member of the open cluster \object{SAI 94} by \citetads{2013MNRAS.434.2238A}, but the association has not been confirmed.
We detected a resolved \texttt{Bound} candidate companion of \object{SX Vel} at a projected separation of 32\,kau, which is probably a solar-like star of spectral type G1V.

\subsubsection{CS Vel}

\object{CS Vel} is not classified as a spectroscopic binary Cepheid. It is possibly a member of the open cluster \object{Ruprecht 79} \citepads{2010Ap&SS.326..219T}, but this association is considered inconclusive by \citetads{2013MNRAS.434.2238A}.
It is also discussed in \citetads{2015MNRAS.446.1268C}, who reach the same conclusion, but based on an incorrect PM vector.
The GDR2 PM vector of \object{CS Vel} is $\mu = [-4.56 \pm 0.06, +3.10 \pm 0.06]$~mas\,a$^{-1}$, which is actually consistent within the mean proper motion of the cluster $\mu =[-5.66 \pm 4.67, +4.04 \pm 4.93]$~mas\,a$^{-1}$.
The GDR2 parallax of \object{CS Vel} ($\varpi = 0.194 \pm 0.030$~mas) is however too small compared to the estimated distance of \object{Ruprecht 79} ($\approx 2$\,kpc; \citeads{2016A&A...585A.101K}).

We detected a \texttt{Bound} candidate companion of \object{CS Vel} (Fig.~\ref{cepheid-field-4}) located at a projected separation of 36\,kau ($8.9\arcsec$), and of spectral type A0V (Table~\ref{cepheids-bound-table2}).

\subsubsection{DK Vel}

This short-period CC ($P=2.48$\,d)  is not a known spectroscopic binary, and we detected a marginal PMa ($\Delta_\mathrm{G2} = 2.1$; Paper~I).
We identified a low mass \texttt{Bound} candidate companion of \object{DK Vel} of spectral type K0V (Table~\ref{cepheids-bound-table2}) located at a large projected separation of 50\,kau (Fig.~\ref{cepheid-field-5}).

\subsubsection{U Vul}

Known since the 19th century \citepads{1899ApJ.....9..179P}, \object{U Vul} is an intermediate period CC ($P=7.99$\,d).
It is classified as a spectroscopic binary \citepads{1991CoKon..96..123S,1996A&AS..116..497I}.
From its strong PMa and the spectroscopic orbital parameters by \citetads{2008A&A...488...25G}, we determined in Paper~I an inclination of the orbital plane of $i=163 \pm 7 \deg$. Our estimate of the mass of the companion  of $2.4 \pm 0.4\,M_\odot$ is consistent with the upper limit of $2.1\,M_\odot$ by \citetads{2015AJ....150...13E} (see also \citeads{1992ApJ...384..220E}).
We did not detect a resolved common proper motion companion of \object{U Vul} in the GDR2 catalog.

\subsection{RR Lyrae stars}

\subsubsection{AT And}

\object{AT And} is an RRab pulsator ($P=0.6169$\,d) that shows the strongest PMa of all the tested RRLs ($\Delta_\mathrm{G2}=16.7$, Paper~I).
This PMa corresponds to a maximum orbital period of around 6\,years (Table~A.7 in Paper~I) and a maximum angular semimajor axis on the order of 7\,mas.
This angular separation is in principle within the capabilities of  the CHARA Array \citepads{2005ApJ...628..453T}.
However, this a relatively faint star ($m_V = 10.60$, $m_K = 9.08$) for the sensitivity of optical interferometers, and the companion may be a very faint compact object that would be difficult to detect.
The most promising way to determine the orbital parameters would be to follow the evolution of its $\gamma$-velocity over $\approx 6$\,years using spectroscopy.
\object{AT And} exhibits the Blazhko effect \citepads{2017A&A...607A..11G}, as well as period change \citepads{1978CoKon..71....1O}.
\citetads{1997A&AS..125..313F} proposed that this star is an anomalous Cepheid, but this was not confirmed by \citetads{2004RMxAA..40...99P}.
We did not detect any additional resolved companion of \object{AT And} from the GDR2 catalog.

\subsubsection{OV And}

\object{OV And} is an RRab star ($P=0.4706$\,d) that does not show the Blazhko effect \citepads{2014A&A...562A..90S}.
This star is absent from the Hipparcos catalog, so we could not test for the presence of a PMa as described in Paper~I.
The PM vector of the first Gaia data release $\mu_\mathrm{G1} = [-5.32_{0.88}, -8.22_{0.47}]$~mas\,a$^{-1}$ \citepads{2016A&A...595A...2G} is however slightly different from the PM vector of the GDR2 $\mu_\mathrm{G2} = [-4.868_{0.057}, -7.672_{0.023}]$~mas\,a$^{-1}$ (uncorrected for frame rotation).
A more sensitive search for the presence of a PMa will be possible using the future Gaia data releases.

The resolved companion of \object{OV And} (Fig.~\ref{rrlyr-field-1}) is a relatively bright and hot star of spectral type F4V (Table~\ref{rrlyrae-bound-table}), with a probable mass around $1.4\,M_\odot$. The presence of such a massive companion orbiting an old population RRL, at a linear projected separation of 3.7\,kau, raises interesting questions on the evolutionary scenario that led to the current state of the system.

\subsubsection{V0363 Cas}

\object{V0363 Cas} is classified as an RR(B) pulsator in the GCVS, but it is listed as an RRc in \citetads{2006MNRAS.368.1757W}, and as a DCEPS(B) in the VSX database \citepads{2006SASS...25...47W}. 
A light curve of \object{V0363 Cas} was obtained by \citetads{2009IBVS.5882....1H} from the Optical Monitoring Camera on board the INTEGRAL satellite\footnote{light curve available from \url{http://sdc.cab.inta-csic.es/omc/var/4014000075.html}}.
They showed that this star is pulsating in two radial modes with a 0.802 period ratio of the first to second overtone periods: $P_1=0.546556$\,d and $P_2=0.438243$\,d. This indicates that \object{V0363 Cas} is probably a very short-period CC, hence its variable type in the VSX database, rather than an RRL.
We detected a strong PMa ($\Delta_\mathrm{G2} = 4.7$; Paper~I) on \object{V0363 Cas}, which confirms that it is a close binary system.

\subsubsection{XZ Cyg}

The RRab pulsator \object{XZ Cyg} has a period of $P=0.4667$\,d and shows the Blazhko effect \citepads{2006MmSAI..77..492S}. This is a particularly remarkable RRL due to its extremely fast rising from minimum ($m_V=10.5$) to maximum ($m_V=8.8$) brightness in only 53 minutes \citepads{2006JRASC.100..130H}. Its light curve presented in \citetads{2004AJ....127.1653L} shows that its Blazhko period ($\approx 57.4$\,d) is changing with time.

We detected a strong PMa ($\Delta_\mathrm{G2}=6.8$; Paper~I), which reveals the presence of a close-in binary companion. Depending on their orbital separation, the presence of this secondary massive body may play a role in the observed change of the Blazhko period of the RRL.

\subsubsection{CS Del}

\object{CS Del} is classified as an RRc pulsator \citepads{1982AJ.....87.1395K} with a period $P=0.366$\,d; however, the light curve reported by \citetads{2014IBVS.6106....1B} is  not conclusive.
The resolved companion of \object{CS Del} that we identified is located at a projected distance of 5\,kau (Fig.~\ref{rrlyr-field-1}) and has a spectral type of G7V (Table~\ref{rrlyrae-bound-table}).

\subsubsection{V0893 Her}

The classification of \object{V0893 Her} as an RR: pulsator ($P=0.4918$\,d; \citeads{2005A&A...442..381M}) appears uncertain, but its absolute magnitude is compatible with this possibility. Its GDR2 parallax of $\varpi = 2.68\pm 0.03$~mas places it among the nearest RRLs. Its resolved companion (Fig.~\ref{rrlyr-field-1}) is a very low mass red dwarf (spectral type M1.5V), located at a projected separation of only 1.9\,kau (Table~\ref{rrlyrae-bound-table}).

\subsubsection{CZ Lac}

\object{CZ Lac} is listed in the GDR2 as a fundamental mode RRL with a period of $P=0.432$\,d.
It was shown by \citetads{2011MNRAS.411.1585S} to exhibit an original multiperiodic Blazhko effect (see also \citeads{2013A&A...554A..46G}).
\object{CZ Lac} shows a strong PMa of $\Delta_\mathrm{G2}=3.9$, which indicates the probable presence of a close-in orbiting companion.
Its K1V resolved \texttt{Near} companion (Fig.~\ref{rrlyr-near}), \object{Gaia DR2 2000976545410515584} is located at an angular separation of $19.4\arcsec$, which corresponds to nearly 25\,kau (Table~\ref{rrlyrae-bound-table}). Its GDR2 parallax $\varpi_\mathrm{G2}=0.773 \pm 0.050$~mas is slightly smaller than that of the RRL $\varpi_\mathrm{G2}=0.852 \pm 0.029$~mas, but the latter may be affected by the detected PMa.
Considering its wide separation, the resolved companion is unlikely to be the cause of the observed PMa, and \object{CZ Lac} is therefore possibly a triple system.

\subsubsection{RR Leo}

\object{RR Leo} is listed in the  \citetads{2014MNRAS.445.1584S} catalog, which gives a mass estimate of $m=0.55 \pm 0.02\,M_\odot$.
It has been observed by the HST (\citeads{2013hst..prop13472F}: Prog ID: 13472, PI: Freedman), but unfortunately the field of view is too small to include the companion that we identified as a \texttt{Bound} candidate, \object{Gaia DR2 630421931138065280} (Fig.~\ref{rrlyr-field-1}).
The proper motion of the two stars is faster than 18\,mas\,a$^{-1}$, which excludes a chance association.
The companion is located at a separation of $9.9\arcsec$ from the RRL (9.3\,kau) and is extremely faint in the visible, with a contrast of $\Delta m_G = 7.1$ (Table~\ref{rrlyrae-bound-table}). The contrast in the infrared is reduced to $\Delta m_K=5.5$, indicating that the companion is a red dwarf of M0V spectral type.
\object{RR Leo} does not show a significant PMa and it is therefore likely a system of only two components.

\subsubsection{V0764 Mon}

The PMa that we detected on \object{V0764 Mon} is the second largest in our sample, at $\Delta_\mathrm{G2} = 7.1$ (Paper~I) and $\Delta_\mathrm{Hip}=3.8$.
It is classified as an RRc pulsator of period $P=0.290$\,d \citepads{2005A&A...442..381M}.
Considering the very strong signature on the PMa detected both at Hipparcos and GDR2 epochs, \object{V0764 Mon} is a certain binary star.
A spectroscopic monitoring of its $\gamma$-velocity should easily reveal its orbital motion.

\subsubsection{SS Oct}

\object{SS Oct} (\object{HIP 108057}) is a $P=0.622$\,day RRL showing the Blazhko effect with a period $P_\mathrm{Blazhko}=144.930$\,d \citepads{2013A&A...549A.101S}.
It is classified in the GDR2 as a fundamental mode pulsator. \object{SS Oct} is also listed in the  \citetads{2014MNRAS.445.1584S} catalog, which gives a mass estimate of $m=0.58 \pm 0.03\,M_\odot$.
Its close companion (Fig.~\ref{rrlyr-field-1}), \object{Gaia DR2 6345324695303800192}, is a red dwarf of spectral type K2V (Table~\ref{rrlyrae-bound-table}), 5.4 magnitudes fainter than the RRL in the visible ($m_G=17.19$).
Their common proper motion is very fast at more than 32\,mas\,a$^{-1}$, making a chance association impossible, and the projected physical separation between the two stars is 2.4\,kau.
SS Oct also presents a marginal PMa at $\Delta_\mathrm{G2}=2.2$, and the future Gaia data releases will confirm whether it has a close companion in addition to the resolved one.

\subsubsection{EY Oph}

\object{EY Oph} is a poorly studied variable, classified as an RR: type star. Its absolute magnitude is faint for an RRL, but may be compatible with this classification if, for example, the adopted color excess $E(B-V)$ is biased.
The candidate \texttt{Bound} companion of \object{EY Oph}, \object{Gaia DR2 6029835295648727168}, is located at a projected separation of 6.6\,kau and has a remarkably low absolute $G$ band magnitude of $\approx +10$. This absolute magnitude corresponds to a very low mass M3V star on the main sequence, but its blue $G_\mathrm{BP}-G_\mathrm{RP}$ color indicates that it is located significantly below the main sequence. This star could therefore be a white dwarf, but in absence of infrared magnitudes, its identification is uncertain.

\subsubsection{V0487 Sco}

\object{V0487 Sco} is a c-type RRL with a pulsation period of $P=0.329$\,d \citepads{2005A&A...442..381M} and a metallicity of $[\mathrm{Fe/H}]=-1.89$ \citepads{2007MNRAS.374.1421M}.
We detected a resolved companion (\object{Gaia DR2 4053551410541112192}) located at a wide projected separation of 28\,kau. It is only $0.6$\,mag fainter than the RRL in the $K$ band, but its $G$ band magnitude is fainter than the RRL by 4.7\,mag. It could be a dusty red giant star with a strong circumstellar absorption differing from that of the RRL.

\subsubsection{AR Ser \label{arser}}

\object{AR Ser} is a fundamental mode \citepads{2009AcA....59..137S} RRab type pulsator ($P=0.5752$\,d) showing the Blazhko effect \citepads{2013A&A...549A.101S,2015IBVS.6132....1B,2016A&A...592A.144S}; however, it is  not listed as a Blazhko star in the SuperWASP catalog  \citepads{2017A&A...607A..11G}.
We detected a strong PMa of $\Delta_\mathrm{G2} = 5.2$, indicative of the presence of an orbiting companion.
The modulation period $P_m = 1325 \pm 60$\,d proposed by \citetads{2016A&A...592A.144S} may correspond to the orbital period of the companion.
We did not detect a resolved common proper motion candidate in the GDR2 catalog.

\subsubsection{TU UMa \label{tuuma}}

\object{TU UMa} is the only RRL member of a binary system whose spectroscopic orbit is available \citepads{1986CoKon..89...57S, 1995IBVS.4205....1K, 1999AJ....118.2442W, 2016A&A...589A..94L}.
The results of the combined fit of its PMa and spectroscopic orbit are presented in Table~3 and Sect.~3.5 of Paper~I. We determined that its companion is likely a massive white dwarf ($m_2 \approx 2\,M_\odot$), orbiting with a semimajor axis of $\approx 11$\,au.
We did not detect a additional resolved companion of \object{TU UMa}.

\subsection{Other variables}

We briefly discuss here the results of our companion search on a selection of stars classified as CCs or RRLs, but that actually do not belong to these classes.

\subsubsection{HM Aql}

\object{HM Aql} is a poorly studied variable classified as an RRL \citepads{1958PZ.....12..277K,2006MNRAS.368.1757W,2014MNRAS.441..715G}, but whose period is unknown. Its absolute magnitude $M_G = +2.62$ appears to be too faint for an RRL (Table~\ref{various-bound-table}).
The resolved \texttt{Bound} candidate companion of \object{HM Aql} (Fig.~\ref{various-field1}) is a low mass dwarf of spectral type K7V located at a large projected separation of 14.7\,kau.

\subsubsection{EN CMi}

The absolute magnitude of \object{EN CMi} in the $G$ band ($M_G=5.34$) is too faint for an RRL, although the light curve presented in \citetads{2004AJ....127.1158V} is similar in shape to that of a $P=0.540$\,day RRab type pulsator.
It is located in a region of low extinction.
Assuming that the GDR2 parallax of \object{EN CMi} is correct, its resolved companion is of comparable brightness and spectral type K1.5V.

\subsubsection{NQ Cyg}

\object{NQ Cyg} is classified as an eclipsing binary by \citetads{2014AJ....147..119C}. Little information is present in the literature on this variable, but its absolute magnitude is too faint to be an RRL. Taichi Kato\footnote{\url{http://ooruri.kusastro.kyoto-u.ac.jp/mailarchive/vsnet-chat/8021}} proposed from its ASAS-SN light curve \citepads{2014ApJ...788...48S,2017PASP..129j4502K} that it is an R-type object (close eclipsing binary with strong reflection lighting effects) with a period $P=0.311592$\,d.

We identified a low mass \texttt{Bound} candidate companion (Fig.~\ref{various-field1}) of spectral type K9V located at a projected separation of 19.4\,kau (Table~\ref{various-bound-table}). Due to their fast proper motion ($\mu > 12$~mas\,a$^{-1}$) and almost perfect PM vector alignment (within $0.35^\circ$), the probability of a chance association is very small.

\subsubsection{V1391 Cyg}

\object{V1391 Cyg} is classified as an eclipsing binary in the catalog by \citetads{2014AJ....147..119C}.
We detected a resolved \texttt{Bound} candidate companion at a projected separation of 19\,kau (Fig.~\ref{various-field1}), with a probable spectral type of M1.5V (Table~\ref{various-bound-table}) based on its $G$-band absolute magnitude. We note,  however, that the Gaia colors $G_\mathrm{BP}$ and $G_\mathrm{RP}$ are not mutually consistent with the $G$ magnitude and are possibly biased.

\subsubsection{V2121 Cyg}

\object{V2121 Cyg} (\object{43 Cyg}, \object{HD 195069}) was identified as a possible RRL from Hipparcos photometry \citepads{1997A&A...323L..49P}.
However, \citetads{2005AJ....129.2815H} established that \object{V2121 Cyg}  is one of the brightest $\gamma$\,Dor non-radial pulsators ($m_G = 5.63$).
\citetads{2003AJ....125.2196F} proposed that it may be a spectroscopic binary star.
It was recently studied in detail by \citetads{2017A&A...608A.103Z} using photometric time series from the BRITE-Constellation nano-satellites.
This is a main sequence F-type star with an effective temperature of around 7000\,K.
We detected a gravitationally bound companion to \object{V2121 Cyg}, \object{Gaia DR2 2084032103278208640} (Fig.~\ref{various-field1}), approximately 6.6 magnitudes fainter than the primary at visible wavelengths ($m_G = 12.27$).
The large parallax ($\varpi \approx 25.6$~mas) and very fast proper motion of the two stars ($\mu \approx 90$~mas\,a$^{-1}$) excludes a chance association.
The companion has an effective temperature $T_\mathrm{eff} = 3700 \pm 100$\,K and a radius $R = 0.50 \pm 0.02\,R_\odot$, which  correspond to a M2V spectral type (Table~\ref{various-bound-table}).
It exhibits a significant X-ray emission (source \object{1RXS J202704.9+492216}; \citeads{2009ApJS..184..138H}, see also \citeads{2016A&A...588A.103B}).

\subsubsection{UU Dor}

\object{UU Dor} is identified as an LMC detached eclipsing binary in the OGLE-III catalog \citepads{2011AcA....61..103G}.
However, its GDR2 parallax implies that it is a Galactic variable, of combined spectral type F6V.
Its resolved \texttt{Bound} candidate companion (Fig.~\ref{various-field1}) is a low mass dwarf of spectral type K3V (Table~\ref{various-bound-table}).

\subsubsection{IW Lib}

\object{IW Lib} is listed in the  \citetads{2012MNRAS.427..343M} catalog, where it does not show infrared excess.
It is classified by \citetads{2002AcA....52..397P} as an eclipsing contact binary with an orbital period of 1.78\,day.
It has been automatically classified as a $\gamma$\,Dor non-radial pulsator by \citetads{2011MNRAS.414.2602D} and more probably as a W UMa eclipsing binary (with a probability of 0.29) by \citetads{2012ApJS..203...32R}, who mostly exclude the possibility of an RRL.
The spectral type of the wide companion is K3V (Table~\ref{various-bound-table}), with an effective temperature $T_\mathrm{eff} = 4800 \pm 200$\,K and a radius $R = 0.71 \pm 0.03\,R_\odot$.
It is 5.3\,mag fainter than the primary in the $G$ band.

\subsubsection{AZ Men}

\object{AZ Men} is classified as an RRc pulsator with a period of 0.318\,day \citepads{2008OEJV...93....1O}; however, its absolute magnitude $M_G = +3.7$ is inconsistent with this class.
Its resolved companion (Fig.~\ref{various-field2}) has an effective temperature $T_\mathrm{eff} = 4900 \pm 200$\,K and a radius $R = 0.63 \pm 0.04\,R_\odot$.
These parameters were derived using the surface brightness--color relations from \citetads{2004A&A...426..297K} and the GDR2 parallax, and match those of a K3V dwarf (Table~\ref{various-bound-table}).

\subsubsection{V1171 Oph}

This star is likely a contact eclipsing binary, of combined spectral type F8V with a mean effective temperature of $T_\mathrm{eff} \approx 6200$\,K. Its resolved companion (Fig.~\ref{various-field2}) is a red dwarf of spectral type K1V.

\subsubsection{V1330 Sgr}

Although its absolute $G$ magnitude is inconsistent with a classification as an RRL ($M_G = +6.9$), \object{V1330 Sgr} is listed as an RRab pulsator with $P=0.597$\,d by \citetads{2011AcA....61....1S}, showing the Blazhko effect.
It is located in the direction of the Galactic Bulge, close to Baade's field \citepads{1973A&A....26..317P}, but its GDR2 parallax ($\varpi_\mathrm{G2} = 1.23 \pm 0.22$~mas) places it at a significantly shorter distance.
Assuming the GDR2 parallax is correct, its resolved companion (Fig.~\ref{various-field2}) is an M0V red dwarf, located at a projected physical separation of 24.7\,kau.

\subsubsection{V1382 Sgr}

\citetads{2014AcA....64..177S} classified \object{V1382 Sgr} as an RRab type pulsator with a period of $P=0.493$\,d with Blazhko effect \citepads{2006ApJ...651..197C}, located in the direction of the Bulge.
However, this is in principle excluded by its very faint absolute magnitude $M_G = +7.3$, unless the GDR2 parallax is incorrect.
Assuming the GDR2 parallax is correct, its resolved common proper motion companion is an M2V red dwarf (Fig.~\ref{various-field2}), approximately 2\,mag fainter than the primary in the $G$ band, and located at a projected physical separation of 11.6\,kau (Table~\ref{various-bound-table}).
It should be noted that the crowding in this region is high and could have caused a bias in the GDR2 parallax estimate.

\subsubsection{V2248 Sgr\label{v2248sgr}}

\object{V2248 Sgr} is classified as a W UMa eclipsing binary by \citetads{2005IBVS.5613....1A}, but it is classified as a DSCT/EC/ESD in the ASAS-3 database \citepads{2002AcA....52..397P} and an RR Lyrae pulsator in the GCVS \citepads{2009yCat....102025S}. Its period is $P=0.315$\,d in the GCVS and 0.158\,d (half of the former) in the ASAS-3 database.
The wide companion (Fig.~\ref{various-field2}) has an effective temperature $T_\mathrm{eff} = 4700 \pm 200$\,K and a radius $R = 0.73 \pm 0.03\,R_\odot$.
These properties correspond to a K3.5V spectral type.

\subsubsection{V3166 Sgr}

The field around \object{V3166 Sgr} shows two candidate bound companions, making it a visual triple star.
The classification as as RRL of this variable star in the GCVS is unlikely to be correct considering its absolute $G$-band magnitude (Table~\ref{various-bound-table}).
The close companion (at a separation of $1.66\arcsec$), considering its absolute magnitude and assuming it is on the main sequence, is most probably a late K-type dwarf.
The more distant companion has $T_\mathrm{eff} = 3500 \pm 100$\,K and $R = 0.79 \pm 0.03 \,R_\odot$, which possibly corresponds to an unusual post-main sequence, ``inflated'' M1 spectral type dwarf.
It could result from the evolution of a $\approx 0.6\,M_\odot$ star, of mid-K spectral type on the main sequence, at an age of $\approx 10$\,Ga \citepads{2012MNRAS.427..127B}.

\subsubsection{HR Sco}

With an absolute $G$ band magnitude of $M_G = +10.2$, \object{HR Sco} is likely a very low mass main sequence star of spectral type M3V.
The candidate resolved companion of \object{HR Sco} (Fig.~\ref{various-field3}) is located at a projected separation of 20\,kau, for a spectral type of M5V (Table~\ref{various-bound-table}).

\subsubsection{EN TrA}

\object{EN TrA} is a binary RV Tauri type variable \citepads{1999A&A...343..202V}. We detected a strong PMa ($\Delta_\mathrm{G2} = 11.7$; Paper~I) that confirms the presence of an orbiting companion.
We did not detect any resolved common proper motion candidate in the GDR2.

\section{Conclusion\label{conclusion}}

We detected a number of new candidate companions of Cepheids and RR Lyrae stars, either from their signature on the proper motion of the targets (Paper~I) or the similarity of their parallax and proper motion to those of the targets (present Paper~II).

Classical Cepheids have long been known to have a high binarity fraction, and our survey of the PM anomalies of bright and nearby CCs indicates that their binarity fraction is likely higher than 80\% (Paper~I), with a significant fraction of triple (e.g., SY Nor, AW Per, W Sgr, V0350 Sgr) or quadruple systems (e.g., $\delta$\,Cep, SY Nor).
The resolved systems that we discovered or confirmed in the present paper (Paper~II) provide interesting fiducial references for the modeling of the evolutionary state of the corresponding CC primaries. They also enable the validation of the GDR2 parallaxes of their CC companions that may be affected by a specific uncertainty due to their photometric and color variability.

The gravitationally bound candidates of RRLs that we present in Table~\ref{rrlyrae-bound-table} are all new discoveries, and provide important constraints on the evolutionary state of their pulsating companions. In Paper~I we find that a small but significant fraction ($\approx 7\%$) of the nearby RRLs shows indications of binarity. The relatively large number of stars in our sample that were misclassified as RRLs is a call for a thorough revision of the RRL catalogs using the Gaia data, and in particular the parallax, to reclassify the other types of variables.

\begin{acknowledgements}
We thank the organizers of the MIAPP workshop ``The extragalactic distance scale in the Gaia era'' held on 11 June--6 July 2018 in Garching (Germany); it  provided a place for  very useful exchanges that enhanced the work presented in the present paper.
We thank Dr. Richard I. Anderson for the interesting discussion on Polaris and $\delta$\,Cep that led to improvements in this paper.
This work has made use of data from the European Space Agency (ESA) mission {\it Gaia} (\url{http://www.cosmos.esa.int/gaia}), processed by the {\it Gaia} Data Processing and Analysis Consortium (DPAC, \url{http://www.cosmos.esa.int/web/gaia/dpac/consortium}).
Funding for the DPAC has been provided by national institutions, in particular the institutions participating in the {\it Gaia} Multilateral Agreement.
The authors acknowledge the support of the French Agence Nationale de la Recherche (ANR), under grant ANR-15-CE31-0012-01 (project UnlockCepheids).
The research leading to these results  has received funding from the European Research Council (ERC) under the European Union's Horizon 2020 research and innovation program (grant agreement No. 695099).
Support was provided to NRE by the Chandra X-ray Center NASA Contract NAS8-03060.
This study was funded by the NKFIH K-115709 grant of the Hungarian National Research and Innovation Office.
W.G. and G.P. gratefully acknowledge financial support for this work from the BASAL Centro de Astrofisica y Tecnologias Afines (CATA) AFB-170002. 
W.G. acknowledges financial support from the Millenium Institute of Astrophysics (MAS) of the Iniciativa Cientifica Milenio del Ministerio de Economia, Fomento y Turismo de Chile, project IC120009.
We acknowledge support from the IdP II 2015 0002 64 grant of the Polish Ministry of Science and Higher Education.
We acknowledge the past financial support to this research program of the ``Programme National de Physique Stellaire'' (PNPS) of CNRS/INSU, France.
This research has made use of Astropy\footnote{Available at \url{http://www.astropy.org/}}, a community-developed core Python package for Astronomy \citepads{2013A&A...558A..33A,2018AJ....156..123A}.
This research has made use of the International Variable Star Index (VSX) database, operated at AAVSO, Cambridge, Massachusetts, USA.
This research was supported by the Munich Institute for Astro- and Particle Physics (MIAPP) of the DFG cluster of excellence ``Origin and Structure of the Universe''.
We used the SIMBAD and VIZIER databases and catalog access tool at the CDS, Strasbourg (France), and  NASA's Astrophysics Data System Bibliographic Services.
The original description of the VizieR service was published in \citetads{2000A&AS..143...23O}.
The Digitized Sky Surveys were produced at the Space Telescope Science Institute under U.S. Government grant NAG W-2166. The images of these surveys are based on photographic data obtained using the Oschin Schmidt Telescope on Palomar Mountain and the UK Schmidt Telescope. The plates were processed into the present compressed digital form with the permission of these institutions.
This research has made use of the Washington Double Star Catalog maintained at the U.S. Naval Observatory.
This publication makes use of data products from the Two Micron All Sky Survey, which is a joint project of the University of Massachusetts and the Infrared Processing and Analysis Center/California Institute of Technology, funded by the National Aeronautics and Space Administration and the National Science Foundation.
\end{acknowledgements}

\bibliographystyle{aa} 
\bibliography{biblioCepheids}

\begin{appendix}

\section{Candidate companions of classical Cepheids \label{cep-tables-appendix}}

\subsection{Tables of candidate companions}

\begin{sidewaystable*}
 \caption{Candidate gravitationally bound companions of Cepheids. The photometry is taken from the GDR2 for the $G_\mathrm{BP}$, $G$ and $G_\mathrm{RP}$ bands \citepads{2018A&A...616A...4E}, from 2MASS for the near infrared $JHK$ bands \citepads{2006AJ....131.1163S} and from WISE for the thermal infrared $W1$, $W2$, $W3$ and $W4$ bands \citepads{2010AJ....140.1868W,2012yCat.2311....0C}.
 $E_\mathrm{BV}$ is the $E(B-V)$ color excess. The negative GDR2 parallaxes of R\,Cru ($-1.094_{0.219}$\,mas) and $\delta$\,Cep ($-1.143_{0.468}$\,mas) have been omitted. Due to their angular proximity, the listed infrared magnitudes of CE Cas AB, V1046 Cyg and S Nor correspond to their combined flux. }
 \label{cepheids-bound-table1}
 \centering
\tiny
 \renewcommand{\arraystretch}{1}
 \setlength\tabcolsep{4.0pt}
 \begin{tabular}{lrrcrrrrrrrrrrrrrrcl}
 \hline
 \hline \noalign{\smallskip}
Cepheid / GDR2 & $\varpi_\mathrm{exp}$ & $\varpi_\mathrm{G2}$ & $\mu_{\alpha,\mathrm{G2}},\,\mu_{\delta,\mathrm{G2}}$ & Sep. & Sep. & $G_\mathrm{BP}$ & $G$ & $G_\mathrm{RP}$ & $J$ & $H$ & $K$ & $W1$ & $W2$ & $W3$ & $W4$ & $E_\mathrm{BV}$ & $M_G$ & $\log T$ & Spectral \\
 & (mas) & (mas) & (mas\,a$^{-1}$) & ($\arcsec$) & (kau) & & & & & & & & & & & & & & type \\
\hline \noalign{\smallskip}
\textbf{\object{TV CMa}} & 0.477 & $ 0.343_{ 0.034}$ & $ -0.69_{ 0.06}$,$ +0.22_{ 0.05}$ & & & 11.03 & 10.24 & 9.34 & 8.16 & 7.74 & 7.48 & 7.33 & 7.34 & 7.32 & 7.11 & 0.583$^a$ & $ -3.48$ & 3.757 & F5 \\
 3044483895574944512 & & $ 0.455_{ 0.050}$ & $ -0.50_{ 0.08}$,$ +0.05_{ 0.07}$ & 6.43 & 13.5 & 16.14 & 15.77 & 14.95 & & & & & & & & & $ +2.51$ & 3.862 & F0V \\
\hline \noalign{\smallskip}
\textbf{\object{ER Car}} & 0.959 & $ 0.825_{ 0.035}$ & $ -9.72_{ 0.06}$,$ +2.48_{ 0.05}$ & & & 7.07 & 6.60 & 5.99 & 5.26 & 4.97 & 4.81 & 4.80 & 4.57 & 4.83 & 4.78 & 0.101$^a$ & $ -4.08$ & 3.749 & F5-F8Ib \\
 5339394048386734336 & & $ 0.918_{ 0.152}$ & $-10.30_{ 0.33}$,$ +3.30_{ 0.25}$ & 35.39 & 36.9 & 18.85 & 18.43 & 17.48 & & & & & & & & & $ +8.00$ & 3.678 & M0V \\
\hline \noalign{\smallskip}
\textbf{\object{CE Cas B}} & 0.343 & $ 0.262_{ 0.030}$ & $ -3.24_{ 0.05}$,$ -2.00_{ 0.06}$ & & & 11.23 & 10.65 & 9.67 & 7.78 & 7.37 & 7.17 & 6.57 & 7.00 & 6.91 & 6.68 & 0.584$^a$ & $ -3.70$ & 3.776 & G1Iab/b \\
 2011892325047232256 & & $ 0.317_{ 0.031}$ & $ -3.32_{ 0.05}$,$ -1.71_{ 0.06}$ & 2.47 & 7.2 & 11.05 & 10.51 & 9.59 &  &  &  &  &  &  &  & & $ -3.46$ & 3.791 & F8I (Ceph.) \\
\hline \noalign{\smallskip}
\textbf{\object{DF Cas}} & 0.462 & $ 0.336_{ 0.028}$ & $ -0.55_{ 0.04}$,$ -0.25_{ 0.06}$ & & & 11.15 & 10.45 & 9.65 & 8.52 & 8.11 & 7.96 & 7.75 & 7.75 & 7.71 & 7.76 & 0.595$^a$ & $ -3.40$ & 3.788 & F6-G4:\\
 465719182408531072 & & $ 0.396_{ 0.080}$ & $ -0.85_{ 0.19}$,$ -0.77_{ 0.18}$ & 16.31 & 35.3 & 18.26 & 17.25 & 16.23 & 14.66 & 13.91 & 13.74 & & & & & & $ +3.90$ & 3.676 & K3IV \\
\hline \noalign{\smallskip}
\textbf{\object{V0659 Cen}} & 1.287 & $ 0.513_{ 0.154}$ & $ -6.00_{ 0.17}$,$ -1.76_{ 0.21}$ & & & 6.90 & 6.47 & 5.86 & 5.22 & 4.94 & 4.68 & 4.57 & 4.33 & 4.56 & 4.41 & 0.134$^e$ & $ -5.32$ & 3.762 & F6/7Ib+B6V \\
 5868451109212716928 & & $ 1.384_{ 0.318}$ & $ -6.14_{ 0.50}$,$ -2.56_{ 0.61}$ & 62.22 & 48.3 & 20.54 & 19.68 & 18.04 & & & & & & & & & $+10.12$ & 2.540 & M3V \\
\hline \noalign{\smallskip}
\textbf{\object{delta Cep}} & 3.755 & $3.71_\mathrm{ 0.12}^{\dagger}$ & $+17.64_{ 0.81}$,$ +3.97_{ 0.73}$ & & & &  & & 2.87 & 2.61 & 2.35 & 0.63 & 0.51 & 2.28 & 2.13 & 0.032$^a$ & $-$ & $-$ & F5Iab: \\
 2200153214212849024 & & $ 3.393_{ 0.049}$ & $+14.09_{ 0.09}$,$ +3.79_{ 0.09}$ & 40.74 & 10.8 & 6.29 & 6.31 & 6.31 & 6.27 & 6.36 & 6.35 & 5.98 & 6.25 & 6.34 & 5.62 & & $ -1.13$ & 4.052 & B7-B8 III-IV+F0V$^g$ \\
\hline \noalign{\smallskip}
\textbf{\object{AX Cir}} & 1.917 & $ 1.774_{ 0.345}$ & $ -4.89_{ 0.44}$,$ -5.08_{ 0.52}$ & & & 6.26 & 5.73 & 5.11 & 4.25 & 3.85 & 3.76 & 3.75 &  3.38 &  3.70 &  3.60 & 0.153$^a$ & $ -3.41$ & 3.749 & F2-G2II+B4 \\
 5874031027625742848 & & $ 1.754_{ 0.374}$ & $ -5.33_{ 0.40}$,$ -4.61_{ 0.70}$ & 81.49 & 42.5 & 20.06 & 19.81 & 18.42 & & & & & & & & & $+10.68$ & 3.586 & M3.5V \\
\hline \noalign{\smallskip}
\textbf{\object{BP Cir}} & 1.700 & $ 1.024_{ 0.040}$ & $ -5.46_{ 0.05}$,$ -3.89_{ 0.07}$ & & & 7.74 & 7.31 & 6.70 & 5.84 & 5.58 & 5.44 & 5.39 & 5.21 & 5.39 & 5.30 & 0.235$^c$ & $ -3.25$ & 3.782 & F2/3II+B6V \\
 5877472464676660480 & & $ 1.525_{ 0.188}$ & $ -5.39_{ 0.31}$,$ -3.61_{ 0.39}$ & 66.30 & 39.0 & 19.90 & 18.90 & 17.77 & & & & & & & & & $ +9.32$ & 3.253 & M2V \\
\hline \noalign{\smallskip}
\textbf{\object{R Cru}} & 1.170 &  & $ -9.47_{ 0.34}$,$ -0.33_{ 0.31}$ & & & &  & & 5.35 & 5.04 & 4.90 & 4.68 & 4.43 & 4.70 & 4.63 & 0.192$^d$ & $-$ & $-$ & F6-G2Ib-II$^j$ \\
 6054935874780531328 & & $ 1.107_{ 0.089}$ & $-11.10_{ 0.28}$,$ -0.04_{ 0.13}$ & 7.70 & 6.6 & 15.88 & 15.69 & 14.57 & & & & & & & & & $ +5.44$ & 3.730 & G8V \\
\hline \noalign{\smallskip}
\textbf{\object{X Cru}} & 0.678 & $ 0.552_{ 0.046}$ & $ -5.89_{ 0.06}$,$ -0.17_{ 0.06}$ & & & 8.69 & 8.14 & 7.41 & 6.58 & 6.14 & 5.95 & 5.88 & 5.82 & 5.86 & 5.76 & 0.286$^a$ & $ -3.86$ & 3.758 & G0/2Ib \\
 6059762524642419968 & & $ 0.638_{ 0.053}$ & $ -6.03_{ 0.08}$,$ -0.20_{ 0.07}$ & 27.23 & 40.2 & 16.48 & 16.04 & 15.36 & 14.58 & 14.14 & 14.13 & & & & & & $ +4.33$ & 3.781 & G1V \\
\hline \noalign{\smallskip}
\textbf{\object{VW Cru}} & 0.710 & $ 0.812_{ 0.045}$ & $ -3.96_{ 0.06}$,$ -1.19_{ 0.06}$ & & & 9.98 & 9.09 & 8.13 & 6.92 & 6.35 & 6.13 & 5.95 & 5.82 & 5.88 & 5.74 & 0.675$^a$ & $ -2.95$ & 3.751 & F6II\\
 6053622508133367680 & & $ 0.708_{ 0.040}$ & $ -3.94_{ 0.04}$,$ -1.13_{ 0.04}$ & 20.43 & 28.8 & 14.58 & 14.08 & 13.40 & 12.47 & 12.22 & 12.07 & & & & & & $ +1.50$ & 3.924 & A2V\\
\hline \noalign{\smallskip}
\textbf{\object{V0532 Cyg}} & 0.727 & $ 0.590_{ 0.032}$ & $ -3.40_{ 0.06}$,$ -3.65_{ 0.06}$ & & & 9.39 & 8.73 & 7.94 & 6.86 & 6.39 & 6.25 & 6.16 & 6.04 & 6.13 & 6.04 & 0.534$^a$ & $ -3.75$ & 3.784 & F8Ib\\
 1971721839529622272 & & $ 0.648_{ 0.040}$ & $ -3.39_{ 0.05}$,$ -3.56_{ 0.05}$ & 21.18 & 29.1 & 15.11 & 14.67 & 14.07 & 13.23 & 12.97 & 12.79 & & & & & & $ +2.28$ & 3.890 & F0V\\
\hline \noalign{\smallskip}
\textbf{\object{V1046 Cyg}} & 0.372 & $ 0.296_{ 0.029}$ & $ -2.98_{ 0.05}$,$ -4.28_{ 0.05}$ & & & 12.71 & 11.65 & 10.57 & 8.98 & 8.34 & 8.10 & 7.79 & 7.77 & 7.78 & 7.81 & 1.143$^e$ & $ -3.72$ & 3.808 & F6Ib\\
 2060460708575795712 & & $ 0.356_{ 0.060}$ & $ -3.34_{ 0.16}$,$ -3.88_{ 0.10}$ & 2.48 & 6.7 & 16.19 & 15.78 & 14.57 &  &  &  &  &  &  &  & & $ +0.49$ & 4.126 & B8V\\
\hline \noalign{\smallskip}
\textbf{\object{CV Mon}} & 0.601 & $ 0.511_{ 0.041}$ & $ +0.43_{ 0.07}$,$ -0.69_{ 0.07}$ & & & 10.49 & 9.62 & 8.71 & 7.37 & 6.77 & 6.52 & 6.46 & 6.31 & 6.45 & 6.37 & 0.722$^b$ & $ -3.57$ & 3.772 & G2Ib/II \\
 3127142224816361600 & & $ 0.567_{ 0.044}$ & $ +0.47_{ 0.06}$,$ -0.84_{ 0.05}$ & 11.07 & 18.4 & 13.96 & 13.57 & 12.97 & 12.14 & 11.91 & 11.73 & & & & & & $ +0.30$ & 4.137 & B8V \\
 3127142327895572352 & & $ 0.537_{0.040}$ & $+0.46_{0.05}, -0.62_{0.05}$ & 14.20 & 23.6 & 13.91 & 13.49 & 12.88 & 12.10 & 11.86 & 11.71 & 11.05 & 11.03 & 10.33 & 8.70 & & $ +0.12$ & 4.095 & B8V\\
\hline \noalign{\smallskip}
\textbf{\object{RS Nor}} & 0.487 & $ 0.450_{ 0.046}$ & $ -2.21_{ 0.07}$,$ -3.45_{ 0.06}$ & & & 10.45 & 9.59 & 8.72 & 7.23 & 6.80 & 6.60 & 6.41 & 6.47 & 6.45 & 6.41 & 0.580$^a$ & $ -3.52$ & 3.750 & F6 \\
 5932812740361508736 & & $ 0.478_{ 0.044}$ & $ -2.48_{ 0.06}$,$ -3.37_{ 0.05}$ & 17.04 & 35.0 & 14.90 & 14.55 & 13.93 & 13.06 & 12.92 & 12.74 & & & & & & $ +1.34$ & 3.969 & A1V \\
\hline \noalign{\smallskip}
\textbf{\object{SY Nor}} & 0.429 & $ 0.429_{ 0.035}$ & $ -1.37_{ 0.06}$,$ -2.10_{ 0.05}$ & & & 9.94 & 9.05 & 8.11 & 6.69 & 6.09 & 5.82 & 5.50 & 5.49 & 5.56 & 5.38 & 0.794$^a$ & $ -4.70$ & 3.779 & F9Ib+B4.5V \\
 5884729035245399424 & & $ 0.443_{ 0.053}$ & $ -2.73_{ 0.16}$,$ -2.45_{ 0.11}$ & 2.46 & 5.7 & 12.33 & 12.11 & 11.45 &  &  &  &  &  &  &  & & $ -1.96$ & 4.418 & B2V \\
 5884729035255068800 & & $ 0.438_{ 0.066}$ & $ -2.65_{ 0.14}$,$ -2.59_{ 0.11}$ & 18.03 & 42.0 & 18.05 & 17.04 & 15.98 & 14.35 & 13.92 & 13.71 & & & & & & $ +3.42$ & 3.739 & F6V \\
\hline
\end{tabular}
\tablefoot{
$^\dagger$ The expected parallax $\varpi_\mathrm{exp}$ of $\delta$\,Cep is the mean value between HST \citepads{2002AJ....124.1695B} and Hipparcos \citepads{2007MNRAS.379..723V}.
}
\tablebib{
(a): \citetads{1990ApJS...72..153F};
(b): \citetads{2013A&A...550A..70G};
(c): \citetads{2015A&A...584A..80M};
(d): \citetads{2003A&A...404..423T};
(e): \citetads{1994ApJ...429..844F};
(f): \citetads{2007A&A...476...73F};
(g): \citetads{2002AJ....124.1695B};
(h): \citetads{2008MNRAS.387L...1U};
(i): \citetads{2011AJ....142...51L};
(j): VSX database \citepads{2006SASS...25...47W};
(k): \citetads{2017A&A...600A.127K}.
}
\end{sidewaystable*}

\begin{sidewaystable*}
 \caption{Continued from Table~\ref{cepheids-bound-table1}, where the references are given.}
 \label{cepheids-bound-table2}
  \centering
\tiny
 \renewcommand{\arraystretch}{1}
 \setlength\tabcolsep{4.5pt}
  \begin{tabular}{lrrcrrrrrrrrrrrrrrcl}
 \hline
 \hline \noalign{\smallskip}
Cepheid / GDR2 & $\varpi_\mathrm{exp}$ & $\varpi_\mathrm{G2}$ & $\mu_{\alpha,\mathrm{G2}},\,\mu_{\delta,\mathrm{G2}}$ & Sep. & Sep. & $G_\mathrm{BP}$ & $G$ & $G_\mathrm{RP}$ & $J$ & $H$ & $K$ & $W1$ & $W2$ & $W3$ & $W4$ & $E_\mathrm{BV}$ & $M_G$ & $\log T$ & Spectral \\
 & (mas) & (mas) & (mas\,a$^{-1}$) & ($\arcsec$) & (kau) & & & & & & & & & & & & & & type \\
\hline \noalign{\smallskip}
\textbf{\object{QZ Nor}} & 0.556 & $ 0.503_{ 0.038}$ & $ -1.84_{ 0.07}$,$ -3.79_{ 0.06}$ & & & 9.14 & 8.64 & 7.96 & 7.00 & 6.70 & 6.54 & 6.48 & 6.47 & 6.49 & 6.41 & 0.253$^f$ & $ -3.49$ & 3.766 & F6Ib/II \\
 5932565899990412672 & & $ 0.481_{ 0.098}$ & $ -1.68_{ 0.20}$,$ -3.34_{ 0.16}$ & 16.55 & 29.8 & 18.13 & 17.92 & 16.84 & & & & & & & & & $ +5.71$ & 3.750 & K1V \\
\hline \noalign{\smallskip}
\textbf{\object{AW Per}} & 1.218 & $ 1.071_{ 0.064}$ & $ +0.29_{ 0.11}$,$ -1.30_{ 0.07}$ & & & 8.57 & 7.05 & 6.60 & 5.10 & 4.79 & 4.63 & 4.55 & 4.30 & 4.51 & 4.43 & 0.489$^f$ & $ -3.90$ & 3.640 & F8I+B6: \\
 174489098011144960 & & $ 1.075_{ 0.249}$ & $ -0.81_{ 0.32}$,$ -3.14_{ 0.40}$ & 10.27 & 8.4 & 17.83 & 17.41 & 16.28 & & & & & & & & & $ +6.37$ & 3.761 & K3.5V \\
\hline \noalign{\smallskip}
\textbf{\object{RS Pup}} & 0.536 & $0.613_{0.026}$ & $-3.55_{0.01}$,$+3.08_{0.02}$ & & & 7.64 & 6.63 & 5.75  & 4.94 & 4.16 & 4.02 &  & 3.374 & 3.438 & 3.246 & 0.496$^k$ & $-5.56$ & 3.681 & F8Iab \\
5546476755539995008 & & $0.532_{0.048}$ & $-3.61_{0.07}$,$+6.22_{0.08}$ & 43.33 & 80.8 & 16.71 & 16.24 & 15.43 & 14.54 & 14.22 & 14.04 & & & & & & $+3.59$ & 3.803 & F6V$^\star$ \\
\hline \noalign{\smallskip}
\textbf{\object{U Sgr}} & 1.669 & $ 1.489_{ 0.045}$ & $ -1.68_{ 0.08}$,$ -6.13_{ 0.07}$ & & & 7.28 & 6.53 & 5.75 & 4.48 & 3.91 & 3.93 & 3.95 & 3.65 & 3.95 & 3.86 & 0.403$^a$ & $ -3.58$ & 3.742 & G1Ib/II \\
 4092905066374437760 & & $ 1.534_{ 0.074}$ & $ -1.75_{ 0.11}$,$ -5.96_{ 0.09}$ & 71.99 & 43.1 & 10.41 & 10.16 & 9.74 & 9.28 & 9.20 & 9.10 & 9.04 & 9.11 & 8.76 & 7.72 & & $ -0.06$ & 3.989 & A0IV-V \\
 4092905203841177856 & & $ 1.490_{ 0.038}$ & $ -1.45_{ 0.07}$,$ -6.18_{ 0.07}$ & 76.96 & 46.1 & 11.40 & 11.15 & 10.73 & 10.23 & 10.09 & 10.02 & 9.86 & 9.90 & 10.10 & 8.75 & & $ +0.87$ & 3.990 & A0V \\
\hline \noalign{\smallskip}
\textbf{\object{V0350 Sgr}} & 1.141 & $ 1.015_{ 0.047}$ & $ -0.28_{ 0.09}$,$ -3.11_{ 0.08}$ & & & 7.86 & 7.30 & 6.59 & 5.69 & 5.27 & 5.12 & 5.07 & 4.91 & 4.99 & 4.92 & 0.299$^f$ & $ -3.42$ & 3.763 & F8Ib/II \\
 4080121319521641344 & & $ 1.044_{ 0.048}$ & $ -1.21_{ 0.08}$,$ -3.62_{ 0.07}$ & 29.85 & 26.2 & 12.45 & 12.28 & 11.94 & 11.54 & 11.43 & 11.35 & 11.05 & 11.00 & 10.96 & 8.36 & & $ +1.50$ & 3.989 & A2V \\
 4080121521343969024 & & $ 1.139_{ 0.145}$ & $ -1.64_{ 0.32}$,$ -3.77_{ 0.30}$ & 34.92 & 30.6 & 17.67 & 17.02 & 16.20 & 15.28 & 14.61 & 14.64 & & & & & & $ +6.58$ & 3.727 & K4.5V \\
\hline \noalign{\smallskip}
\textbf{\object{V0950 Sco}} & 1.073 & $ 0.869_{ 0.052}$ & $ -0.48_{ 0.10}$,$ -1.66_{ 0.09}$ & & & 7.49 & 7.06 & 6.44 & 5.72 & 5.44 & 5.24 & 5.18 & 5.03 & 5.17 & 5.09 & 0.254$^j$ & $ -3.91$ & 3.784 & F6Ib/II \\
 5960623340819000192 & & $ 0.922_{ 0.058}$ & $ -0.45_{ 0.10}$,$ -1.88_{ 0.09}$ & 16.07 & 15.0 & 15.53 & 15.28 & 14.50 & 12.81 & 13.35 & 13.18 & & & & & & $ +4.44$ & 3.786 & G1V \\
\hline \noalign{\smallskip}
\textbf{\object{CM Sct}} & 0.518 & $ 0.405_{ 0.065}$ & $ -1.01_{ 0.10}$,$ -1.41_{ 0.09}$ & & & 11.40 & 10.51 & 9.59 & 8.22 & 7.75 & 7.49 & 7.40 & 7.43 & 7.13 & 5.92 & 0.795$^a$ & $ -3.37$ & 3.783 & F8Ib/II \\
 4253603428053877504 & & $ 0.547_{ 0.050}$ & $ -1.48_{ 0.08}$,$ -1.56_{ 0.08}$ & 25.40 & 49.0 & 15.38 & 14.73 & 13.90 & 12.83 & 12.46 & 11.56 & 11.18 & 11.39 & 9.09 & 6.77 & & $ +1.36$ & 3.863 & A1V \\
\hline \noalign{\smallskip}
\textbf{\object{EV Sct}} & 0.556 & $ 0.526_{ 0.054}$ & $ -0.22_{ 0.09}$,$ -2.70_{ 0.08}$ & & & 10.45 & 9.66 & 8.82 & 7.55 & 7.13 & 6.96 & 6.42 & 6.78 & 6.71 & 6.50 & 0.679$^a$ & $ -3.41$ & 3.786 & G0II \\
 4156513016572003840 & & $ 0.510_{ 0.043}$ & $ -0.04_{ 0.06}$,$ -2.66_{ 0.06}$ & 24.71 & 44.4 & 13.96 & 13.62 & 13.08 & 12.36 & 11.22 & 12.00 & & & & & & $ +0.21$ & 4.189 & B9V \\
\hline \noalign{\smallskip}
\textbf{\object{Polaris}} ($\alpha$ UMi) & 7.540$^{\ddagger}$ & & $+44.48_{0.11}^{\ddagger}$,$-11.85_{0.13}^{\ddagger}$ & & & &  & & 0.80 & 0.46 & 0.46 & -1.94 & -1.47 & 0.58 & 0.57 & 0.003$^f$ & $-$ & $-$ & F7Ib-II: \\
 576402619921510144 & & $ 7.321_{ 0.028}$ & $+41.95_{ 0.06}$,$-13.67_{ 0.06}$ & 17.67 & 2.3 & 8.82 & 8.64 & 8.32 & & & & & & & & & $ +2.95$ & 3.810 & F3V$^h$ \\
\hline \noalign{\smallskip}
\textbf{\object{SX Vel}} & 0.490 & $ 0.438_{ 0.041}$ & $ -4.50_{ 0.07}$,$ +4.82_{ 0.07}$ & & & 8.66 & 8.12 & 7.42 & 6.54 & 6.11 & 5.95 & 5.87 & 5.71 & 5.83 & 5.81 & 0.349$^a$ & $ -4.56$ & 3.776 & F8II \\
 5329838158460399488 & & $ 0.461_{ 0.068}$ & $ -4.83_{ 0.15}$,$ +4.50_{ 0.16}$ & 15.64 & 31.9 & 17.47 & 17.01 & 16.33 & 15.39 & 15.15 & 14.94 & & & & & & $ +4.43$ & 3.791 & G1V \\
\hline \noalign{\smallskip}
\textbf{\object{CS Vel}} & 0.292 & $ 0.194_{ 0.030}$ & $ -4.56_{ 0.06}$,$ +3.10_{ 0.06}$ & & & 12.04 & 11.19 & 10.21 & 8.79 & 8.30 & 8.07 & 7.81 & 7.81 & 7.80 & 7.37 & 0.737$^i$ & $ -4.13$ & 3.767 & G1I \\
 5308893046071732096 & & $ 0.251_{ 0.048}$ & $ -4.61_{ 0.10}$,$ +3.16_{ 0.08}$ & 10.40 & 35.6 & 16.62 & 16.20 & 15.54 & 14.66 & 14.42 & 14.26 & & & & & & $ +1.14$ & 4.070 & A0V \\
\hline \noalign{\smallskip}
\textbf{\object{DK Vel}} & 0.431 & $ 0.238_{ 0.025}$ & $ -4.07_{ 0.05}$,$ +3.62_{ 0.05}$ & & & 10.91 & 10.40 & 9.74 & 8.87 & 8.51 & 8.40 & 8.31 & 8.31 & 8.30 & 8.38 & 0.287$^i$ & $ -3.45$ & 3.774 & F8II \\
 5311599390863537408 & & $ 0.402_{ 0.089}$ & $ -3.97_{ 0.21}$,$ +3.63_{ 0.22}$ & 21.32 & 49.5 & 18.57 & 18.08 & 17.39 & & & & & & & & & $ +5.37$ & 3.771 & K0V \\
 \hline
\end{tabular}
\tablefoot{$^\ddagger$ The expected parallax $\varpi_\mathrm{exp}$ and the proper motion of Polaris are from \citetads{2007ASSL..350.....V}. $^\star$ The companion of \object{RS Pup} is an unbound field star embedded in the circumstellar nebula of the Cepheid (Sect.~\ref{rspup}).}
\end{sidewaystable*}

\begin{sidewaystable*}
 \caption{Candidate \texttt{Near} companions of Cepheids, up to a maximum projected separation of 50\,kau.
Part of the listed targets also host a candidate \texttt{Bound} companion, listed in Table~\ref{cepheids-bound-table1}, where the references for the photometry are also listed.
The adopted color excesses $E(B-V)$ are taken essentially from \citetads{1990ApJS...72..153F,1994ApJ...429..844F}.}
 \label{cepheids-near-table1}
 \centering
\tiny
 \renewcommand{\arraystretch}{1}
 \setlength\tabcolsep{4.0pt}
 \begin{tabular}{lrrcrrrrrrrrrrrrrrcl}
 \hline
 \hline \noalign{\smallskip}
Cepheid / GDR2 & $\varpi_\mathrm{exp}$ & $\varpi_\mathrm{G2}$ & $\mu_{\alpha,\mathrm{G2}},\,\mu_{\delta,\mathrm{G2}}$ & Sep. & Sep. & $G_\mathrm{BP}$ & $G$ & $G_\mathrm{RP}$ & $J$ & $H$ & $K$ & $W1$ & $W2$ & $W3$ & $W4$ & $E_\mathrm{BV}$ & $M_G$ & $\log T$ &  \\
 & (mas) & (mas) & (mas\,a$^{-1}$) & ($\arcsec$) & (kau) & & & & & & & & & & & & & & \\
\hline \noalign{\smallskip}
\textbf{\object{FF Aql}} & 2.048 & $ 1.839_{ 0.107}$ & $ -1.01_{ 0.16}$,$ -9.80_{ 0.17}$ & & & 5.73 & 5.18 & 4.58 & 4.03 & 3.74 & 3.53 & 3.51 & 2.94 & 3.42 & 3.38 & 0.224 & $ -4.07$ & 3.765 & \\
 4514145597469864192 & & $ 2.082_{ 0.449}$ & $ -3.76_{ 1.35}$,$ -4.46_{ 1.62}$ & 95.96 & 46.9 & 20.45 & 20.33 & 19.18 & & & & & & & & & $+11.37$ & 3.745 & \\
\hline \noalign{\smallskip}
\textbf{\object{V0916 Aql}} & 0.325 & $ 0.506_{ 0.050}$ & $ -1.65_{ 0.08}$,$ -5.06_{ 0.07}$ & & & 11.17 & 10.06 & 8.98 & 7.56 & 6.95 & 6.72 & 5.80 & 6.17 & 6.23 & 6.23 & 0.761 & $ -3.11$ & 3.694 & \\
 4313179507877107072 & & $ 0.347_{ 0.048}$ & $ -2.27_{ 0.13}$,$ -5.89_{ 0.12}$ & 3.45 & 10.6 & 14.43 & 13.50 & 12.30 & & & & & & & & & $ -0.51$ & 3.711 & \\
\hline \noalign{\smallskip}
\textbf{\object{eta Aql}} & 3.755 & $ 2.641_{ 0.485}$ & $ +7.67_{ 0.84}$,$ -8.84_{ 0.78}$ & & & & 7.01 & & 2.35 & 1.89 & 1.79 & 0.04 & -0.30 & 1.89 & 1.82 & 0.035 &  & & \\
 4240272815933813888 & & $ 3.361_{ 0.785}$ & $ +3.86_{ 3.00}$,$ +0.73_{ 2.47}$ & 135.78 & 36.2 & 19.92 & 20.49 & 19.40 & & & & & & & & & $+13.02$ & 3.817 & \\
\hline \noalign{\smallskip}
\textbf{\object{CK Cam}} & 1.959 & $ 1.301_{ 0.035}$ & $ +4.27_{ 0.05}$,$ -3.03_{ 0.05}$ & & & 7.76 & 7.18 & 6.45 & 5.44 & 5.12 & 4.96 & 5.01 & 4.81 & 4.95 & 4.89 & 0.501 & $ -3.53$ & 3.798 & \\
 279381953248087808 & & $ 1.670_{ 0.144}$ & $ +9.37_{ 0.27}$,$ -8.69_{ 0.25}$ & 76.53 & 39.1 & 19.89 & 18.46 & 17.22 & 15.45 & 14.71 & 14.51 & 14.20 & 14.31 & 12.76 & 9.05 & & $ +8.58$ & 3.006 & \\
\hline \noalign{\smallskip}
\textbf{\object{Y Car}} & 0.695 & $ 0.330_{ 0.036}$ & $ -9.10_{ 0.07}$,$ +2.43_{ 0.07}$ & & & 8.38 & 8.04 & 7.52 & 6.80 & 6.57 & 6.48 & 6.41 & 6.37 & 6.39 & 6.19 & 0.178 & $ -4.84$ & 3.796 & \\
 5351428787238024576 & & $ 0.718_{ 0.144}$ & $ -5.88_{ 0.34}$,$ +2.75_{ 0.31}$ & 22.47 & 32.3 & 19.08 & 18.47 & 17.49 & & & & & & & & & $ +7.34$ & 3.632 & \\
\hline \noalign{\smallskip}
\textbf{\object{UX Car}} & 0.710 & $ 0.590_{ 0.038}$ & $ -7.08_{ 0.06}$,$ +2.40_{ 0.06}$ & & & 8.65 & 8.27 & 7.71 & 7.10 & 6.87 & 6.71 & 6.48 & 6.44 & 6.49 & 6.11 & 0.123 & $ -3.20$ & 3.772 & \\
 5351721944542867072 & & $ 0.775_{ 0.086}$ & $ -8.45_{ 0.17}$,$ +3.13_{ 0.16}$ & 26.63 & 37.5 & 18.20 & 17.54 & 16.71 & 15.04 & 14.24 & 14.79 & & & & & & $ +6.70$ & 3.642 & \\
\hline \noalign{\smallskip}
\textbf{\object{UZ Car}} & 0.437 & $ 0.403_{ 0.028}$ & $ -6.23_{ 0.05}$,$ +2.92_{ 0.06}$ & & & 10.87 & 9.11 & 8.45 & 7.57 & 7.22 & 7.08 & 6.95 & 7.08 & 7.01 & 6.76 & 0.187 & $ -3.24$ & 2.788 & \\
 5254298479854443392 & & $ 0.484_{ 0.112}$ & $ -4.37_{ 0.26}$,$ +1.91_{ 0.30}$ & 20.85 & 47.7 & 18.55 & 18.29 & 17.48 & & & & & & & & & $ +6.24$ & 3.767 & \\
\hline \noalign{\smallskip}
\textbf{\object{XY Car}} & 0.366 & $ 0.359_{ 0.027}$ & $ -5.18_{ 0.05}$,$ +3.28_{ 0.05}$ & & & 9.65 & 8.97 & 8.17 & 6.96 & 6.44 & 6.17 & 6.05 & 6.01 & 6.01 & 5.92 & 0.417 & $ -4.28$ & 3.755 & \\
 5240441472232303744 & & $ 0.371_{ 0.045}$ & $ -6.39_{ 0.07}$,$ +3.49_{ 0.06}$ & 17.84 & 48.7 & 16.45 & 16.02 & 15.41 & 14.46 & 14.40 & 14.23 & & & & & & $ +2.75$ & 3.829 & \\
\hline \noalign{\smallskip}
\textbf{\object{ER Car}} & 0.959 & $ 0.825_{ 0.035}$ & $ -9.72_{ 0.06}$,$ +2.48_{ 0.05}$ & & & 7.07 & 6.60 & 5.99 & 5.26 & 4.97 & 4.81 & 4.80 & 4.57 & 4.83 & 4.78 & 0.101 & $ -4.08$ & 3.749 & \\
 5339393979667360512 & & $ 0.842_{ 0.114}$ & $ -6.15_{ 0.22}$,$ +2.08_{ 0.19}$ & 12.94 & 13.5 & 17.82 & 17.68 & 16.64 & & & & & & & & & $ +7.05$ & 3.730 & \\
\hline \noalign{\smallskip}
\textbf{\object{EY Car}} & 0.446 & $ 0.360_{ 0.025}$ & $ -6.26_{ 0.05}$,$ +3.27_{ 0.05}$ & & & 10.67 & 10.10 & 9.38 & 8.28 & 7.94 & 7.80 & 7.70 & 7.71 & 7.66 & 7.57 & 0.352 & $ -3.01$ & 3.770 & \\
 5254070090673424768 & & $ 0.395_{ 0.046}$ & $ -5.62_{ 0.08}$,$ +3.30_{ 0.07}$ & 18.39 & 41.2 & 16.58 & 16.14 & 15.48 & 14.80 & 14.32 & 14.34 & & & & & & $ +3.20$ & 3.796 & \\
\hline \noalign{\smallskip}
\textbf{\object{DD Cas}} & 0.322 & $ 0.250_{ 0.029}$ & $ -1.77_{ 0.06}$,$ -1.20_{ 0.05}$ & & & 10.07 & 9.43 & 8.63 & 7.46 & 7.02 & 6.89 & 6.84 & 6.79 & 6.81 & 6.75 & 0.501 & $ -4.83$ & 3.778 & \\
 2013029941628289920 & & $ 0.335_{ 0.061}$ & $ -2.72_{ 0.14}$,$ -1.66_{ 0.09}$ & 15.82 & 49.1 & 17.50 & 16.99 & 16.31 & 15.33 & 15.04 & 14.95 & & & & & & $ +3.30$ & 3.822 & \\
\hline \noalign{\smallskip}
\textbf{\object{VW Cen}} & 0.256 & $ 0.249_{ 0.031}$ & $ -6.03_{ 0.05}$,$ -1.17_{ 0.05}$ & & & 10.78 & 9.87 & 8.98 & 7.70 & 7.08 & 6.82 & 6.62 & 6.86 & 6.79 & 6.08 & 0.448 & $ -4.18$ & 3.691 & \\
 5864955727374022656 & & $ 0.243_{ 0.057}$ & $ -6.84_{ 0.08}$,$ -1.22_{ 0.09}$ & 10.41 & 40.7 & 17.21 & 16.79 & 16.07 & & & & & & & & & $ +2.54$ & 3.816 & \\
\hline \noalign{\smallskip}
\textbf{\object{XX Cen}} & 0.589 & $ 0.566_{ 0.042}$ & $ -4.11_{ 0.07}$,$ -1.06_{ 0.08}$ & & & 8.13 & 7.50 & 6.85 & 5.85 & 5.53 & 5.39 & 5.37 & 5.23 & 5.34 & 5.27 & 0.260 & $ -4.38$ & 3.752 & \\
 5871922301788975232 & & $ 0.646_{ 0.131}$ & $ -2.21_{ 0.30}$,$ -1.33_{ 0.37}$ & 27.82 & 47.2 & 18.70 & 18.35 & 17.45 & 16.51 & 15.82 & 15.65 & & & & & & $ +6.76$ & 3.757 & \\
\hline \noalign{\smallskip}
\textbf{\object{AY Cen}} & 0.639 & $ 0.579_{ 0.032}$ & $ -6.75_{ 0.06}$,$ +2.12_{ 0.05}$ & & & 9.14 & 8.55 & 7.83 & 6.78 & 6.45 & 6.28 & 6.22 & 6.15 & 6.23 & 6.24 & 0.310 & $ -3.41$ & 3.759 & \\
 5334506135103632768 & & $ 0.551_{ 0.107}$ & $ -5.66_{ 0.26}$,$ +1.38_{ 0.18}$ & 27.34 & 42.8 & 18.55 & 18.05 & 17.20 & & & & & & & & & $ +5.99$ & 3.753 & \\
\hline \noalign{\smallskip}
\textbf{\object{AV Cir}} & 1.802 & $ 1.045_{ 0.036}$ & $ -3.42_{ 0.04}$,$ -2.22_{ 0.06}$ & & & 7.67 & 7.14 & 6.46 & 5.58 & 5.23 & 5.11 & 5.03 & 4.81 & 5.02 & 4.95 & 0.397 & $ -3.79$ & 3.791 & \\
 5848499882282952960 & & $ 1.806_{ 0.041}$ & $-10.14_{ 0.04}$,$ -6.50_{ 0.05}$ & 89.64 & 49.7 & 14.89 & 14.43 & 13.80 & 13.02 & 12.61 & 12.51 & & & & & & $ +4.67$ & 3.811 & \\
\hline \noalign{\smallskip}
\textbf{\object{AX Cir}} & 1.917 & $ 1.774_{ 0.345}$ & $ -4.89_{ 0.44}$,$ -5.08_{ 0.52}$ & & & 6.26 & 5.73 & 5.11 & 4.25 & 3.85 & 3.76 & 3.75 & 3.38 & 3.70 & 3.60 & 0.153 & $ -3.41$ & 3.749 & \\
 5874030958906259072 & & $ 1.857_{ 0.406}$ & $ -7.45_{ 0.49}$,$ -4.66_{ 0.82}$ & 44.15 & 23.0 & 18.95 & 18.44 & 17.22 & & & & & & & & & $ +9.43$ & 3.532 & \\
\hline \noalign{\smallskip}
\textbf{\object{BP Cir}} & 1.700 & $ 1.024_{ 0.040}$ & $ -5.46_{ 0.05}$,$ -3.89_{ 0.07}$ & & & 7.74 & 7.31 & 6.70 & 5.84 & 5.58 & 5.44 & 5.39 & 5.21 & 5.39 & 5.30 & 0.235 & $ -3.25$ & 3.782 & \\
 5877460610633466496 & & $ 1.552_{ 0.260}$ & $ -3.61_{ 0.45}$,$ -3.54_{ 0.56}$ & 82.78 & 48.7 & 19.93 & 19.39 & 17.95 & & & & & & & & & $ +9.83$ & 3.400 & \\
 5877460782359164928 & & $ 1.837_{ 0.364}$ & $ -5.35_{ 0.63}$,$ -1.70_{ 0.81}$ & 56.99 & 33.5 & 20.20 & 19.70 & 18.36 & & & & & & & & & $+10.50$ & 3.521 & \\
 5877472430316940672 & & $ 1.506_{ 0.172}$ & $ -3.02_{ 0.29}$,$ -2.59_{ 0.35}$ & 49.45 & 29.1 & 19.34 & 18.62 & 17.54 & 15.54 & 13.87 & 13.76 & 12.24 & 12.44 & 11.93 & 8.56 & & $ +8.98$ & 3.555 & \\
\hline \noalign{\smallskip}
\textbf{\object{T Cru}} & 1.306 & $ 1.169_{ 0.028}$ & $-10.96_{ 0.05}$,$ -0.54_{ 0.05}$ & & & 6.97 & 6.41 & 5.76 & 4.93 & 4.62 & 4.44 & 4.26 & 4.09 & 4.35 & 4.24 & 0.193 & $ -3.74$ & 3.749 & \\
 6054829771915831808 & & $ 1.254_{ 0.050}$ & $-14.26_{ 0.09}$,$ -1.33_{ 0.07}$ & 46.43 & 35.6 & 17.10 & 16.46 & 15.67 & 14.70 & 14.09 & 14.16 & & & & & & $ +6.50$ & 3.698 & \\
\hline \noalign{\smallskip}
\textbf{\object{X Cru}} & 0.678 & $ 0.552_{ 0.046}$ & $ -5.89_{ 0.06}$,$ -0.17_{ 0.06}$ & & & 8.69 & 8.14 & 7.41 & 6.58 & 6.14 & 5.95 & 5.88 & 5.82 & 5.86 & 5.76 & 0.286 & $ -3.86$ & 3.758 & \\
 6059763997807862656 & & $ 0.677_{ 0.128}$ & $ -6.06_{ 0.21}$,$ +0.42_{ 0.23}$ & 21.68 & 32.0 & 18.75 & 18.22 & 17.46 & & & & & & & & & $ +6.66$ & 3.757 & \\
\hline \noalign{\smallskip}
\textbf{\object{SU Cyg}} & 1.234 & $ 1.198_{ 0.052}$ & $ -1.79_{ 0.10}$,$ -3.28_{ 0.08}$ & & & 7.22 & 6.85 & 6.28 & 5.79 & 5.49 & 5.42 & 5.26 & 5.10 & 5.22 & 5.16 & 0.096 & $ -3.01$ & 3.767 & \\
 2031776236946397696 & & $ 1.212_{ 0.239}$ & $ -0.93_{ 0.58}$,$ -3.76_{ 0.59}$ & 42.43 & 34.4 & 19.64 & 19.23 & 18.16 & & & & & & & & & $ +9.42$ & 3.630 & \\
\hline
\end{tabular}
\end{sidewaystable*}

\begin{sidewaystable*}
 \caption{Continued from Table~\ref{cepheids-near-table1}. The negative parallax of Y Sgr ($\varpi_\mathrm{G2} = -0.441 \pm 0.280$\,mas) has been omitted.}
 \label{cepheids-near-table2}
  \centering
\tiny
 \renewcommand{\arraystretch}{1}
 \setlength\tabcolsep{4.5pt}
  \begin{tabular}{lrrcrrrrrrrrrrrrrrcl}
 \hline
 \hline \noalign{\smallskip}
Cepheid / GDR2 & $\varpi_\mathrm{exp}$ & $\varpi_\mathrm{G2}$ & $\mu_{\alpha,\mathrm{G2}},\,\mu_{\delta,\mathrm{G2}}$ & Sep. & Sep. & $G_\mathrm{BP}$ & $G$ & $G_\mathrm{RP}$ & $J$ & $H$ & $K$ & $W1$ & $W2$ & $W3$ & $W4$ & $E_\mathrm{BV}$ & $M_G$ & $\log T$ & \\
 & (mas) & (mas) & (mas\,a$^{-1}$) & ($\arcsec$) & (kau) & & & & & & & & & & & & & & \\
\hline \noalign{\smallskip}
\textbf{\object{SZ Cyg}} & 0.419 & $ 0.391_{ 0.026}$ & $ -2.20_{ 0.05}$,$ -5.56_{ 0.06}$ & & & 9.89 & 9.03 & 8.03 & 6.39 & 5.89 & 5.71 & 5.49 & 5.45 & 5.50 & 5.60 & 0.631 & $ -4.48$ & 3.737 & \\
 2071433765909165952 & & $ 0.398_{ 0.065}$ & $ -1.98_{ 0.14}$,$ -4.54_{ 0.17}$ & 19.60 & 46.8 & 18.50 & 17.16 & 15.96 & 13.74 & 12.86 & 12.59 & & & & & & $ +3.86$ & 3.342 & \\
\hline \noalign{\smallskip}
\textbf{\object{V1334 Cyg}} & 1.502 & $ 1.180_{ 0.066}$ & $ +3.77_{ 0.10}$,$ +0.18_{ 0.12}$ & & & 6.02 & 5.75 & 5.32 & 5.17 & 4.66 & 4.46 & 4.38 & 4.06 & 4.45 & 4.38 & 0.074 & $ -4.09$ & 3.796 & \\
 1964855939153629312 & & $ 1.409_{ 0.318}$ & $ -0.37_{ 0.66}$,$ -3.27_{ 0.68}$ & 65.67 & 43.7 & 20.51 & 19.66 & 18.53 & 16.36 & 15.41 & 15.28 & & & & & & $+10.25$ & 3.184 & \\
\hline \noalign{\smallskip}
\textbf{\object{RR Lac}} & 0.515 & $ 0.352_{ 0.033}$ & $ -3.65_{ 0.06}$,$ -3.21_{ 0.06}$ & & & 9.26 & 8.73 & 7.98 & 7.14 & 6.71 & 6.58 & 6.36 & 6.31 & 6.37 & 6.33 & 0.353 & $ -4.43$ & 3.772 & \\
 2006989121667183872 & & $ 0.490_{ 0.041}$ & $ -4.78_{ 0.05}$,$ -4.53_{ 0.05}$ & 18.46 & 35.8 & 15.39 & 15.03 & 14.47 & 13.73 & 13.44 & 13.45 & & & & & & $ +2.53$ & 3.837 & \\
\hline \noalign{\smallskip}
\textbf{\object{CV Mon}} & 0.601 & $ 0.511_{ 0.041}$ & $ +0.43_{ 0.07}$,$ -0.69_{ 0.07}$ & & & 10.49 & 9.62 & 8.71 & 7.37 & 6.77 & 6.52 & 6.46 & 6.31 & 6.45 & 6.37 & 0.722 & $ -3.57$ & 3.772 & \\
 3127142220517515392 & & $ 0.594_{ 0.062}$ & $ +0.12_{ 0.12}$,$ -0.77_{ 0.12}$ & 17.70 & 29.4 & 17.19 & 16.48 & 15.65 & & & & & & & & & $ +3.52$ & 3.814 & \\
 3127142327895572352 & & $ 0.537_{ 0.040}$ & $ +0.46_{ 0.05}$,$ -0.62_{ 0.05}$ & 14.20 & 23.6 & 13.91 & 13.49 & 12.88 & 12.10 & 11.86 & 11.71 & 11.05 & 11.03 & 10.33 & 8.70 & & $ +0.12$ & 4.095 & \\
\hline \noalign{\smallskip}
\textbf{\object{S Nor}} & 1.126 & $ 1.091_{ 0.042}$ & $ -1.51_{ 0.07}$,$ -2.32_{ 0.06}$ & & & 6.82 & 6.18 & 5.55 & 4.87 & 4.50 & 4.22 & 4.08 & 3.82 & 4.01 & 3.95 & 0.189 & $ -4.10$ & 3.737 & \\
 5835124121463456896 & & $ 0.977_{ 0.052}$ & $ -1.45_{ 0.09}$,$ -2.50_{ 0.08}$ & 46.32 & 41.1 & 16.69 & 16.04 & 15.24 & 14.19 & 13.59 & 13.36 & 12.54 & 12.59 & 12.87 & 9.21 & & $ +5.54$ & 3.691 & \\
 5835124121533801088 & & $ 1.171_{ 0.097}$ & $ -2.02_{ 0.32}$,$ -1.85_{ 0.16}$ & 43.66 & 38.8 & & 17.09 & & 13.03 & 12.42 & 12.26 & 11.53 & 11.58 & 12.25 & 9.11 & &  & & \\
\hline \noalign{\smallskip}
\textbf{\object{U Nor}} & 0.699 & $ 0.641_{ 0.050}$ & $ -2.01_{ 0.10}$,$ -2.79_{ 0.09}$ & & & 9.49 & 8.43 & 7.43 & 5.77 & 5.18 & 4.91 & 4.64 & 4.44 & 4.80 & 4.67 & 0.892 & $ -4.62$ & 3.762 & \\
 5884368494207976960 & & $ 0.751_{ 0.170}$ & $ -4.36_{ 0.54}$,$ -2.47_{ 0.41}$ & 16.47 & 23.6 & 20.10 & 18.60 & 17.17 & 13.53 & 13.93 & 13.41 & & & & & & $ +6.21$ & 3.266 & \\
\hline \noalign{\smallskip}
\textbf{\object{RS Nor}} & 0.487 & $ 0.450_{ 0.046}$ & $ -2.21_{ 0.07}$,$ -3.45_{ 0.06}$ & & & 10.45 & 9.59 & 8.72 & 7.23 & 6.80 & 6.60 & 6.41 & 6.47 & 6.45 & 6.41 & 0.580 & $ -3.52$ & 3.750 & \\
 5932812431123859712 & & $ 0.480_{ 0.059}$ & $ -1.03_{ 0.09}$,$ -1.92_{ 0.07}$ & 7.55 & 15.5 & 16.31 & 15.63 & 14.67 & 11.52 & 12.82 & 11.09 & & & & & & $ +2.63$ & 3.765 & \\
\hline \noalign{\smallskip}
\textbf{\object{QZ Nor}} & 0.556 & $ 0.503_{ 0.038}$ & $ -1.84_{ 0.07}$,$ -3.79_{ 0.06}$ & & & 9.14 & 8.64 & 7.96 & 7.00 & 6.70 & 6.54 & 6.48 & 6.47 & 6.49 & 6.41 & 0.253 & $ -3.49$ & 3.766 & \\
 5932565968801321344 & & $ 0.509_{ 0.070}$ & $ -1.23_{ 0.14}$,$ -2.49_{ 0.11}$ & 27.43 & 49.3 & 17.78 & 17.23 & 16.47 & 15.80 & 16.08 & 15.94 & & & & & & $ +5.13$ & 3.745 & \\
 5932565900081828480 & & $ 0.554_{ 0.045}$ & $ -2.35_{ 0.06}$,$ -4.72_{ 0.06}$ & 16.54 & 29.7 & 13.73 & 13.36 & 12.80 & 12.07 & 11.81 & 11.79 & & & & & & $ +1.40$ & 3.801 & \\
\hline \noalign{\smallskip}
\textbf{\object{Y Oph}} & 1.850 & $ 1.390_{ 0.083}$ & $ -3.21_{ 0.14}$,$ -4.73_{ 0.11}$ & & & 7.08 & 5.60 & 4.73 & 3.67 & 3.08 & 2.82 & 1.36 & 1.17 & 2.58 & 2.47 & 0.655 & $ -5.09$ & 3.536 & \\
 4175017591096420352 & & $ 1.914_{ 0.424}$ & $ +0.30_{ 1.09}$,$ -1.00_{ 1.17}$ & 81.31 & 44.0 & 20.76 & 19.80 & 18.61 & & & & & & & & & $ +9.75$ & 3.657 & \\
\hline \noalign{\smallskip}
\textbf{\object{AQ Pup}} & 0.312 & $ 0.319_{ 0.034}$ & $ -2.57_{ 0.06}$,$ +2.57_{ 0.06}$ & & & 9.31 & 8.37 & 7.44 & 5.88 & 5.33 & 5.09 & 4.62 & 4.65 & 4.99 & 4.88 & 0.512 & $ -5.29$ & 3.694 & \\
 5597379745861712512 & & $ 0.339_{ 0.060}$ & $ -2.77_{ 0.09}$,$ +2.27_{ 0.10}$ & 9.88 & 31.7 & 16.84 & 16.45 & 15.79 & & & & & & & & & $ +2.72$ & 3.874 & \\
\hline \noalign{\smallskip}
\textbf{\object{U Sgr}} & 1.669 & $ 1.489_{ 0.045}$ & $ -1.68_{ 0.08}$,$ -6.13_{ 0.07}$ & & & 7.28 & 6.53 & 5.75 & 4.48 & 3.91 & 3.93 & 3.95 & 3.65 & 3.95 & 3.86 & 0.403 & $ -3.58$ & 3.742 & \\
 4092904791496145664 & & $ 1.562_{ 0.101}$ & $ -0.88_{ 0.24}$,$ -5.99_{ 0.22}$ & 66.82 & 40.0 & 18.53 & 17.64 & 16.55 & & & & & & & & & $ +7.71$ & 3.576 & \\
 4092904993291428864 & & $ 1.512_{ 0.286}$ & $ -4.86_{ 0.68}$,$ -6.52_{ 0.62}$ & 35.35 & 21.2 & 19.05 & 18.70 & 17.13 & 15.58 & 13.47 & 13.36 & & & & & & $ +8.70$ & 3.609 & \\
\hline \noalign{\smallskip}
\textbf{\object{Y Sgr}} & 2.146 & $-$ & $ -3.19_{ 0.37}$,$ -8.13_{ 0.28}$ & & & 7.60 & 8.23 & 6.01 & 3.92 & 3.56 & 3.50 & 3.61 & 3.34 & 3.62 & 3.61 & 0.205 & $-$ & 3.647 & \\
 4096107737550993664 & & $ 1.891_{ 0.370}$ & $ +0.30_{ 1.68}$,$ -4.49_{ 1.65}$ & 71.37 & 33.3 & 20.09 & 19.62 & 18.11 & & & & & & & & & $+10.56$ & 3.369 & \\
 4096107874989915648 & & $ 1.991_{ 0.367}$ & $ -1.90_{ 1.28}$,$ -2.40_{ 1.24}$ & 95.02 & 44.3 & & 19.58 & & & & & & & & & &  & & \\
\hline \noalign{\smallskip}
\textbf{\object{XX Sgr}} & 0.733 & $ 0.731_{ 0.059}$ & $ -0.05_{ 0.10}$,$ -0.48_{ 0.09}$ & & & 9.23 & 8.54 & 7.68 & 6.46 & 5.94 & 5.72 & 5.57 & 5.39 & 5.63 & 5.71 & 0.543 & $ -3.47$ & 3.770 & \\
 4096979650283681408 & & $ 0.697_{ 0.070}$ & $ +0.68_{ 0.50}$,$ -1.85_{ 0.35}$ & 7.87 & 10.7 & & 15.24 & & & & & & & & & &  & & \\
\hline \noalign{\smallskip}
\textbf{\object{AP Sgr}} & 1.287 & $ 1.148_{ 0.053}$ & $ +0.88_{ 0.16}$,$ -3.85_{ 0.13}$ & & & 7.46 & 6.90 & 6.24 & 5.27 & 4.98 & 4.78 & 4.77 & 4.71 & 4.86 & 4.84 & 0.192 & $ -3.27$ & 3.747 & \\
 4066429101224454144 & & $ 1.174_{ 0.187}$ & $ -1.86_{ 0.69}$,$ -3.42_{ 0.65}$ & 63.02 & 49.0 & 19.43 & 18.73 & 17.26 & & & & & & & & & $ +8.68$ & 3.141 & \\
 4066428998145128064 & & $ 1.249_{ 0.284}$ & $ -0.83_{ 0.82}$,$ -6.67_{ 0.76}$ & 36.28 & 28.2 & & 19.37 & & & & & & & & & &  & & \\
\hline \noalign{\smallskip}
\textbf{\object{V0950 Sco}} & 1.073 & $ 0.869_{ 0.052}$ & $ -0.48_{ 0.10}$,$ -1.66_{ 0.09}$ & & & 7.49 & 7.06 & 6.44 & 5.72 & 5.44 & 5.24 & 5.18 & 5.03 & 5.17 & 5.09 & 0.254 & $ -3.91$ & 3.784 & \\
 5960623340819585792 & & $ 1.149_{ 0.371}$ & $ +3.13_{ 1.00}$,$ -1.53_{ 0.94}$ & 16.93 & 15.8 & 17.93 & 17.79 & 16.73 & 12.81 & 13.35 & 13.18 & & & & & & $ +7.45$ & 3.764 & \\
 5960623340819009024 & & $ 1.148_{ 0.141}$ & $ -1.65_{ 0.38}$,$ -6.85_{ 0.32}$ & 29.33 & 27.3 & 18.83 & 18.40 & 17.24 & & & & & & & & & $ +8.11$ & 3.673 & \\
\hline \noalign{\smallskip}
\textbf{\object{Y Sct}} & 0.581 & $ 0.458_{ 0.067}$ & $ -0.68_{ 0.10}$,$ -3.08_{ 0.08}$ & & & 9.83 & 8.86 & 7.85 & 6.48 & 5.85 & 5.59 & 5.49 & 5.39 & 5.61 & 5.35 & 0.619 & $ -4.25$ & 3.702 & \\
 4156450099579515904 & & $ 0.663_{ 0.123}$ & $ -1.97_{ 0.29}$,$ -3.61_{ 0.26}$ & 28.38 & 48.8 & 18.40 & 17.23 & 16.08 & 14.49 & 14.06 & 13.67 & & & & & & $ +5.02$ & 3.530 & \\
\hline \noalign{\smallskip}
\textbf{\object{R TrA}} & 1.733 & $ 1.504_{ 0.037}$ & $ -5.23_{ 0.04}$,$ -8.31_{ 0.07}$ & & & 6.90 & 6.44 & 5.88 & 5.34 & 5.01 & 4.85 & 4.76 & 4.55 & 4.73 & 4.70 & 0.172 & $ -3.11$ & 3.772 & \\
 5824464631144871424 & & $ 1.567_{ 0.364}$ & $ -5.10_{ 0.52}$,$ -6.33_{ 0.77}$ & 56.57 & 32.6 & 19.82 & 19.31 & 18.45 & & & & & & & & & $ +9.87$ & 3.708 & \\
\hline \noalign{\smallskip}
\textbf{\object{LR TrA}} & 1.048 & $ 0.918_{ 0.032}$ & $ -4.86_{ 0.04}$,$ -7.87_{ 0.05}$ & & & 8.01 & 7.60 & 7.02 & 6.31 & 6.02 & 5.88 & 5.85 & 5.77 & 5.84 & 5.79 & 0.154 & $ -2.98$ & 3.772 & \\
 5824226655600814976 & & $ 1.005_{ 0.187}$ & $ -5.77_{ 0.24}$,$ -8.84_{ 0.39}$ & 37.56 & 35.8 & 19.94 & 19.00 & 17.96 & 16.64 & 15.43 & 15.01 & & & & & & $ +8.68$ & 3.303 & \\
   \hline
\end{tabular}
\end{sidewaystable*}

\subsection{Field charts\label{fieldcharts-cep}}

The fields surrounding the CCs with detected \texttt{Bound} candidate companions, and a selection of CCs with \texttt{Near} candidate companions are presented in Fig.~\ref{cepheid-field-1} to \ref{cepheids-near3}.
$\delta$\,Cep and Polaris are discussed separately in Sects.~\ref{delcep} and \ref{polaris}.
The target position is shown with a magenta $+$ symbol, and the tested field stars (with a parallax within $\pm 15\%$ of that of the target) are represented with cyan $\times$ symbols.
The identified \texttt{Near} candidates are represented in orange, the \texttt{LowV} candidates are shown in red, and the \texttt{Bound} candidates are indicated with a yellow star symbol. Their proper motion vectors, and that of the target, are shown.
The GDR2 proper motion vector $\vec{\mu_\mathrm{DR2}}$ is shown in magenta for each target star.
When available, the mean proper motion vector of the target $\vec{\mu_\mathrm{HG}}$ estimated from the Hipparcos and GDR2 positions (see Paper~I) is also displayed in cyan.
The expected parallax of the target is indicated in the upper left corner of each panel, together with the GDR2 value.
For CCs, this is the renormalized Leavitt law parallax from the catalog by \citetads{2000A&AS..143..211B} (see Sect.~\ref{parallax}).
The background images are taken from the DSS2-Red.

\begin{figure*}[h]
\centering
\includegraphics[width=9cm]{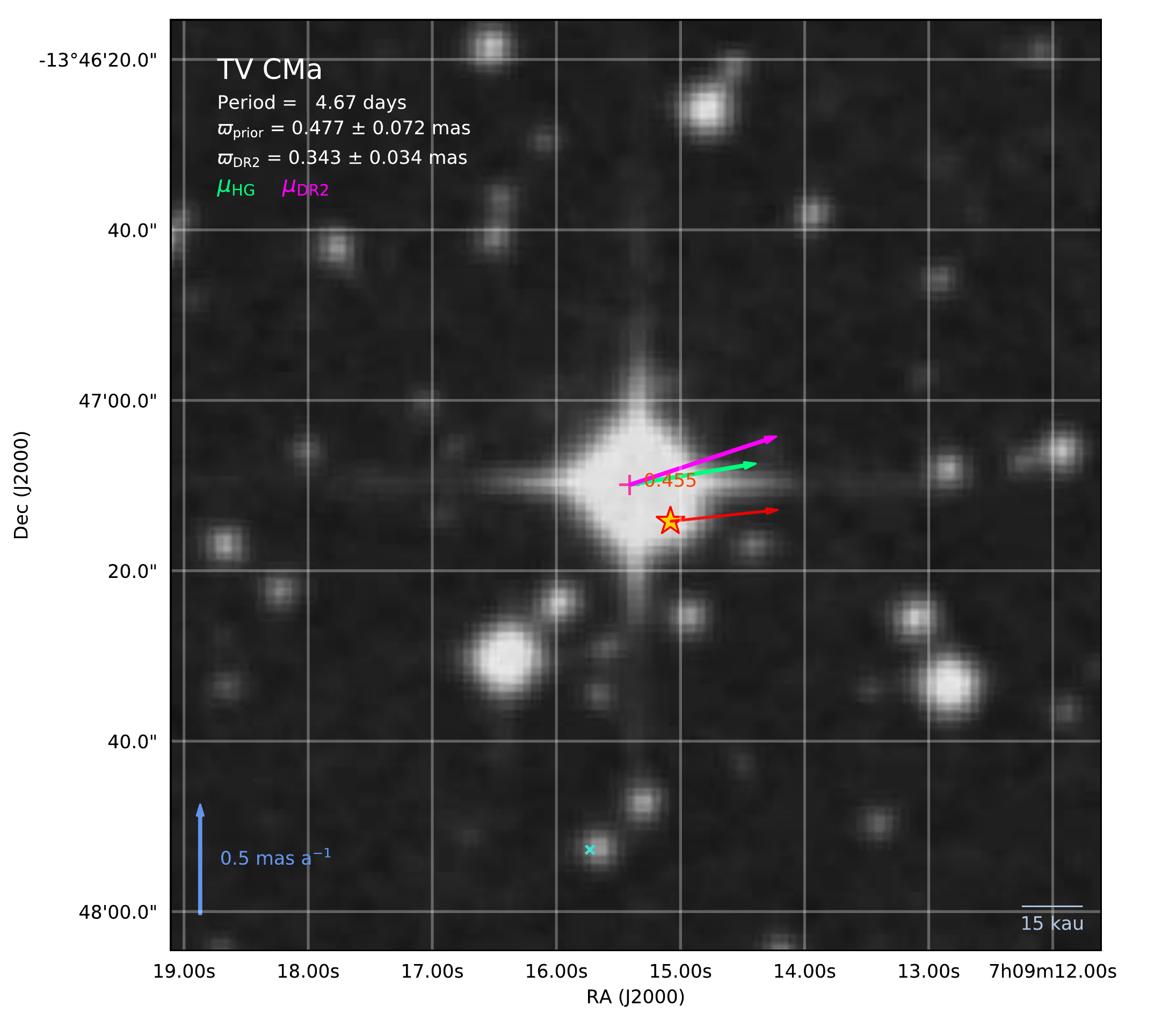}
\includegraphics[width=9cm]{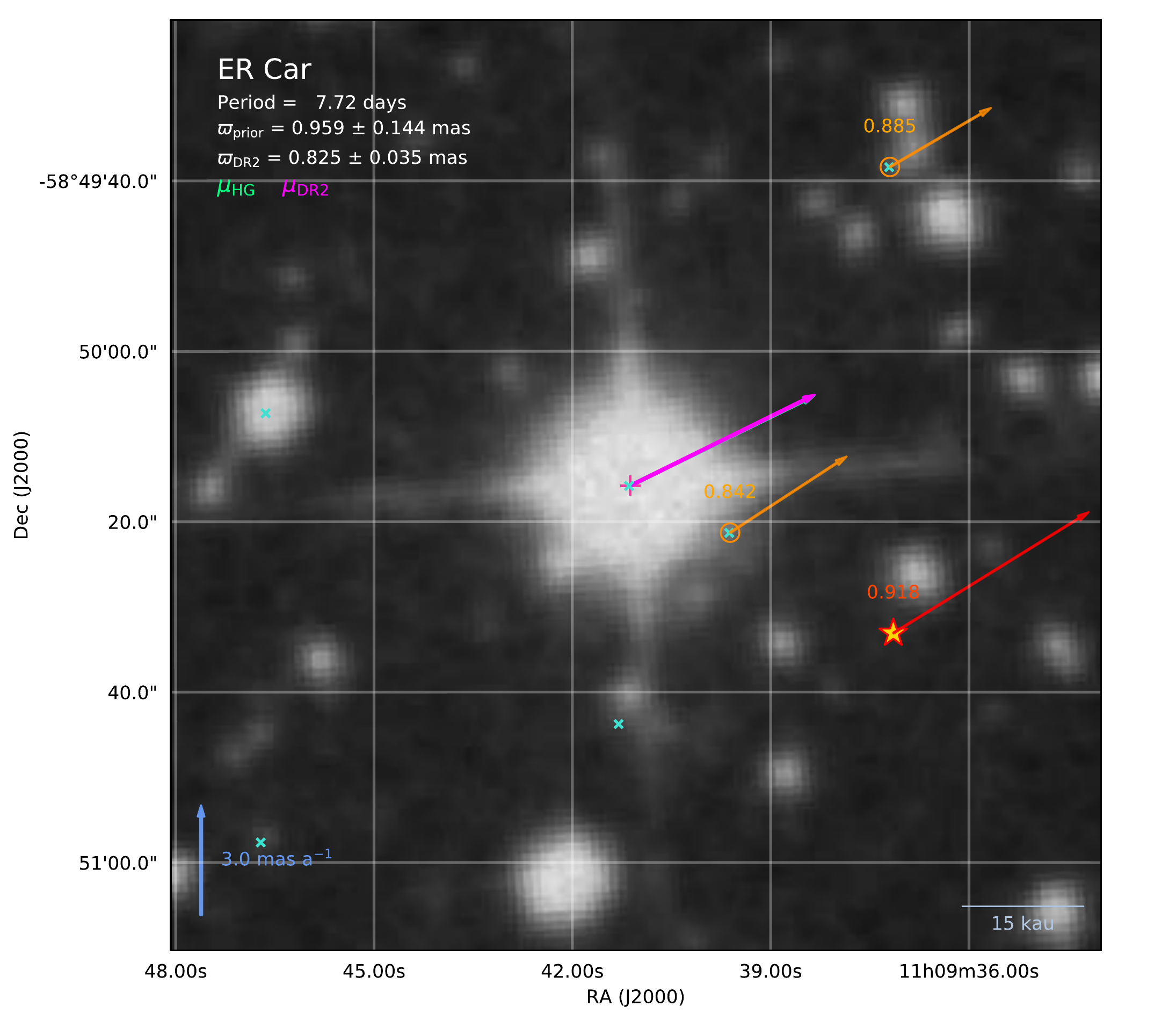}
\includegraphics[width=9cm]{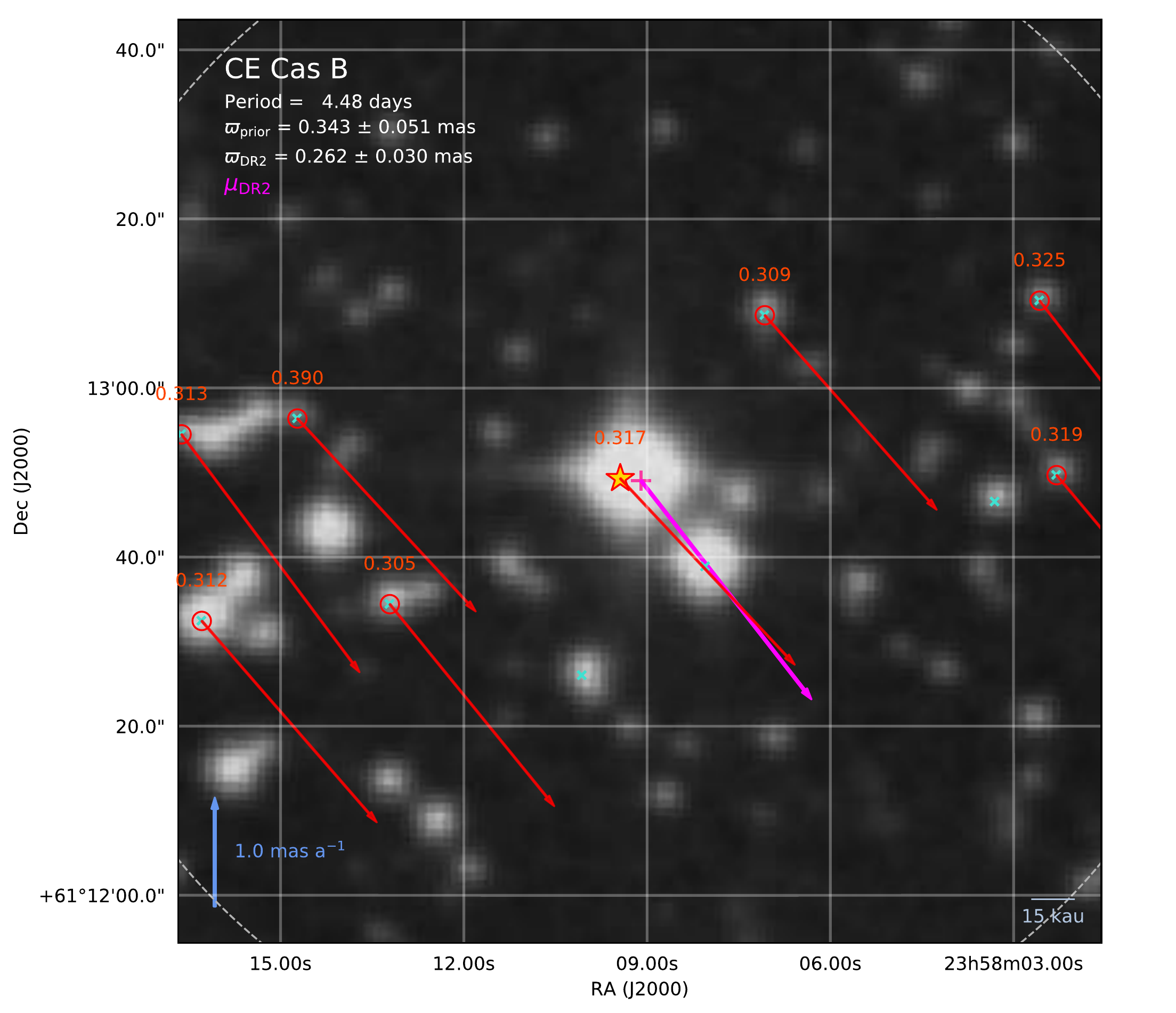}
\includegraphics[width=9cm]{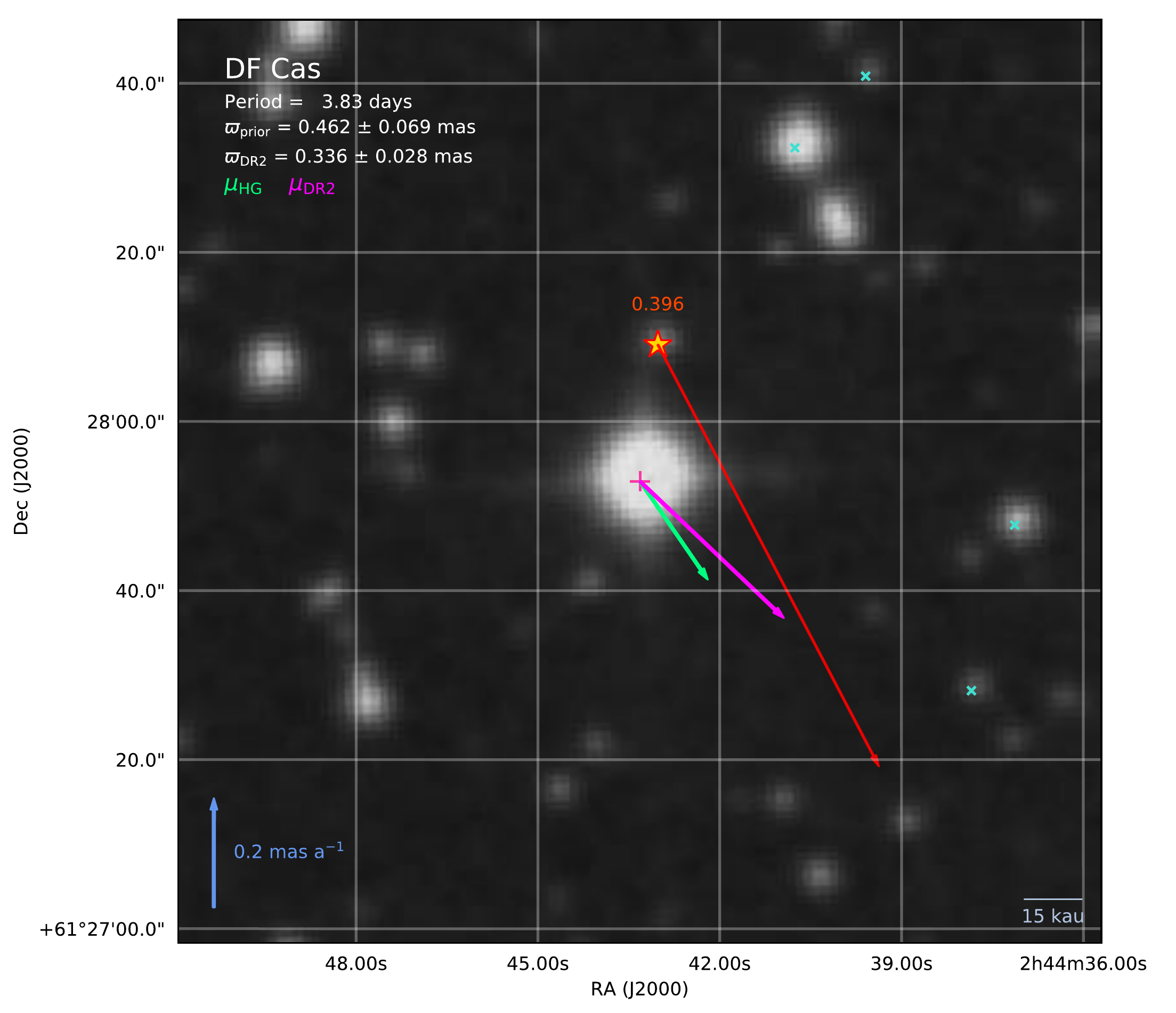}
\includegraphics[width=9cm]{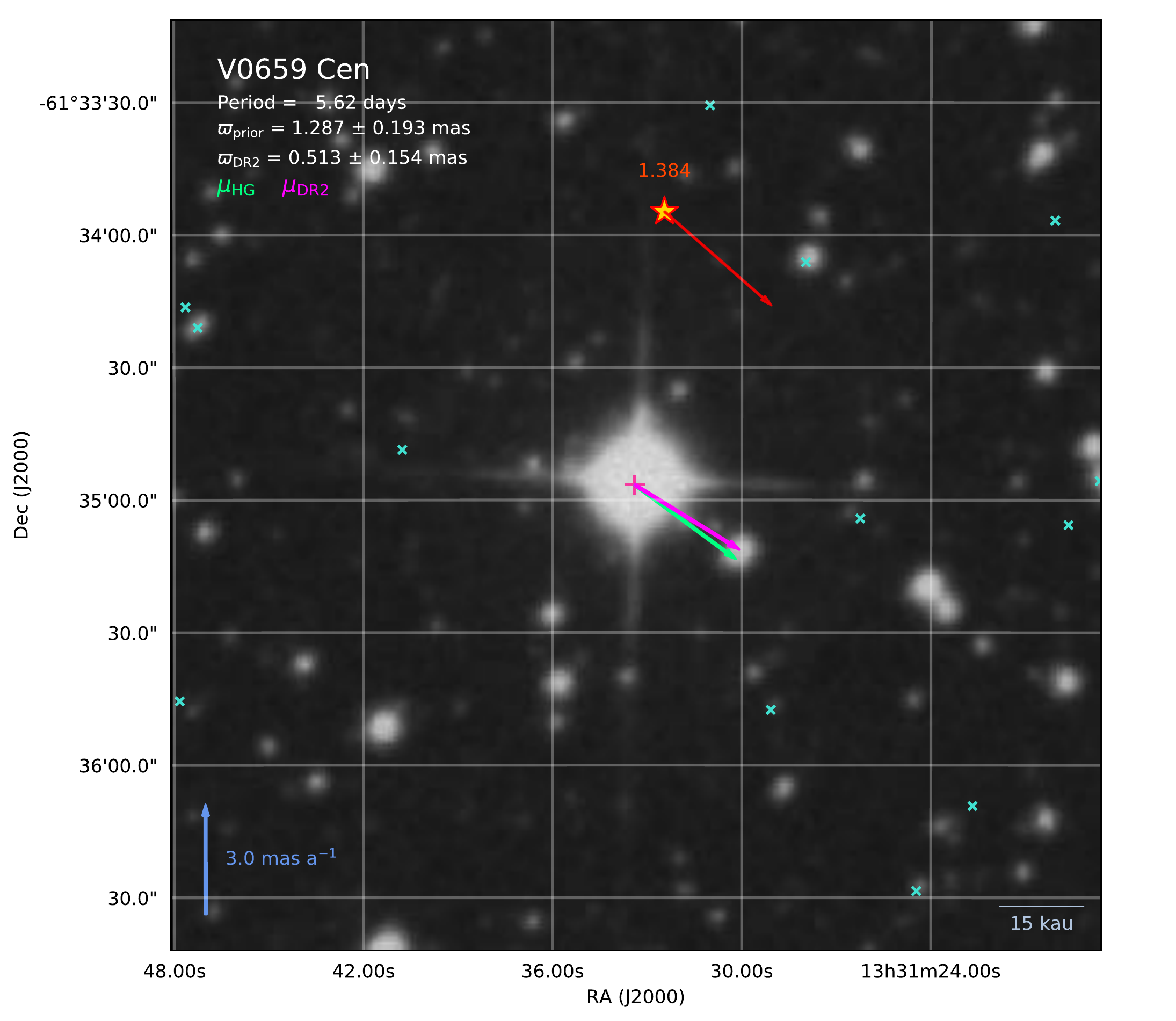}
\includegraphics[width=9cm]{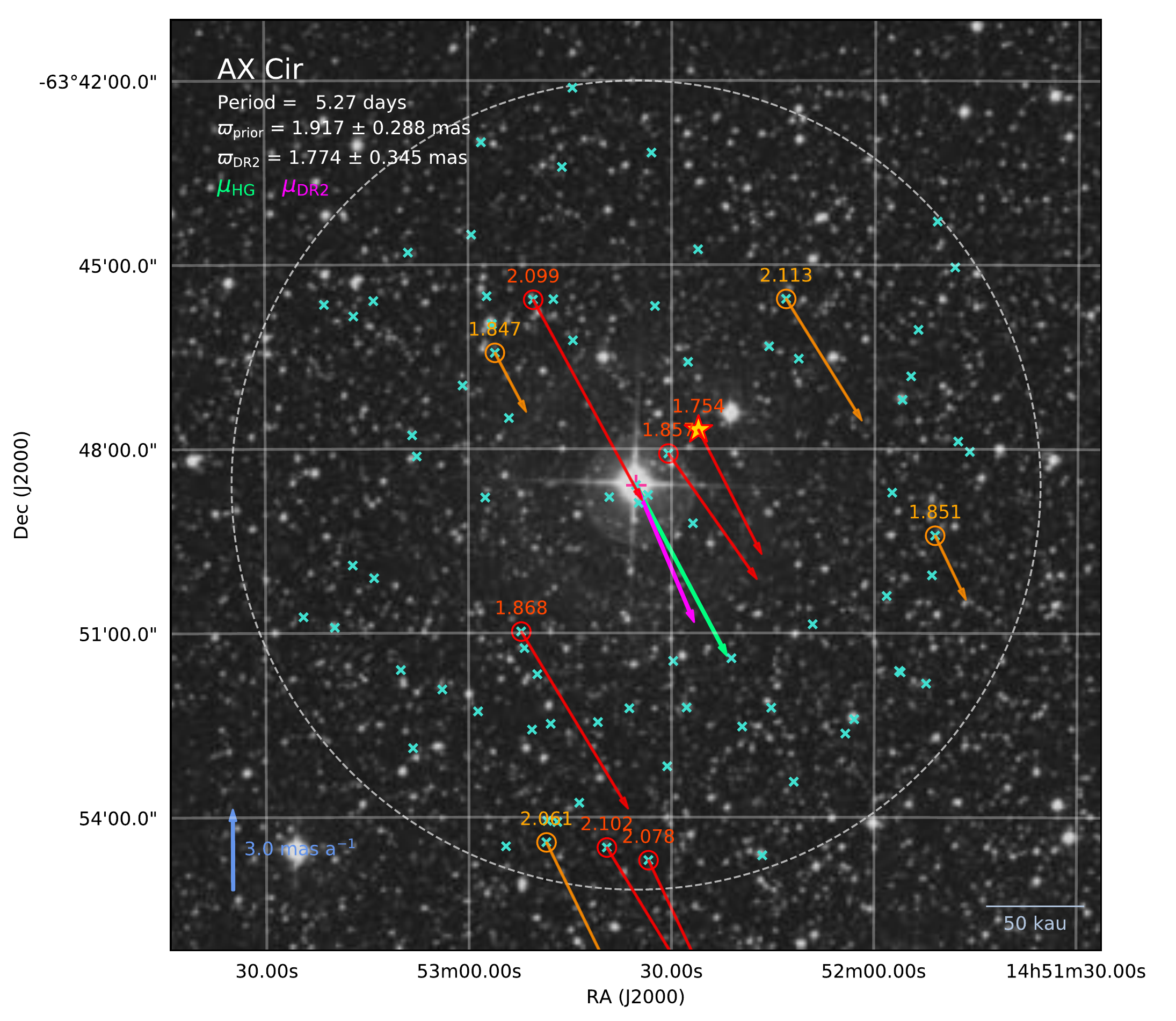}
\caption{Cepheids with \texttt{Bound} candidate companions.
\label{cepheid-field-1}}
\end{figure*}

\begin{figure*}[h]
\centering
\includegraphics[width=9cm]{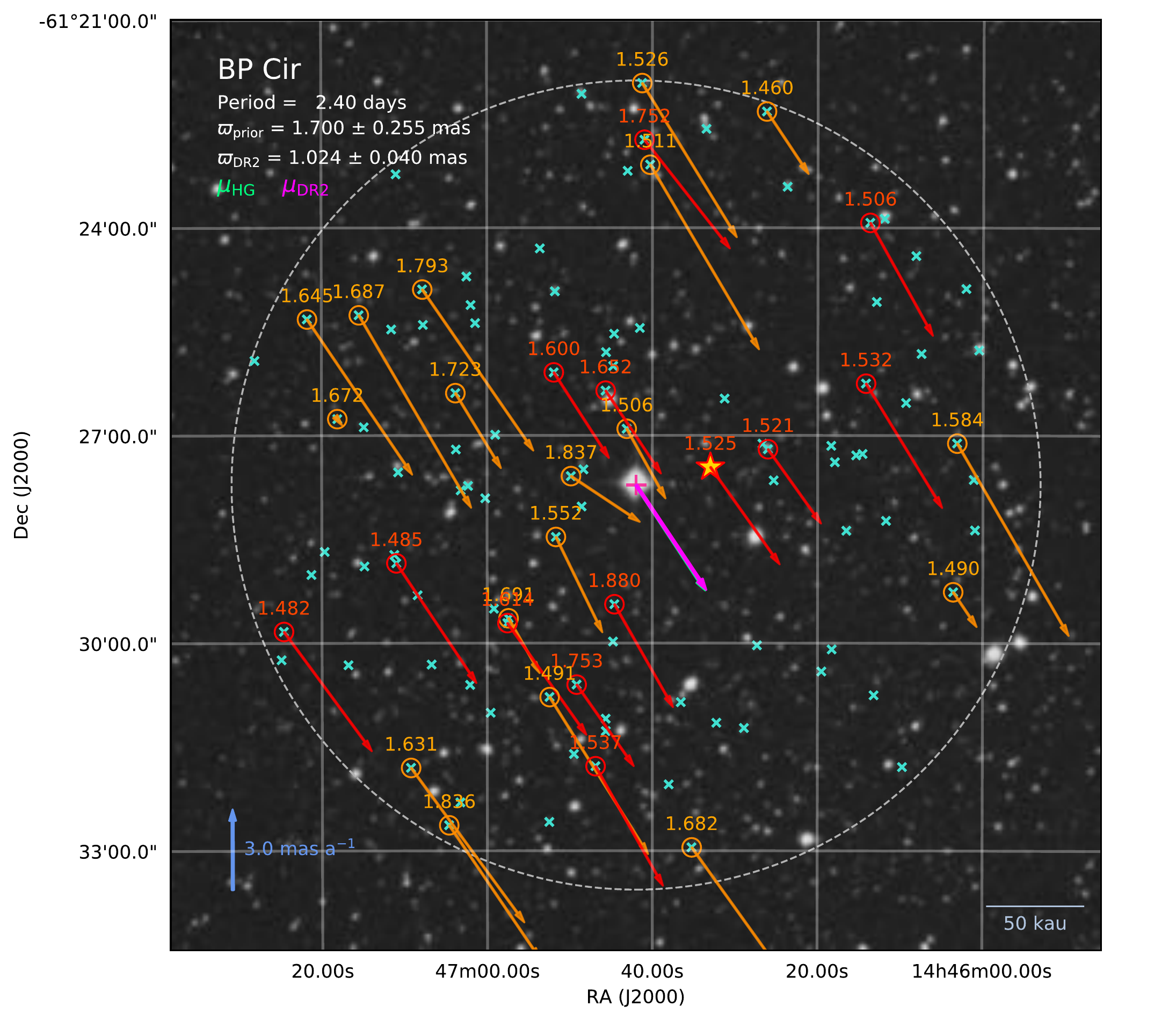}
\includegraphics[width=9cm]{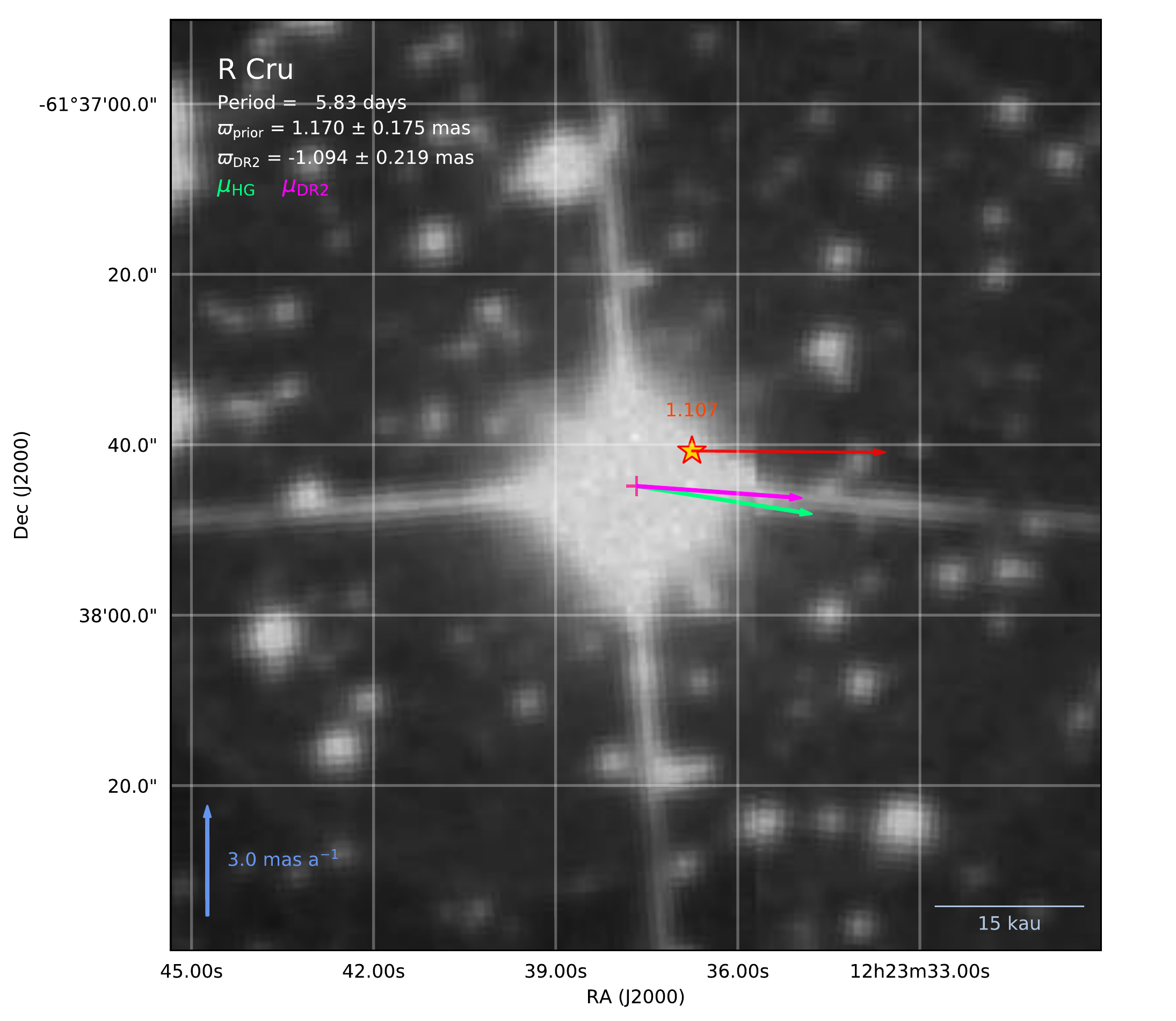}
\includegraphics[width=9cm]{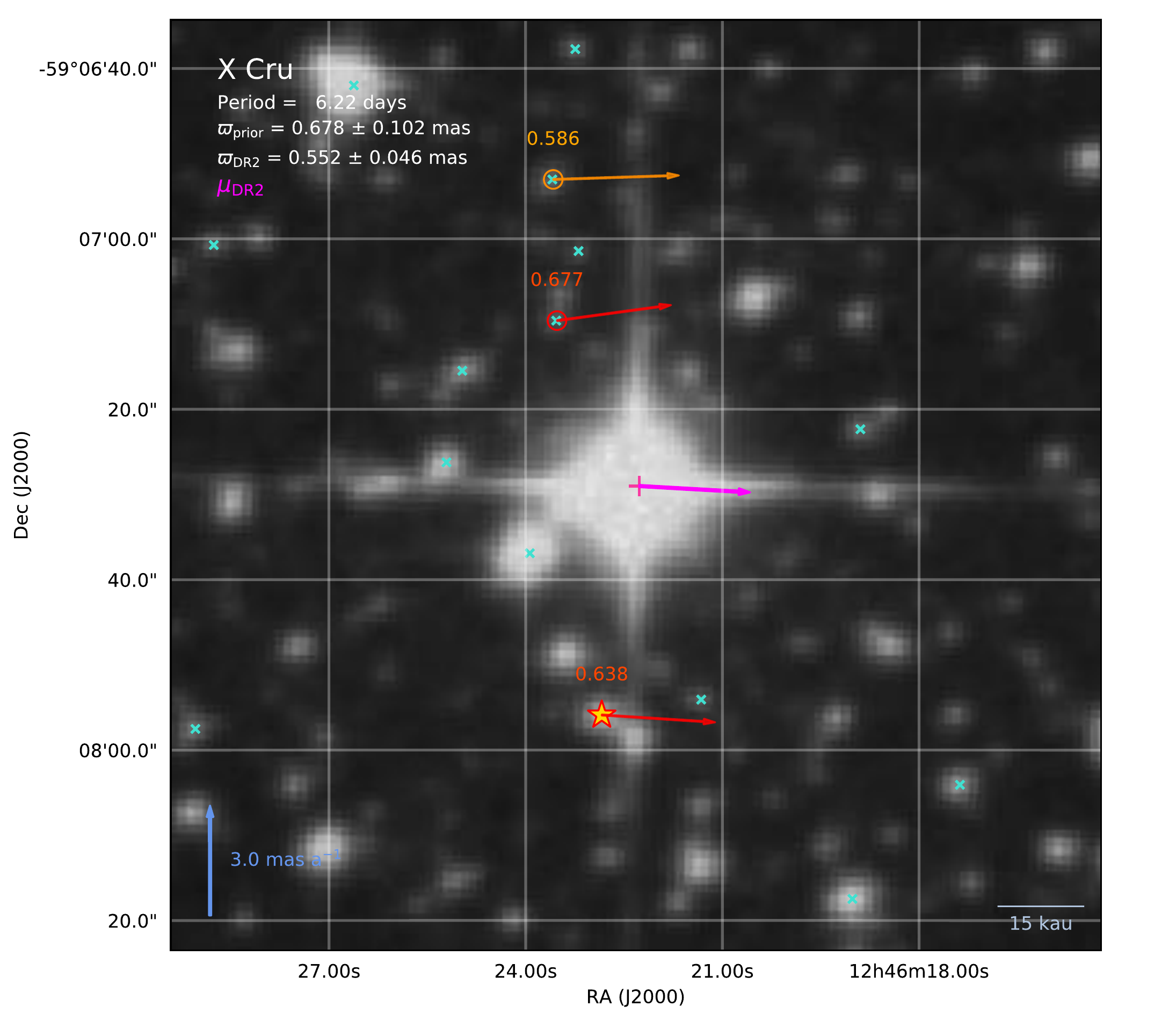}
\includegraphics[width=9cm]{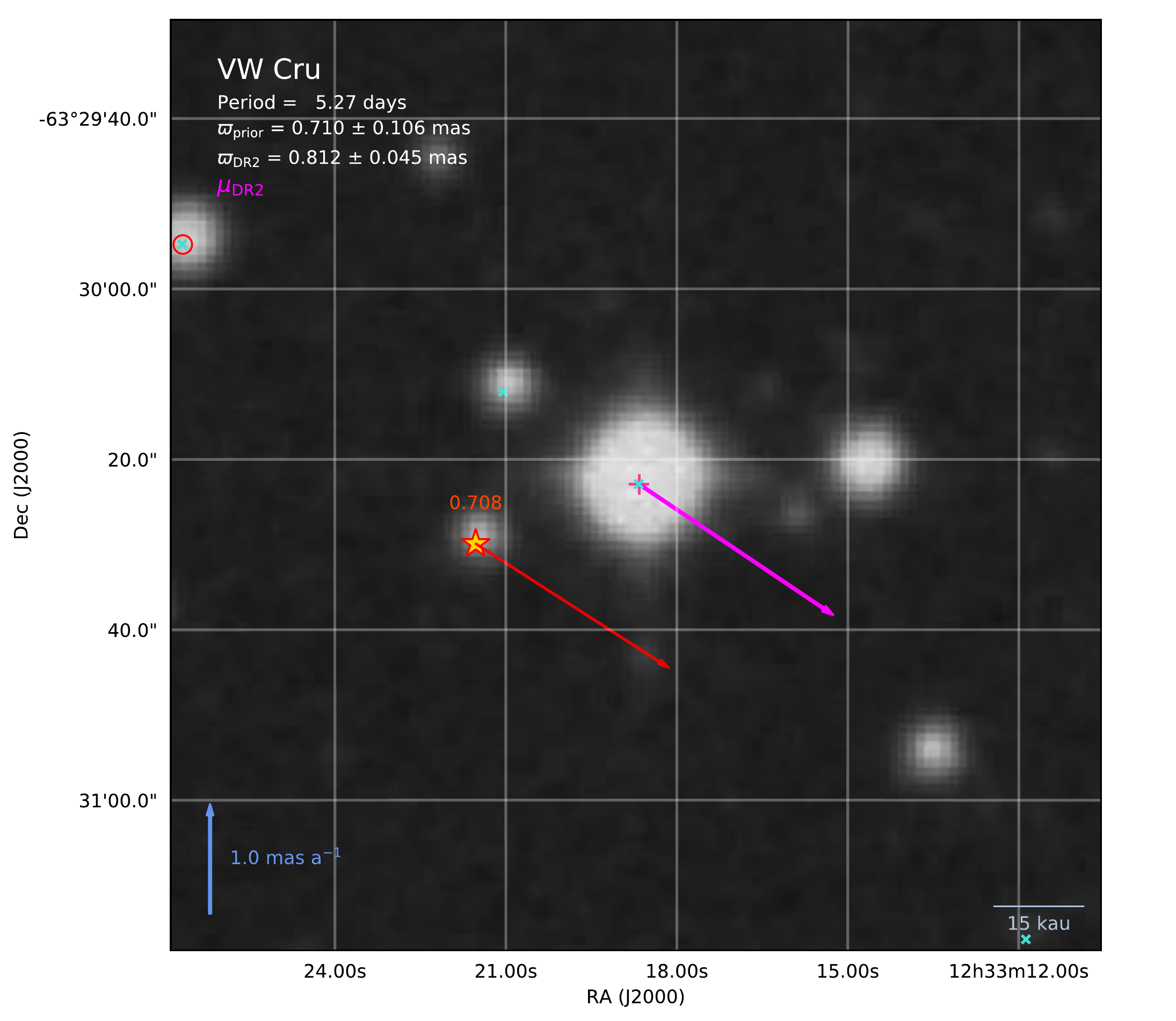}
\includegraphics[width=9cm]{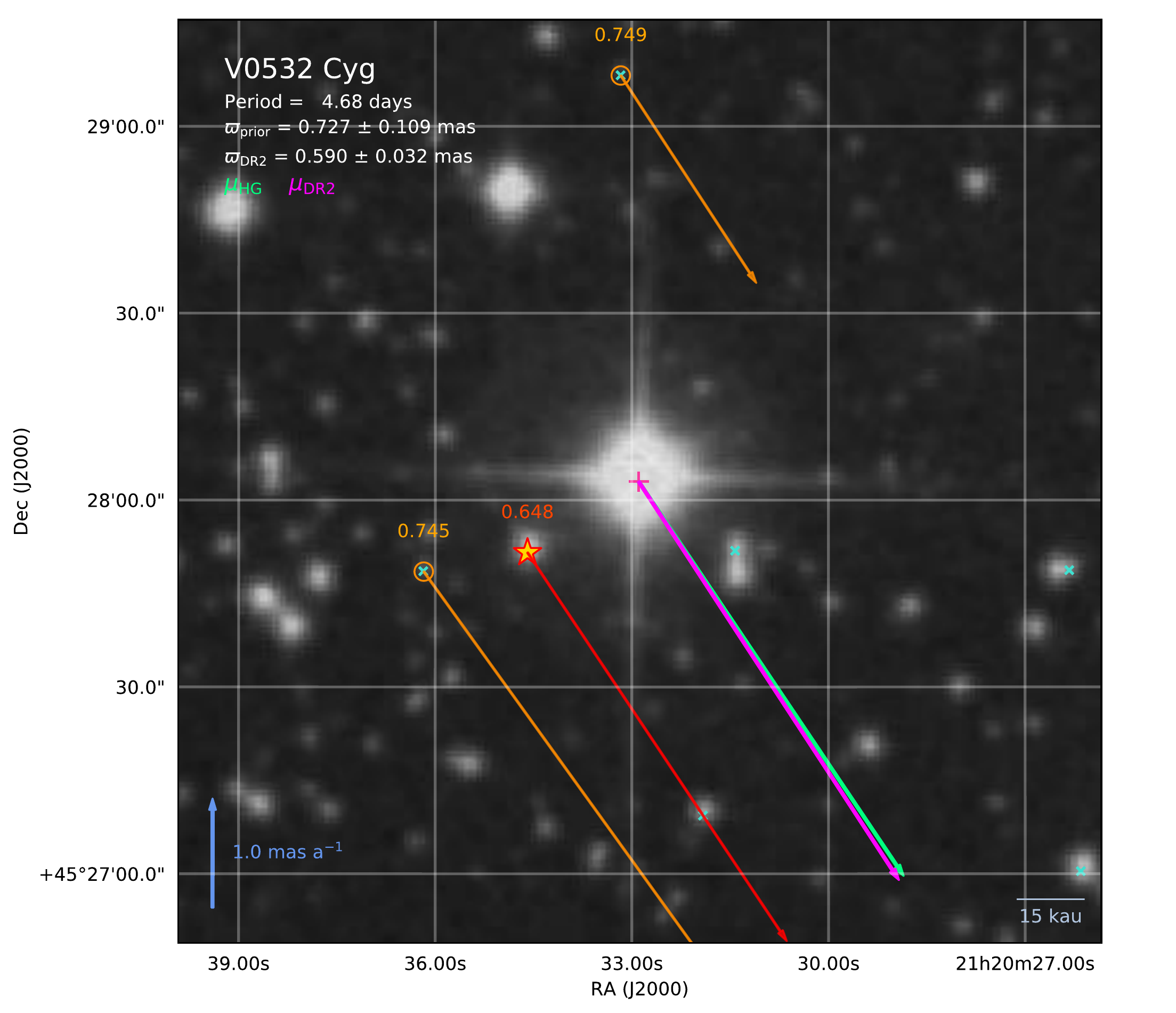}
\includegraphics[width=9cm]{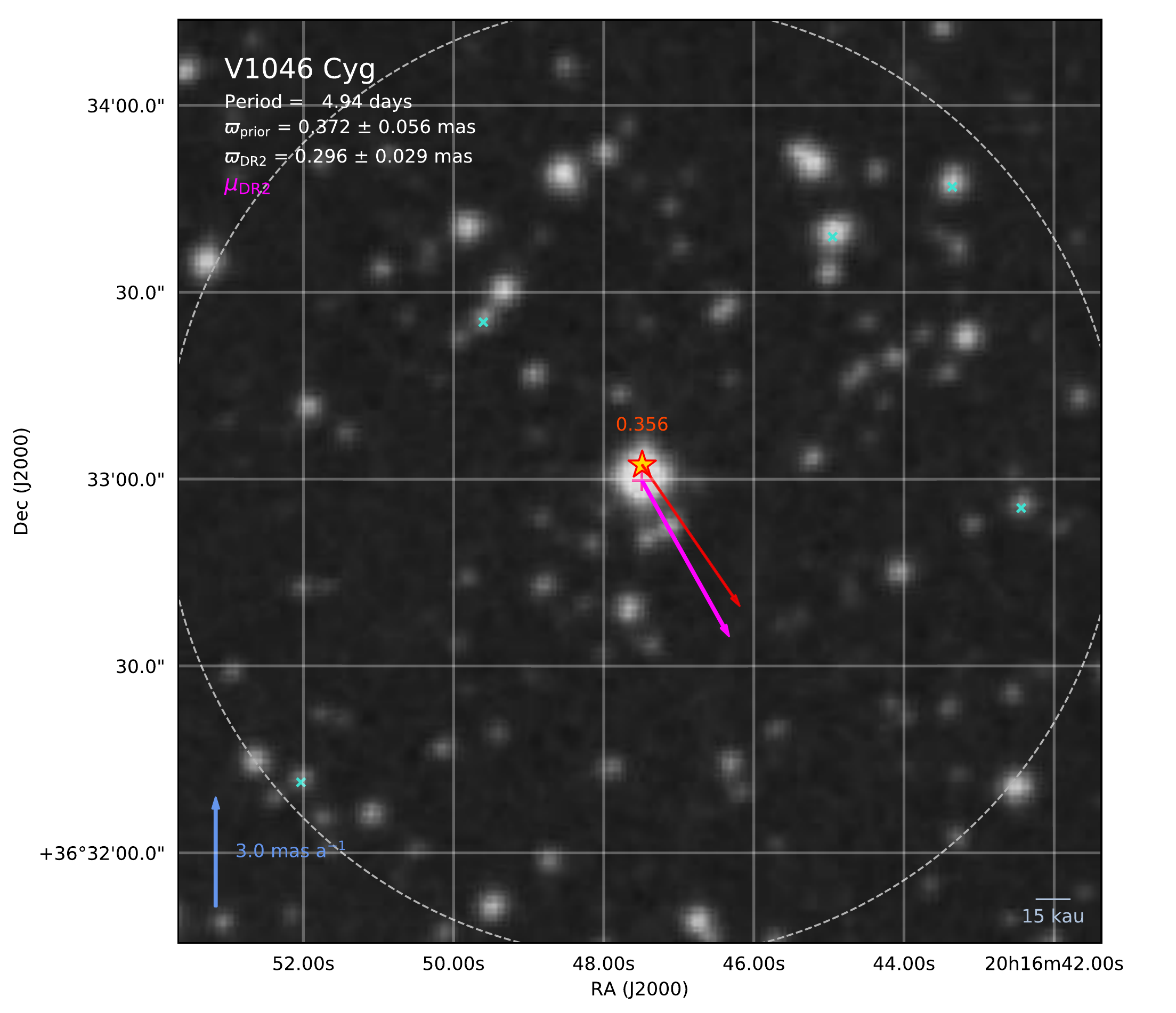}
\caption{Continued from Fig.~\ref{cepheid-field-1}.\label{cepheid-field-2}}
\end{figure*}

\begin{figure*}[]
\centering
\includegraphics[width=9cm]{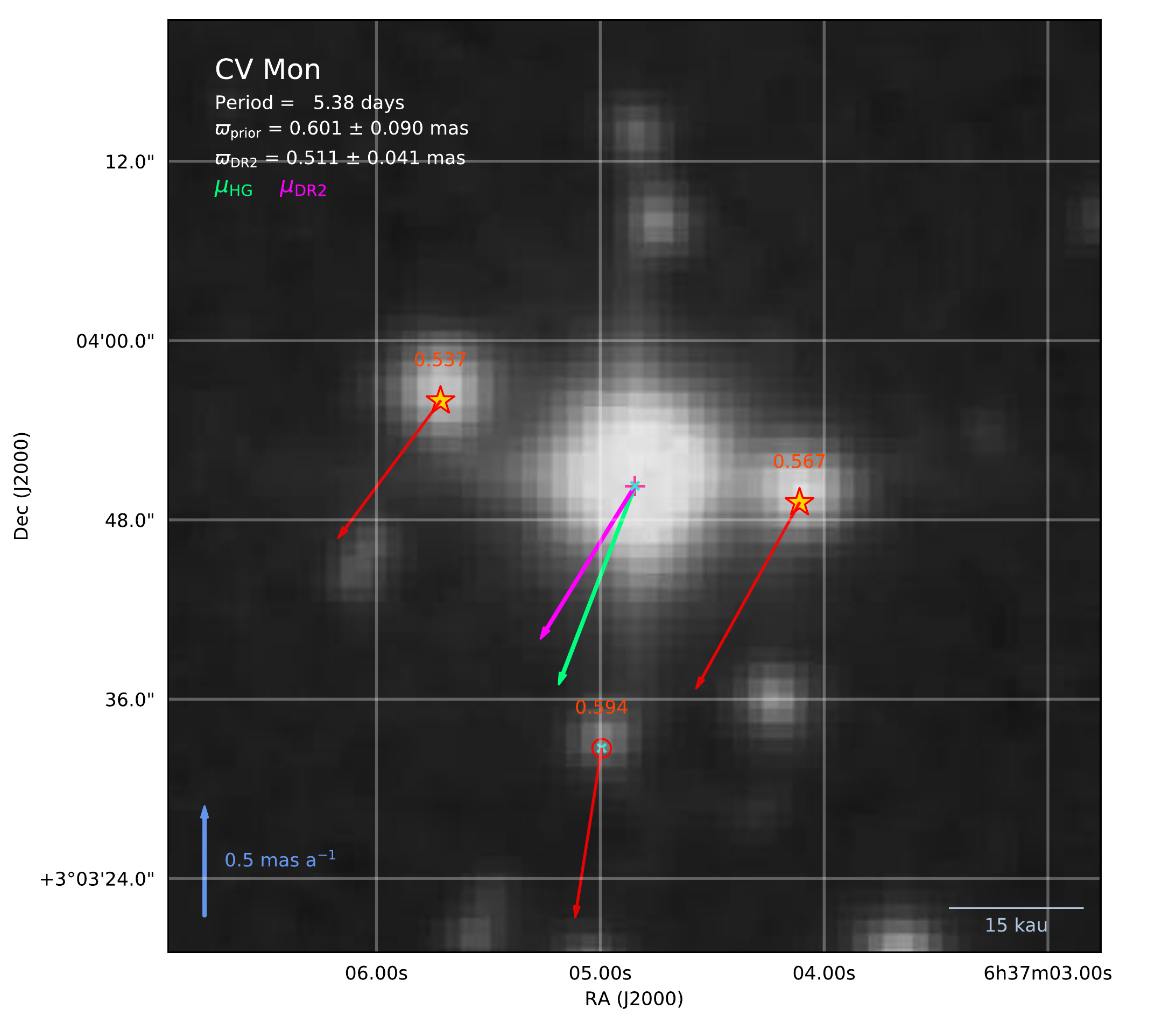}
\includegraphics[width=9cm]{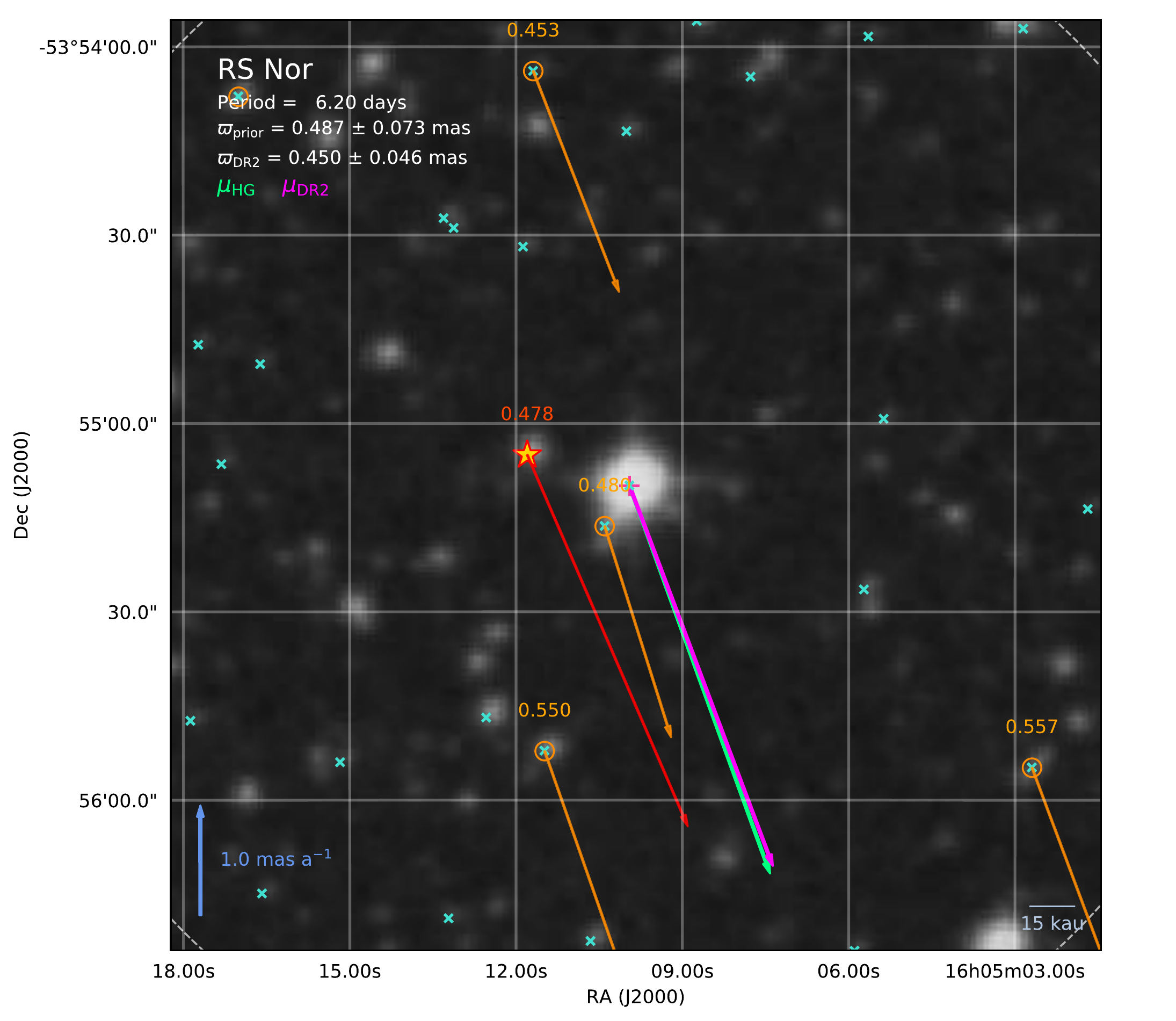}
\includegraphics[width=9cm]{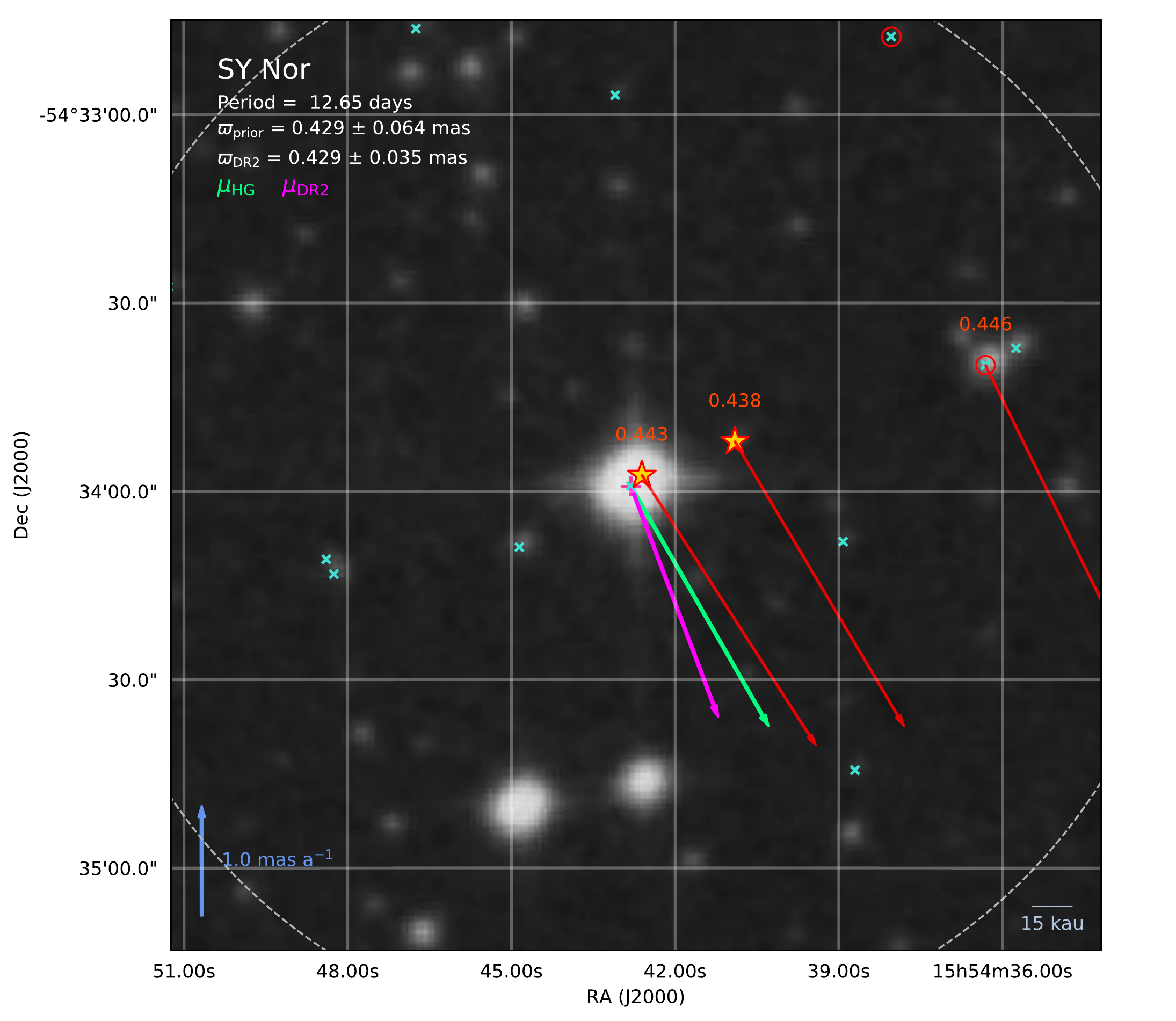}
\includegraphics[width=9cm]{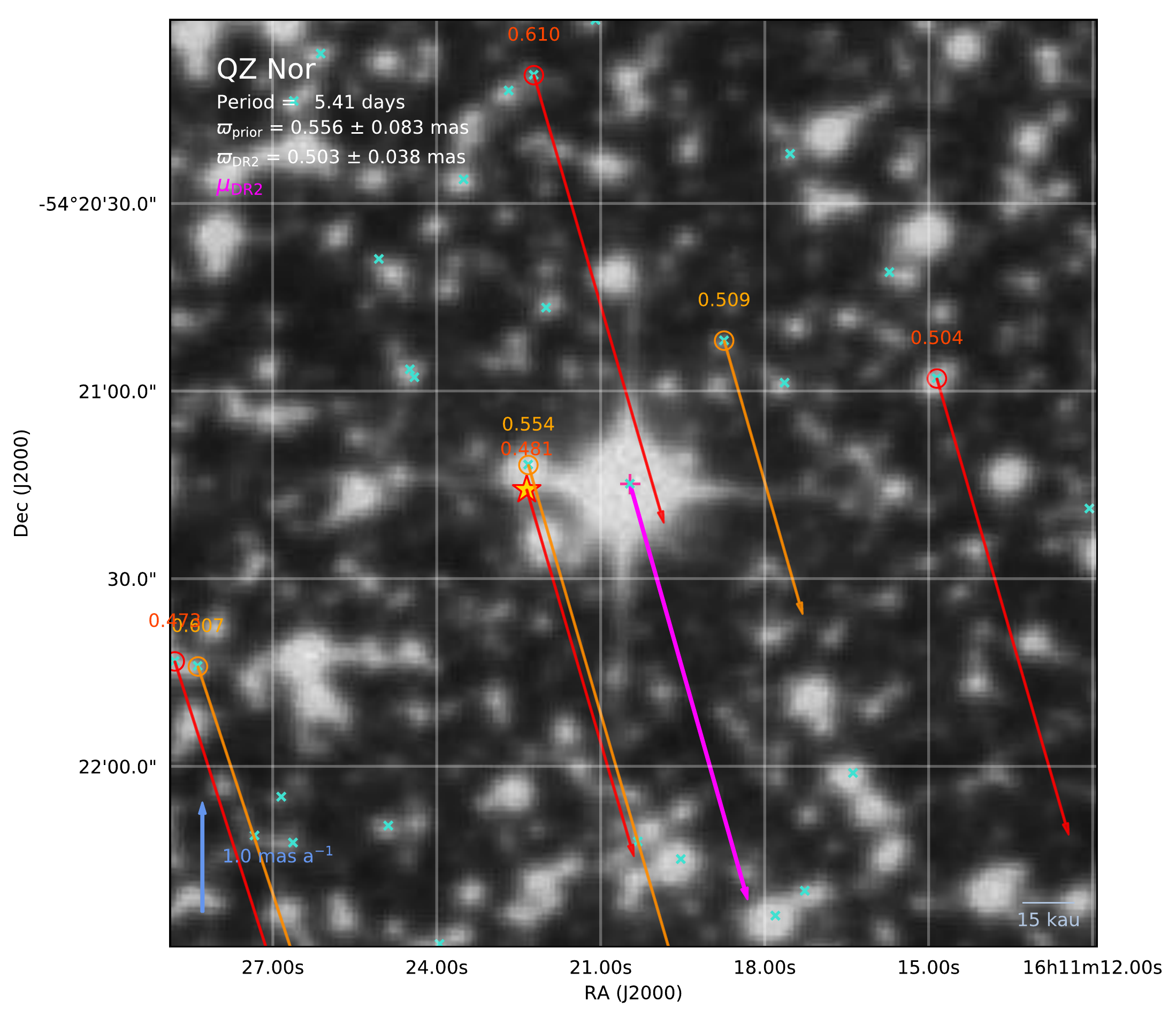}
\includegraphics[width=9cm]{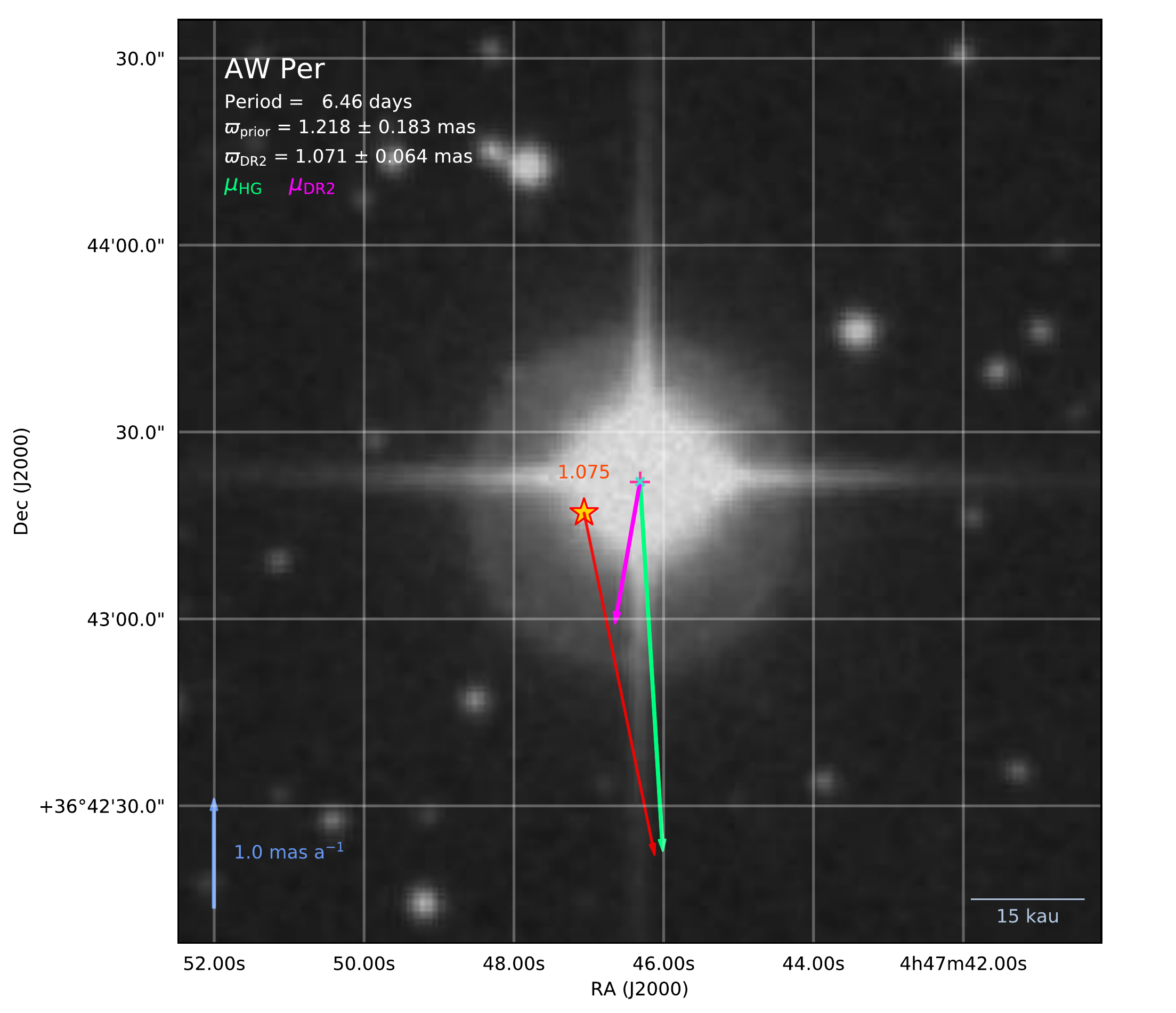}
\includegraphics[width=9cm]{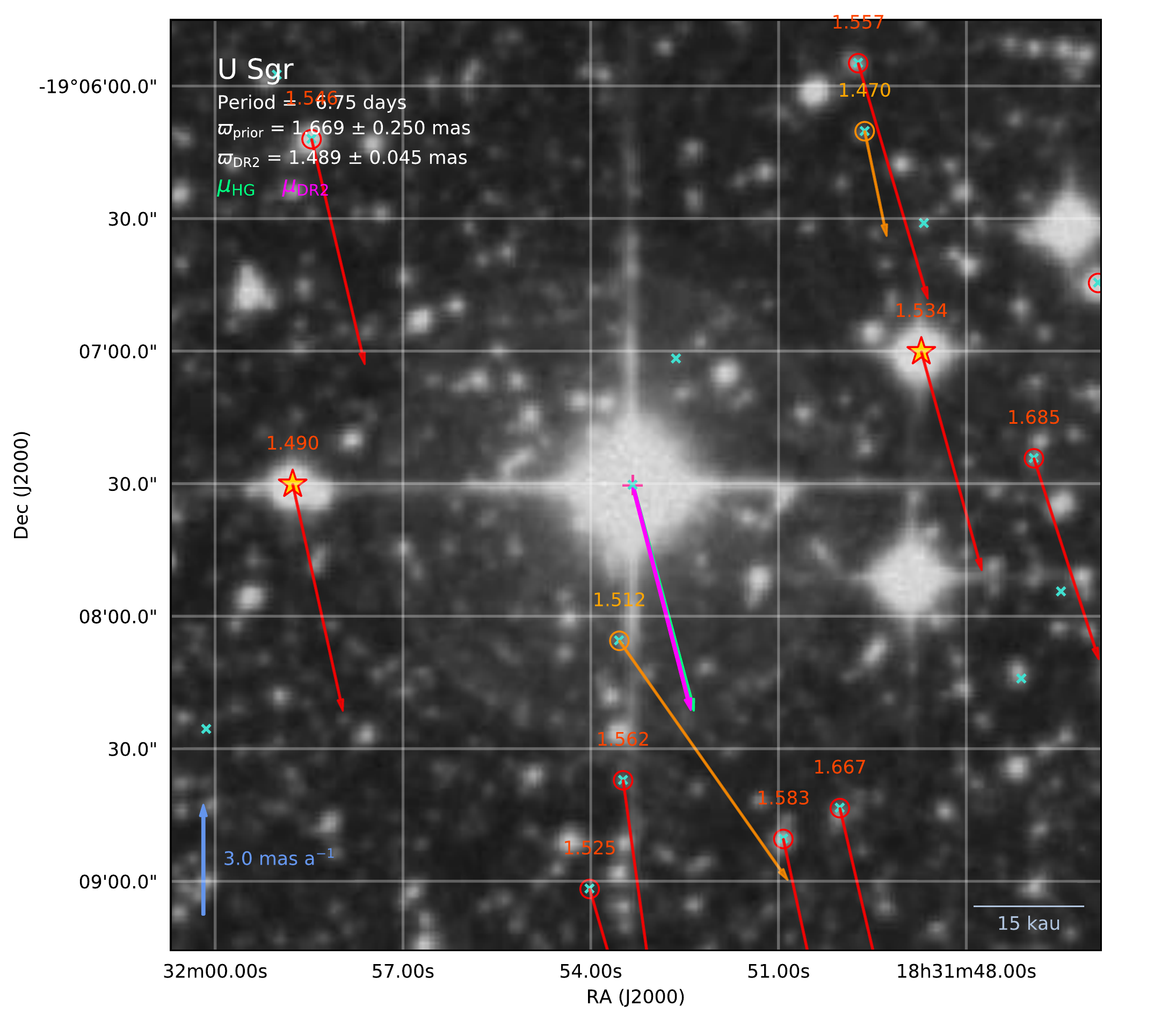}
\caption{Continued from Fig.~\ref{cepheid-field-2}.\label{cepheid-field-3}}
\end{figure*}

\begin{figure*}[]
\centering
\includegraphics[width=9cm]{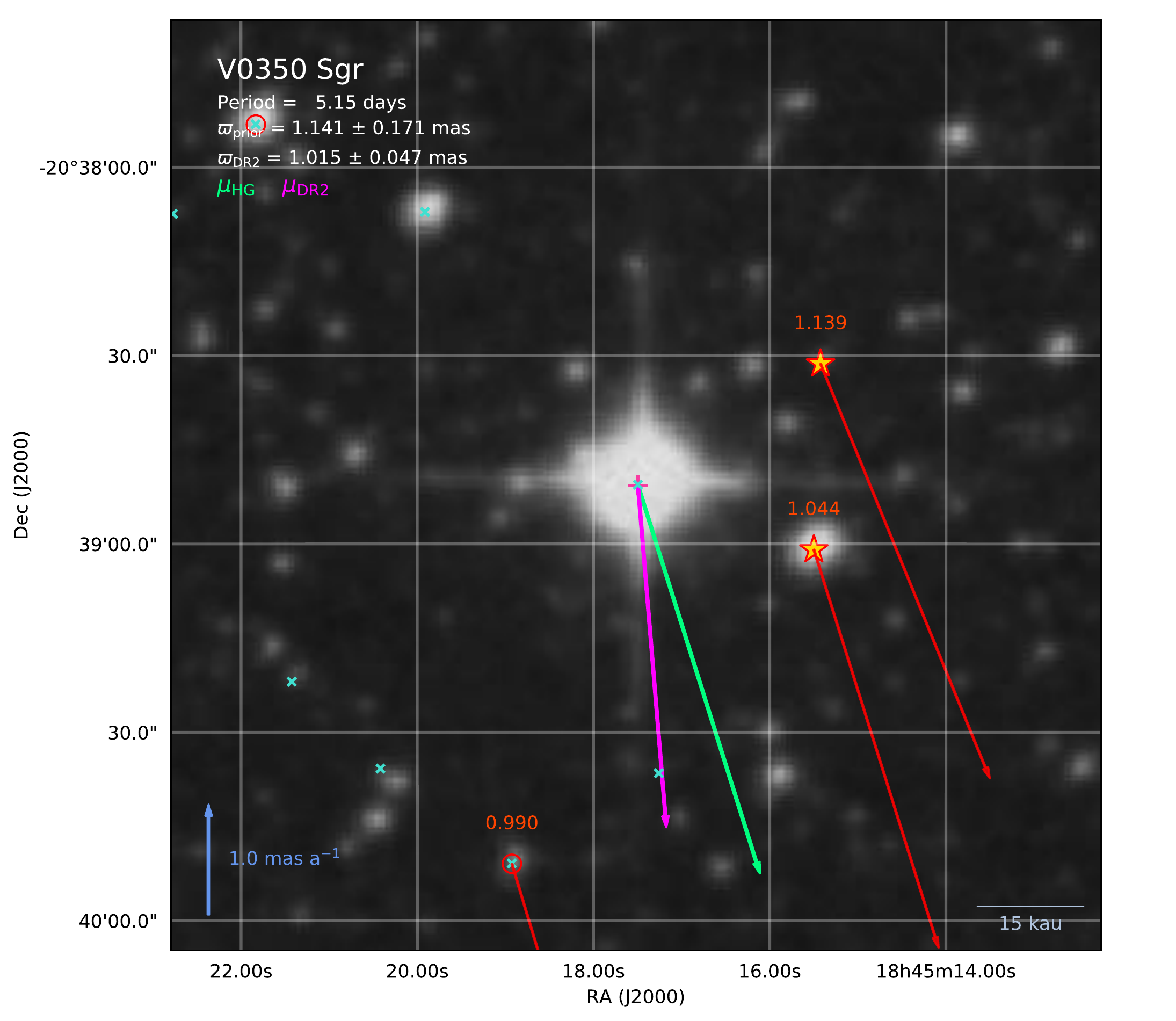}
\includegraphics[width=9cm]{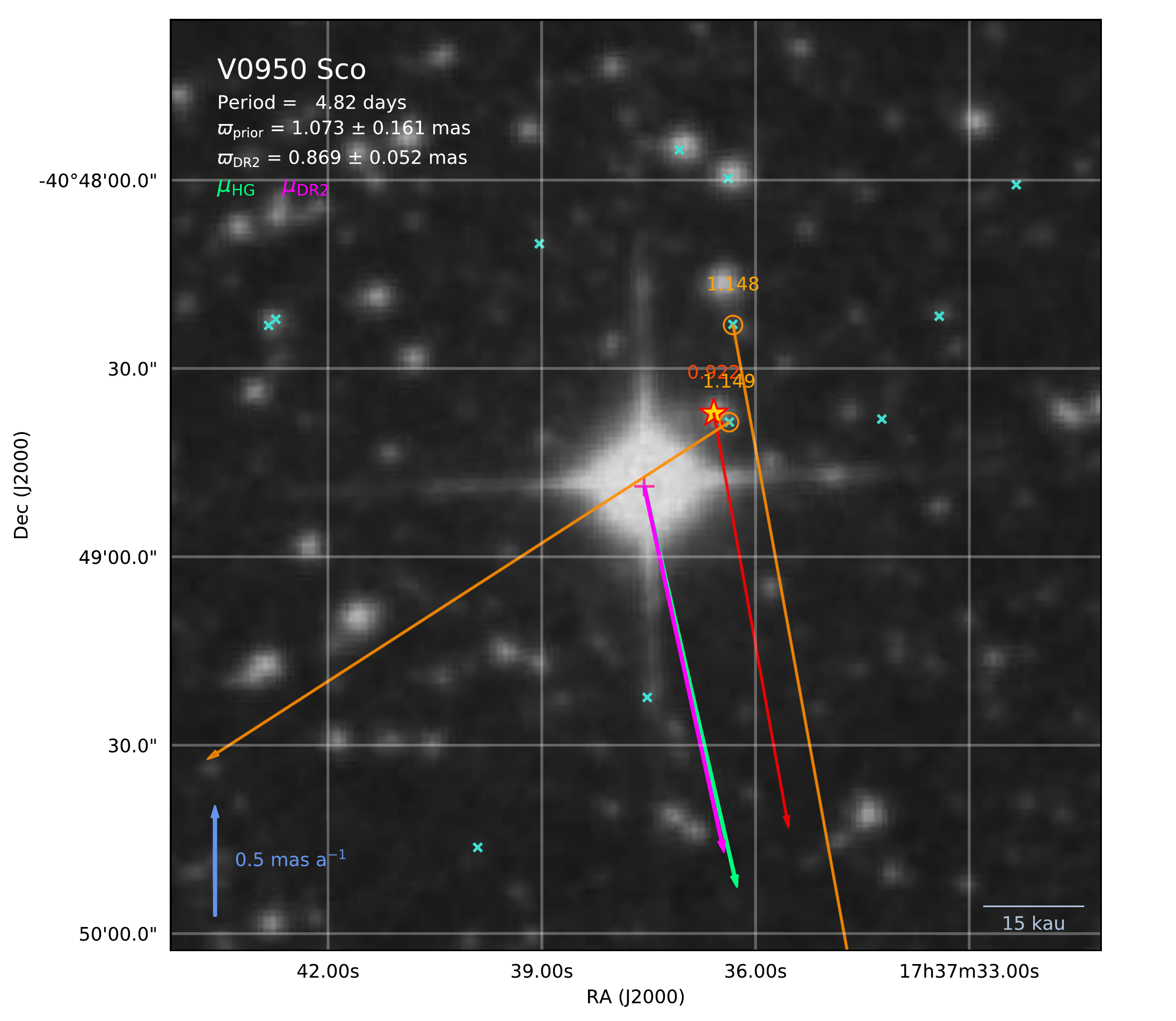}
\includegraphics[width=9cm]{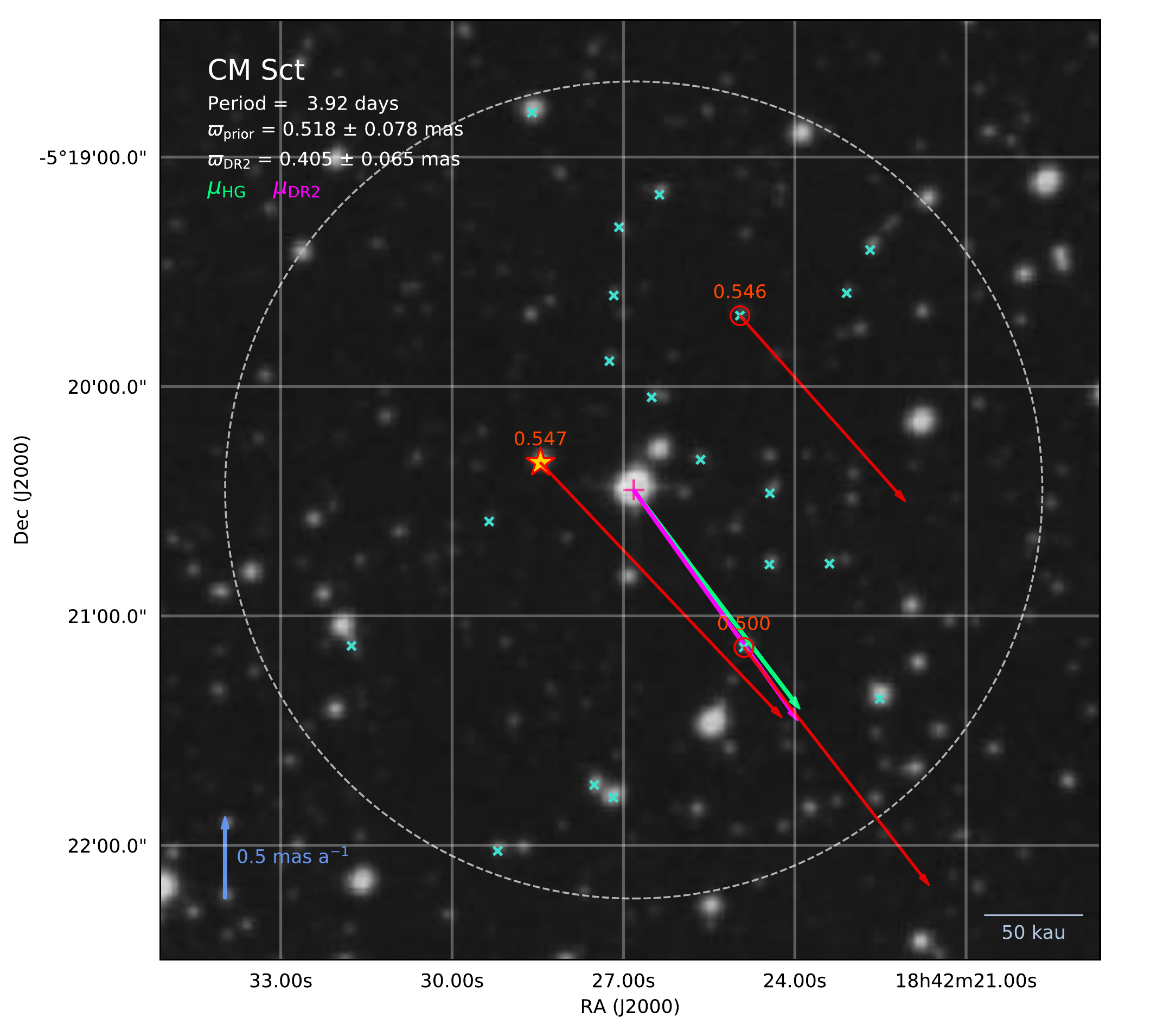}
\includegraphics[width=9cm]{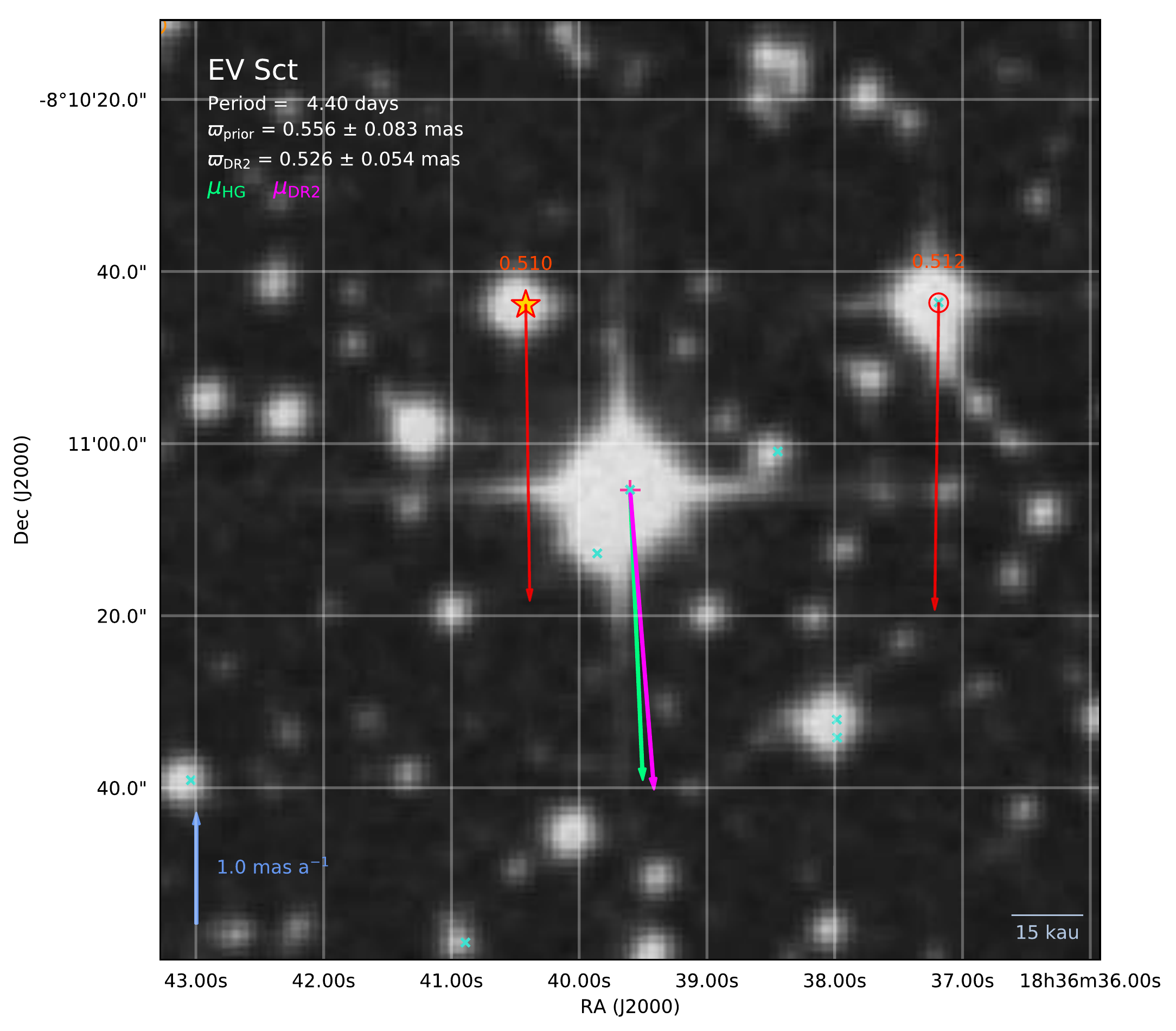}
\includegraphics[width=9cm]{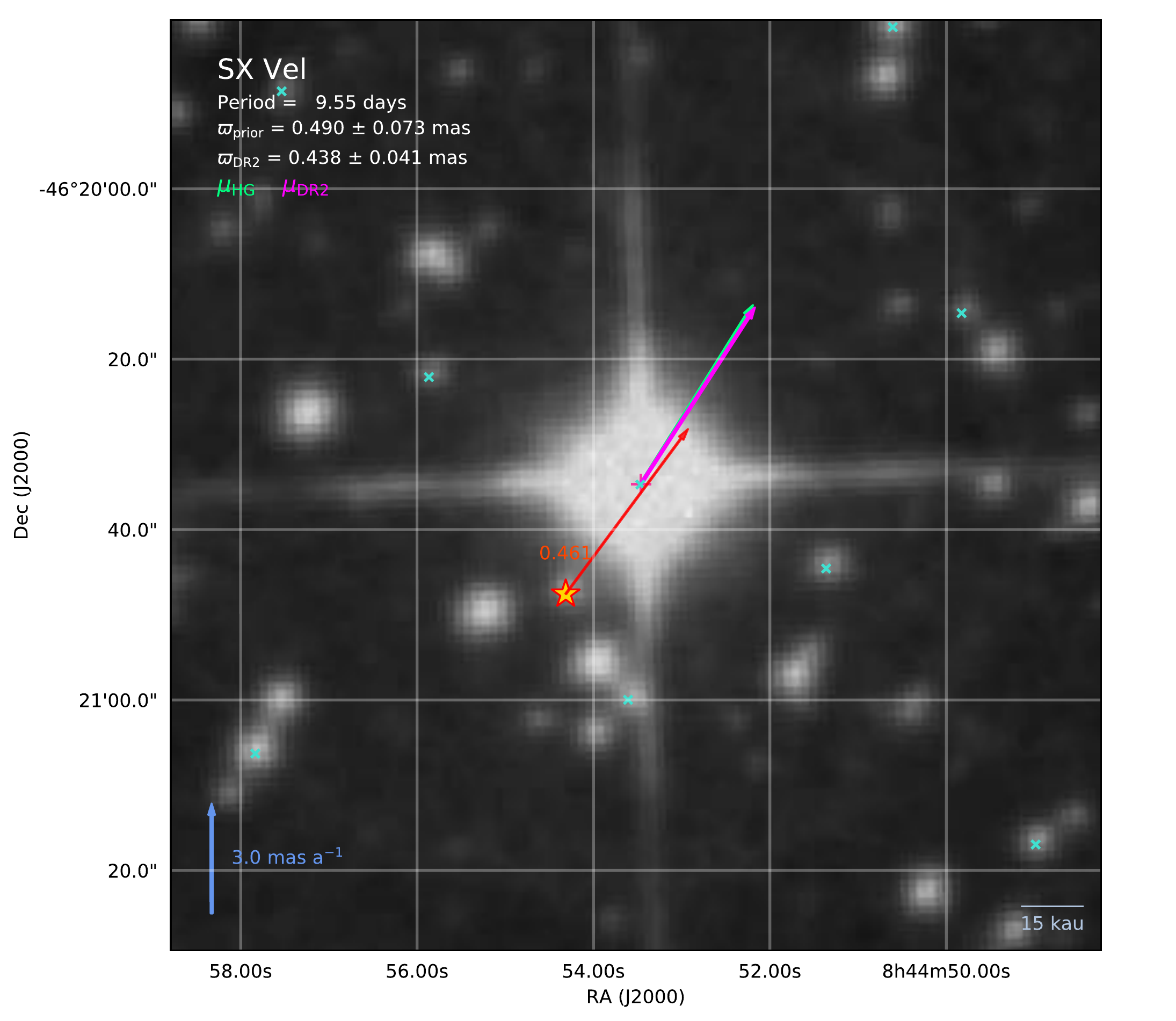}
\includegraphics[width=9cm]{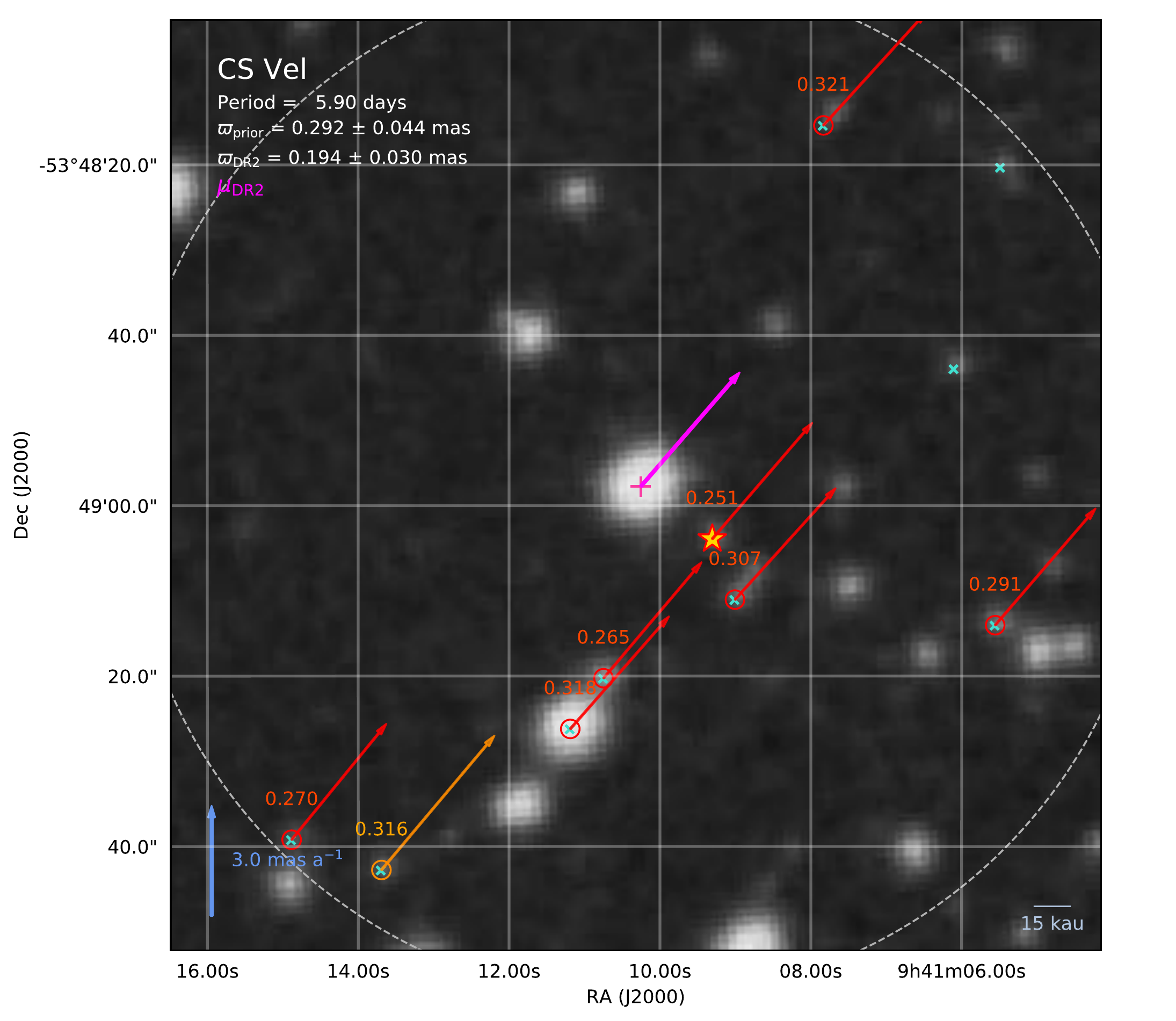}
\caption{Continued from Fig.~\ref{cepheid-field-3}.\label{cepheid-field-4}}
\end{figure*}

\begin{figure}[]
\centering
\includegraphics[width=9cm]{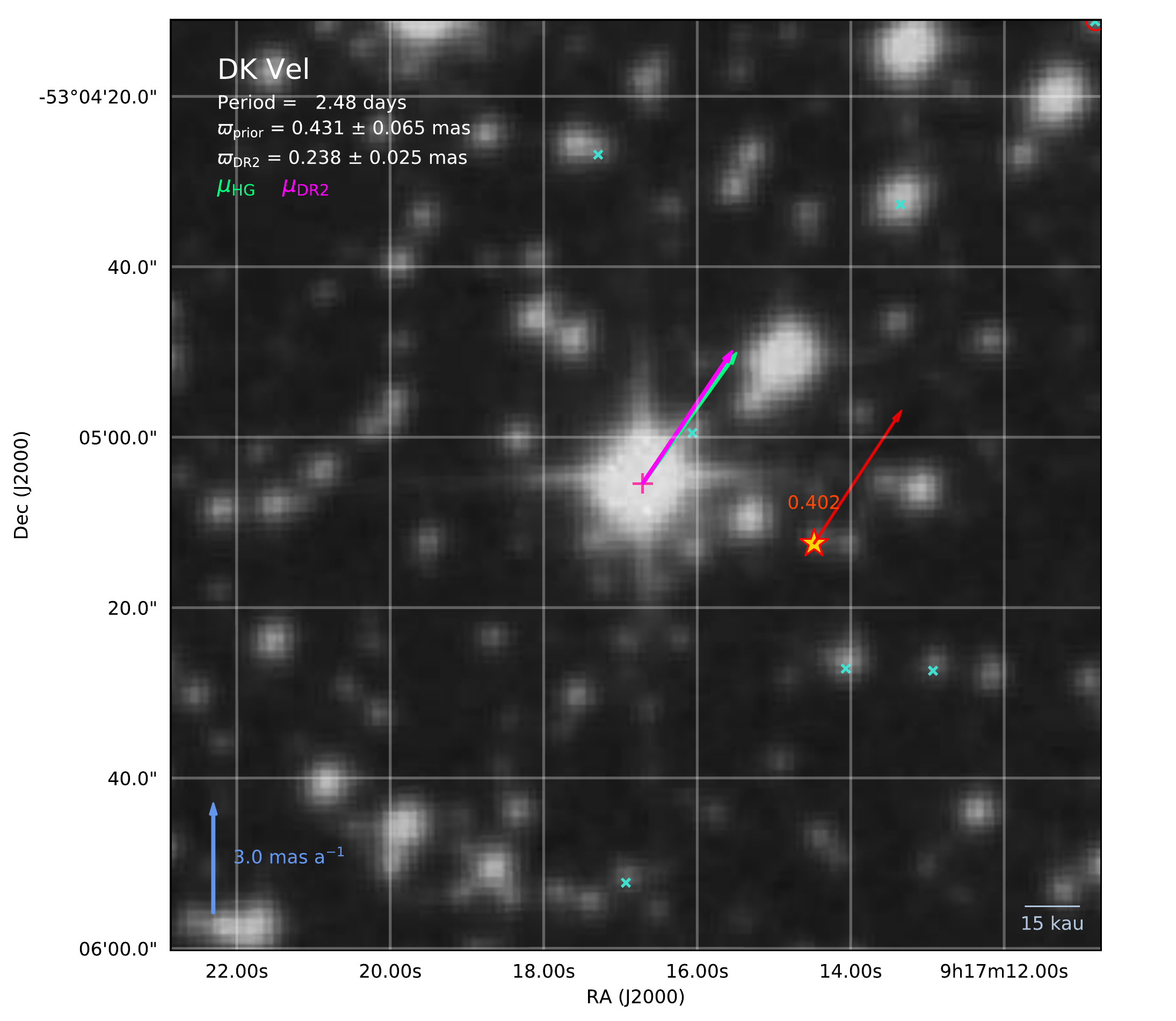}
\caption{Continued from Fig.~\ref{cepheid-field-4}.\label{cepheid-field-5}}
\end{figure}

\begin{figure*}[]
\centering
\includegraphics[width=9cm]{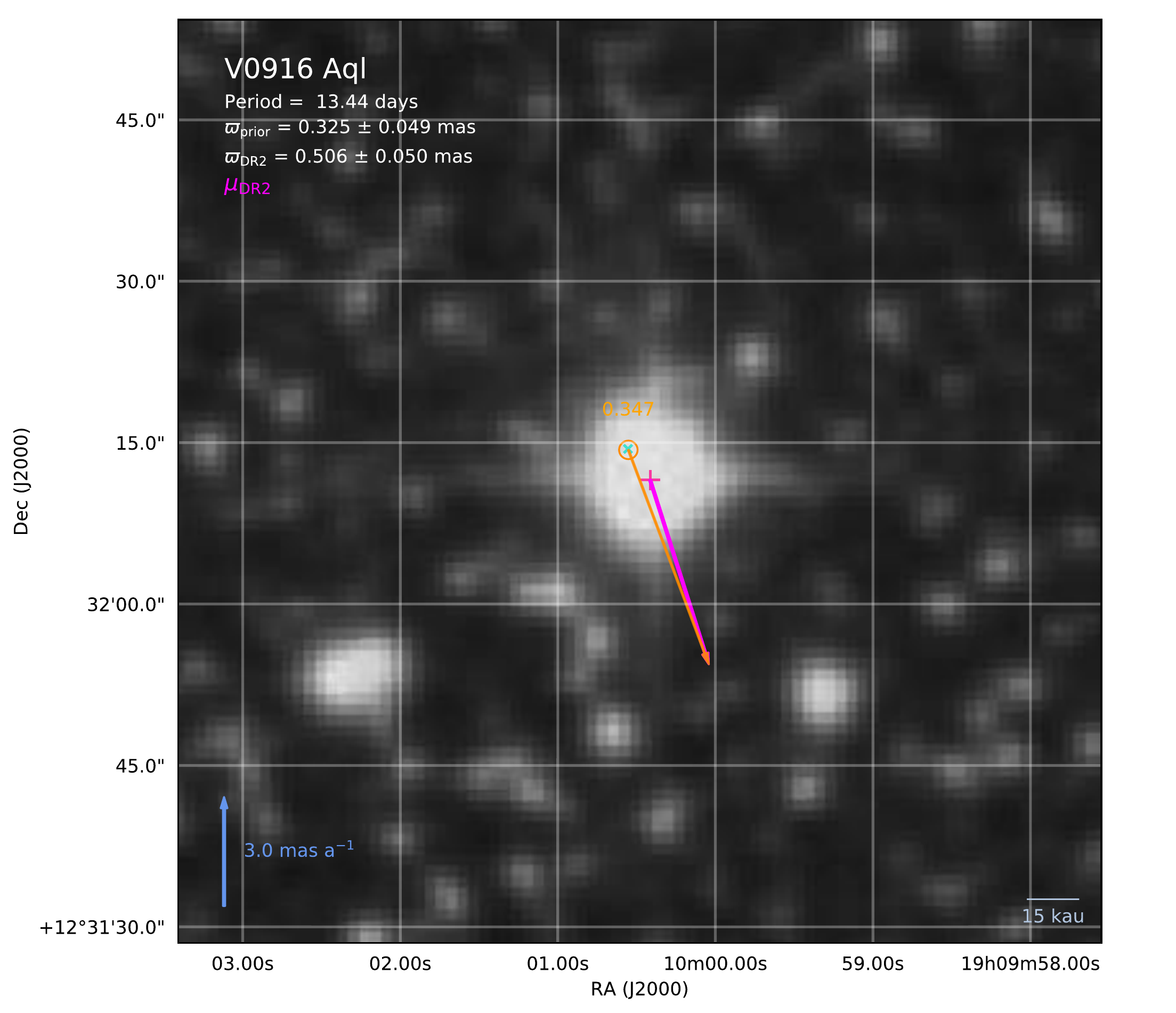}
\includegraphics[width=9cm]{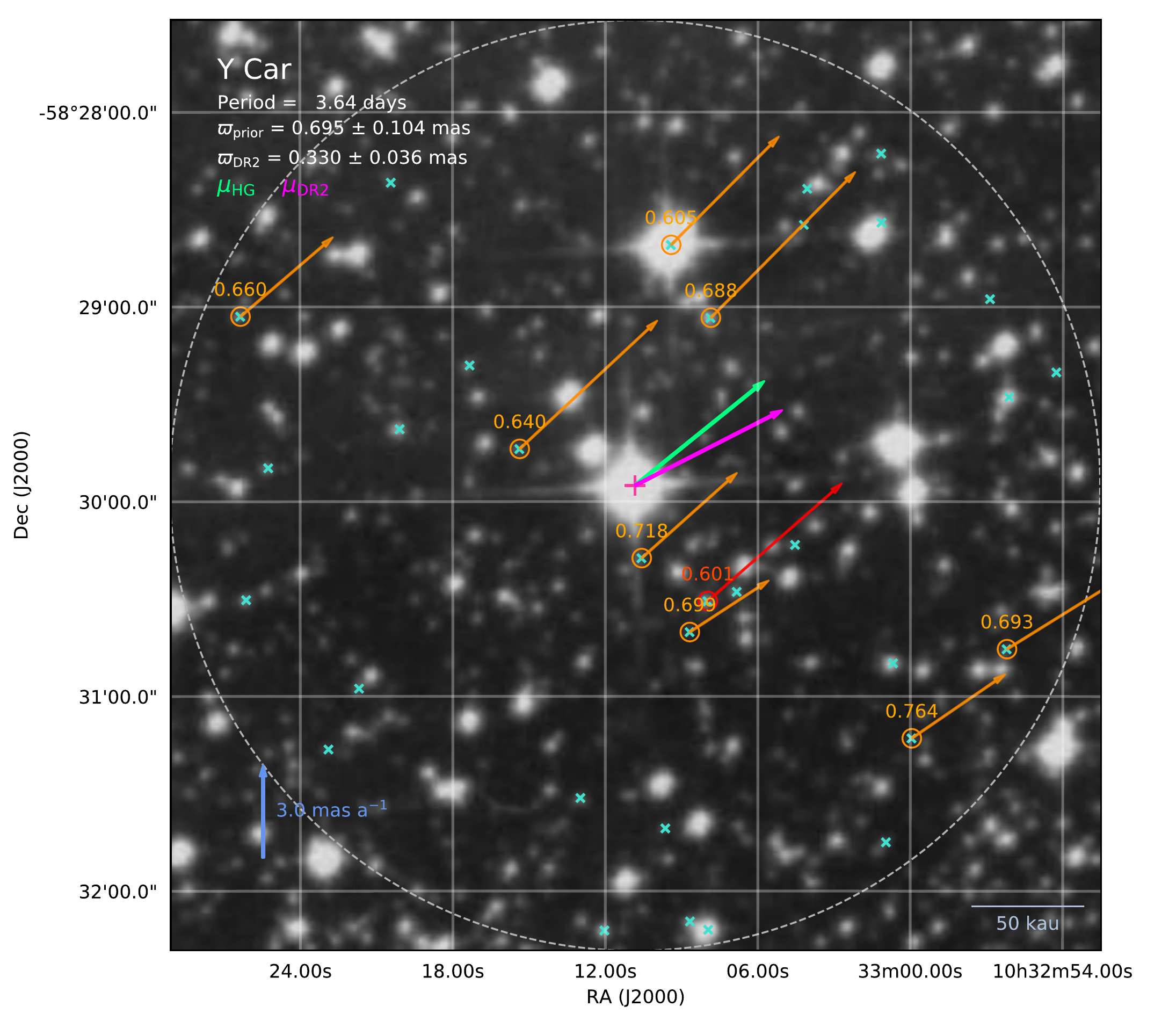}
\includegraphics[width=9cm]{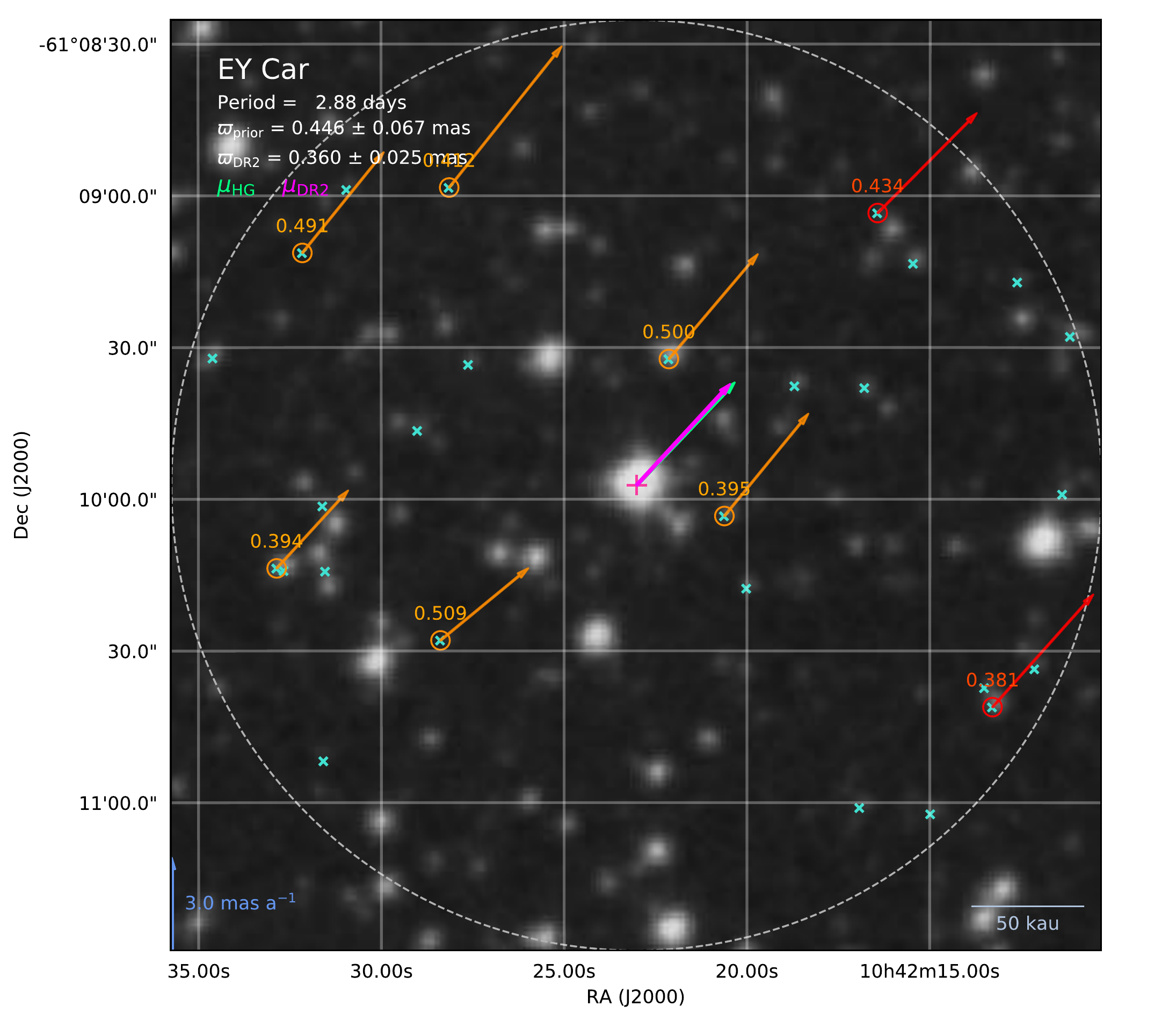}
\includegraphics[width=9cm]{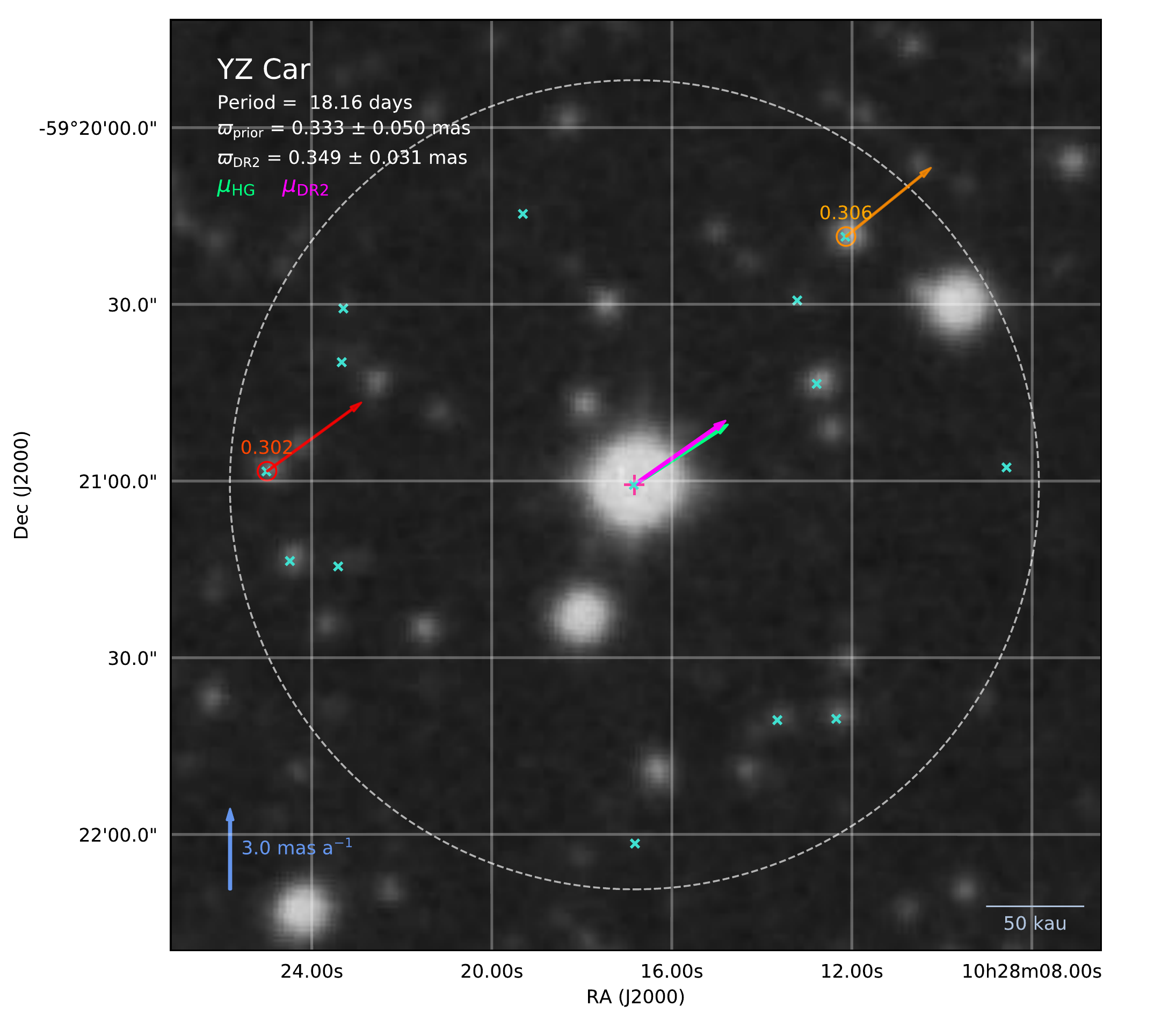}
\includegraphics[width=9cm]{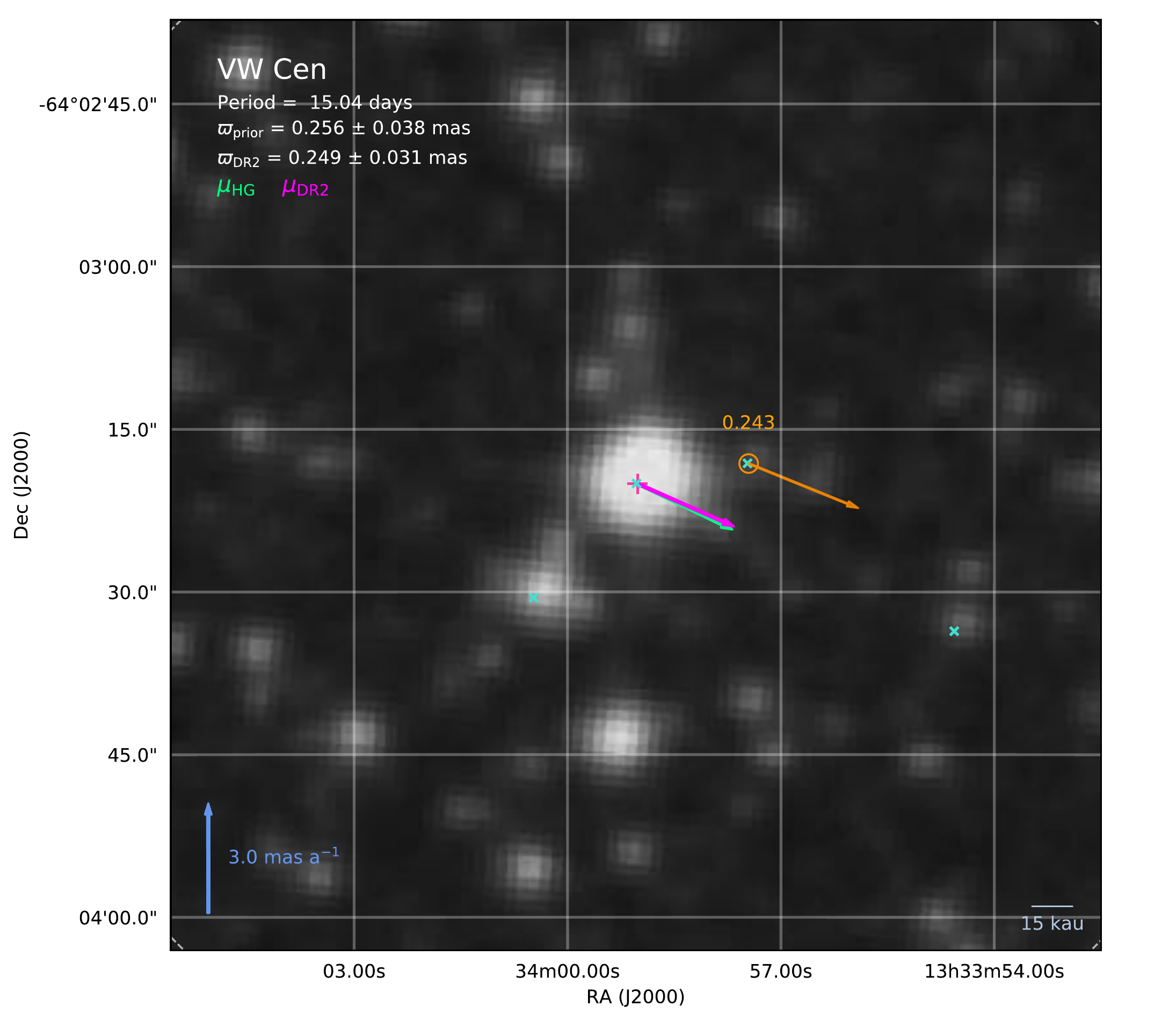}
\includegraphics[width=9cm]{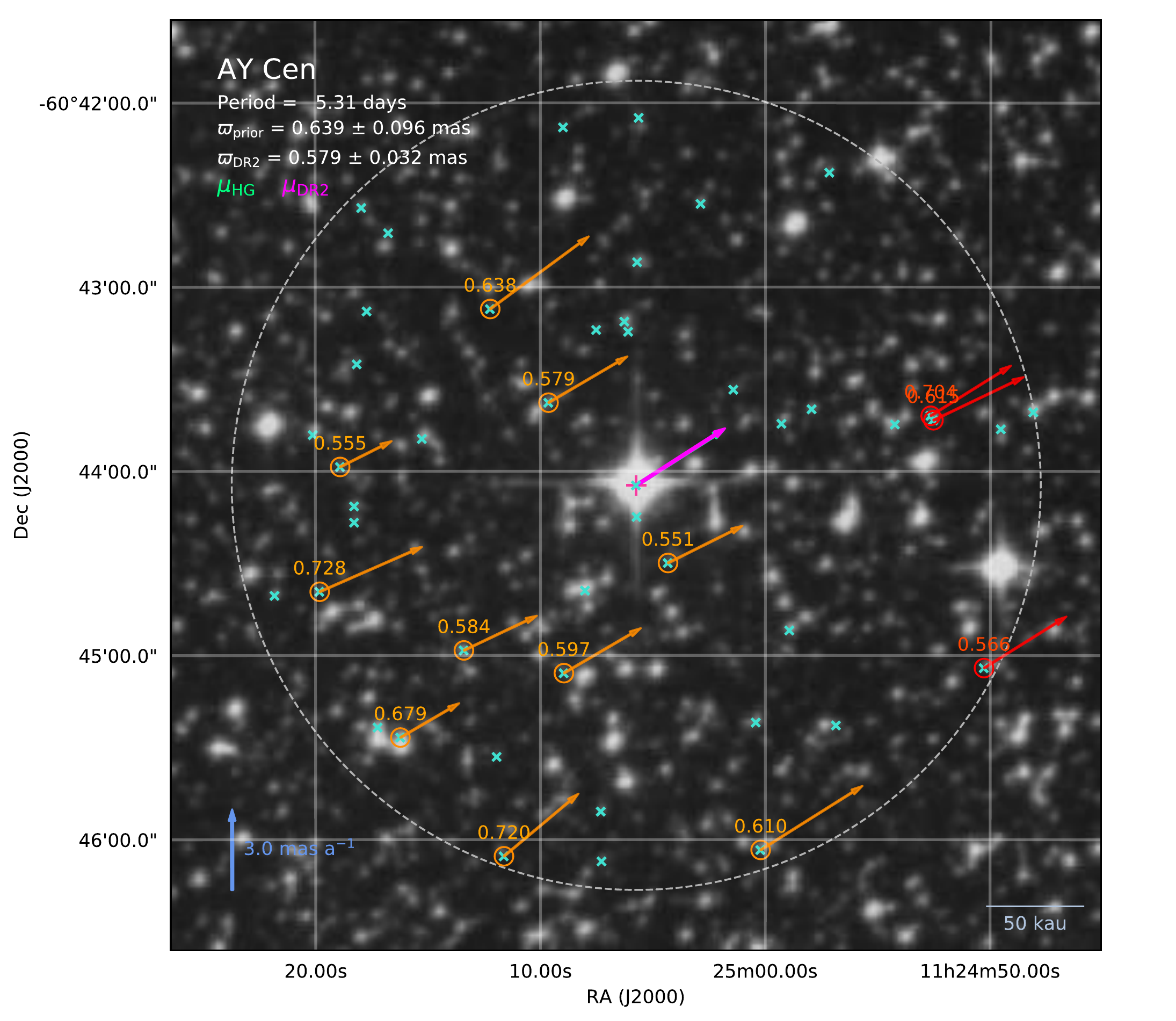}
\caption{Fields around selected Cepheids with \texttt{Near} candidate companions.\label{cepheids-near1}}
\end{figure*}

\begin{figure*}[]
\centering
\includegraphics[width=9cm]{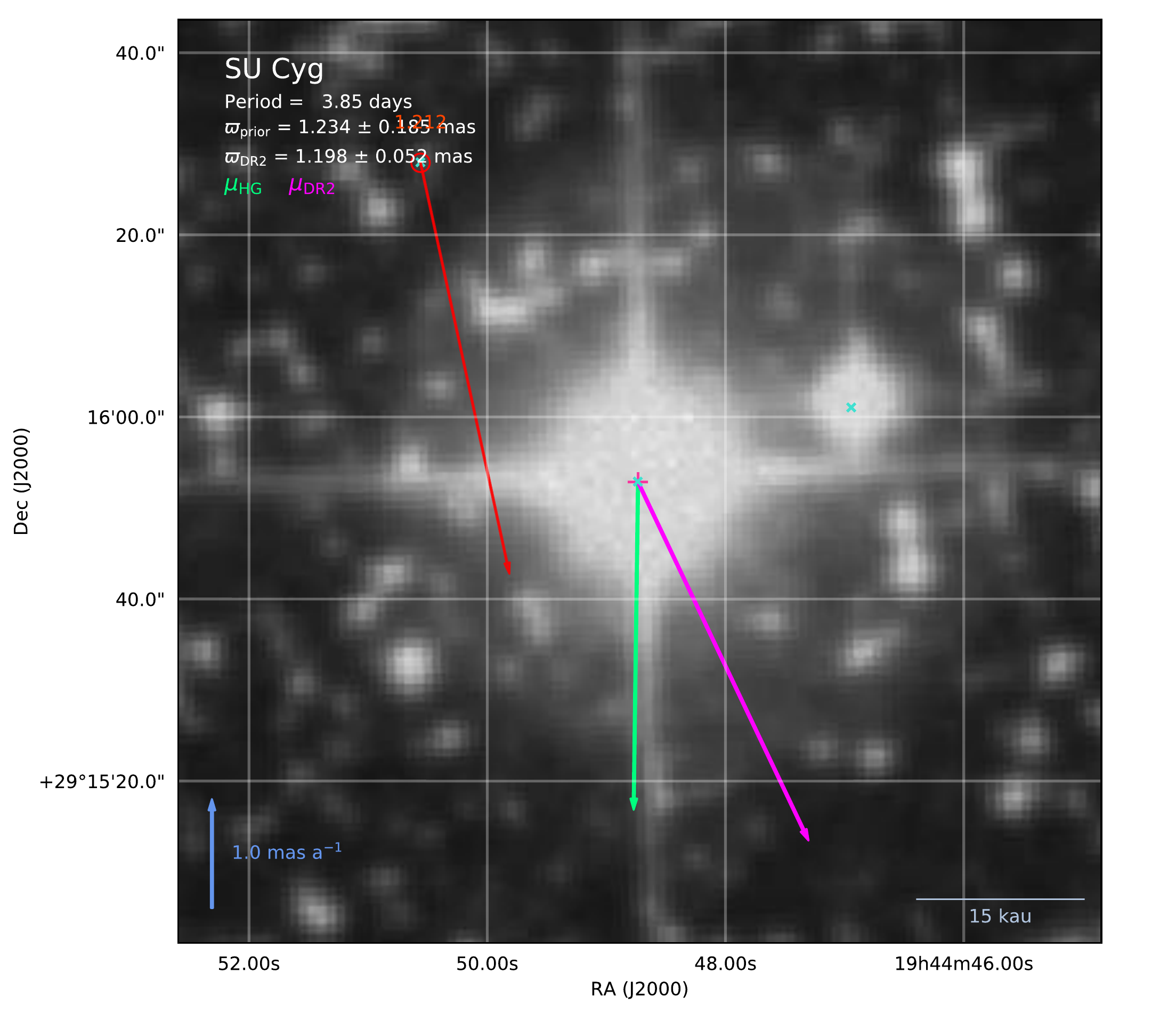}
\includegraphics[width=9cm]{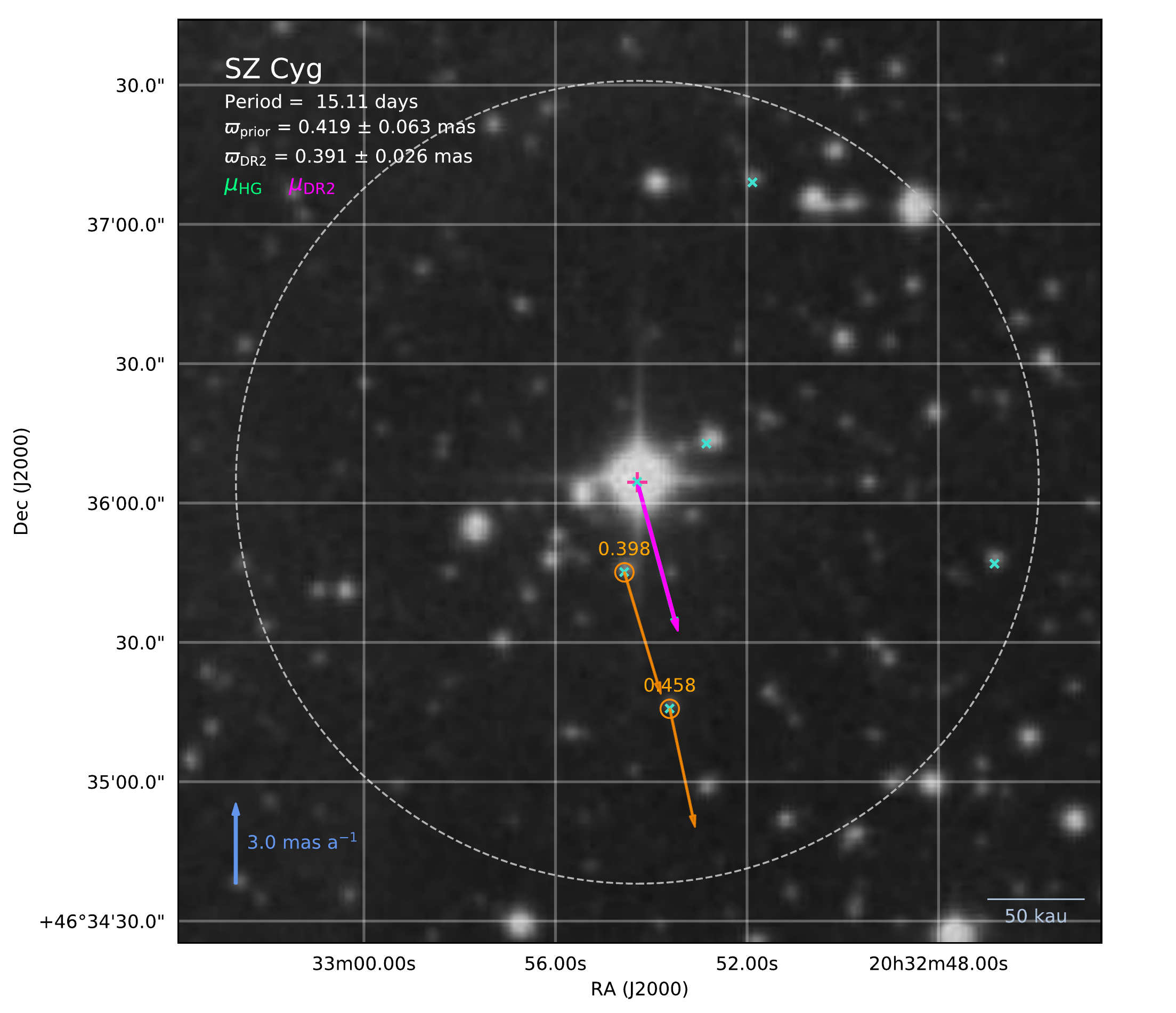}
\includegraphics[width=9cm]{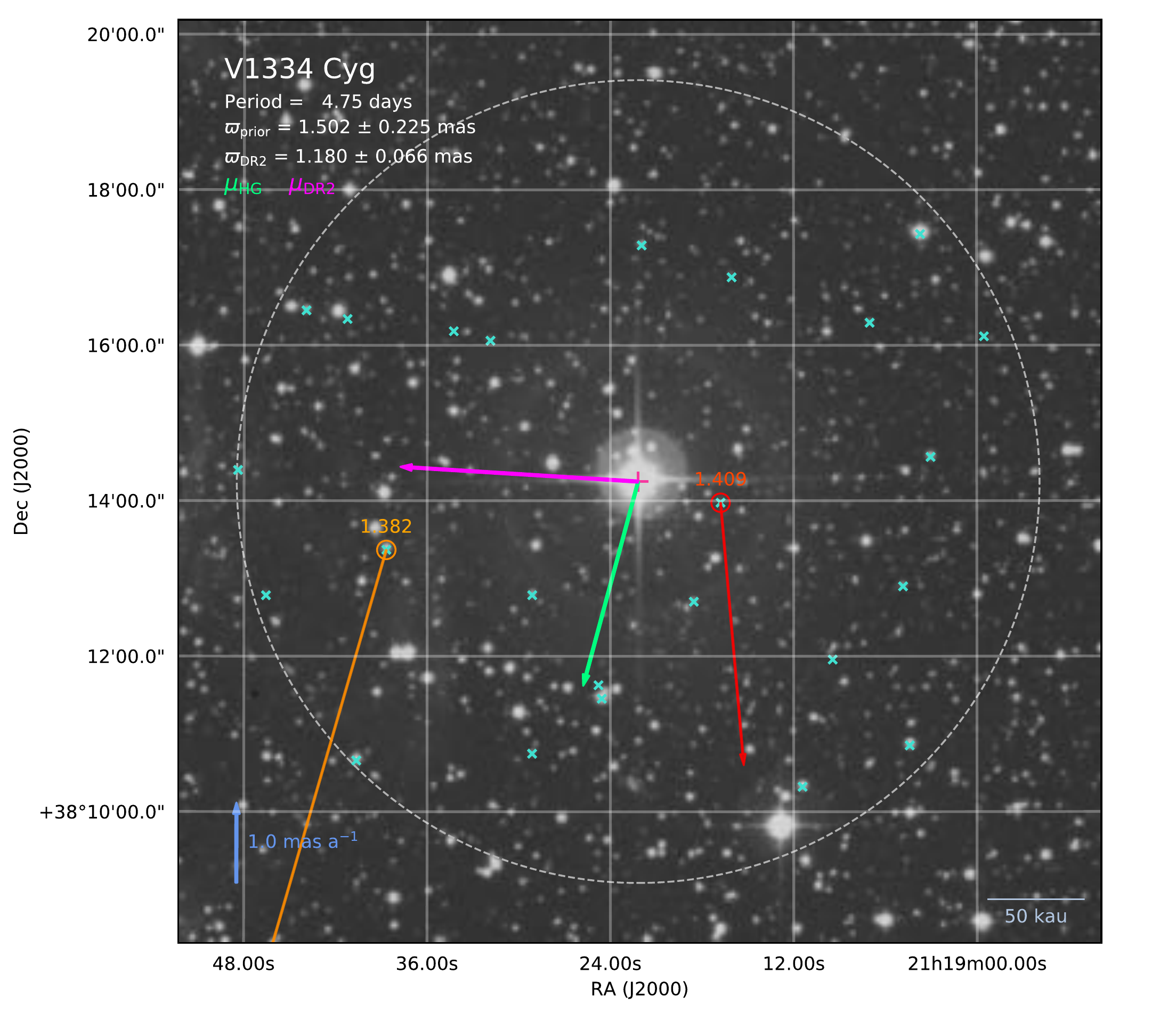}
\includegraphics[width=9cm]{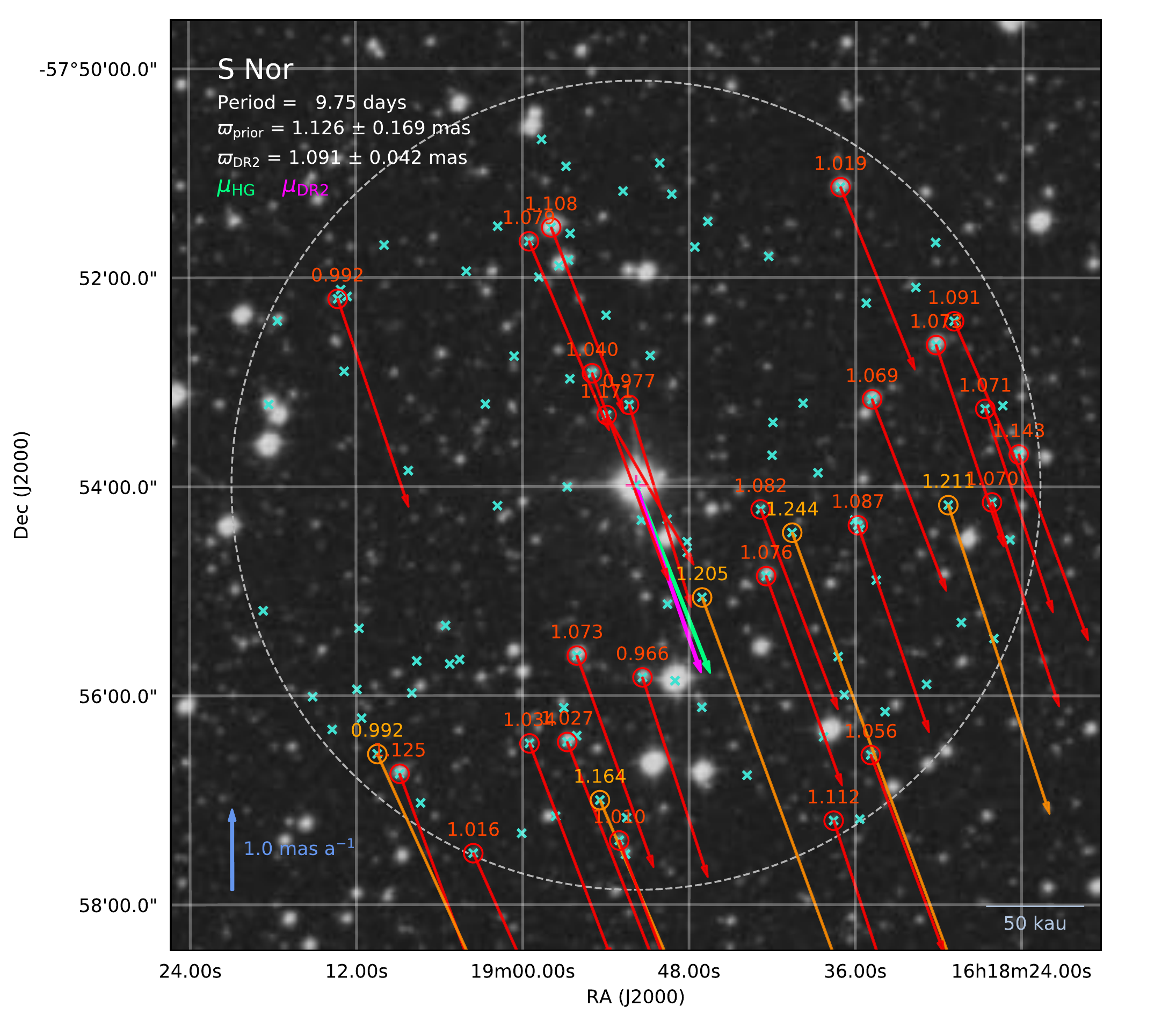}
\includegraphics[width=9cm]{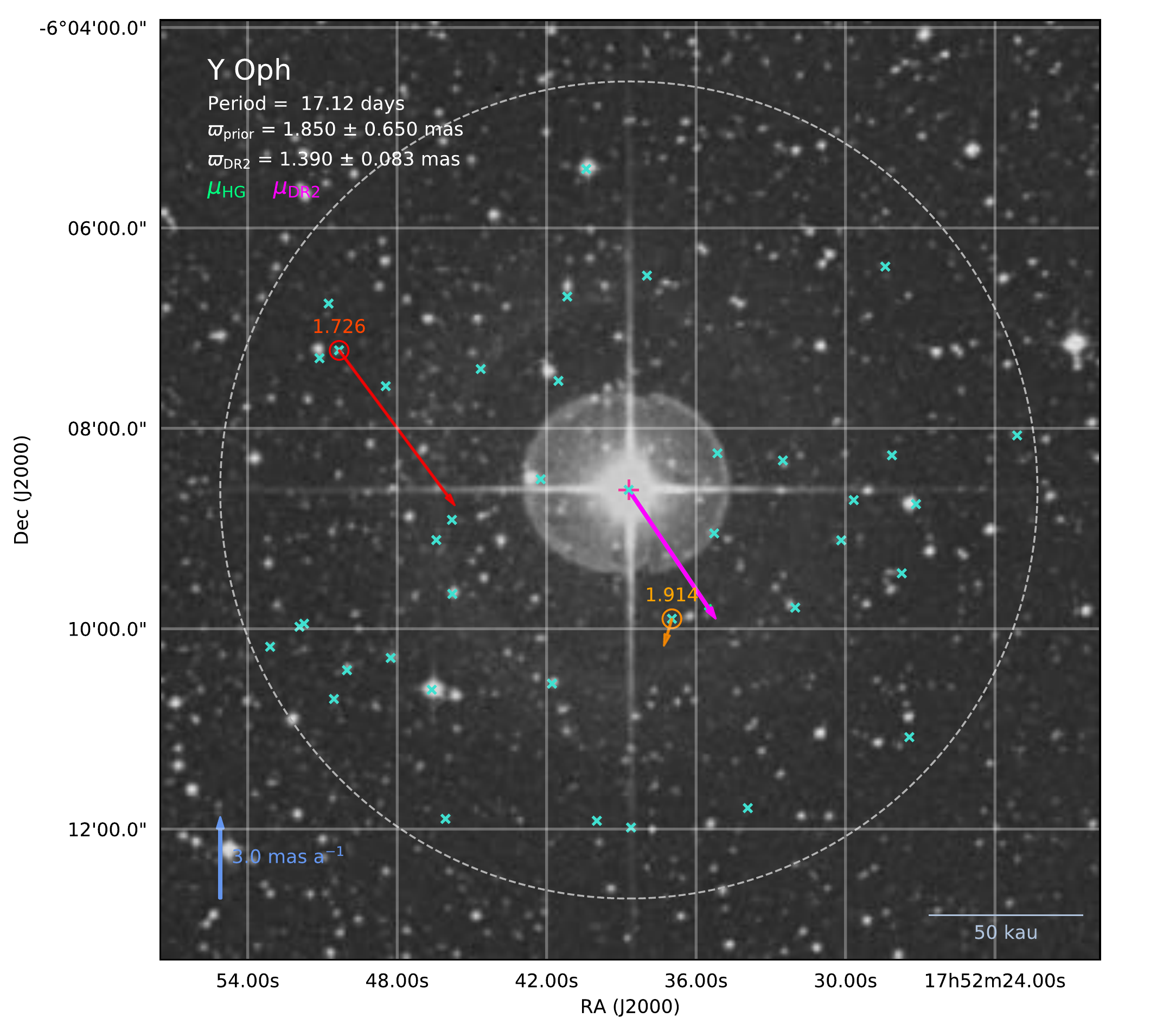}
\includegraphics[width=9cm]{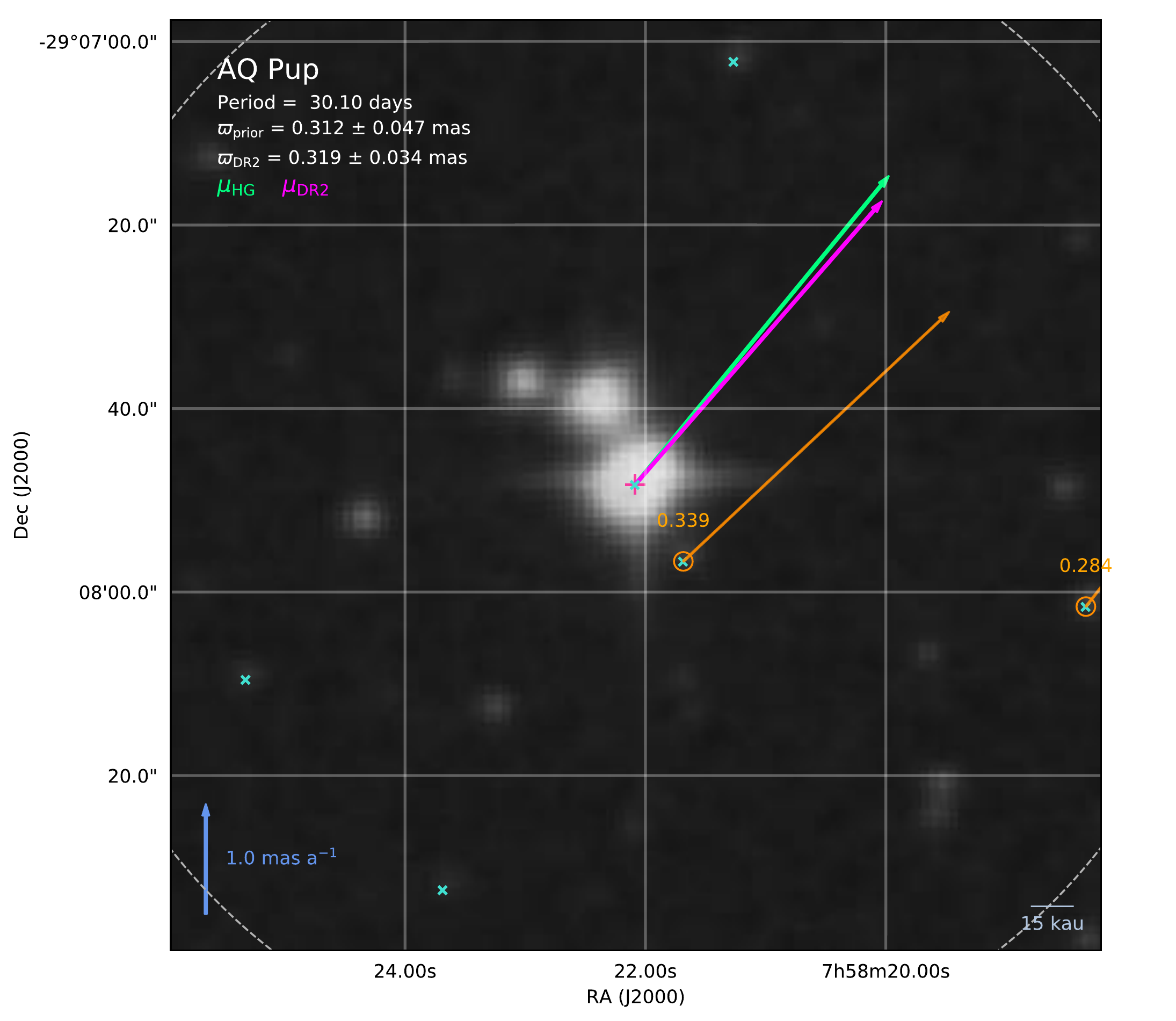}
\caption{Continued from Fig.~\ref{cepheids-near1}.\label{cepheids-near2}}
\end{figure*}

\begin{figure*}[]
\centering
\includegraphics[width=9cm]{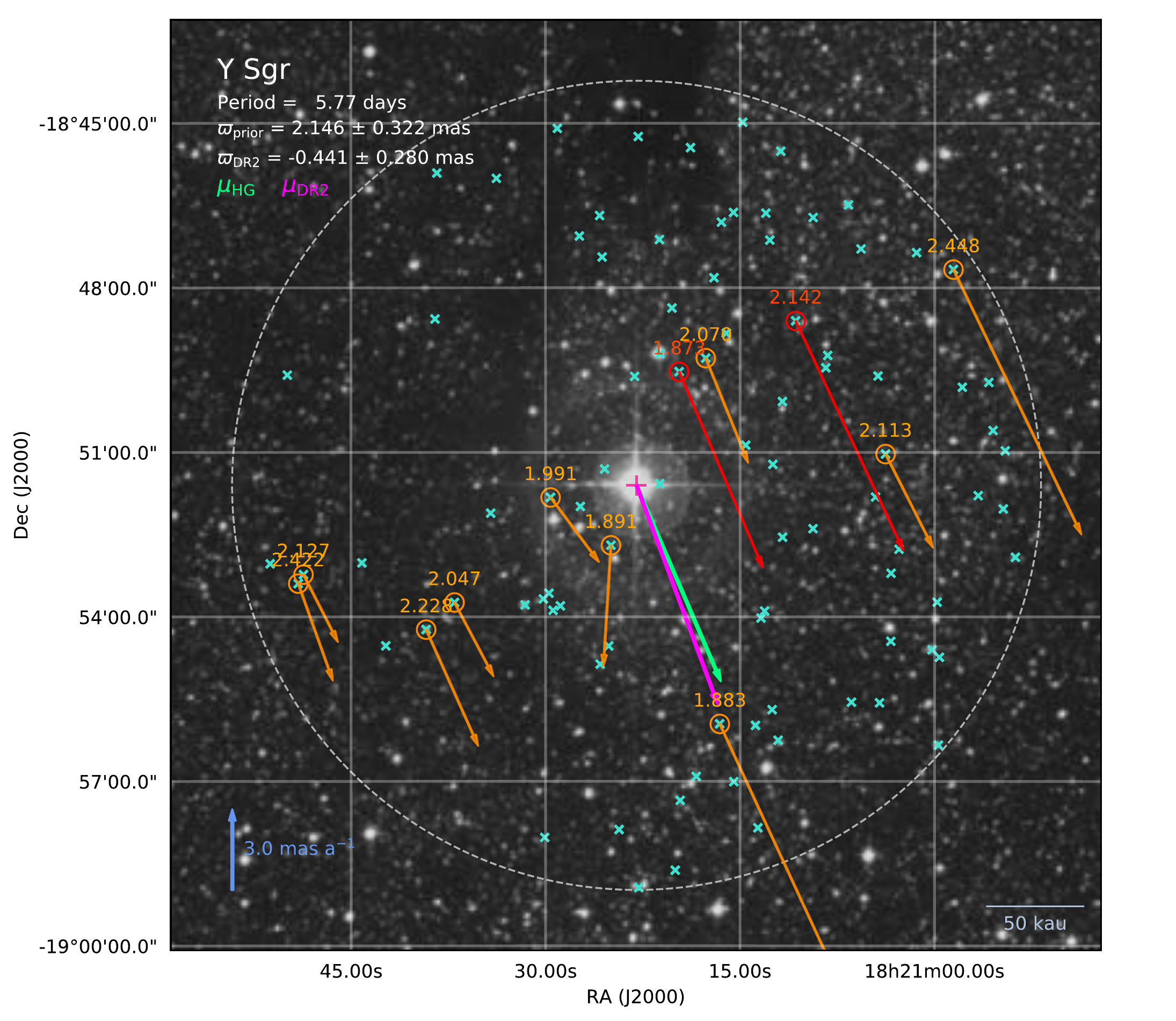}
\includegraphics[width=9cm]{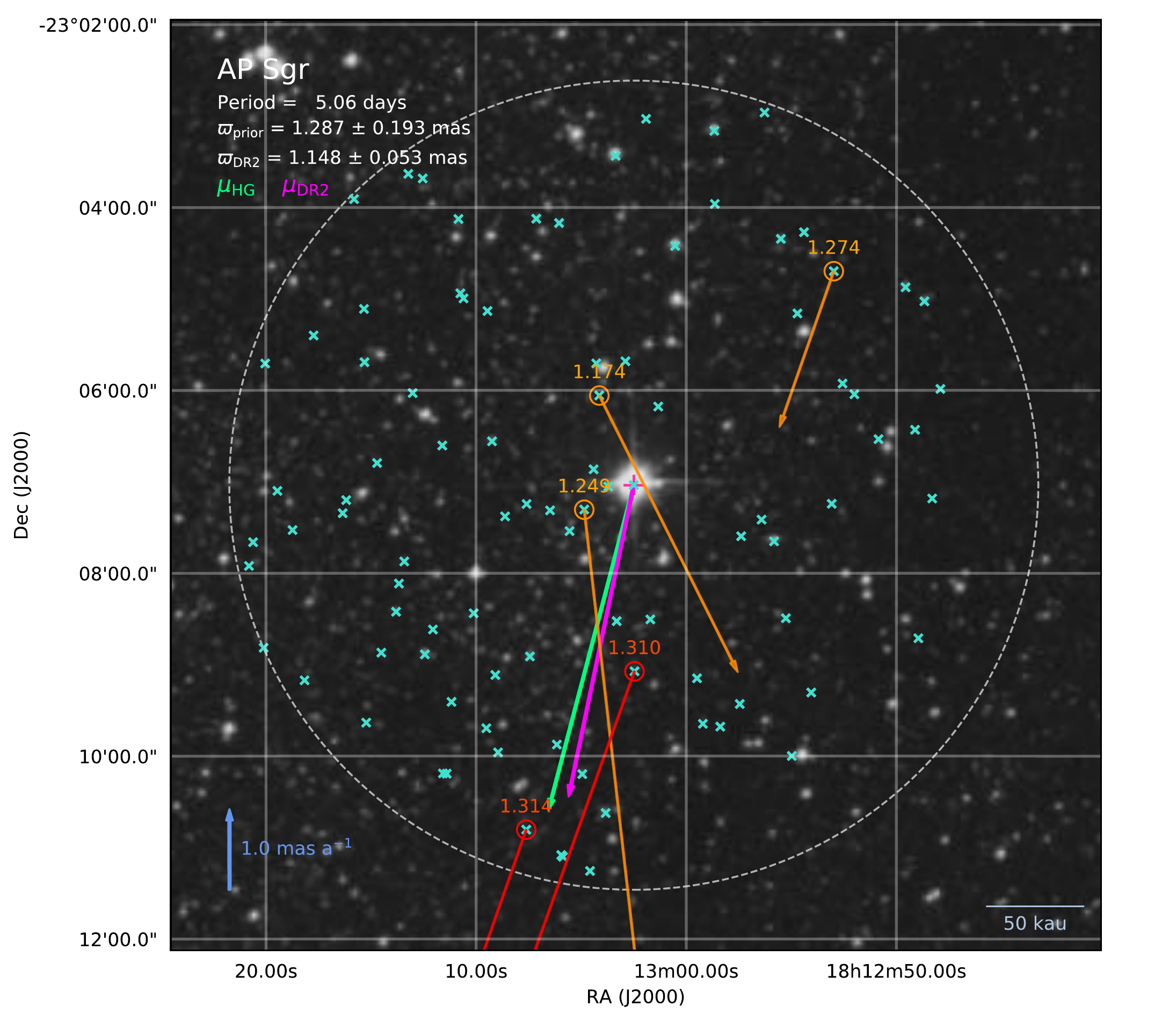}
\includegraphics[width=9cm]{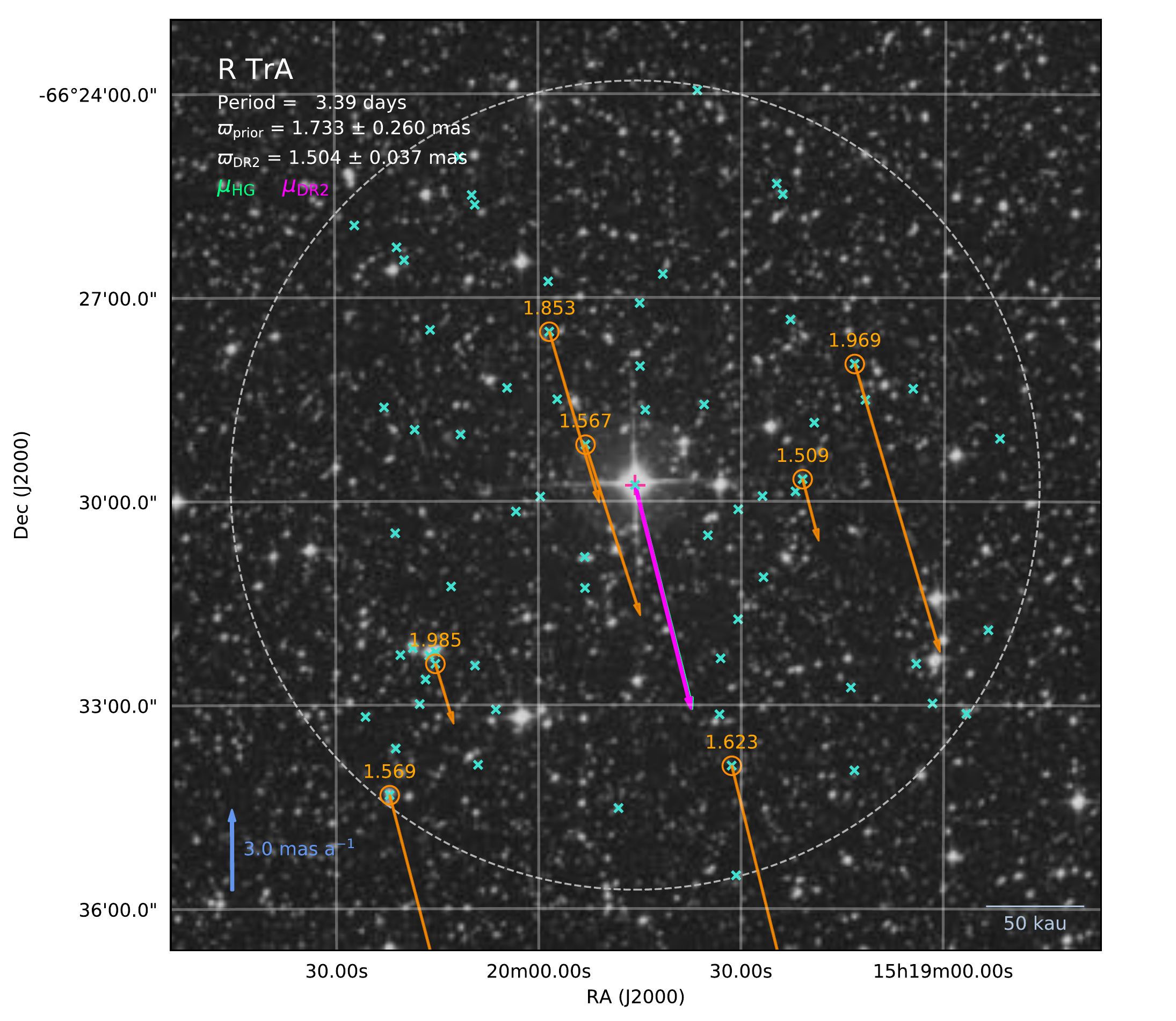}
\caption{Continued from Fig.~\ref{cepheids-near2}.\label{cepheids-near3}}
\end{figure*}

\section{Candidate companions of RR Lyrae stars \label{rrl-tables-appendix}}

\subsection{Tables of candidate companions}

\begin{sidewaystable*}
\tiny
 \caption{Candidate gravitationally bound companions of RR Lyrae stars.
 The RR subtype is taken from the VSX database \citepads{2006SASS...25...47W}.
 The references for the photometry are given in the caption of Table~\ref{cepheids-bound-table1}.
 }
 \label{rrlyrae-bound-table}
 \centering
 \renewcommand{\arraystretch}{1}
\setlength\tabcolsep{4.5pt}
 \begin{tabular}{lrcrrrrrrrrrrrrrrcll}
 \hline
\hline \noalign{\smallskip}
RRL name / GDR2 & $\varpi_\mathrm{G2}$ & $\mu_{\alpha,\mathrm{G2}},\,\mu_{\delta,\mathrm{G2}}$ & Sep. & Sep. & $G_\mathrm{BP}$ & $G$ & $G_\mathrm{RP}$ & $J$ & $H$ & $K$ & $W1$ & $W2$ & $W3$ & $W4$ & $E(B-V)$ & $M_G$ & $\mathrm{Log}\,T$ & RR \\
 & (mas) & (mas\,a$^{-1}$) & ($\arcsec$) & (kau) & & & & & & & & & & & & & & type \\
\hline \noalign{\smallskip}
\textbf{\object{OV And}} & $ 0.938_{ 0.038}$ & $ -4.90_{ 0.06}$,$ -7.78_{ 0.03}$  &  &  & 11.93 & 11.34 & 10.64 &  9.68 &  9.15 &  9.00 &  8.91 &  8.99 &  8.91 &  8.53 & 0.053$^a$ & $ +1.08$ & 3.683 & RRab \\
 380489851377596928 &  $0.909_{ 0.039}$ & $ -4.43_{ 0.04}$,$ -7.71_{ 0.03}$ &   3.43 &   3.7 & 13.79 & 13.54 & 13.02 &  &  &  &  &  &  &  &  & $ +3.20$ & 3.781 & F4V \\
\hline \noalign{\smallskip}
\textbf{\object{CS Del}} & $0.635_{0.031}$ & $ -0.32_{ 0.05}$,$ +2.16_{ 0.06}$ & & & 13.05 & 12.85 & 12.50 & 12.02 & 11.88 & 11.83 & 11.74 & 11.79 & 11.78 & 9.07 & 0.095$^a$ & $+1.60$ & 3.830 & RRc \\
 1805526185917504512 & $0.644_{0.056}$ & $ -0.28_{ 0.10}$,$ +2.19_{ 0.10}$ & 3.17 & 5.0 & 16.32 & 16.14 & 15.42 & & & & & & & & & $+4.93$ & 3.774 & G7V \\
\hline \noalign{\smallskip}
\textbf{\object{V0893 Her}} & $2.679_{0.028}$ & $ -6.75_{ 0.05}$,$+11.54_{ 0.06}$ & & & 9.33 & 9.16 & 8.84 & 8.45 & 8.32 & 8.26 & 8.25 & 8.26 & 8.22 & 8.02 & 0.022$^a$ & $+1.24$ & 3.818 & RR: \\
 1385661079389302144 & $2.905_{0.443}$ & $ -9.68_{ 1.14}$,$+14.08_{ 1.20}$ & 5.16 & 1.9 & 16.19 & 16.65 & 15.09 & & & & & & & & & $+8.91$ & 3.725 & M1.5V \\
\hline \noalign{\smallskip}
\textbf{\object{RR Leo}} & $1.032_{0.088}$ & $-15.29_{ 0.10}$,$-10.16_{ 0.10}$ & & & 11.12 & 10.91 & 10.48 & 9.92 & 9.68 & 9.64 & 9.63 & 9.63 & 9.58 & 8.25 & 0.037$^b$ & $+0.87$ & 3.795 & RRab \\
 630421931138065280 & $1.062_{0.163}$ & $-16.30_{ 0.44}$,$-10.15_{ 0.32}$ & 9.64 & 9.3 & 18.74 & 18.05 & 17.16 & 15.97 & 15.31 & 15.11 & & & & & & $+8.10$ & 3.523 & M0V \\
\hline \noalign{\smallskip}
\textbf{\object{SS Oct}} & $0.862_{0.023}$ & $ -1.90_{ 0.05}$,$-32.70_{ 0.05}$ & & & 12.17 & 11.79 & 11.11 & 10.05 & 9.83 & 9.75 & 9.73 & 9.71 & 9.66 & 9.27 & 0.285$^b$ & $+0.73$ & 3.789 & RRab \\
 6345324695303800192 & $0.782_{0.114}$ & $ -2.22_{ 0.32}$,$-32.85_{ 0.24}$ & 2.67 & 3.1 & 16.80 & 17.18 & 15.80 & & & & & & & & & $+5.90$ & 3.797 & K2V \\
\hline \noalign{\smallskip}
\textbf{\object{EY Oph}} & $ 0.658_{ 0.032}$ & $ -2.64_{ 0.06}$,$ -1.75_{ 0.04}$  &  &  & 14.00 & 13.55 & 12.92 & 12.48 & 12.13 & 12.07 & 11.41 & 11.49 & 11.57 &  8.49 & 0.200$^a$ & $ +2.12$ & 3.769 & RR: \\
 6029835295648727168 &  $ 1.597_{ 0.341}$ & $ -4.52_{ 0.81}$,$ -4.26_{ 0.52}$ &  12.21 &   6.6 & 19.45 & 19.43 & 18.16 &  &  &  &  &  &  &  &  & $ +9.95$ & 3.738 & M3V (WD?)  \\
\hline \noalign{\smallskip}
\textbf{\object{V0487 Sco}} & $ 1.069_{ 0.075}$ & $ -9.83_{ 0.15}$,$-12.30_{ 0.12}$  &  &  & 11.54 & 11.22 & 10.67 &  9.90 &  9.65 &  9.60 &  8.14 &  8.20 &  8.30 &  8.67 & 0.252$^b$ & $+0.69$ & 3.811 & RRc \\
 4053551410541112192 &  $ 0.942_{ 0.060}$ & $-10.60_{ 0.12}$,$-12.61_{ 0.10}$ &  29.59 &  27.7 & 16.85 & 15.95 & 14.90 & 12.04 & 10.62 & 10.16 &  9.85 &  9.98 &  9.92 &  8.68 &  & $+5.26$ & 3.458 & dusty M giant ? \\
\hline
\end{tabular}
\tablebib{(a): 3D map \texttt{Bayestar17} by \citetads{2018MNRAS.478..651G}; (b): \citetads{2008MNRAS.386.2115F}. }
\end{sidewaystable*}

\begin{sidewaystable*}
 \caption{Candidate \texttt{Near} companions of stars of RR Lyrae type in the GCVS \citepads{2009yCat....102025S}.
 EY Oph also has a candidate \texttt{Bound} companion listed in Table~\ref{rrlyrae-bound-table}.}
 \label{rrl-near-table1}
 \centering
\tiny
 \renewcommand{\arraystretch}{1}
\setlength\tabcolsep{4.5pt}
 \begin{tabular}{lrcrrrrrrrrrrrrrrcll}
 \hline
\hline \noalign{\smallskip}
RRL name / GDR2 & $\varpi_\mathrm{G2}$ & $\mu_{\alpha,\mathrm{G2}},\,\mu_{\delta,\mathrm{G2}}$ & Sep. & Sep. & $G_\mathrm{BP}$ & $G$ & $G_\mathrm{RP}$ & $J$ & $H$ & $K$ & $W1$ & $W2$ & $W3$ & $W4$ & $E(B-V)$ & $M_G$ & $\mathrm{Log}\,T$ & RR \\
 & (mas) & (mas\,a$^{-1}$) & ($\arcsec$) & (kau) & & & & & & & & & & & & & & type \\
\hline \noalign{\smallskip}
\textbf{\object{V0830 Cyg}} & $ 0.618_{ 0.019}$ & $ -1.59_{ 0.04}$,$ -1.98_{ 0.04}$ & & & 13.71 & 13.24 & 12.59 & 11.69 & 11.39 & 11.27 & 11.26 & 11.26 & 11.17 & 8.89 & 0.363 & $ +1.25$ & 3.796 & AB \\
 1873027163251912192 & $ 0.600_{ 0.124}$ & $ -3.35_{ 0.30}$,$ -4.12_{ 0.30}$ & 11.55 & 18.7 & 19.27 & 18.63 & 17.69 & 16.28 & 15.81 & 16.06 & & & & & & $ +6.65$ & 3.722 & \\
\hline \noalign{\smallskip}
\textbf{\object{CZ Lac}} & $ 0.852_{ 0.029}$ & $ -2.55_{ 0.06}$,$ -1.59_{ 0.05}$ & & & 11.95 & 11.69 & 11.18 & 10.82 & 10.60 & 10.53 & 10.28 & 10.27 & 10.20 & 8.98 & 0.065 & $ +1.17$ & 3.784 & AB \\
 2000976545410515584 & $ 0.773_{ 0.050}$ & $ -1.20_{ 0.09}$,$ -2.33_{ 0.08}$ & 21.05 & 24.7 & 16.94 & 16.42 & 15.78 & 15.05 & 14.56 & 14.49 & 14.22 & 14.33 & 12.72 & 9.24 & & $ +5.70$ & 3.722 & \\
\hline \noalign{\smallskip}
\textbf{\object{V0424 Lyr}} & $ 1.579_{ 0.187}$ & $ -0.41_{ 0.35}$,$ -4.72_{ 0.41}$ & & & 15.61 & 15.26 & 14.63 & 14.00 & 13.66 & 13.51 & 13.36 & 13.38 & 12.86 & 9.19 & 0.141 & $ +5.89$ & 3.771 & AB \\
 2036210631733553152 & $ 1.459_{ 0.209}$ & $ +1.37_{ 0.37}$,$ -4.99_{ 0.57}$ & 31.17 & 19.7 & 20.51 & 19.26 & 18.16 & 16.46 & 16.18 & 15.59 & 15.61 & 16.40 & 12.32 & 9.10 & & $ +9.80$ & 2.802 & \\
\hline \noalign{\smallskip}
\textbf{\object{AG Nor}} & $ 0.978_{ 0.093}$ & $ -4.49_{ 0.14}$,$ -5.71_{ 0.14}$ & & & 12.16 & 9.57 & 8.21 & 5.47 & 4.40 & 4.04 & 3.89 & 3.82 & 3.67 & 3.28 & & & & AB \\
 5835953600010583296 & $ 0.991_{ 0.198}$ & $ -4.59_{ 0.37}$,$ -4.83_{ 0.43}$ & 23.27 & 23.8 & 19.20 & 18.67 & 17.57 & & & & & & & & & & & \\
 5835953703089735552 & $ 1.100_{ 0.255}$ & $ -5.79_{ 0.47}$,$ -3.92_{ 0.55}$ & 12.59 & 12.9 & 19.45 & 18.67 & 16.83 & & & & & & & & & & & \\
 5835953668729965568 & $ 0.966_{ 0.154}$ & $ -2.96_{ 0.31}$,$ -4.91_{ 0.34}$ & 17.33 & 17.7 & 18.62 & 18.24 & 17.20 & 14.62 & 14.44 & 14.02 & & & & & & & & \\
\hline \noalign{\smallskip}
\textbf{\object{KP Nor}} & $ 0.520_{ 0.054}$ & $ -1.76_{ 0.09}$,$ -3.92_{ 0.08}$ & & & 16.77 & 16.37 & 15.70 & 14.86 & 14.59 & 14.43 & 13.83 & 14.14 & 12.32 & 8.44 & & & & \\
 5831240057986282624 & $ 0.585_{ 0.124}$ & $ -1.03_{ 0.25}$,$ -2.53_{ 0.23}$ & 3.96 & 7.6 & 18.94 & 18.30 & 17.45 & 16.29 & 14.64 & 14.43 & & & & & & & & \\
\hline \noalign{\smallskip}
\textbf{\object{EY Oph}} & $ 0.658_{ 0.032}$ & $ -2.64_{  0.06}$,$ -1.75_{  0.04}$  &  &  & 14.00 & 13.55 & 12.92 & 12.48 & 12.13 & 12.07 & 11.41 & 11.49 & 11.57 &  8.49 & 0.200 & $ +2.12$ & 3.769 & : \\
 6029835364429606784 &  $ 1.801_{ 0.419}$ & $ -7.11_{  0.81}$,$ -3.89_{  0.58}$ &  47.01 &  25.4 & 19.29 & 19.24 & 18.35 &    &    &    &    &    &    &    &  & $ +9.99$ & 3.787 &  \\
\hline \noalign{\smallskip}
\textbf{\object{IT Oph}} & $ 6.206_{ 0.045}$ & $ -0.92_{ 0.10}$,$-19.56_{ 0.06}$ & & & 13.58 & 12.75 & 11.81 & 10.27 & 9.30 & 8.59 & 7.85 & 7.28 & 5.61 & 3.64 & 0.032 & $ +6.64$ & 3.349 & \\
 4108442162053920640 & $ 5.602_{ 0.813}$ & $ -0.73_{ 2.87}$,$ -5.72_{ 1.93}$ & 150.54 & 24.3 & & 20.06 & & & & & & & & & & & & \\
\hline \noalign{\smallskip}
\textbf{\object{MS Oph}} & $ 0.658_{ 0.146}$ & $ +3.36_{ 0.24}$,$ -0.60_{ 0.16}$ & & & 14.33 & 11.33 & 9.86 & 7.01 & 6.02 & 5.59 & 5.48 & 5.35 & 0.00 & 3.72 & 0.148 & $ +0.19$ & -1.535 & : \\
 6029618245128099200 & $ 0.604_{ 0.070}$ & $ +4.91_{ 0.13}$,$ -0.20_{ 0.08}$ & 8.16 & 12.4 & 17.04 & 16.56 & 15.84 & & & & & & & & & $ +5.09$ & 3.739 & \\
\hline \noalign{\smallskip}
\textbf{\object{V1693 Oph}} & $ 0.677_{ 0.340}$ & $ +0.26_{ 1.30}$,$ -7.16_{ 0.99}$ & & & & 19.61 & & 15.75 & 15.19 & 14.98 & & & & & 0.219 & & & AB \\
 4134311639114110464 & $ 0.592_{ 0.071}$ & $ +0.59_{ 0.19}$,$ -2.04_{ 0.12}$ & 19.58 & 28.9 & 17.37 & 16.82 & 16.10 & 15.17 & 14.67 & 14.54 & 14.18 & 14.16 & 11.96 & 8.81 & & $ +5.14$ & 3.744 & \\
\hline \noalign{\smallskip}
\textbf{\object{UY Ori}} & $ 2.840_{ 0.082}$ & $ +1.78_{ 0.18}$,$ +0.96_{ 0.12}$ & & & 12.49 & 12.40 & 12.13 & 11.48 & 10.51 & 9.33 & 7.86 & 6.94 & 3.11 & 1.24 & 0.164 & $ +4.19$ & 3.945 & : \\
 3209513031759316736 & $ 2.593_{ 0.449}$ & $ +5.73_{ 1.38}$,$ -2.05_{ 1.22}$ & 24.77 & 8.7 & 20.93 & 19.97 & 18.58 & 16.92 & 15.81 & 15.15 & & & & & & $+11.71$ & 2.855 & \\
\hline \noalign{\smallskip}
\textbf{\object{V1154 Ori}} & $ 1.087_{ 0.054}$ & $ -1.47_{ 0.10}$,$ -2.21_{ 0.09}$ & & & 10.94 & 10.87 & 10.71 & 10.48 & 10.44 & 10.37 & 10.31 & 10.34 & 10.08 & 5.70 & 0.267 & $ +0.25$ & 4.181 & C \\
 3345190391512281600 & $ 0.993_{ 0.045}$ & $ -0.81_{ 0.08}$,$ -1.53_{ 0.08}$ & 28.28 & 26.0 & 11.82 & 11.77 & 11.60 & 11.39 & 11.39 & 11.33 & 11.25 & 11.30 & 10.62 & 5.40 & & $ +0.94$ & 4.186 & \\
\hline \noalign{\smallskip}
\textbf{\object{V0701 Sgr}} & $ 2.000_{ 0.302}$ & $ +2.84_{ 0.69}$,$ +0.43_{ 0.63}$ & & & 13.94 & 13.65 & 13.08 & 12.25 & 11.99 & 11.90 & 11.88 & 11.94 & 11.38 & 8.04 & & & & AB \\
 4038120868676971008 & $ 2.117_{ 0.349}$ & $ +1.59_{ 0.85}$,$ -3.41_{ 0.78}$ & 54.71 & 27.4 & 19.54 & 19.35 & 17.91 & 16.02 & 15.58 & 15.39 & 15.05 & 16.84 & 11.85 & 8.91 & & & & \\
 4038120971756176128 & $ 2.036_{ 0.224}$ & $ +1.15_{ 0.54}$,$ -0.98_{ 0.49}$ & 43.74 & 21.9 & 19.07 & 18.94 & 17.91 & & & & & & & & & & & \\
\hline \noalign{\smallskip}
\textbf{\object{V2481 Sgr}} & $ 0.910_{ 0.197}$ & $ +3.27_{ 0.65}$,$ -1.10_{ 0.54}$ & & & 15.60 & 14.72 & 13.56 & 12.15 & 11.14 & 11.02 & & & & & 0.107 & $ +4.29$ & 3.161 & C: \\
 4062366092483595776 & $ 0.974_{ 0.150}$ & $ -0.05_{ 0.49}$,$ -0.70_{ 0.41}$ & 25.76 & 28.3 & 16.89 & 16.68 & 15.64 & & & & & & & & & $ +6.36$ & 3.717 & \\
\hline \noalign{\smallskip}
\textbf{\object{V2626 Sgr}} & $ 1.363_{ 0.047}$ & $ +4.15_{ 0.08}$,$ -1.40_{ 0.07}$ & & & 15.39 & 14.94 & 14.31 & 13.57 & 13.18 & 13.12 & 12.86 & 13.04 & 11.78 & 8.25 & & & & AB \\
 4039178186528038400 & $ 1.203_{ 0.245}$ & $ +3.46_{ 0.50}$,$ -0.66_{ 0.45}$ & 22.13 & 16.2 & & 18.90 & & 15.95 & 15.10 & 15.08 & & & & & & & & \\
\hline \noalign{\smallskip}
\textbf{\object{V3531 Sgr}} & $ 0.766_{ 0.462}$ & $ -1.14_{ 0.96}$,$ -3.00_{ 0.81}$ & & & 17.19 & 17.01 & 16.26 & 15.43 & 15.03 & 15.04 & & & & & & & & AB \\
 4046831577743784192 & $ 0.748_{ 0.150}$ & $ +2.36_{ 0.31}$,$ -2.71_{ 0.26}$ & 10.82 & 14.1 & 17.91 & 17.63 & 16.87 & 16.17 & 15.80 & 15.07 & & & & & & & & \\
\hline \noalign{\smallskip}
\textbf{\object{V4107 Sgr}} & $ 1.265_{ 0.142}$ & $ -3.28_{ 0.40}$,$ -1.74_{ 0.34}$ & & & 16.81 & 16.61 & 15.55 & & & & & & & & 0.113 & $ +6.84$ & 3.715 & AB \\
 4050225564037689472 & $ 1.404_{ 0.254}$ & $ -4.61_{ 0.53}$,$ -4.06_{ 0.49}$ & 36.78 & 29.1 & 18.08 & 18.12 & 16.57 & & & & & & & & & $ +8.59$ & 3.630 & \\
\hline \noalign{\smallskip}
\textbf{\object{V4313 Sgr}} & $ 1.700_{ 0.282}$ & $ -3.98_{ 0.68}$,$ -1.83_{ 0.57}$ & & & 16.20 & 16.22 & 15.31 & 14.32 & 14.04 & 14.00 & & & & & 0.087 & $ +7.14$ & 3.772 & AB \\
 4049184322231156736 & $ 1.702_{ 0.230}$ & $ -3.80_{ 0.58}$,$ +0.79_{ 0.48}$ & 47.85 & 28.2 & 17.90 & 17.48 & 16.43 & & & & & & & & & $ +8.43$ & 3.629 & \\
 4049184425310375808 & $ 1.869_{ 0.326}$ & $ -1.86_{ 0.72}$,$ -3.94_{ 0.72}$ & 28.12 & 16.5 & & 18.91 & & 13.15 & 12.53 & 12.34 & 10.90 & 11.17 & 10.93 & 8.06 & & & & \\
\hline
\end{tabular}
\end{sidewaystable*}

\begin{sidewaystable*}
 \caption{Continued from Table~\ref{rrl-near-table1}.}
 \label{rrl-near-table2}
  \centering
\tiny
 \renewcommand{\arraystretch}{1}
 \setlength\tabcolsep{4.5pt}
\begin{tabular}{lrcrrrrrrrrrrrrrrcll}
 \hline
\hline \noalign{\smallskip}
RRL name / GDR2 & $\varpi_\mathrm{G2}$ & $\mu_{\alpha,\mathrm{G2}},\,\mu_{\delta,\mathrm{G2}}$ & Sep. & Sep. & $G_\mathrm{BP}$ & $G$ & $G_\mathrm{RP}$ & $J$ & $H$ & $K$ & $W1$ & $W2$ & $W3$ & $W4$ & $E(B-V)$ & $M_G$ & $\mathrm{Log}\,T$ & RR \\
 & (mas) & (mas\,a$^{-1}$) & ($\arcsec$) & (kau) & & & & & & & & & & & & & & type \\
\hline \noalign{\smallskip}
\textbf{\object{V4355 Sgr}} & $ 2.433_{ 0.313}$ & $ +0.13_{ 0.69}$,$ -3.87_{ 0.61}$ & & & 16.48 & 16.20 & 15.42 & 14.60 & 14.27 & 14.05 & & & & & & & & AB \\
 4048847360538237696 & $ 2.357_{ 0.538}$ & $ -3.29_{ 1.41}$,$ -3.73_{ 1.27}$ & 69.70 & 28.6 & & 19.60 & & 15.19 & 14.69 & 14.42 & & & & & & & & \\
 4048847704135531136 & $ 2.484_{ 0.404}$ & $ -1.82_{ 1.17}$,$ -6.79_{ 1.03}$ & 34.25 & 14.1 & 19.43 & 19.31 & 18.15 & & & & & & & & & & & \\
 4048847772895728000 & $ 2.131_{ 0.415}$ & $ -3.41_{ 0.89}$,$ -3.50_{ 1.00}$ & 47.67 & 19.6 & 19.07 & 19.19 & 17.66 & & & & & & & & & & & \\
 4048847734320589312 & $ 2.100_{ 0.359}$ & $ -2.61_{ 0.99}$,$ -4.08_{ 0.97}$ & 40.85 & 16.8 & & 19.14 & & & & & & & & & & & & \\
 4048847803040124416 & $ 2.691_{ 0.434}$ & $ -3.75_{ 1.11}$,$ -3.50_{ 1.04}$ & 64.61 & 26.6 & 18.62 & 19.25 & 17.45 & & & & & & & & & & & \\
\hline \noalign{\smallskip}
\textbf{\object{V4591 Sgr}} & $ 1.837_{ 0.179}$ & $ -0.34_{ 0.46}$,$ -2.21_{ 0.38}$ & & & 17.67 & 17.29 & 16.34 & 15.51 & 14.80 & 14.62 & & & & & & & & AB \\
 4048827462005964416 & $ 1.821_{ 0.303}$ & $ -1.50_{ 0.67}$,$ -6.42_{ 0.60}$ & 26.75 & 14.6 & 19.23 & 18.93 & 17.68 & & & & & & & & & & & \\
\hline \noalign{\smallskip}
\textbf{\object{IY Sco}} & $ 2.530_{ 0.042}$ & $ -8.94_{ 0.06}$,$ -4.25_{ 0.05}$ & & & 14.07 & 13.59 & 12.95 & 12.17 & 11.78 & 11.66 & 11.58 & 11.67 & 11.51 & 8.82 & & & & \\
 6028393767161904000 & $ 2.276_{ 0.516}$ & $ -8.19_{ 1.38}$,$ -1.00_{ 0.88}$ & 67.23 & 26.6 & 19.93 & 19.81 & 18.72 & & & & & & & & & & & \\
 6028394007680341248 & $ 2.641_{ 0.513}$ & $ -3.55_{ 1.70}$,$ -0.40_{ 1.15}$ & 30.23 & 11.9 & 19.05 & 19.87 & 18.58 & & & & & & & & & & & \\
\hline \noalign{\smallskip}
\textbf{\object{KN Sco}} & $ 0.959_{ 0.043}$ & $ -0.60_{ 0.07}$,$ -2.42_{ 0.05}$ & & & 14.22 & 13.79 & 13.21 & 12.34 & 12.09 & 11.98 & 11.74 & 11.77 & 12.05 & 8.47 & 0.423 & $ +2.56$ & 3.841 & \\
 6028937883678549888 & $ 1.077_{ 0.199}$ & $ +0.02_{ 0.48}$,$ -4.26_{ 0.32}$ & 27.43 & 28.6 & 19.34 & 18.66 & 17.71 & & & & & & & & & $ +7.82$ & 3.728 & \\
\hline \noalign{\smallskip}
\textbf{\object{V0828 Sco}} & $ 0.877_{ 0.435}$ & $ -6.58_{ 1.89}$,$ -8.06_{ 1.77}$ & & & & 19.07 & & 14.62 & 14.28 & 14.42 & & & & & & & & \\
 5954643058331900160 & $ 0.871_{ 0.124}$ & $ -8.88_{ 0.33}$,$ -5.36_{ 0.30}$ & 22.86 & 26.1 & 18.70 & 18.06 & 17.34 & 16.41 & 15.51 & 15.56 & & & & & & & & \\
\hline \noalign{\smallskip}
\textbf{\object{V0348 Sct}} & $ 1.249_{ 0.042}$ & $ -1.36_{ 0.06}$,$ +0.51_{ 0.05}$ & & & 13.56 & 13.17 & 12.61 & 11.94 & 11.66 & 11.57 & 11.51 & 11.51 & 11.97 & 8.74 & 0.109 & $ +3.37$ & 3.768 & \\
 4203792424572067968 & $ 1.096_{ 0.161}$ & $ -3.73_{ 0.52}$,$ +1.42_{ 0.43}$ & 26.44 & 21.2 & 18.94 & 18.16 & 17.24 & 15.91 & 15.36 & 15.03 & & & & & & $ +8.11$ & 3.507 & \\
\hline
\end{tabular}
\end{sidewaystable*}

\subsection{Field charts}

\begin{figure*}[h]
\centering
\includegraphics[width=9cm]{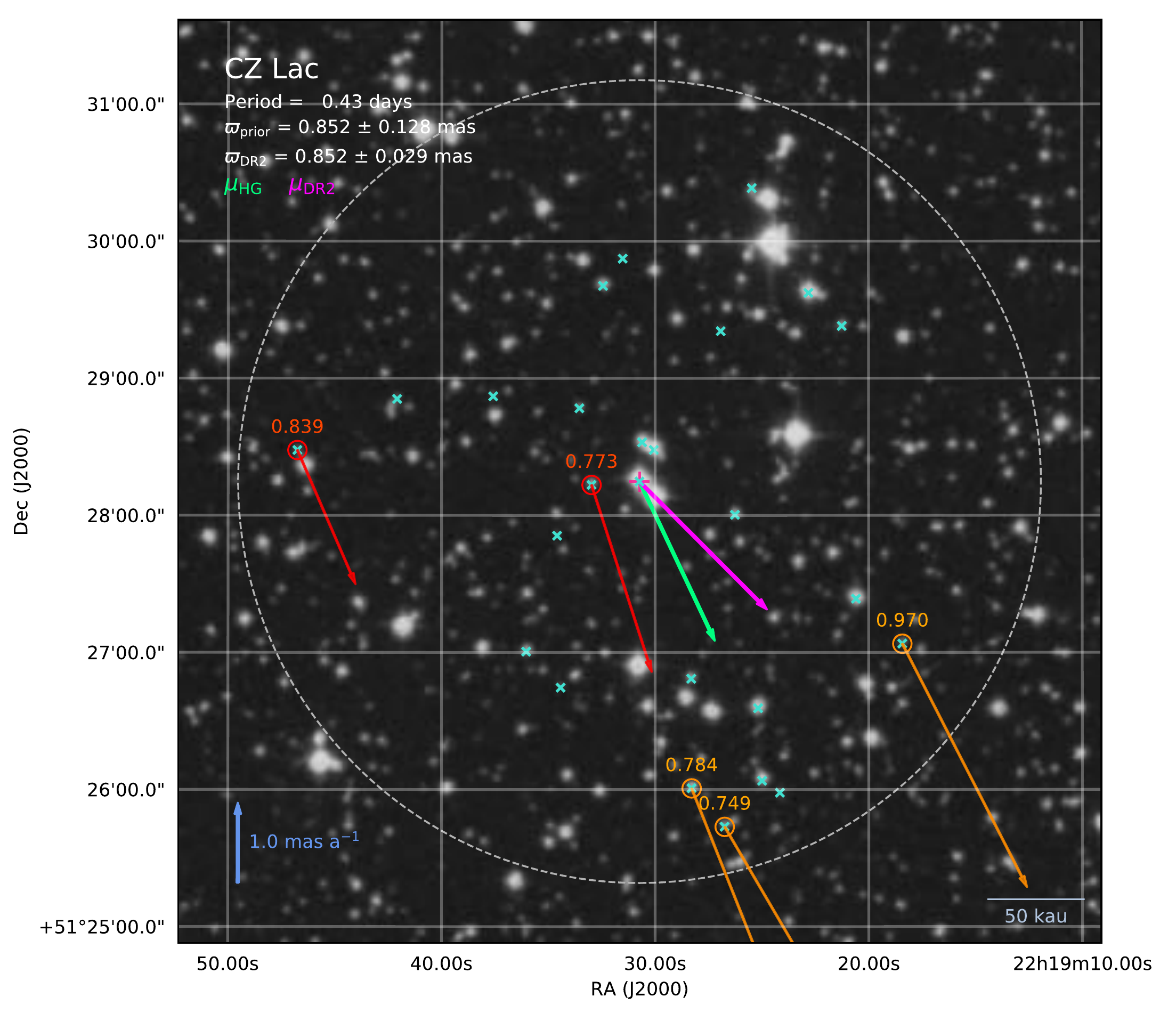}
\includegraphics[width=9cm]{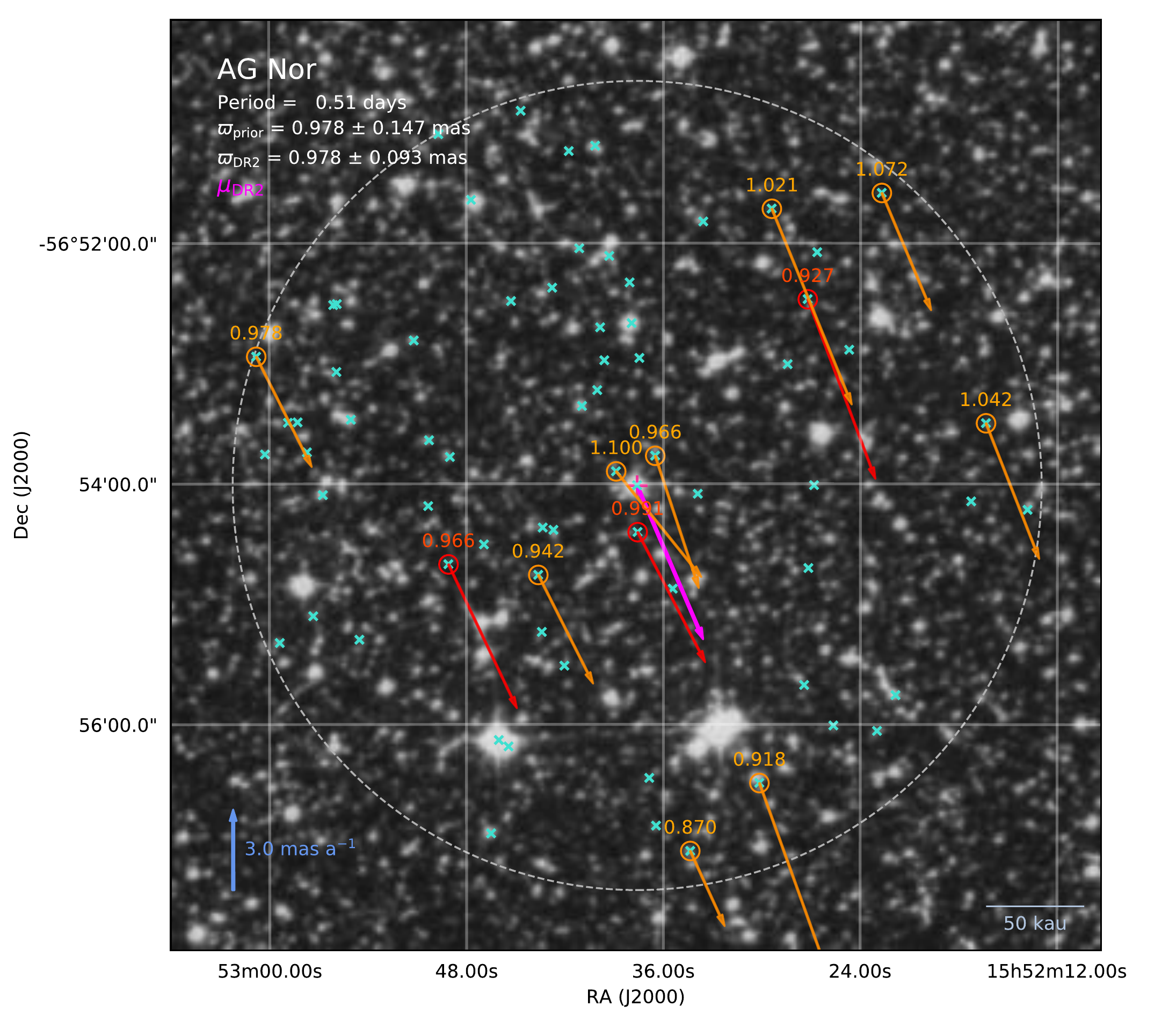}
\includegraphics[width=9cm]{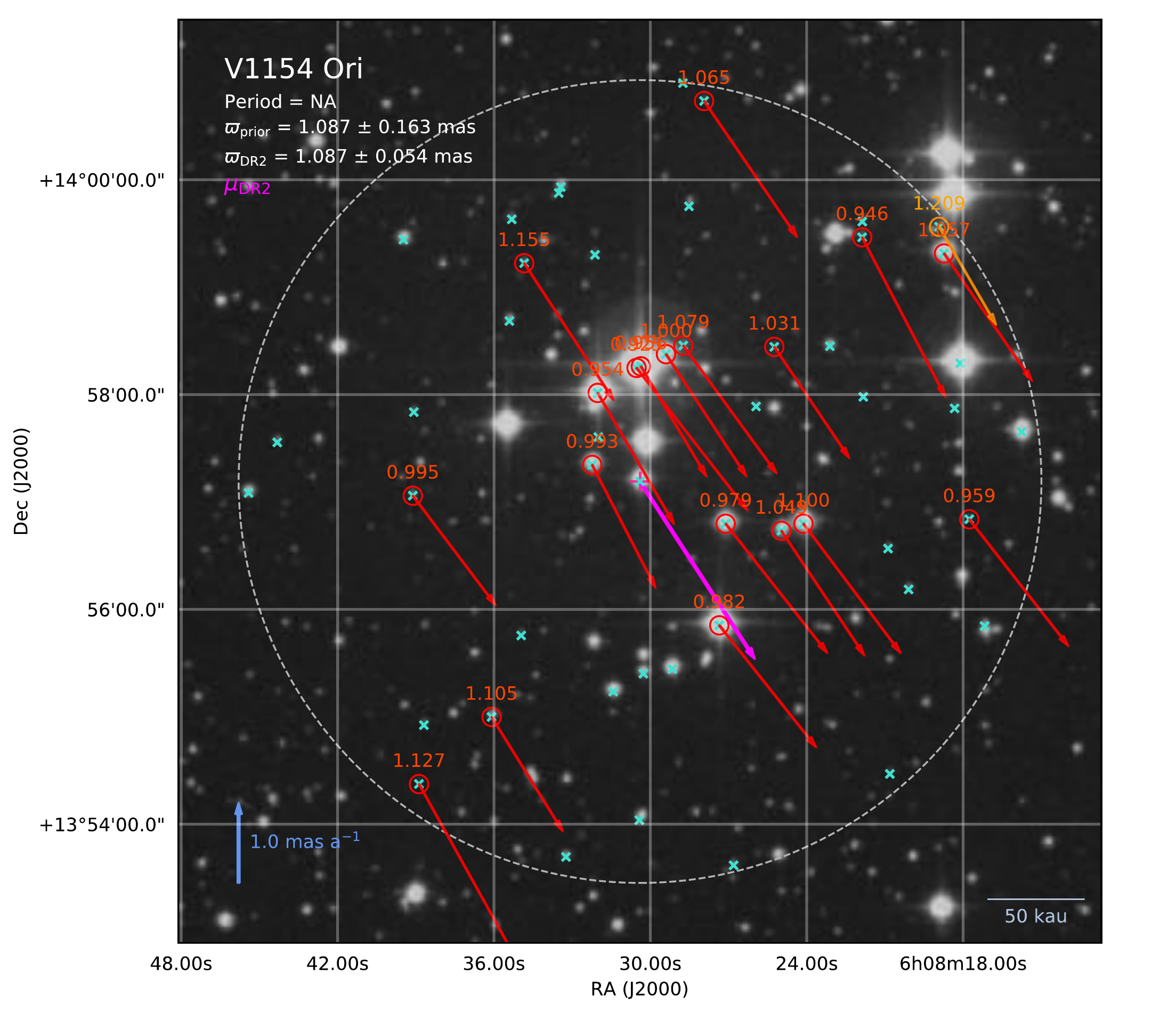}
\includegraphics[width=9cm]{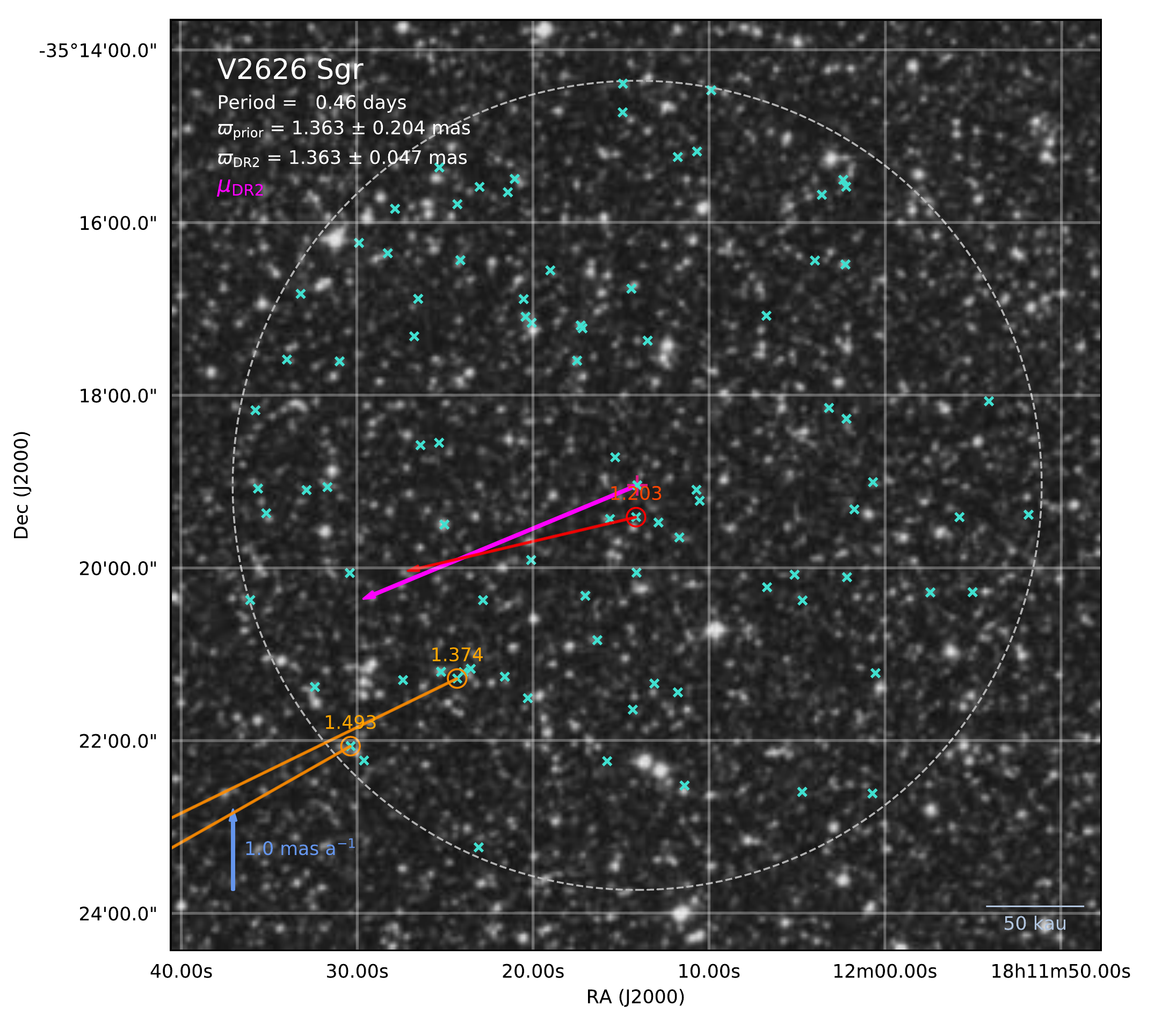}
\includegraphics[width=9cm]{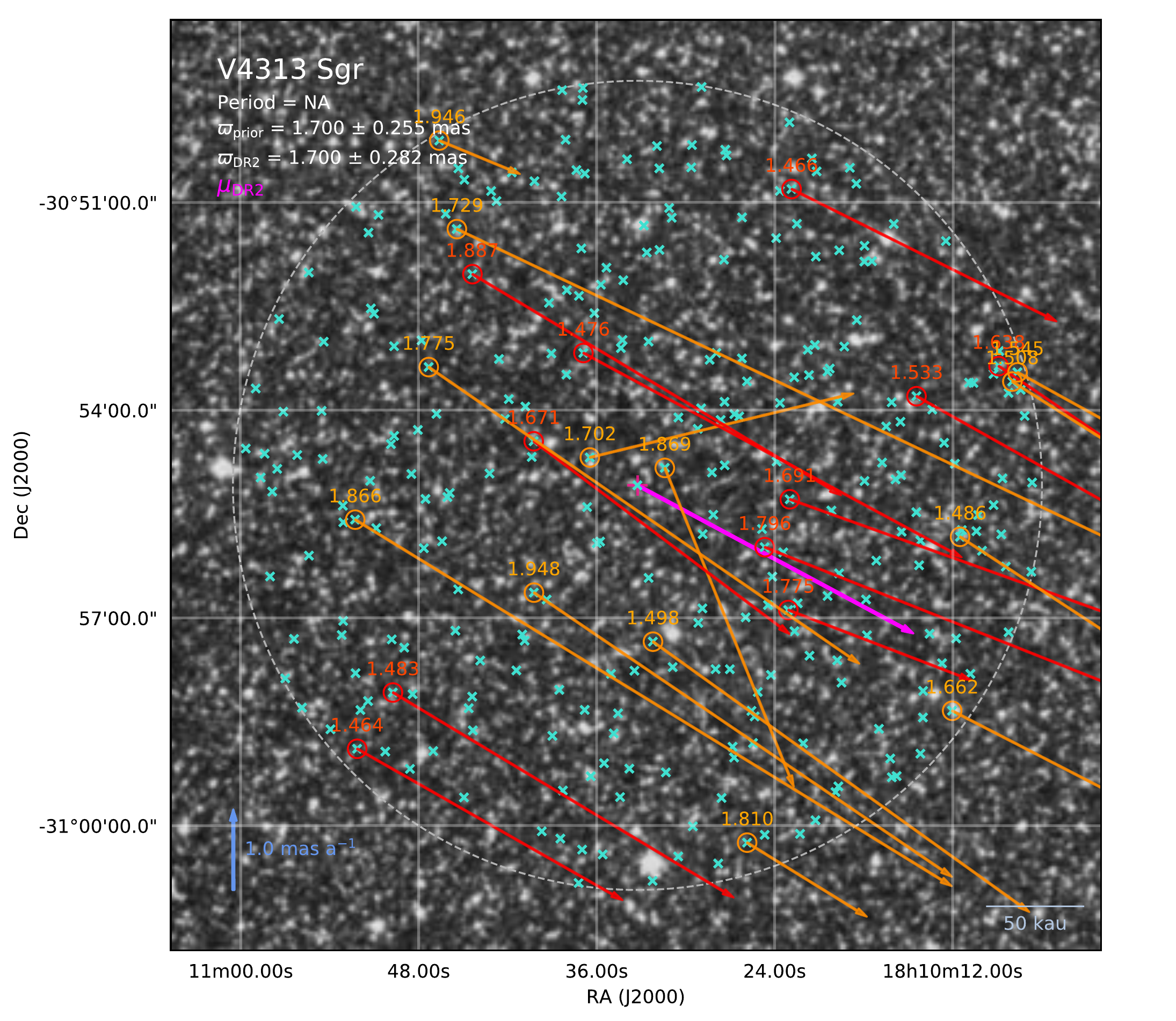}
\caption{Selected RR Lyrae stars with \texttt{Near} candidate companions.\label{rrlyr-near}}
\end{figure*}

The fields surrounding the RRLs with detected \texttt{Bound} candidate companions, and a selection of RRLs with \texttt{Near} candidate companions are presented in Fig.~\ref{rrlyr-field-1} to \ref{rrlyr-near}. The adopted symbols are described in Sect.~\ref{fieldcharts-cep}.

\begin{figure*}[h]
\centering
\includegraphics[width=9cm]{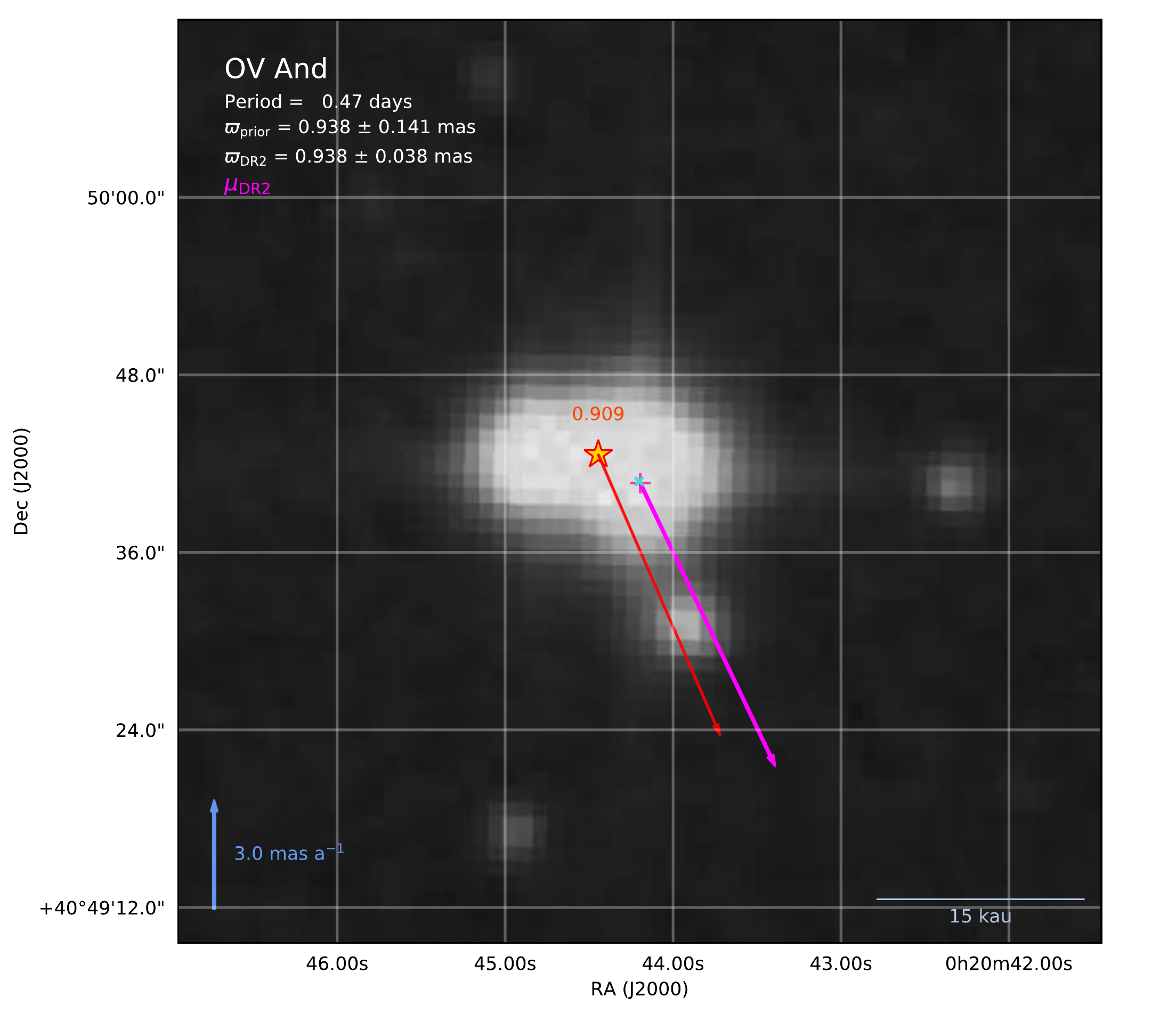}
\includegraphics[width=9cm]{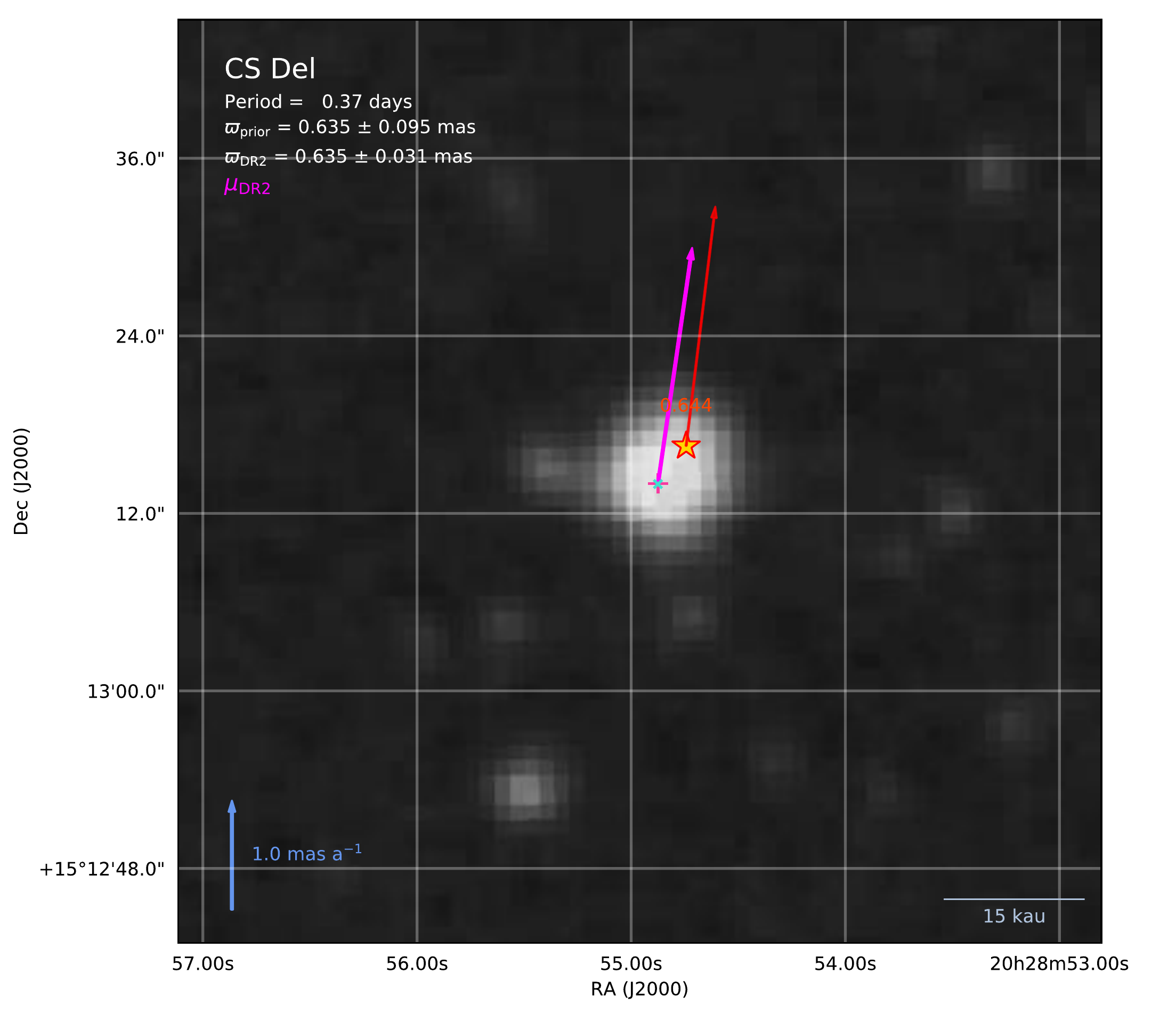}
\includegraphics[width=9cm]{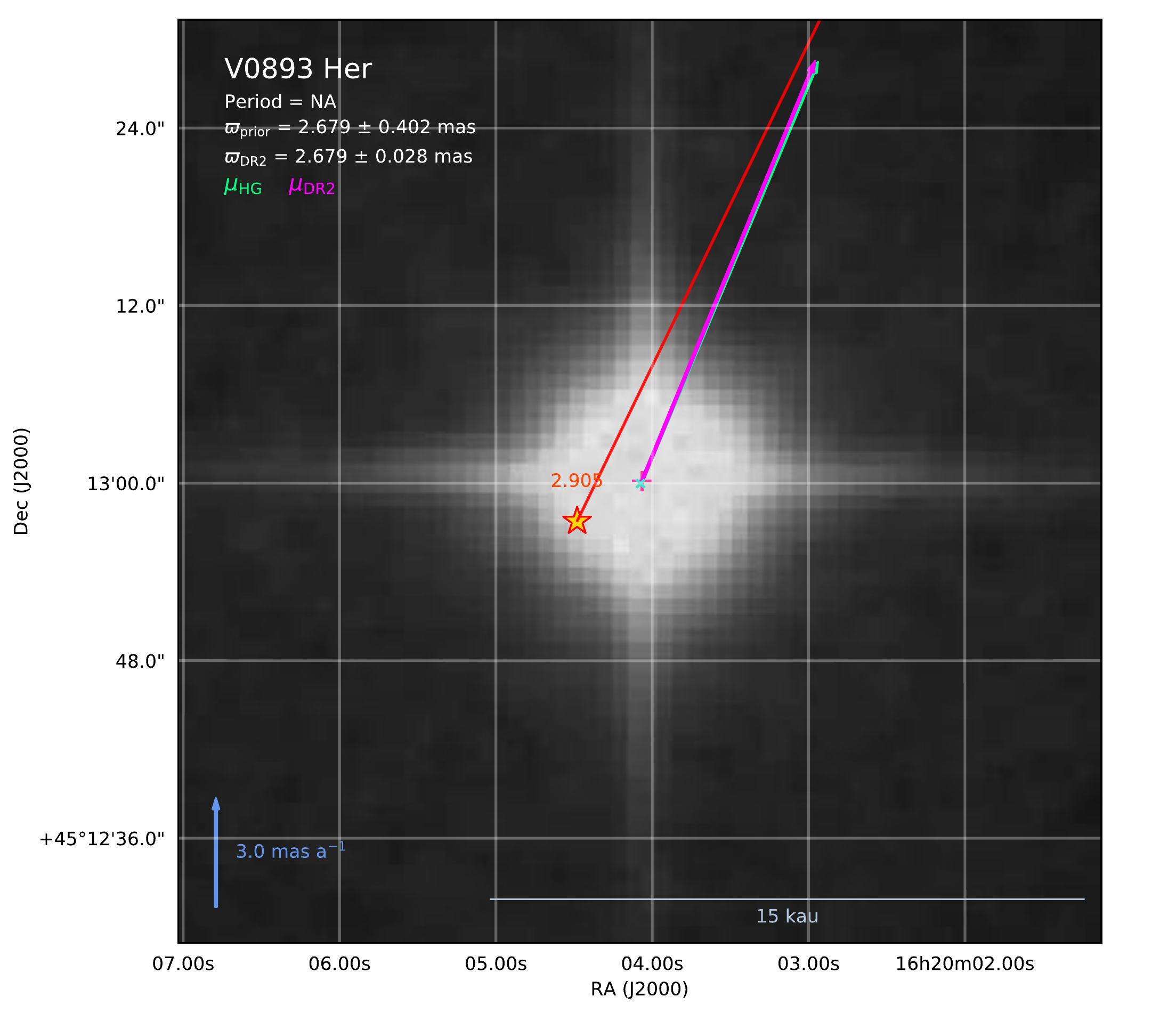}
\includegraphics[width=9cm]{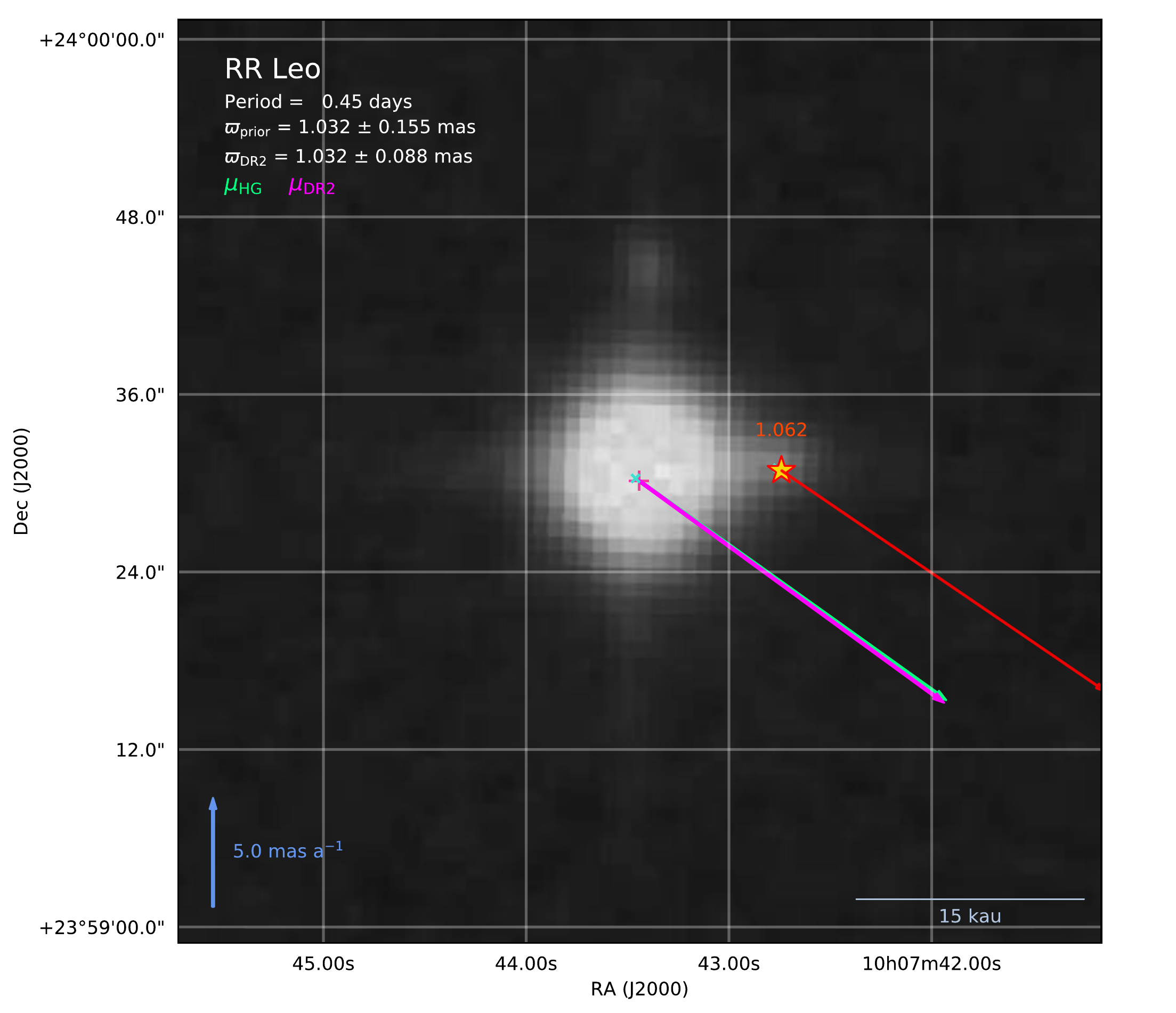}
\includegraphics[width=9cm]{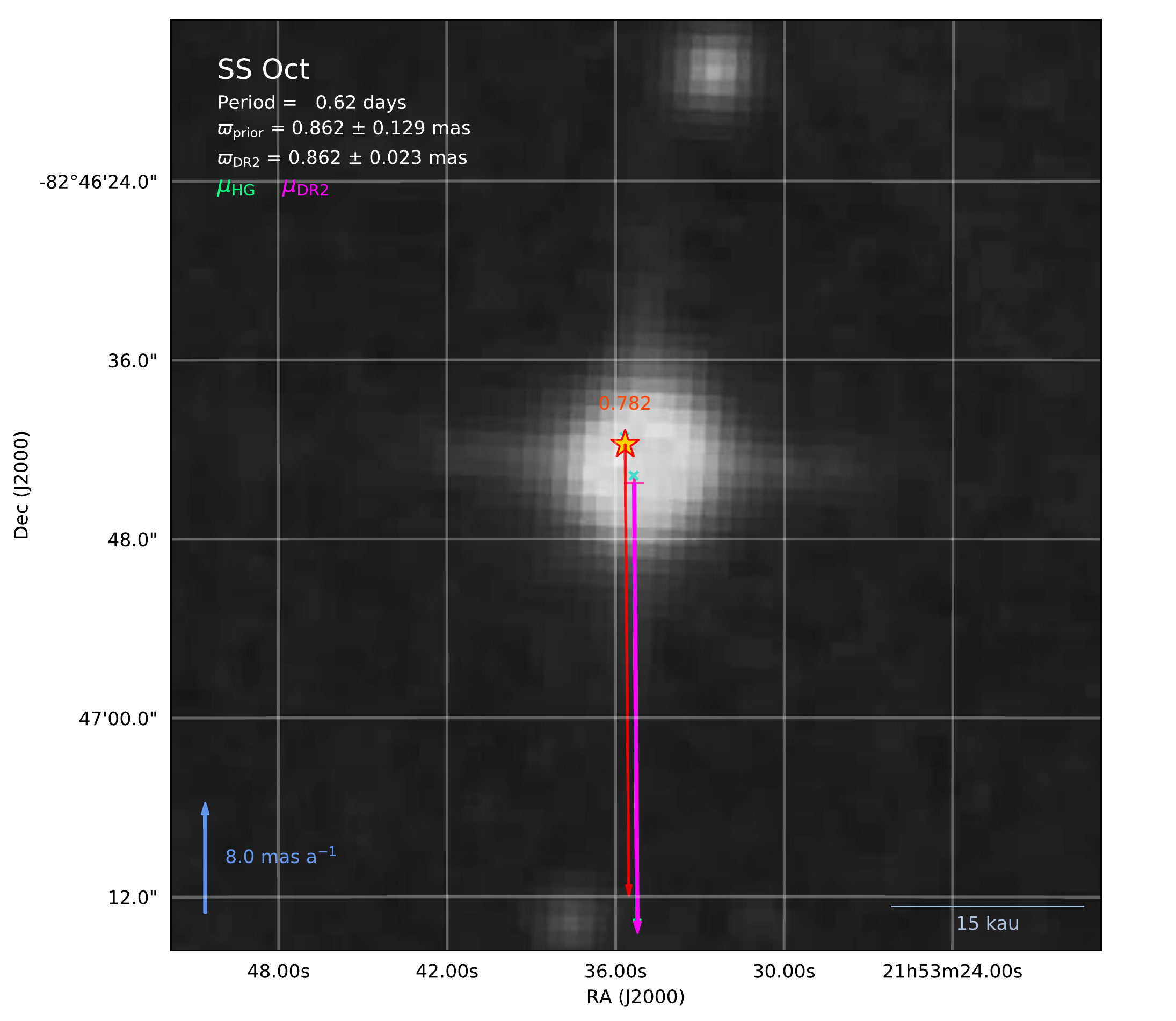}
\includegraphics[width=9cm]{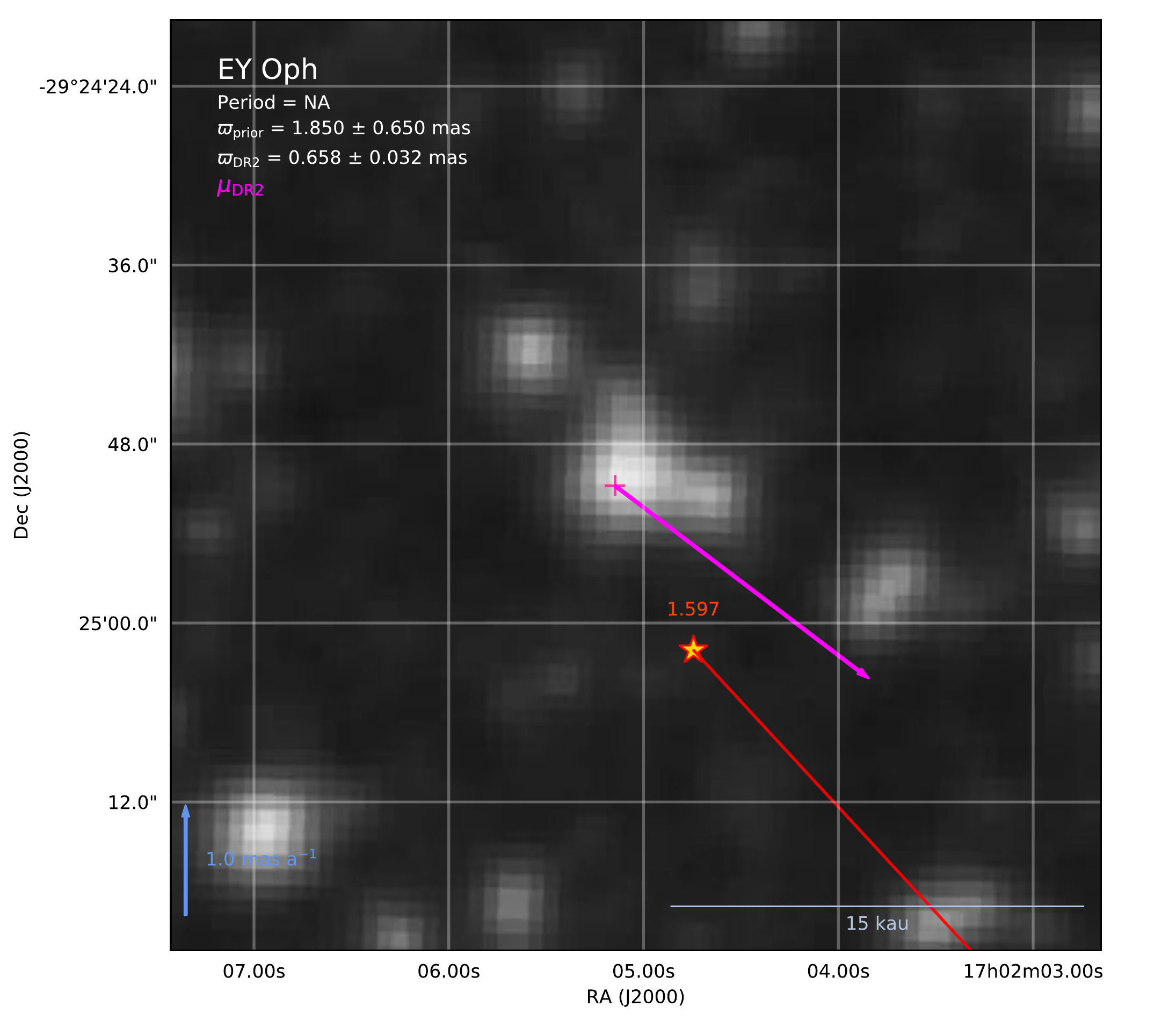}
\caption{RR Lyrae stars with \texttt{Bound} candidate companions.\label{rrlyr-field-1}}
\end{figure*}

\begin{figure}[h]
\centering
\includegraphics[width=9cm]{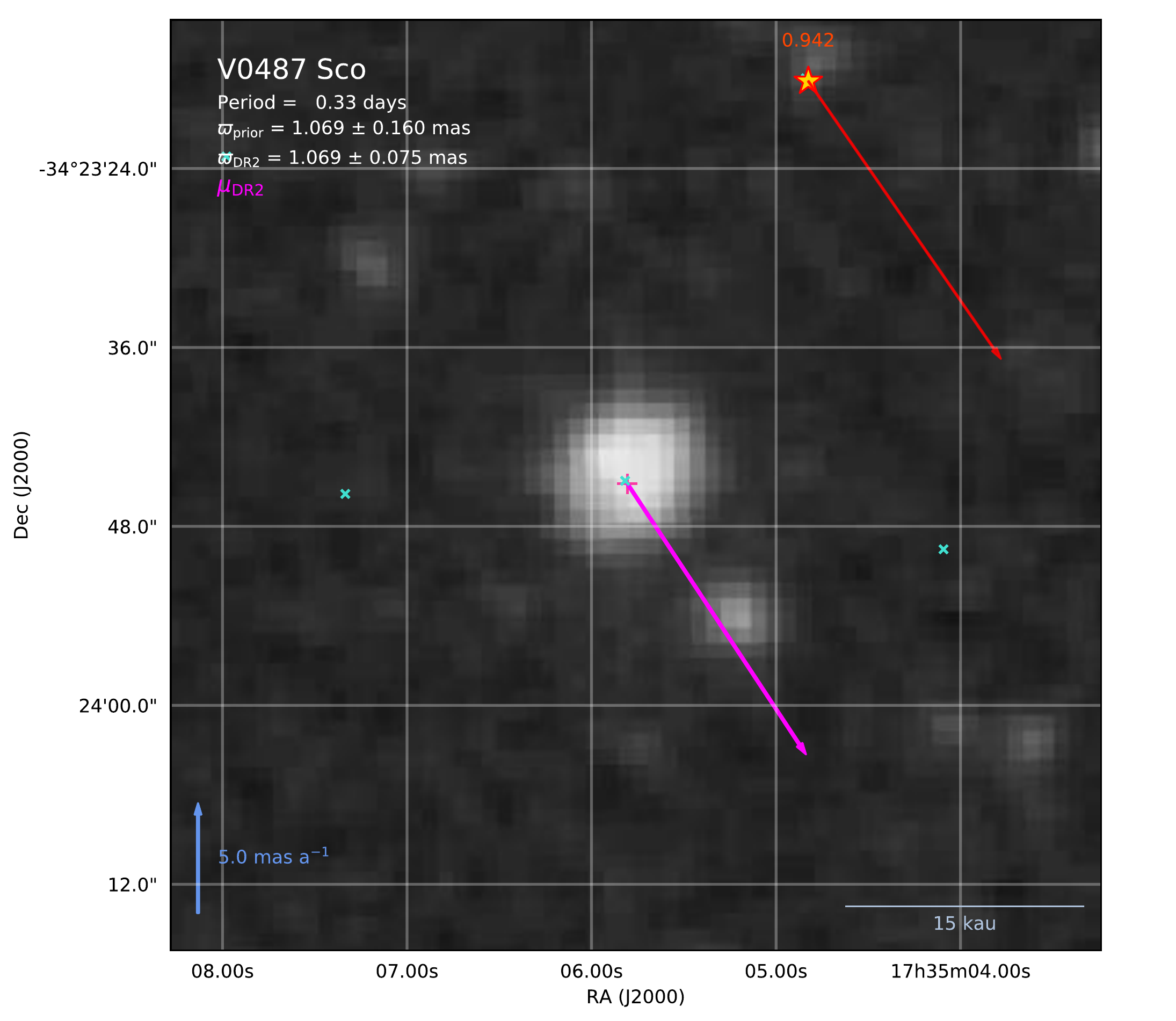}
\caption{Continued from Fig.~\ref{rrlyr-field-1}.\label{rrlyr-field-2}}
\end{figure}

\section{Candidate \texttt{Bound} companions of various variable stars \label{var-tables-appendix}}

\subsection{Table of candidate companions}

\begin{sidewaystable*}
\caption{Candidate gravitationally bound companions of various variable stars. The references for the photometry are given in the caption of Table~\ref{cepheids-bound-table1}.}
\label{various-bound-table}
\centering
\tiny
\renewcommand{\arraystretch}{1}
\setlength\tabcolsep{4.5pt}
\begin{tabular}{lrcrrrrrrrrrrrrrrcll}
\hline
\hline \noalign{\smallskip}
Star name / GDR2 & $\varpi_\mathrm{G2}$ & $\mu_{\alpha,\mathrm{G2}},\,\mu_{\delta,\mathrm{G2}}$ & Sep. & Sep. & $G_\mathrm{BP}$ & $G$ & $G_\mathrm{RP}$ & $J$ & $H$ & $K$ & $W1$ & $W2$ & $W3$ & $W4$ & $E(B-V)$ & $M_G$ & $\mathrm{Log}\,T$ & Type \\
 & (mas) & (mas\,a$^{-1}$) & ($\arcsec$) & (kau) & & & & & & & & & & & & &  \\
\hline \noalign{\smallskip}
\textbf{\object{HM Aql}} & $1.829_{0.037}$ & $ +0.65_{ 0.07}$,$ -5.85_{ 0.06}$ & & & 13.54 & 12.85 & 12.02 & 10.85 & 10.49 & 10.34 & & & & & 0.616$^a$ & $+2.62$ & 3.789 & F1V, ? \\
 4259109541759198848 & $1.709_{0.105}$  & $ +0.42_{ 0.23}$,$ -6.37_{ 0.21}$& 26.84 & 14.7 & 19.09 & 17.49 & 16.24 & 14.06 & 13.12 & 12.81 & & & & & & $+7.44$ & (2.971) & K7V \\
\hline \noalign{\smallskip}
\textbf{\object{EN CMi}} & $ 0.569_{0.066}$ & $ -3.70_{ 0.13}$,$ +2.24_{ 0.08}$ & & & 17.10 & 16.66 & 16.06 & 15.39 & 15.08 & 14.86 & 14.68 & 14.78 & 12.38 & 8.22 & 0.037$^a$ & $ +5.34$ & 3.741 & G9V, ? \\
 3083253225327313024 & $ 0.519_{0.091}$ & $ -4.30_{ 0.20}$,$ +2.72_{ 0.11}$ & 13.84 & 24.3 & 17.57 & 17.16 & 16.62 & 16.10 & 15.58 & 15.22 & 15.58 & 15.76 & 12.51 & 8.71 & & $ +5.64$ & 3.755 & K1.5V \\
\hline \noalign{\smallskip}
\textbf{\object{NQ Cyg}} & $ 0.938_{0.093}$ & $ -5.86_{ 0.21}$,$-11.20_{ 0.20}$ & & & 16.33 & 16.37 & 16.05 & 15.14 & 14.92 & 14.78 & & & & & 0.082$^a$ & $ +5.99$ & 3.916 & K2.5V, Ecl. binary$^c$ \\
 2072926589158425856 & $ 0.932_{0.106}$ & $ -5.25_{ 0.25}$,$ -9.94_{ 0.25}$ & 18.20 & 19.4 & 18.93 & 18.17 & 17.34 & 16.35 & 15.69 & 15.41 & & & & & & $ +7.83$ & 3.556 & K9V \\
\hline \noalign{\smallskip}
\textbf{\object{V1391 Cyg}} & $1.232_{0.056}$ & $ -1.17_{ 0.11}$,$ -3.76_{ 0.12}$ & & & 16.47 & 15.42 & 14.38 & 12.68 & 11.86 & 11.56 & 11.21 & 10.91 & 8.16 & 5.84 & 0.216$^a$ & $+5.42$ & 3.268 & K0V, Ecl. binary$^d$ \\
 2067355638620630912 & $1.057_{0.183}$ & $ -1.32_{ 0.42}$,$ -3.90_{ 0.56}$ & 3.57 & 2.9 & 20.50 & 19.22 & 17.63 & & & & & & & & & $+8.94$ & (2.077) & M1.5V \\
\hline \noalign{\smallskip}
\textbf{\object{V2121 Cyg}} & $25.597_{0.064}$ & $+68.69_{ 0.11}$,$+57.11_{ 0.13}$ & & & 5.82 & 5.66 & 5.37 & 5.35 & 5.01 & 4.91 & 4.91 & 4.73 & 4.96 & 4.91 & 0.004$^a$ & $+2.69$ & 3.822 & F1V, $\gamma$\,Dor$^e$ \\
 2084032103278208640 & $25.652_{0.029}$ & $+72.29_{ 0.06}$,$+53.72_{ 0.05}$ & 55.64 & 2.2 & 13.59 & 12.27 & 11.13 & 9.58 & 8.90 & 8.70 & 8.57 & 8.48 & 8.44 & 9.09 & & $+9.31$ & 2.279 & M2V \\
\hline \noalign{\smallskip}
\textbf{\object{UU Dor}} & $0.856_{0.060}$ & $ -2.26_{ 0.12}$,$ -0.75_{ 0.19}$ & & & 14.61 & 14.39 & 13.84 & 13.41 & 13.11 & 13.03 & 12.73 & 12.75 & 11.67 & 9.58 & 0.159$^c$ & $+3.62$ & 3.805 & F6V, Ecl. binary$^f$ \\
 4657579638104003968 & $0.857_{0.098}$ & $ -1.55_{ 0.48}$,$ -0.87_{ 0.32}$ & 0.92 & 1.1 & & 16.90 & &  &  &  &  &  &  &  & & $+6.13$ & & K3V \\
\hline \noalign{\smallskip}
\textbf{\object{IW Lib}} & $3.571_{0.112}$ & $+10.95_{ 0.11}$,$ -5.18_{ 0.08}$ & & & 9.59 & 9.22 & 8.66 & 8.01 & 7.70 & 7.63 & 7.57 & 7.55 & 7.58 & 7.43 & 0.231$^a$ & $+1.37$ & 3.753 & A1V, W\,UMa ecl.$^g$ \\
 6234240520157591424 & $3.417_{0.063}$ & $+11.69_{ 0.08}$,$ -5.23_{ 0.06}$ & 17.27 & 4.8 & 15.26 & 14.48 & 13.62 & 12.43 & 11.77 & 11.69 & 11.33 & 11.32 & 11.29 & 8.92 & & $+6.62$ & 3.639 & K3V \\
\hline \noalign{\smallskip}
\textbf{\object{AZ Men}} & $1.088_{0.037}$ & $ -1.56_{ 0.04}$,$ -5.57_{ 0.04}$ & & & 14.15 & 13.81 & 13.30 & 12.61 & 12.33 & 12.22 & 12.29 & 12.30 & 12.10 & 9.28 & 0.114$^c$ & $+3.69$ & 3.782 & ?F7V,  \\
 4628608777985655296 & $1.159_{0.053}$ & $ -1.60_{ 0.10}$,$ -5.50_{ 0.11}$ & 16.71 & 15.4 & 17.46 & 16.77 & 15.98 & 14.97 & 14.39 & 14.29 & 14.17 & 14.22 & 13.19 & 9.63 & & $+6.82$ & 3.638 & K5V \\
\hline \noalign{\smallskip}
\textbf{\object{V1171 Oph}} & $1.098_{0.048}$ & $ -1.97_{ 0.07}$,$ -0.78_{ 0.05}$ & & & 14.09 & 13.70 & 13.13 & 12.40 & 12.08 & 11.96 & 11.81 & 11.84 & 11.46 & 8.86 & 0.237$^a$ & $+3.28$ & 3.792 & F8V, ? \\
 4129610367918498432 & $0.985_{0.071}$ & $ -2.17_{ 0.14}$,$ -1.10_{ 0.09}$ & 3.94 & 3.6 & 16.82 & 16.27 & 15.50 & 14.46 & 14.07 & 13.81 & & & & & & $+5.65$ & 3.740 & K1V \\
\hline \noalign{\smallskip}
\textbf{\object{V1330 Sgr}} & $ 1.229_{0.224}$ & $ -5.80_{ 0.45}$,$ -2.23_{ 0.37}$ & & & 17.32 & 16.92 & 15.70 & 14.12 & 13.27 & 13.00 & & & & & 0.191$^a$ & $ +6.93$ & 3.624 & K5V, ? \\
 4050169523277366656 & $ 1.398_{0.320}$ & $ -3.57_{ 0.61}$,$ -1.36_{ 0.51}$ & 30.35 & 24.7 & 18.33 & 17.86 & 16.42 & & & & & & & & & $ +8.17$ & 3.413 & M0V \\
\hline \noalign{\smallskip}
\textbf{\object{V1382 Sgr}} & $1.611_{0.280}$ & $ -3.95_{ 0.40}$,$ -5.42_{ 0.33}$ & & & 16.86 & 16.51 & 15.08 & 13.71 & 13.02 & 12.79 & & & & & 0.119$^a$ & $+7.28$ & 3.450 & K6V, ? \\
 4050171683873801216 & $1.719_{0.287}$ & $ -2.96_{ 0.54}$,$ -5.13_{ 0.46}$ & 18.71 & 11.6 & 18.49 & 18.33 & 16.77 & & & & & & & & & $+9.24$ & 3.500 & M2V \\
\hline \noalign{\smallskip}
\textbf{\object{V2248 Sgr}} & $1.948_{0.040}$ & $ +5.50_{ 0.05}$,$-11.43_{ 0.04}$ & & & 14.06 & 13.57 & 12.95 & 12.18 & 11.77 & 11.70 & 11.56 & 11.60 & 11.15 & 8.80 & 0.043$^b$ & $+4.92$ & 3.729 & G7V, W\,UMa ecl.$^h$ \\
 6694128163462947200 & $1.923_{0.048}$ & $ +5.39_{ 0.07}$,$-11.52_{ 0.06}$ & 53.54 & 27.5 & 16.01 & 15.35 & 14.59 & 13.66 & 13.05 & 12.89 & 12.87 & 12.89 & 12.39 & 8.85 & & $+6.67$ & 3.622 & K3.5V \\
\hline \noalign{\smallskip}
\textbf{\object{V3166 Sgr}} & $6.646_{0.070}$ & $ -0.97_{ 0.15}$,$-26.78_{ 0.12}$ & & & & 14.34 & & 10.60 & 9.78 & 9.37 & 8.67 & 8.20 & 6.61 & 4.50 & 0.030$^b$ & $+8.48$ & 3.488 & M0.5V, ? \\
 4044760647636888576 & $6.705_{0.053}$ & $ -3.27_{ 0.11}$,$-26.91_{ 0.09}$ & 1.64 & 0.2 & 14.86 & 13.73 & 12.43 &  &  &  &  &  &  &  & & $+7.81$ & (2.387) & K8V \\
 4044760643280597504 & $6.883_{0.098}$ & $ -2.31_{ 0.17}$,$-25.97_{ 0.14}$ & 9.00 & 1.4 & 16.57 & 14.73 & 13.42 & 11.46 & 10.82 & 10.60 & 10.37 & 10.15 & 10.20 & 8.11 & & $+8.87$ & 3.547 & M1V \\
\hline \noalign{\smallskip}
\textbf{\object{HR Sco}} & $7.370_{0.576}$ & $ -1.32_{ 1.20}$,$ +7.59_{ 0.70}$ & & & 16.13 & 15.89 & 14.94 & 13.91 & 13.50 & 13.57 & & & & & 0.010$^a$ & $+10.20$ & 3.683 & M3V, ? \\
 6029291827664185600 & $7.052_{1.490}$ & $ -2.30_{ 2.07}$,$ +5.53_{ 1.20}$ & 155.4 & 21.1 & 19.66 & 20.38 & 18.59 & & & & & & & & & $+14.60$ & 3.728 & M5V \\
\hline
\end{tabular}
\tablebib{(a): \citetads{2018MNRAS.478..651G};
(b): \citetads{2018A&A...616A.132L};
(c): \citetads{2017AstL...43..472G}, converted using $E(B-V)=1.655 \, E(J-K_S)$.
(d) \citetads{2014AJ....147..119C};
(e) \citetads{2005AJ....129.2815H};
(f) \citetads{2011AcA....61..103G};
(g) \citetads{2012ApJS..203...32R};
(h) \citetads{2005IBVS.5613....1A}.}
\end{sidewaystable*}

\subsection{Field charts}

The fields surrounding the variable stars of various classes with detected \texttt{Bound} candidate companions are presented in Fig.~\ref{various-field1} to \ref{various-field3}. The adopted symbols are described in Sect.~\ref{fieldcharts-cep}.

\begin{figure*}[h]
\centering
\includegraphics[width=9cm]{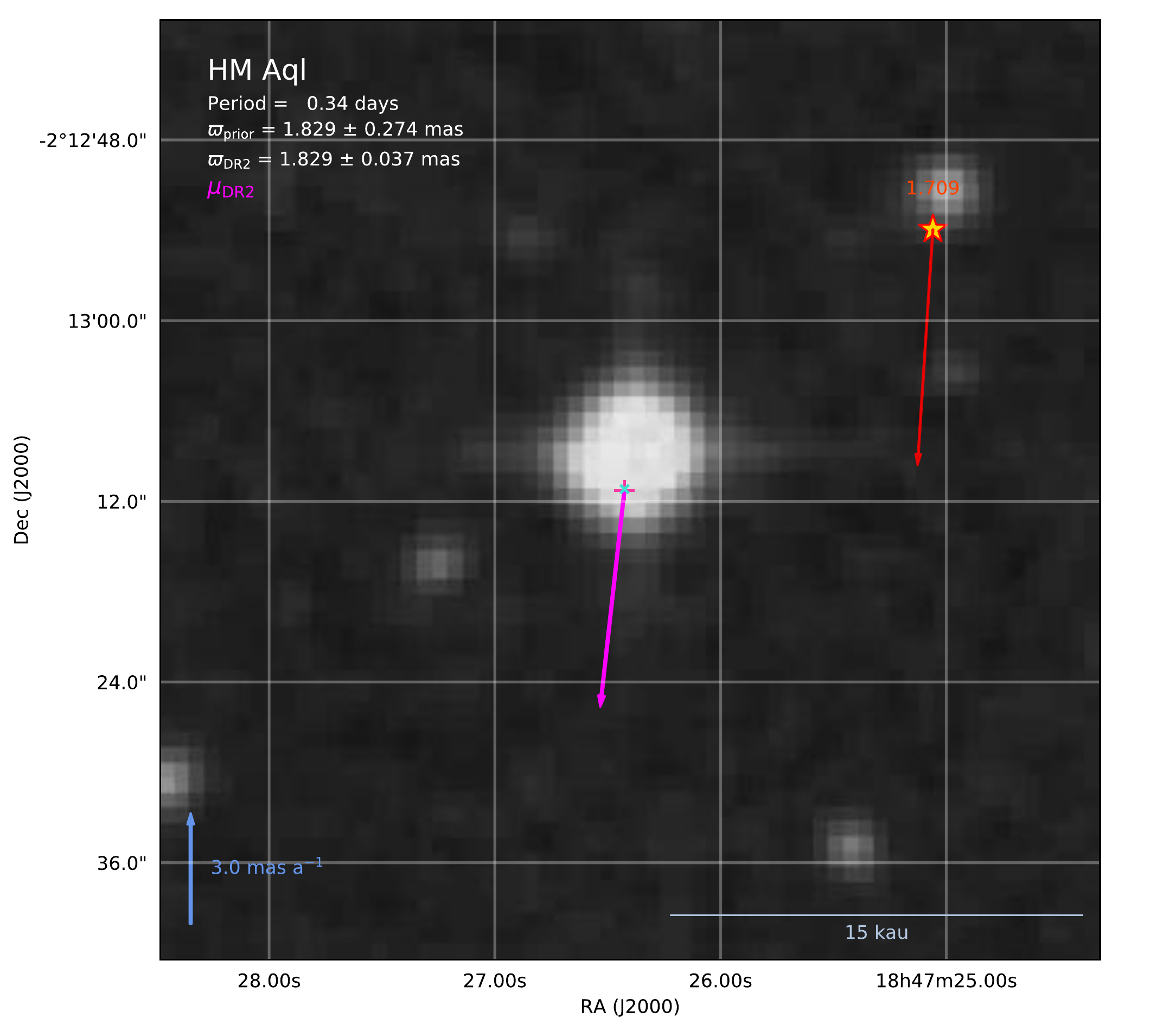}
\includegraphics[width=9cm]{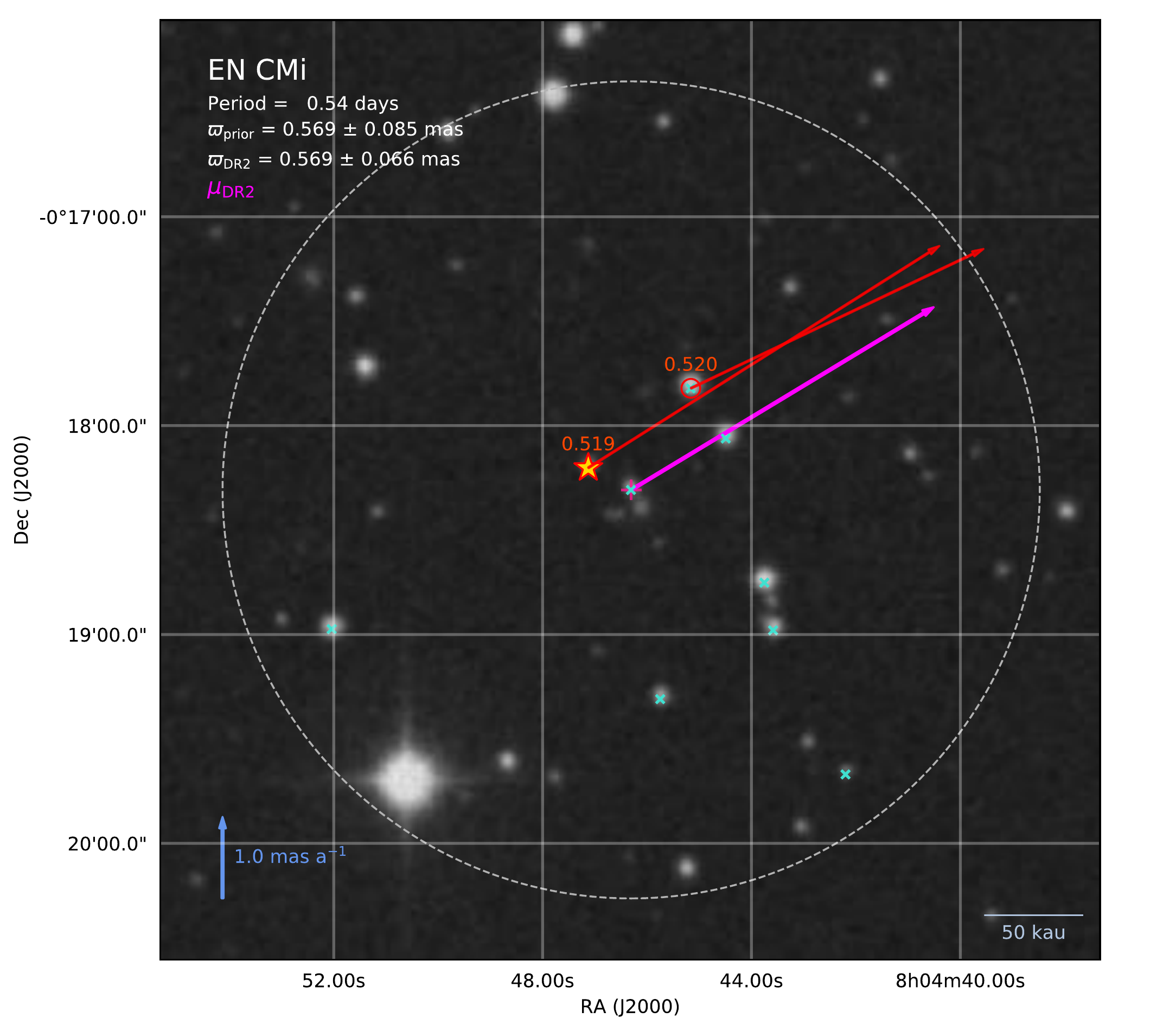}
\includegraphics[width=9cm]{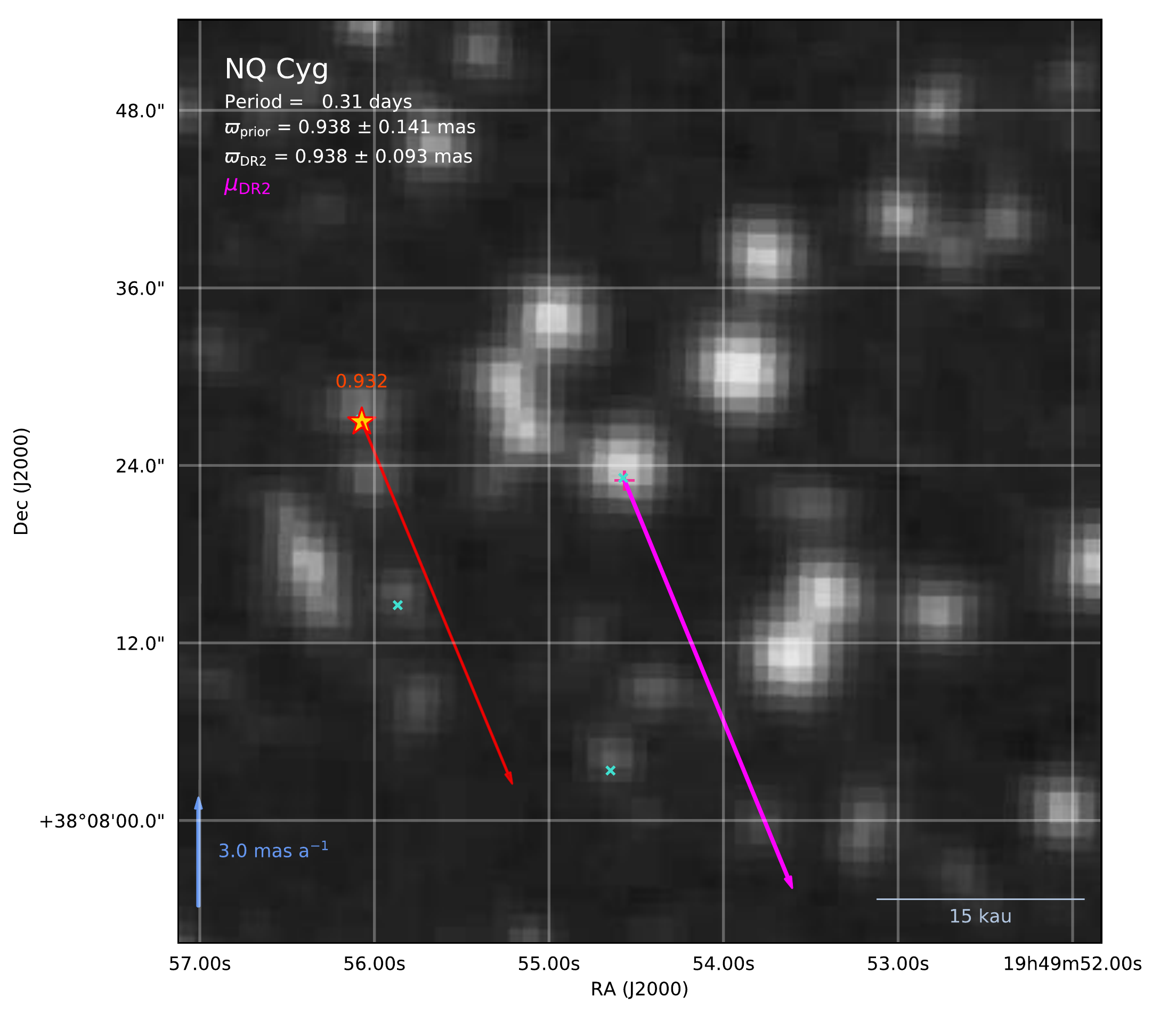}
\includegraphics[width=9cm]{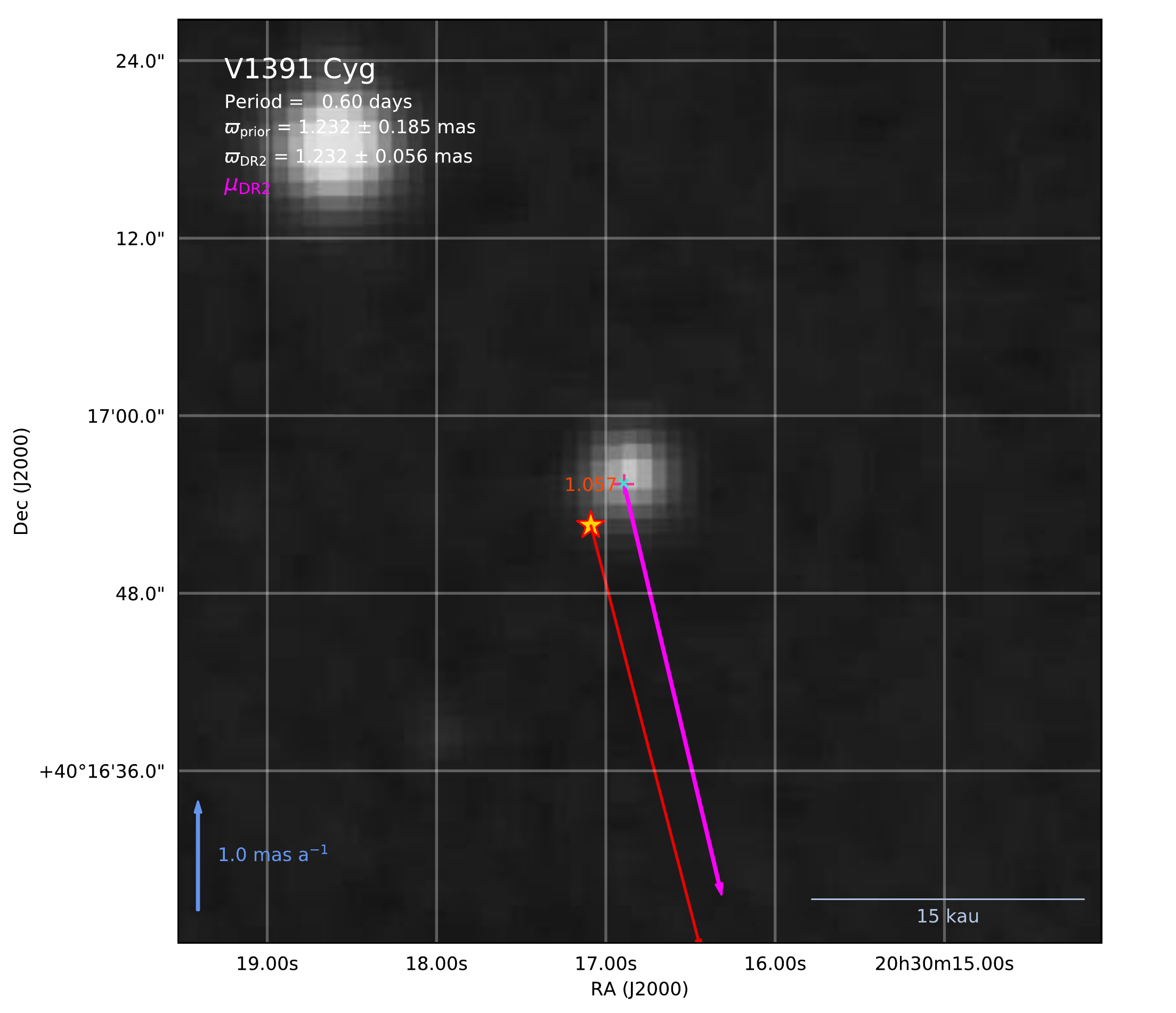}
\includegraphics[width=9cm]{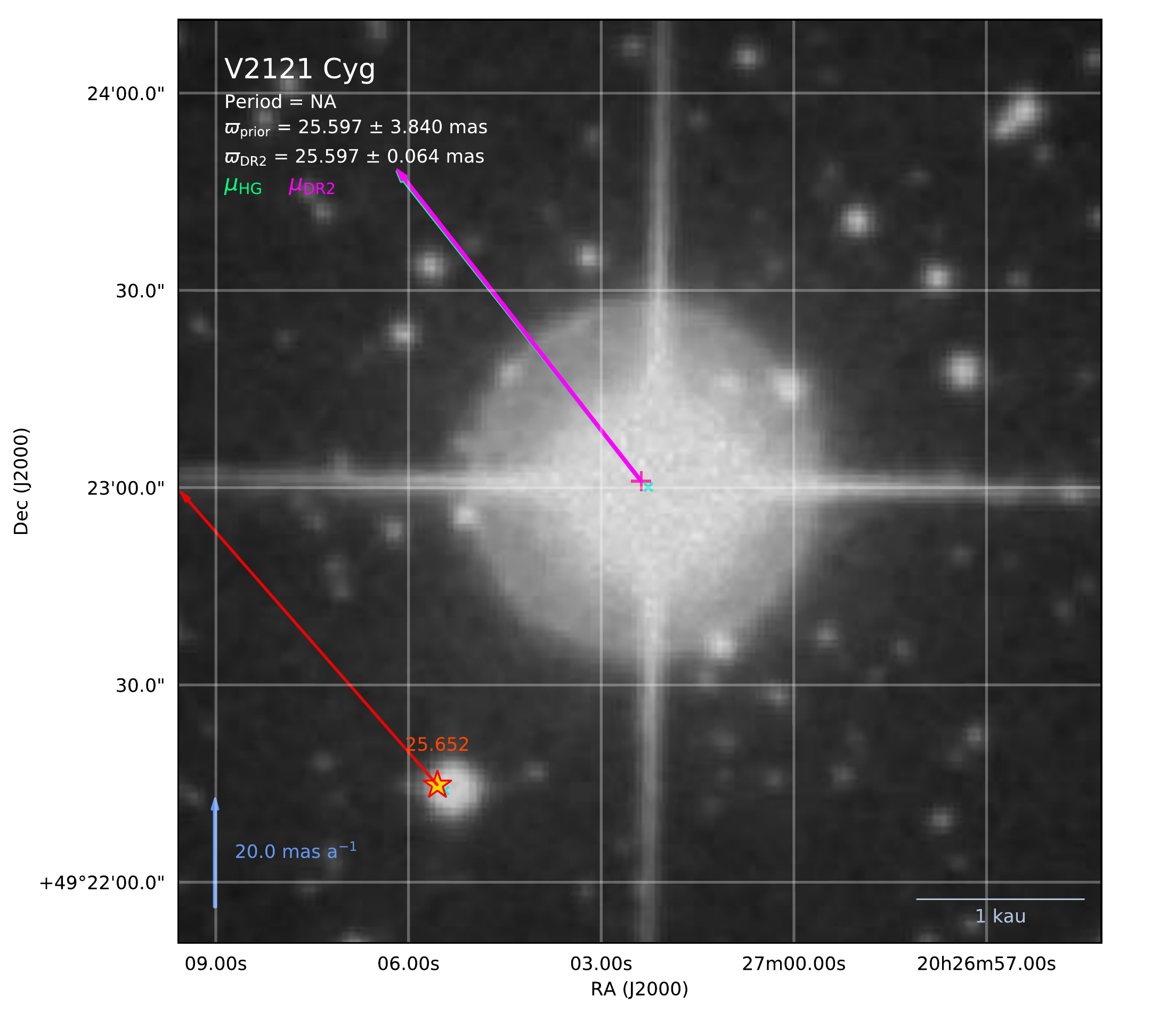}
\includegraphics[width=9cm]{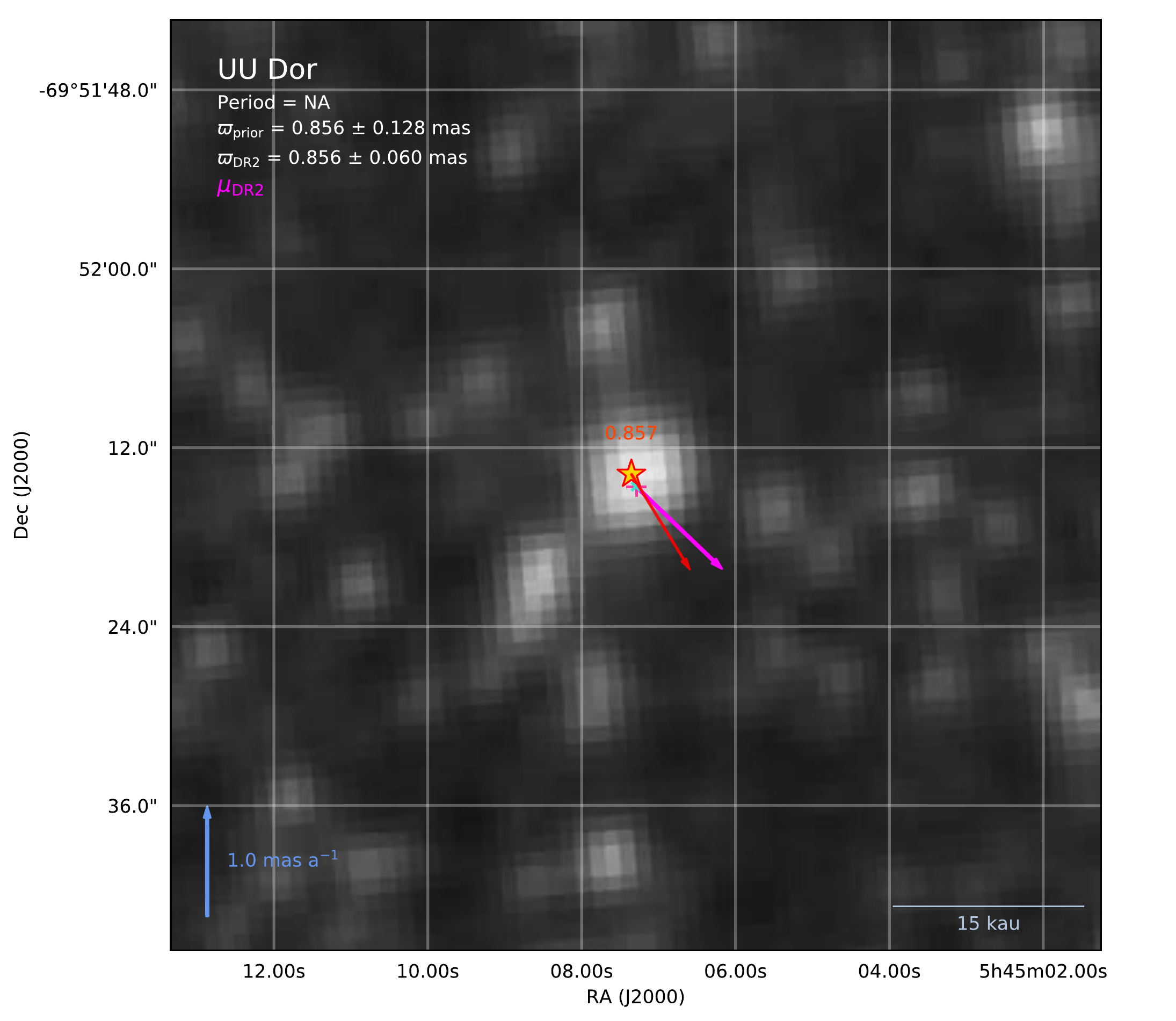}
\caption{Variable stars of various classes with candidate \texttt{Bound} companions.\label{various-field1}}
\end{figure*}

\begin{figure*}[h]
\centering
\includegraphics[width=9cm]{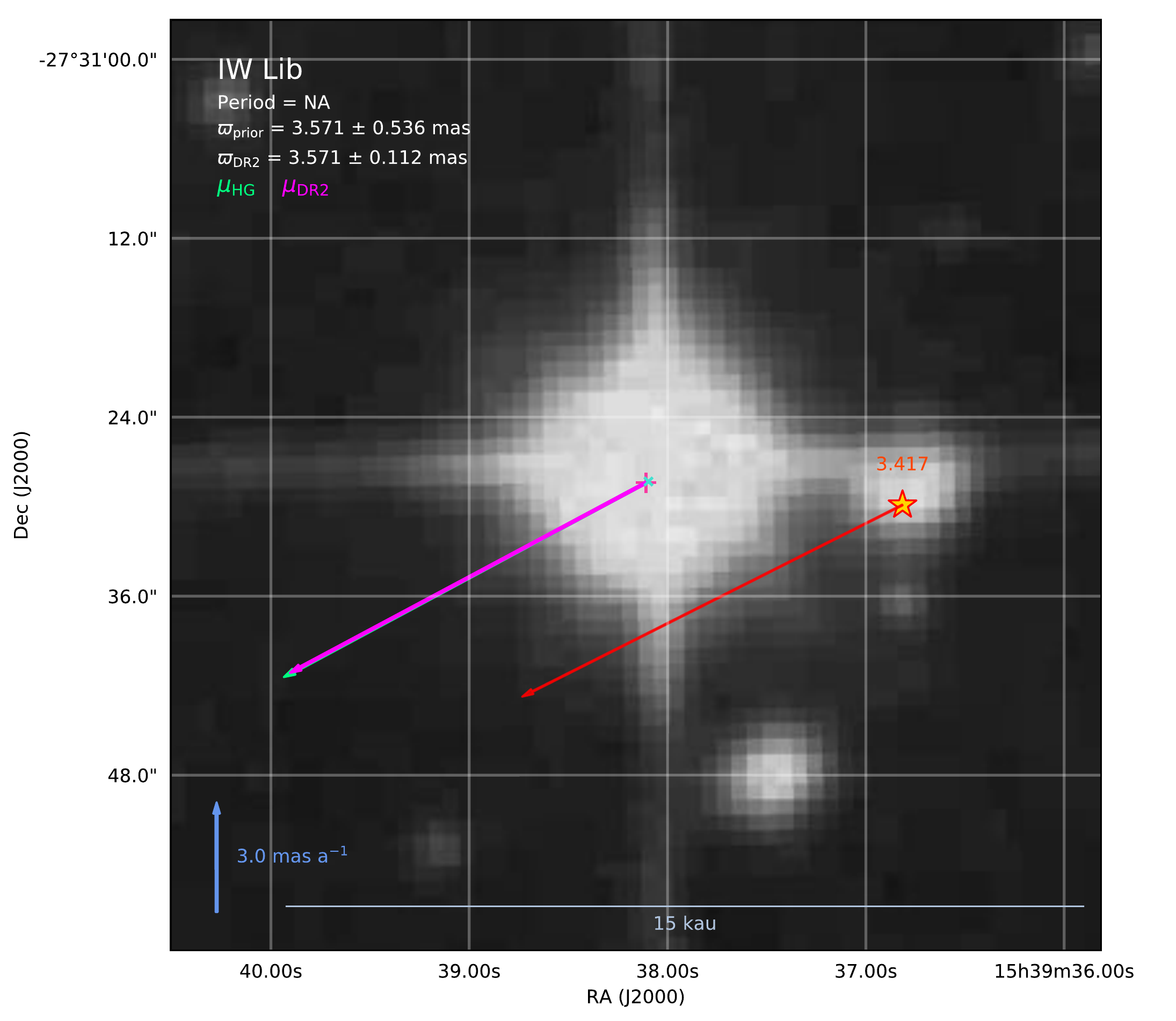}
\includegraphics[width=9cm]{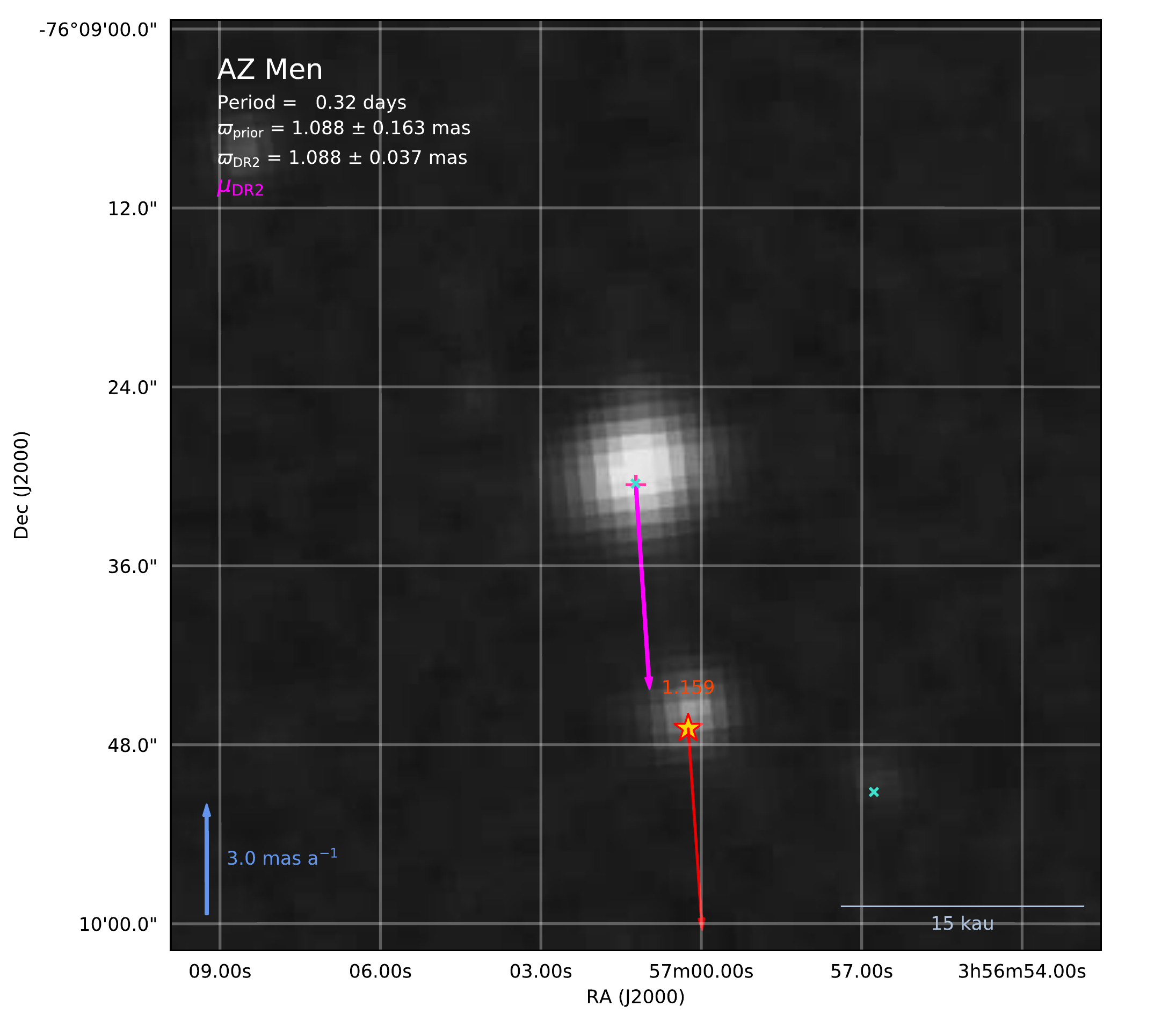}
\includegraphics[width=9cm]{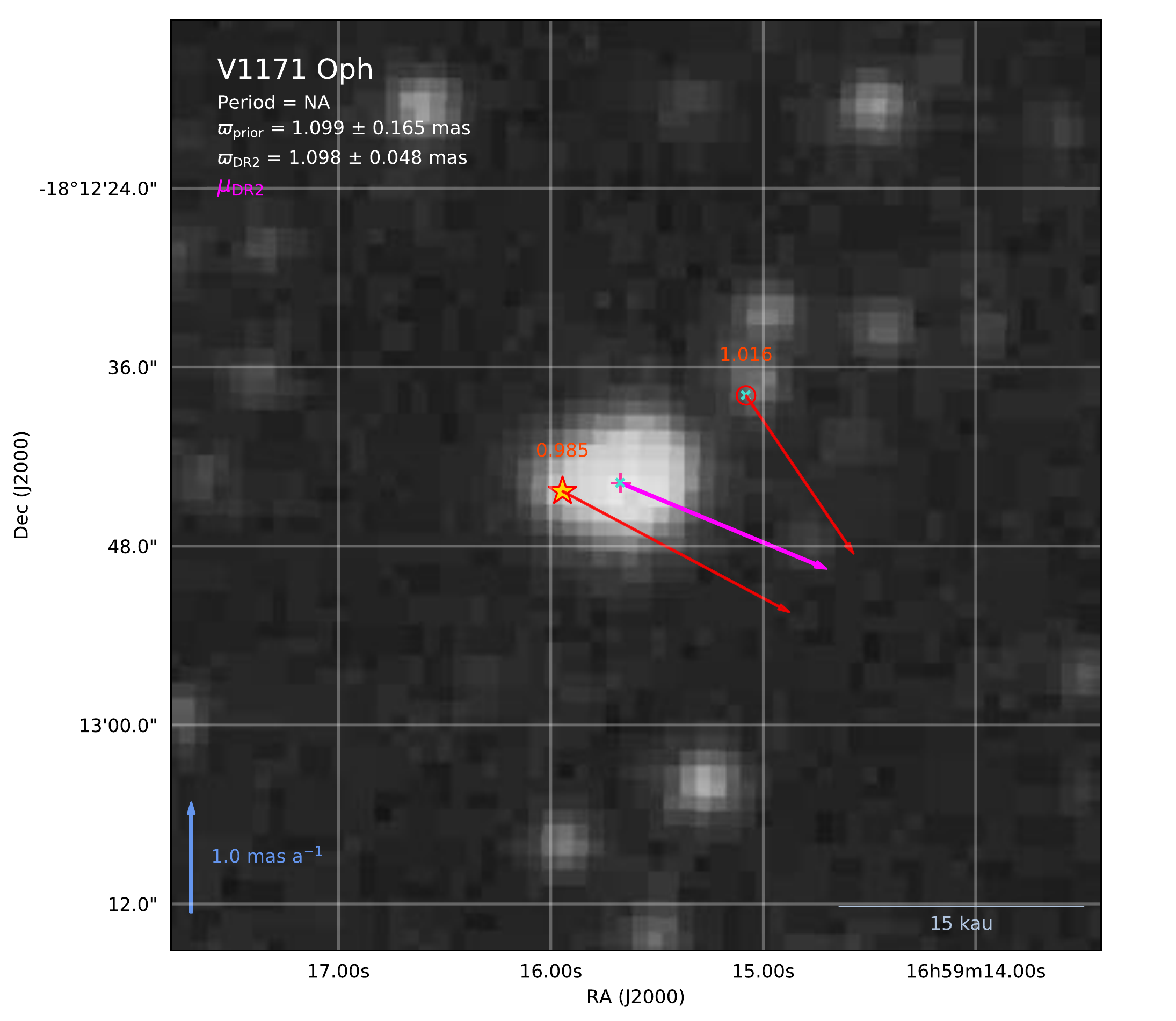}
\includegraphics[width=9cm]{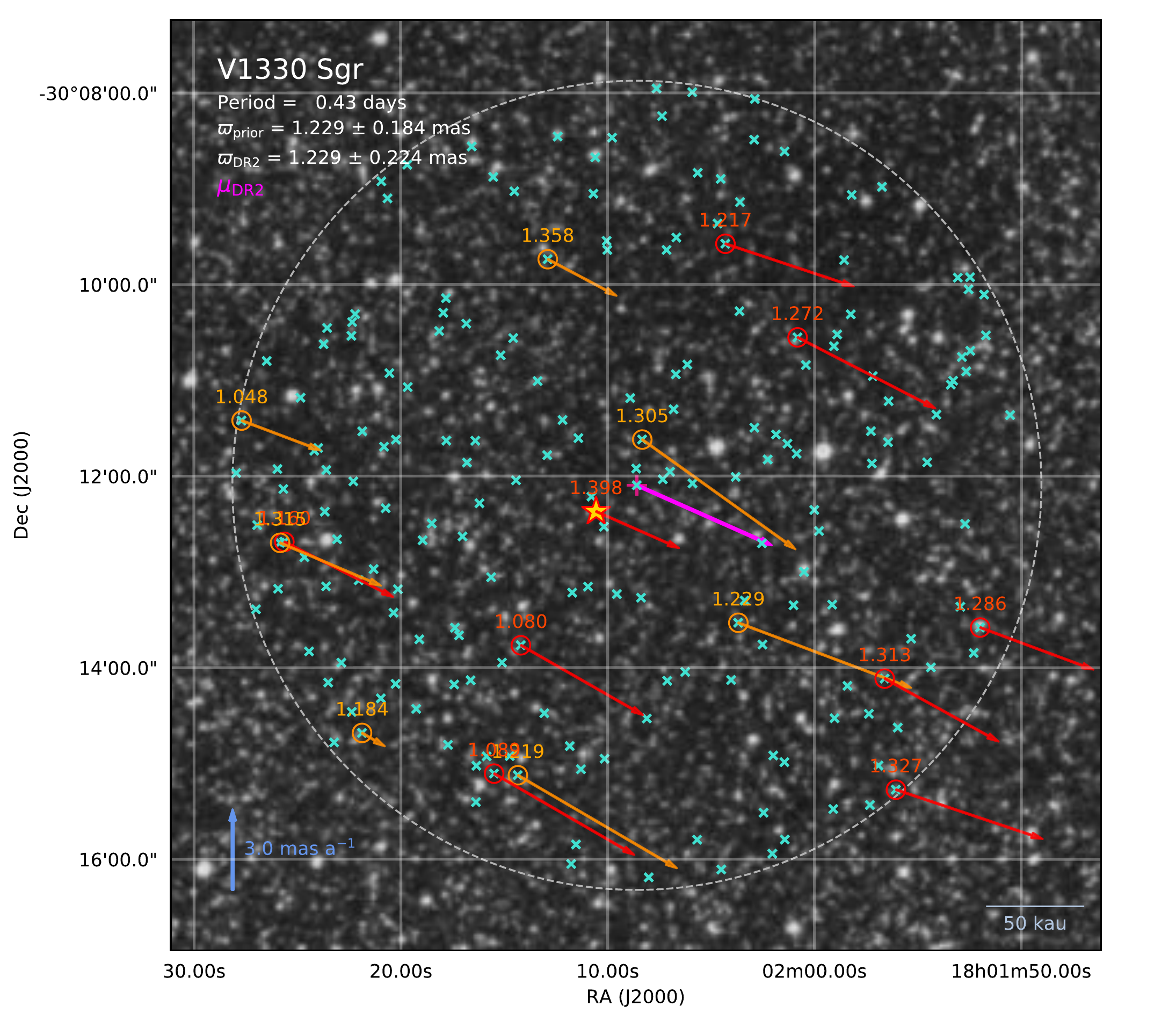}
\includegraphics[width=9cm]{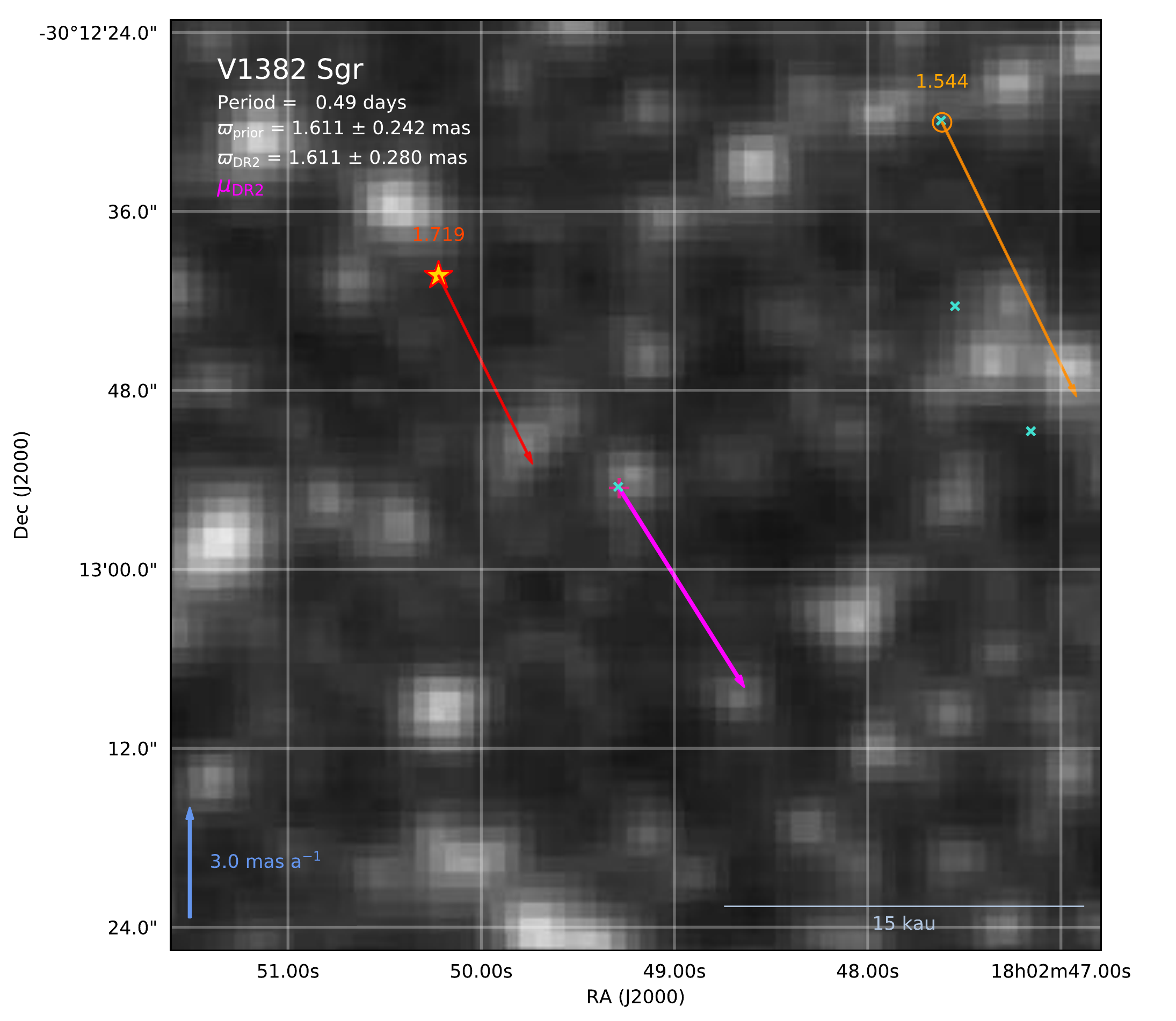}
\includegraphics[width=9cm]{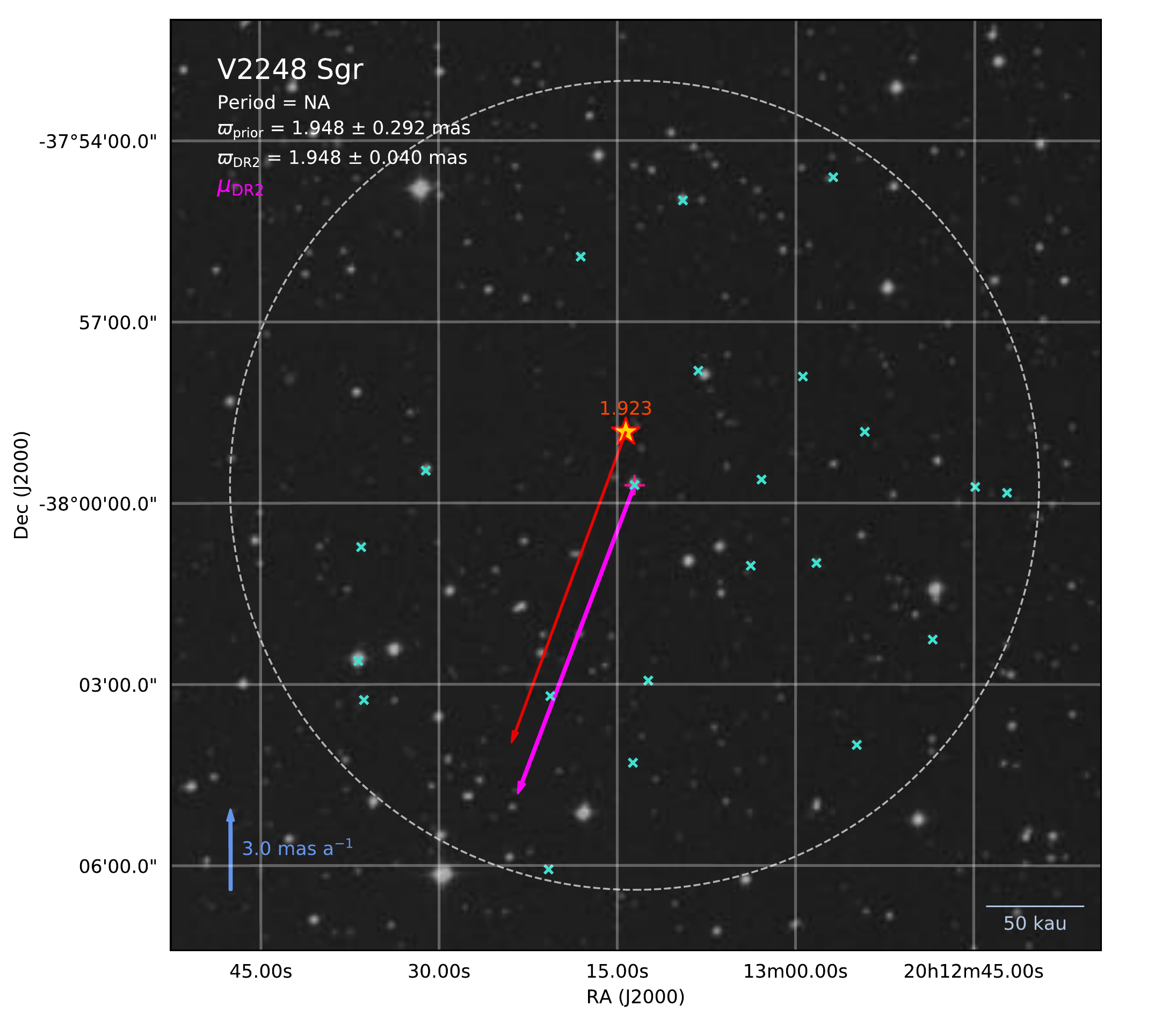}
\caption{Continued from Fig.~\ref{various-field1}.\label{various-field2}}
\end{figure*}

\begin{figure*}[h]
\centering
\includegraphics[width=9cm]{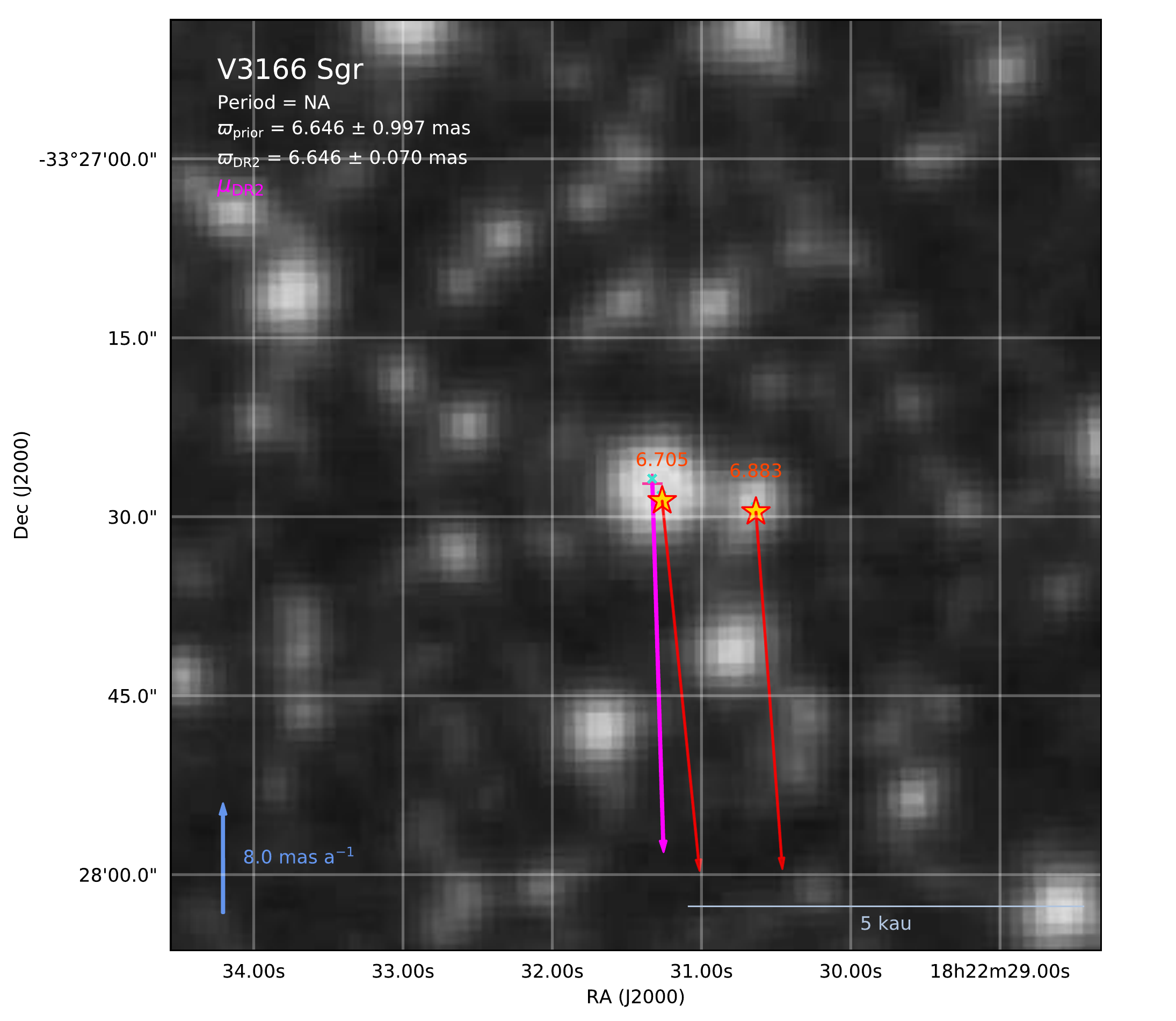}
\includegraphics[width=9cm]{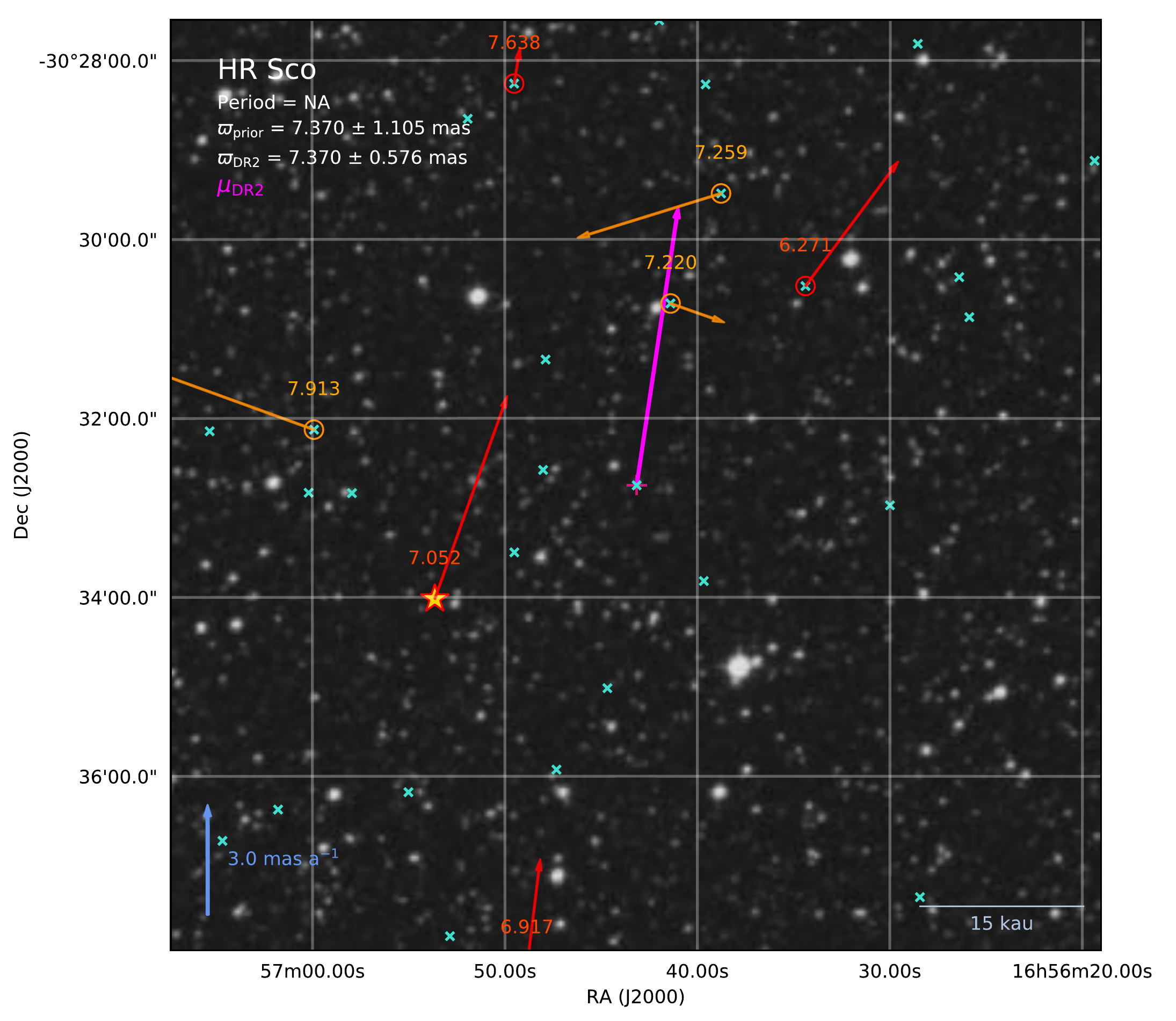}
\caption{Continued from Fig.~\ref{various-field2}.\label{various-field3}}
\end{figure*}

\end{appendix}

\end{document}